# Technical Design Report for the $\overline{\text{P}}$ANDA Endcap Disc DIRC

**(AntiProton Annihilations at Darmstadt)**

**Strong Interaction Studies with Antiprotons**

$\overline{\text{P}}$ANDA Collaboration

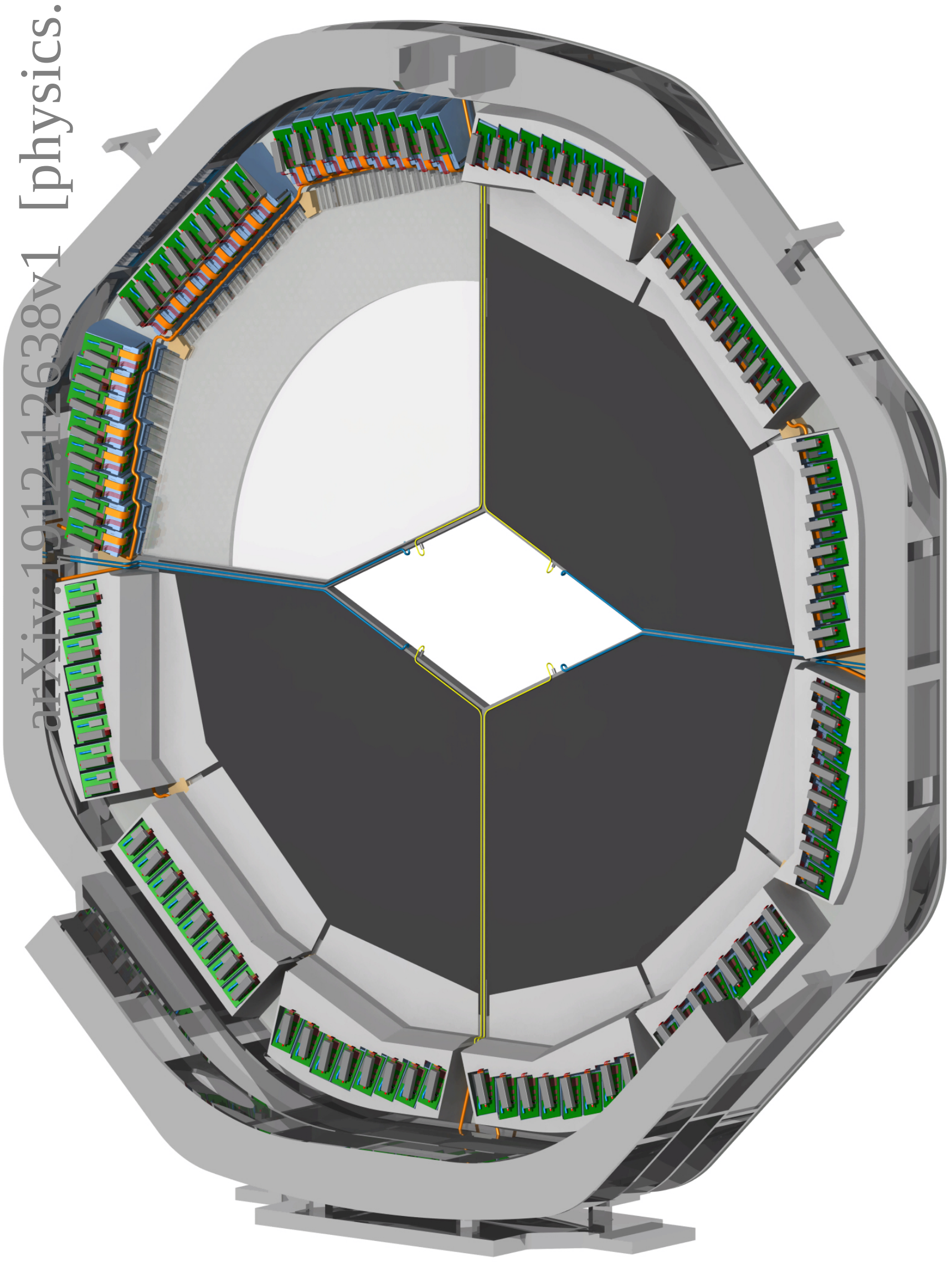

# The $\overline{\text{P}}$ANDA Collaboration

Università Politecnica delle Marche-Ancona, **Ancona**, Italy
F. Davì

Universität Basel, **Basel**, Switzerland
W. Erni, B. Krusche, M. Steinacher, N. Walford

Institute of High Energy Physics, Chinese Academy of Sciences, **Beijing**, China
H. Liu, Z. Liu, B. Liu, X. Shen, C. Wang, J. Zhao

Ruhr-Universität Bochum, Institut für Experimentalphysik I, **Bochum**, Germany
M. Albrecht, T. Erlen, F. Feldbauer, M. Fink, V. Freudenreich, M. Fritsch, F.H. Heinsius, T. Held, T. Holtmann, I. Keshk, H. Koch, B. Kopf, M. Kuhlmann, M. Kümmel, S. Leiber, P. Musiol, A. Mustafa, M. Pelizäus, A. Pitka, G. Reicherz, M. Richter, C. Schnier, T. Schröder, S. Sersin, L. Sohl, C. Sowa, M. Steinke, T. Triffterer, U. Wiedner

Rheinische Friedrich-Wilhelms-Universität Bonn, **Bonn**, Germany
R. Beck, C. Hammann, J. Hartmann, B. Ketzer, M. Kube, M. Rossbach, C. Schmidt, R. Schmitz, U. Thoma, M. Urban

Università di Brescia, **Brescia**, Italy
A. Bianconi

Institutul National de C&D pentru Fizica si Inginerie Nucleara "Horia Hulubei", **Bukarest-Magurele**, Romania
M. Bragadireanu, D. Pantea

University of Technology, Institute of Applied Informatics, **Cracow**, Poland
W. Czyzycki, M. Domagala, G. Filo, J. Jaworowski, M. Krawczyk, E. Lisowski, F. Lisowski, M. Michałek, J. Płażek

IFJ, Institute of Nuclear Physics PAN, **Cracow**, Poland
K. Korcyl, A. Kozela, P. Kulessa, P. Lebiedowicz, K. Pysz, W. Schäfer, A. Szczurek

AGH, University of Science and Technology, **Cracow**, Poland
T. Fiutowski, M. Idzik, B. Mindur, K. Swientek

Instytut Fizyki, Uniwersytet Jagiellonski, **Cracow**, Poland
J. Biernat, B. Kamys, S. Kistryn, G. Korcyl, W. Krzemien, A. Magiera, P. Moskal, W. Przygoda, Z. Rudy, P. Salabura, J. Smyrski, P. Strzempek, A. Wronska

FAIR, Facility for Antiproton and Ion Research in Europe, **Darmstadt**, Germany
I. Augustin, R. Böhm, I. Lehmann, D. Nicmorus Marinescu, L. Schmitt, V. Varentsov

GSI Helmholtzzentrum für Schwerionenforschung GmbH, **Darmstadt**, Germany
M. Al-Turany, A. Belias, H. Deppe, N. Divani Veis, R. Dzhygadlo, H. Flemming, A. Gerhardt, K. Götzen, R. Karabowicz, U. Kurilla, D. Lehmann, S. Löchner, J. Lühning, U. Lynen, S. Nakhoul, H. Orth, K. Peters, T. Saito, G. Schepers, C. J. Schmidt, C. Schwarz, J. Schwiening, A. Täschner, M. Traxler, B. Voss, P. Wieczorek, A. Wilms

Joint Institute for Nuclear Research, **Dubna**, Russia
V. Abazov, G. Alexeev, V. A. Arefiev, V. Astakhov, M. Yu. Barabanov, B. V. Batyunya, V. Kh. Dodokhov, A. Efremov, A. Fechtchenko, A. Galoyan, G. Golovanov, E. K. Koshurnikov, Y. Yu. Lobanov, V. I. Lobanov, V. Malyshev, A. G. Olshevskiy, A. A. Piskun, A. Samartsev, M. G. Sapozhnikov, N. B. Skachkov, A. N. Skachkova, E. A. Strokovsky, V. Tokmenin, V. Uzhinsky, A. Verkheev, A. Vodopianov, N. I. Zhuravlev, A. Zinchenko

University of Edinburgh, **Edinburgh**, United Kingdom
D. Branford, D. Glazier, D. Watts

Friedrich Alexander Universität Erlangen-Nürnberg, **Erlangen**, Germany
M. Böhm, W. Eyrich, A. Lehmann, D. Miehling, M. Pfaffinger, S. Stelter, F. Uhlig

Northwestern University, **Evanston**, U.S.A.
S. Dobbs, K. Seth, A. Tomaradze, T. Xiao

Università di Ferrara and INFN Sezione di Ferrara, **Ferrara**, Italy
D. Bettoni

Goethe Universität, Institut für Kernphysik, **Frankfurt**, Germany
A. Ali, A. Hamdi, M. Krebs, F. Nerling

Frankfurt Institute for Advanced Studies, **Frankfurt**, Germany
V. Akishina, S. Gorbunov, I. Kisel, G. Kozlov, M. Pugach, M. Zyzak

INFN Laboratori Nazionali di Frascati, **Frascati**, Italy
N. Bianchi, P. Gianotti, C. Guaraldo, V. Lucherini

Dept of Physics, University of Genova and INFN-Genova, **Genova**, Italy
G. Bracco

Justus Liebig-Universität Gießen II. Physikalisches Institut, **Gießen**, Germany
S. Bodenschatz, K.T. Brinkmann, V. Di Pietro, S. Diehl, V. Dormenev, M. Düren, E. Etzelmüller, K. Föhl, M. Galuska, T. Geßler, E. Gutz, C. Hahn, A. Hayrapetyan, M. Kesselkaul, W. Kühn, T. Kuske, J. S. Lange, Y. Liang, O. Merle, V. Metag, M. Moritz, M. Nanova, R. Novotny, T. Quagli, A. Riccardi, J. Rieke, M. Schmidt, R. Schnell, H. Stenzel, M. Strickert, U. Thöring, T. Wasem, B. Wohlfahrt, H.G. Zaunick

IRFU, CEA, Université Paris-Saclay, **Gif-sur-Yvette Cedex**, France
E. Tomasi-Gustafsson

University of Glasgow, **Glasgow**, United Kingdom
D. Ireland, G. Rosner, B. Seitz

Birla Institute of Technology and Science, Pilani, K K Birla Goa Campus, **Goa**, India
P.N. Deepak, A. Kulkarni

KVI-Center for Advanced Radiation Technology (CART), University of Groningen, **Groningen**, Netherlands
A. Apostolou, M. Babai, M. Kavatsyuk, H. Loehner, J. Messchendorp, P. Schakel, M. Tiemens, J. C. van der Weele, S. Vejdani

Gauhati University, Physics Department, **Guwahati**, India
K. Dutta, K. Kalita

Fachhochschule Südwestfalen, **Iserlohn**, Germany
H. Sohlbach

Forschungszentrum Jülich, Institut für Kernphysik, **Jülich**, Germany
M. Bai, L. Bianchi, M. Büscher, A. Derichs, R. Dosdall, A. Erven, V. Fracassi, A. Gillitzer, F. Goldenbaum, D. Grunwald, L. Jokhovets, G. Kemmerling, H. Kleines, A. Lai, A. Lehrach, M. Mikirtychyants, S. Orfanitski, D. Prasuhn, E. Prencipe, J. Pütz, J. Ritman, E. Rosenthal, S. Schadmand, T. Sefzick, V. Serdyuk, G. Sterzenbach, T. Stockmanns, P. Wintz, P. Wüstner, H. Xu, Y. Zhou

Chinese Academy of Science, Institute of Modern Physics, **Lanzhou**, China
Z. Li, X. Ma, H. Xu

INFN Laboratori Nazionali di Legnaro, **Legnaro**, Italy
V. Rigato

Lunds Universitet, Department of Physics, **Lund**, Sweden
L. Isaksson

Johannes Gutenberg-Universität, Institut für Kernphysik, **Mainz**, Germany
P. Achenbach, A. Aycock, O. Corell, A. Denig, M. Distler, M. Hoek, W. Lauth, Z. Liu, H. Merkel, U. Müller, J. Pochodzalla, S. Sanchez, S. Schlimme, C. Sfienti, M. Thiel, M. Zambrana

Helmholtz-Institut Mainz, **Mainz**, Germany
H. Ahmadi, S. Ahmed , S. Bleser, L. Capozza, M. Cardinali, A. Dbeyssi, A. Ehret, B. Fröhlich, P. Grasemann, S. Haasler, D. Izard, J. Jorge, D. Khaneft, R. Klasen, R. Kliemt, J. Köhler, H. H. Leithoff, D. Lin, F. Maas, S. Maldaner, M. Michel, M. C. Mora Espí, C. Morales Morales, C. Motzko, O. Noll, S. Pflüger, D. Rodríguez Piñeiro, M. Steinen, E. Walaa, S. Wolff, I. Zimmermann

Research Institute for Nuclear Problems, Belarus State University, **Minsk**, Belarus
A. Fedorov, M. Korzhik, O. Missevitch

Institute for Theoretical and Experimental Physics, **Moscow**, Russia
P. Balanutsa, V. Chernetsky, A. Demekhin, A. Dolgolenko, P. Fedorets, A. Gerasimov, V. Goryachev, D. Y. Kirin, V. A. Matveev, A. V. Stavinskiy

Moscow Power Engineering Institute, **Moscow**, Russia
A. Balashoff, A. Boukharov, O. Malyshev, I. Marishev

Nuclear Physics Division, Bhabha Atomic Research Centre, **Mumbai**, India
V. Chandratre, V. Datar, V. Jha, H. Kumawat, A.K. Mohanty, A. Parmar, A. K. Rai, B. Roy, G. Sonika

Westfälische Wilhelms-Universität Münster, **Münster**, Germany
C. Fritzsch, S. Grieser, A.K. Hergemöller, B. Hetz, N. Hüsken, A. Khoukaz, J. P. Wessels

Suranaree University of Technology, **Nakhon Ratchasima**, Thailand
C. Herold, K. Khosonthongkee, C. Kobdaj, A. Limphirat, P. Srisawad, Y. Yan

Novosibirsk State University, **Novosibirsk**, Russia
A. E. Blinov, S. Kononov, E. A. Kravchenko

Budker Institute of Nuclear Physics, **Novosibirsk**, Russia
E. Antokhin, M. Barnyakov, A. Yu. Barnyakov, K. Beloborodov, V. E. Blinov, V. S. Bobrovnikov, I. A. Kuyanov, A. P. Onuchin, S. Pivovarov, E. Pyata, S. Serednyakov, Y. Tikhonov

Institut de Physique Nucléaire, CNRS-IN2P3, Univ. Paris-Sud, Université Paris-Saclay, 91406, **Orsay cedex**, France
R. Kunne, D. Marchand, B. Ramstein, J. van de Wiele, Y. Wang

Dipartimento di Fisica, Università di Pavia, INFN Sezione di Pavia, **Pavia**, Italy
G. Boca

Charles University, Faculty of Mathematics and Physics, **Prague**, Czech Republic
V. Burian, M. Finger, M. Finger, A. Nikolovova, M. Pesek, M. Peskova, M. Pfeffer, I. Prochazka, M. Slunecka

Czech Technical University, Faculty of Nuclear Sciences and Physical Engineering, **Prague**, Czech Republic
P. Gallus, V. Jary, J. Novy, M. Tomasek, M. Virius, V. Vrba

Institute for High Energy Physics, **Protvino**, Russia
V. Abramov, N. Belikov, S. Bukreeva, A. Davidenko, A. Derevschikov, Y. Goncharenko, V. Grishin, V. Kachanov, V. Kormilitsin, A. Levin, Y. Melnik, N. Minaev, V. Mochalov, D. Morozov, L. Nogach, S. Poslavskiy, A. Ryazantsev, S. Ryzhikov, P. Semenov, I. Shein, A. Uzunian, A. Vasiliev, A. Yakutin

Sikaha-Bhavana, Visva-Bharati, WB, **Santiniketan**, India
U. Roy

University of Sidney, School of Physics, **Sidney**, Australia
B. Yabsley

National Research Centre "Kurchatov Institute" B. P. Konstantinov Petersburg Nuclear Physics Institute, Gatchina, **St. Petersburg**, Russia
S. Belostotski, G. Gavrilov, A. Izotov, S. Manaenkov, O. Miklukho, D. Veretennikov, A. Zhdanov

Kungliga Tekniska Högskolan, **Stockholm**, Sweden
T. Bäck, B. Cederwall

Stockholms Universitet, **Stockholm**, Sweden

K. Makonyi, M. Preston, P.E. Tegner, D. Wölbing

Veer Narmad South Gujarat University, Department of Physics, **Surat**, India
S. Godre

Università di Torino and INFN Sezione di Torino, **Torino**, Italy
M. P. Bussa, S. Marcello, S. Spataro

Politecnico di Torino and INFN Sezione di Torino, **Torino**, Italy
F. Iazzi, R. Introzzi, A. Lavagno

INFN Sezione di Torino, **Torino**, Italy
D. Calvo, P. De Remigis, A. Filippi, G. Mazza, A. Rivetti, R. Wheadon

Università di Trieste and INFN Sezione di Trieste, **Trieste**, Italy
A. Martin

Uppsala Universitet, Institutionen för fysik och astronomi, **Uppsala**, Sweden
H. Calen, W. Ikegami Andersson, T. Johansson, A. Kupsc, P. Marciniewski, M. Papenbrock, J. Pettersson, J. Regina, K. Schönning, M. Wolke

Instituto de Física Corpuscular, Universidad de Valencia-CSIC, **Valencia**, Spain
J. Diaz

Sardar Patel University, Physics Department, **Vallabh Vidynagar**, India
V. Pothodi Chackara

National Centre for Nuclear Research, **Warsaw**, Poland
A. Chlopik, G. Kesik, D. Melnychuk, B. Slowinski, A. Trzcinski, M. Wojciechowski, S. Wronka, B. Zwieglinski

Österreichische Akademie der Wissenschaften, Stefan Meyer Institut für Subatomare Physik, **Wien**, Austria
P. Bühler, J. Marton, D. Steinschaden, K. Suzuki, E. Widmann, S. Zimmermann, J. Zmeskal

Editors:

| | |
|---|---|
| Michael Düren | Email: michael.dueren@exp2.physik.uni-giessen.de |
| Erik Etzelmüller | Email: erik.etzelmueller@exp2.physik.uni-giessen.de |
| Klaus Föhl | Email: klaus.foehl@exp2.physik.uni-giessen.de |
| Avetik Hayrapetyan | Email: avetik.hayrapetyan@exp2.physik.uni-giessen.de |
| Matthias Hoek | Email: hoek@uni-mainz.de |
| Albert Lehmann | Email: albert.lehmann@physik.uni-erlangen.de |
| Julian Rieke | Email: julian.rieke@physik.uni-giessen.de |
| Mustafa Schmidt | Email: mustafa.schmidt@exp2.physik.uni-giessen.de |

Technical Coordinator:

| | |
|---|---|
| Lars Schmitt | Email: l.schmitt@gsi.de |

Spokesperson:

| | |
|---|---|
| Klaus Peters | Email: k.peters@gsi.de |

Deputy:

| | |
|---|---|
| Tord Johansson | Email: tord.johansson@physics.uu.se |

# Preface

PANDA (anti-Proton ANnihiliation at DArmstadt) is planned to be one of the four main experiments at the future international accelerator complex FAIR (Facility for Antiproton and Ion Research) in Darmstadt, Germany. It is going to address fundamental questions of hadron physics and quantum chromodynamics using cooled antiproton beams with a high intensity and and momenta between 1.5 and 15 GeV/c. PANDA is designed to reach a maximum luminosity of $2 \times 10^{32}\,\mathrm{cm}^{-2}\,\mathrm{s}$. Most of the physics programs require an excellent particle identification (PID). The PID of hadronic states at the forward endcap of the target spectrometer will be done by a fast and compact Cherenkov detector that uses the detection of internally reflected Cherenkov light (DIRC) principle. It is designed to cover the polar angle range from 5° to 22° and to provide a separation power for the separation of charged pions and kaons up to 3 standard deviations (s.d.) for particle momenta up to 4 GeV/c in order to cover the important particle phase space. This document describes the technical design and the expected performance of the novel PANDA Disc DIRC detector that has not been used in any other high energy physics experiment (HEP) before. The performance has been studied with Monte-Carlo simulations and various beam tests at DESY and CERN. The final design meets all PANDA requirements and guarantees sufficient safety margins.

# Contents

# 1 Executive Summary

## The $\overline{\mathrm{P}}$ANDA Experiment

$\overline{\mathrm{P}}$ANDA [1] will be one of the main experiments for the new international accelerator complex FAIR (Facility for Antiproton and Ion Research) in Darmstadt, Germany. A stored and cooled antiproton beam of 1.5 - 15 GeV/c momentum, colliding with a fixed proton or nuclear target, will allow for high luminosity hadron production and formation experiments. The annihilating baryons will produce new particles at high rate and with various quantum numbers, giving insight into the properties of mesons and baryons and the strong interactions of quarks and gluons. Recently discovered new hadronic states, especially the $Z^+$ (4430) [2], which is most likely an exotic state, will be mapped out in detail and searches for other exotic matter, especially glueballs, will be performed.

To cope with high production rates and complex hadronic final states, a $4\pi$ detector is designed with excellent tracking, calorimetry, and particle identification. It will be read out in a triggerless mode. The importance of the charm quark sector for the study of exotic states requires the use of the $\overline{\mathrm{P}}$ANDA electromagnetic calorimeter for photon detection, the micro vertex detector for the measurement of sub-mm decay lengths, and the positive identification of charged kaons.

The identification of charged particles is implemented in three angular regions: The region of the barrel EMC (polar angles 22° - 140°) will be equipped with a Barrel DIRC (where DIRC stands for Detection of Internally Reflected Cherenkov light) [3]; the endcap EMC region (5° - 22°) with the Endcap Disc DIRC (EDD) described here, and the very forward region ($< 5°$) with an aerogel RICH detector downstream of the forward spectrometer. In addition time-of-flight counters are provided for low momentum particle identification. Muons will be identified by a dedicated muon detection system which also serves as a hadronic calorimeter.

## A compact, fast Disc DIRC

This document describes the proposed EDD for $\overline{\mathrm{P}}$ANDA. The requirement is to positively identify kaons in the momentum range of 1 - 4 GeV/c, i.e. to separate kaons from pions and protons. Muons are identified in muon chambers.

DIRC detectors make use of internally reflected Cherenkov light in a solid state radiator. Here a transparent fused silica plate of 20 mm thickness is employed. The photons are read out at the rim of the plate. This design makes DIRC detectors very thin, usually just a few centimeters, whereas other RICH detectors typically occupy an order of magnitude more space along the path of the detected particles. The EDD is significantly more economic than a RICH design as the forward endcap EMC and the complete forward spectrometer can be moved closer to the interaction point. Therefore, the transverse size of the subsequent detectors can be reduced.

More than 10 years of R&D have been invested into the current, novel design of a Disc DIRC. Several versions have been investigated in order to minimize cost and to maximize performance. The EDD will be the first Disc DIRC in a physics experiment. The $\overline{\mathrm{P}}$ANDA requirements of the EDD are at the limit of today's technology. A major challenge is the readout of individual Cherenkov photons with high detection efficiency and at high rates in an environment with large background, high radiation dose, tight spatial constraints, and a large magnetic field. Only recently the technology of microchannel-plate photomultipliers (MCP-PMTs) has advanced such that the proposed EDD becomes feasible.

The EDD design uses a very limited number of component types, has a modular structure, a rigid optical setup, and can be fabricated by external companies to a large extend. This way, we aim to achieve a stable and reliable operation even with a limited manpower during construction, operation, and maintenance. The EDD consists of four identical quadrants, each of them is a fully functional PID detector that is mechanically, optically, and electronically separated from the other quadrants. A common holding cross aligns the quadrants mechanically. It is mounted on the holding structure of the forward endcap EMC.

## The Quadrant

Each quadrant consists of a radiation hard, precision polished synthetic fused silica plate that forms the active area of the detector. Charged particles that traverse the radiator produce Cherenkov light. About half of the Cherenkov photons undergo multiple total reflections off the surfaces until they leave the plate at the outer rim. Attached to the rim are light guide bars and focusing elements (FELs) which guide and focus the photons onto position-sensitive MCP-PMTs. The position registered by the photon sensor in the focal area of the FEL is used to extract the Cherenkov angle and, thus, the velocity of the particle that traverses the radiator. The precise timing of the MCP-PMT signal relative to the timing of other photons is used to separate the signal from background and to disentangle the photon pattern with different numbers of reflections at the rim of the radiator plate.

The radiator plate is an irregular hexagon and has three outer edges which are equipped with eight optical readout modules (ROMs) each. A ROM consists of an MCP-PMT, three FELs, and three connecting light guide bars. The radiator, the light guide bar, and the focusing element are optically bonded with high precision and form a rigid mechanical unit.

### The Readout Module

Each of the 96 ROMs of the EDD is a separate readout unit. The MCP-PMT which is attached to each ROM has an anode structure with at least $3 \times 100$ pixels. The pixels with a pitch of not more than 0.5 mm measure the Cherenkov angle. The three columns are assigned to the three light guide bars and constitute the azimuthal resolution of the EDD. Each MCP-PMT anode surface is attached to a small PCB that holds several ASICs (Application Specific Integrated Circuits) which discriminate and digitize the analog MCP-PMT signals. A front-end board contains the low voltage control and concentrates the data from the ASICs onto an optical fiber link which brings the signals out of the experimental area to the $\overline{\text{P}}$ANDA DAQ system.

### Major Components

The detector consists of 4 identical rigid optical units (radiator and FELs) in a common rigid mechanical frame and 96 identical opto-electronic readout modules calles ROMs (consisting of three FELs, MCP-PMT with front-end electronics, and an optical fibre). The specifications of the elements present a challenge. The critical items are the optical elements, the MCP-PMTs, and the readout electronics.

The optical elements will be manufactured by suitable optical companies according to the specifications. The assembly and the quality control will partly be done by the Cherenkov group in dedicated optical labs and by companies.

Until recently, MCP-PMTs had a very limited lifetime due to degradation effects of the photocathode after large exposure. New developments by several companies solved the lifetime problem to the extent that today's MCP-PMTs can stand integrated anode currents of more than $15\,\text{C}/cm^2$ [4]. In our design the integrated charge will be limited by adding an optical band or long-pass filter in front of the MCP-PMTs to reduce the incident photon rate. Current developments also provide an option where the sensitivity of the MCP-PMT photocathodes is shifted towards higher wavelengths. The drawback of the reduced photon statistics is partially compensated by a reduction of the dispersion effects on the photon patterns. Currently, three companies are developing MCP-PMTs with the required spatial resolution. First prototypes have been developed and are currently being evaluated.

The heart of the front-end electronics is a modified version of the TOFPET ASIC [5]. This commercially available ASIC is suitable for our application concerning charge sensitivity, rate, and time resolution. In a common effort of several $\overline{\text{P}}$ANDA groups the chip is modified for $\overline{\text{P}}$ANDA applications. The same chip is planned to be used for the Barrel Time-of-Flight detector[6] and a similar version of the chip is planned for the MVD[7]. In all cases the digital part of the chip needs modifications which are currently being designed and tested.

### Performance

Different versions of the EDD have been studied in detailed simulations and a sophisticated pattern reconstruction program has been developed where the particle type can be determined from time-ordered photon signals of the MCP-PMT anodes using likelihood methods. All known effects of the optical transmission including focusing, scattering, absorption, quantum efficiencies of the photocathode, and degradation of the surfaces are included in the simulation. Minimum bias events at full $\overline{\text{P}}$ANDA luminosity and dark count rates of the MCP-PMTs have been included. The optimized version of the EDD shows a performance that meets the requirements of $\overline{\text{P}}$ANDA. The design will deliver the required separation power between pions and kaons in most of the phase space for momenta up to 4 GeV/c.

### Lab and Beam Tests

The individual optical and electronic components of the EDD have been tested intensively in the lab. Small-scale prototypes have been evaluated success-fully in test beam experiments. At mixed hadron beams at CERN the capabilities of the EDD have been demonstrated by providing a clean separation for pions and protons at 3 GeV/c on a single photon level. Furthermore, it was shown that the measured single photon resolutions and the number of detected Cherenkov photons agree with Monte Carlo simulations.

## References

[1] PANDA Collaboration. Physics Performance Report for PANDA: Strong Interaction Studies with Antiprotons. 2009.

[2] S.-K. Choi et al. Observation of a Resonance-like Structure in the $\pi^{+-}\psi'$ Mass Distribution in exclusive $B \to k\pi^{+-}\psi'$ Decays. *Phys. Rev. Lett.*, 100:142001, 2008.

[3] B. Singh et al. Technical design report for the PANDA Barrel DIRC detector. *Journal of Physics G: Nuclear and Particle Physics*, 46(4):045001, mar 2019.

[4] A. Lehmann et al. Lifetime of MCP-PMTs. *J. Instrum.*, 11:C05009, 2016.

[5] M.D. Rolo et al. TOFPET ASIC for PET applications. *J. Instrum.*, 8:C02050, 2013.

[6] B. Singh and others (PANDA Collaboration). Technical Design Report for the PANDA Barrel Time-of-Flight. *Internal Report FAIR, Darmstadt, https://fair-center.eu/en/for-users/publications/*, 2018.

[7] PANDA Collaboration. Technical Design Report for the: PANDA Micro Vertex Detector. arXiv:1207.6581, 2011.

# 2 The $\overline{\text{P}}$ANDA Experiment and its PID Concept

## 2.1 The $\overline{\text{P}}$ANDA Experiment

### 2.1.1 The Scientific Program

The $\overline{\text{P}}$ANDA (anti-Proton ANnihiliation at DArmstadt) collaboration [1] envisages a physics core program [2] that comprises

- charmonium spectroscopy with precision measurements of mass, width and decay branches;
- the investigation of states that are assumed to have more exotic configurations like multiquark states, charmed hybrids and glueballs;
- the search for medium modifications of charmed hadrons in nuclear matter;
- the $\gamma$-ray spectroscopy of hypernuclei, in particular double-$\Lambda$ states.

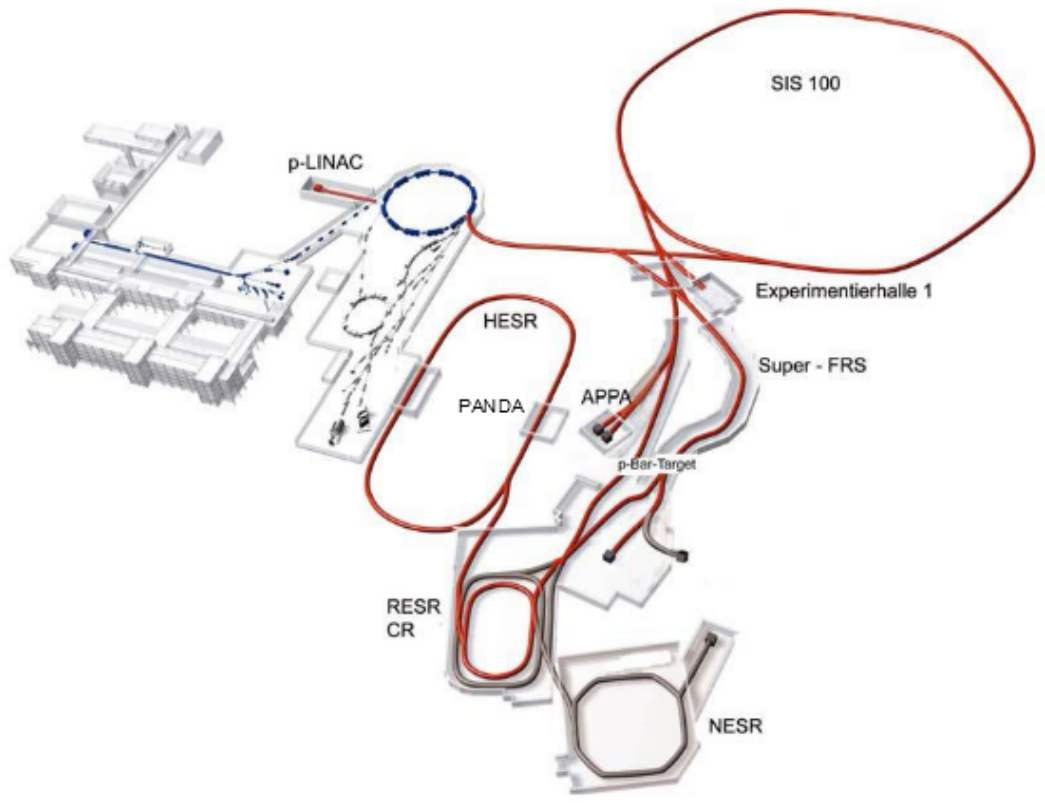

**Figure 2.1:** Schematic of the future FAIR layout incorporating the current GSI installations on the left; on the right the future installations, among them the SIS 100 synchrotron, the storage and cooler ring complex including CR and HESR, and the Super FRS experiments (reproduced from [3]).

In charmonium spectroscopy, for example, theoretical calculations differ significantly above the $\mathrm{D\bar{D}}$ threshold and do not properly predict several recently discovered states. Experimentally, a fair number of states and their properties at higher energy are not well established.

Charmonia states have been observed in $\mathrm{e^+e^-}$ interactions as well, where only $1^{--}$ states are directly formed, which limits the number of resonances that can be accurately scanned by ramping the accelerator energy. In $\mathrm{\bar{p}p}$ reactions more meson states can be directly formed, and an excellent mass resolution of $\sigma = 20\,\mathrm{keV}$ allows an accurate determination of the width of a resonance.

Further states (spin exotics) can be studied using the production mechanism.

### 2.1.2 High Energy Storage Ring

The combination of the High Energy Storage Ring (HESR) and $\overline{\text{P}}$ANDA aims at both high reaction rates and high resolution to be able to study rare production processes and small branching ratios. With up to $10^{11}$ stored antiprotons for beam momenta from 1.5 to 15 GeV/c and high density targets the anticipated antiproton production rate of $2{\cdot}10^7\,\mathrm{s^{-1}}$ governs the experiment interaction rate in the order of cycle-averaged $1{\cdot}10^7\,\mathrm{s^{-1}}$. The stored antiprotons do not have a bunch structure, and with 10–20% allocated to a barrier bucket, the antiprotons are continuously spread over 90–80% of the HESR circumference.

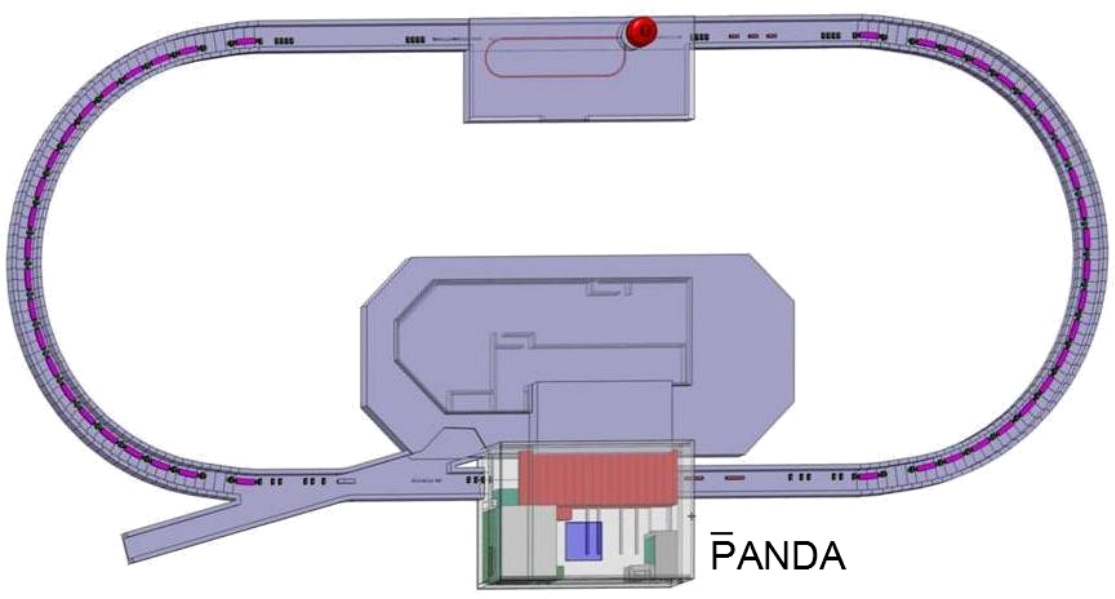

**Figure 2.2:** The HESR ring with the $\overline{\text{P}}$ANDA experimental area at the bottom (marked in red) and the electron cooler installation at the top. Standard operation has the antiproton injection from RESR (during the modularized startup phase from CR) from the left. Protons may be incected at reversed field polarities.

Two complementary operating modes are planned, named *high luminosity* and *high resolution*. The high luminosity mode with a momentum resolution of $\Delta p/p = 10^{-4}$, stochastic cooling and a pellet target density of $4 \cdot 10^{15}\,\mathrm{cm^{-2}}$ will have a luminosity of $L = 2 \cdot 10^{32}\,\text{cm}^{-2}\text{s}^{-1}$. For the high resolution mode $\Delta p/p = 4 \cdot 10^{-5}$ will be achieved with electron cooling for momenta up to $p = 8.9\,\text{GeV/c}$. This mode will operate in conjunction with a cluster jet target to limit the energy broadening caused by the target. The cycle-averaged luminosity is expected to be $L = 1 \cdot 10^{31}\,\text{cm}^{-2}\text{s}^{-1}$.

### 2.1.3 Targets

The PANDA Target Spectrometer is designed to allow the installation of different targets [1, 4]. For hydrogen as target material both Cluster Jet and Pellet Targets are being prepared. A major technical challenge is the distance of 2.5 m between the target generation and the accelerator beam and the distance of 5 m between the target generation and the target dump.

The cluster jet should give a constant thickness as a function of time whereas a pellet beam with its granular structure may give density variations on the $10 - 100\,\mu$s timescale. The cluster jet gives a larger interaction region and is anticipated to cause more background gas than the pellets for similar target thickness, the latter according to results from comparative studies made at CELSIUS and at COSY.

An extension of the targets to heavier gases such as deuterium, nitrogen or argon is planned for complementary studies with nuclear targets.

### 2.1.4 Luminosity Considerations

The luminosity is proportional to the number of stored antiprotons and is limited by the antiproton production rate. To first approximation the cycle averaged reaction rate should equal the antiproton production rate. Due to injection time and dumping of beam particles at the end of a cycle the time averaged reaction rate will be lower. Figure 2.3 illustrates beam preparation periods with target off and data taking periods with target on. The red curve shows the luminosity at constant target thickness which is proportional to the decreasing number of antiprotons during data taking. Measures to implement a target density increasing with time are studied in order to achieve constant luminosity.

Some variations of the instantaneous luminosity on short time scales of milliseconds and below will occur due to the target. The newly developed PANDA cluster jet target is expected to give small variations. In the case of a pellet target, variations of the instantaneous luminosity may occur. These are depending on antiproton beam profile, pellet size and distributions of pellet velocity and spacing between pellets.

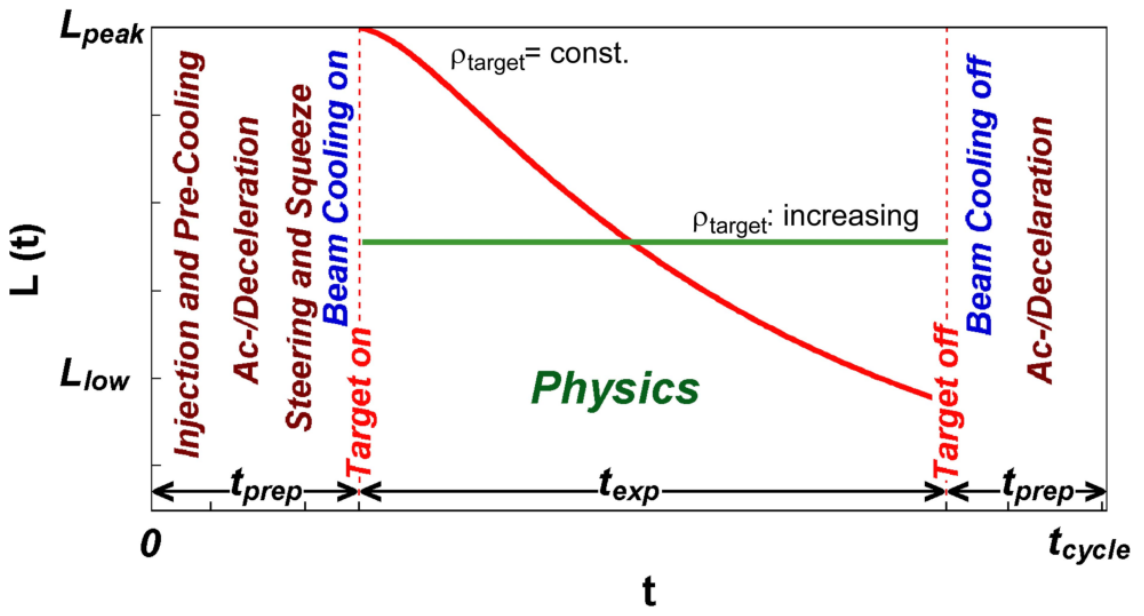

**Figure 2.3:** Time dependent macroscopic luminosity profile $L(t)$ in one operation cycle for constant (solid red) and increasing (green dotted) target density $\rho_{target}$ [5]. Different measures for beam preparation are indicated. Pre-cooling is performed at 3.8 GeV/c. A maximum ramp of 25 mT/s is specified for beam ac-/deceleration.

The design goal of the pellet target in high luminosity (PHL) mode is to always have more than 10 small pellets (with diameters ¡ $10\,\mu$m) in the beam region. In this mode, the velocity spread is not very crucial and one could have a stochastic time distribution of pellets. This would cause luminosity variations below 30% [4].

In pellet tracking (PTR) mode the pellet velocity spread should be below the per cent level for highly effective tracking. In this mode there will most of the time be only one pellet in the interaction region and its position will be known with high accuracy. At the pellet tracking development station (UPTS), the velocity spread is a few per mille and for this case only occasionally there would be two or more pellets in the beam region [4].

Target density fluctuations and background level will be taken into account when optimizing target settings in order to avoid that the maximum total interaction rate will exceed the 10 MHz mark.

## 2.2 The $\overline{\text{P}}$ANDA Detector

Figure 2.4 shows the $\overline{\text{P}}$ANDA detector as a partial cut-out. $\overline{\text{P}}$ANDA is a fixed target experiment and consists of two parts, the Target Spectrometer and the Forward Spectrometer. The antiproton beam is scattered off a pellet or cluster jet target. $\overline{\text{P}}$ANDA will be measuring $p\bar{p}$ reactions comprehensively and exclusively, which requires simultaneous measurements of leptons and photons as well as charged

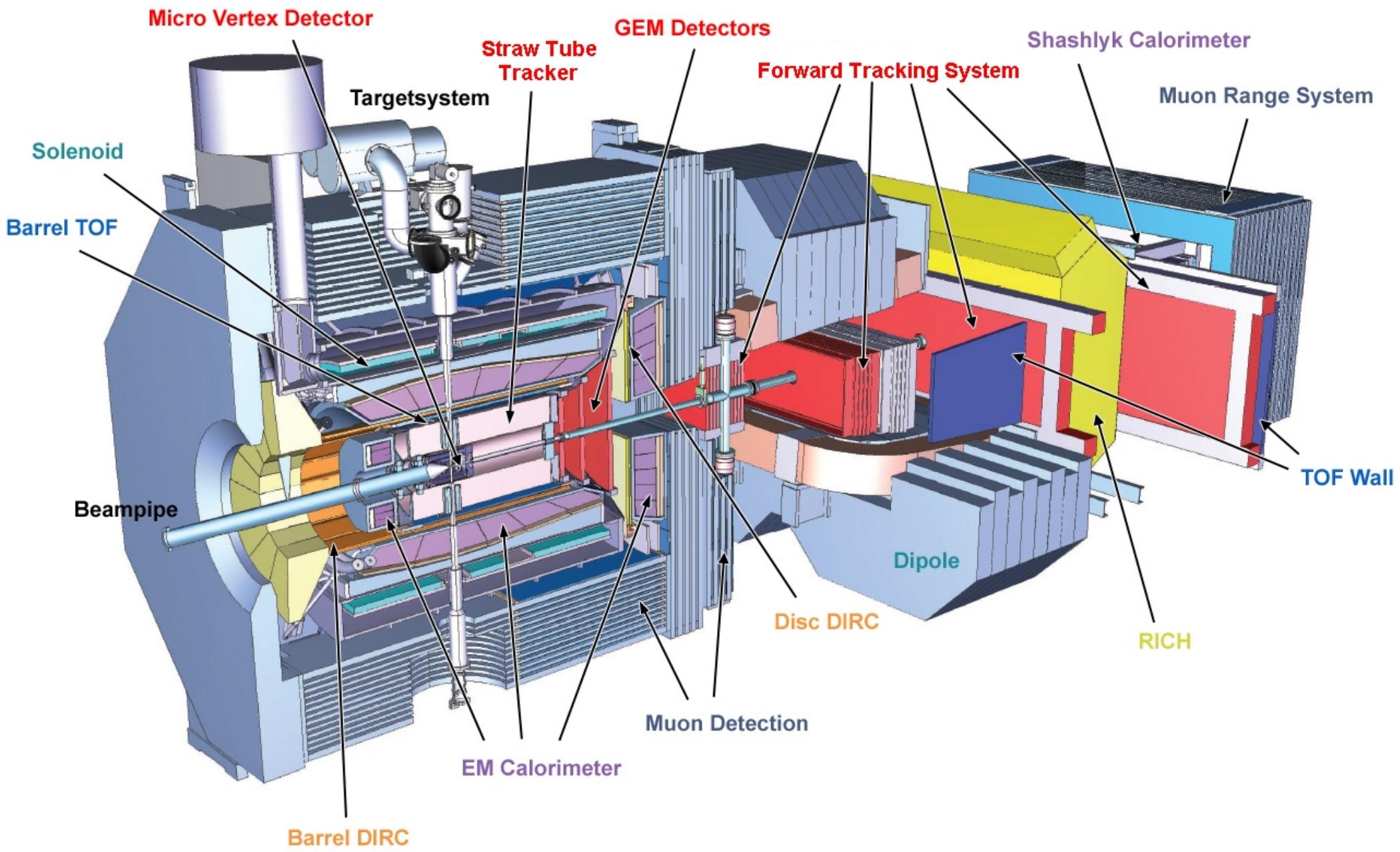

**Figure 2.4:** Side view of $\overline{\text{P}}$ANDA with the target spectrometer (TS) on the left side, and the forward spectrometer (FS) starting with the Dipole magnet center and right. The antiproton beam enters from the left.

and neutral hadrons, with high multiplicities.

The physics requirements for the detectors are:

- cover the full solid angle of the released particles,
- detect the momenta of the charged reaction products,
- measure the energy of photons,
- measure decay vertices of strange and charm particles,
- identify particle types over the full range of momenta of the reaction products.

### 2.2.1 Target Spectrometer

The Target Spectrometer components, subdivided into backward endcap, barrel and forward endcap regions, are housed within the yoke of a superconducting solenoid magnet [6]. The Target Spectrometer is almost hermetically sealed to avoid solid angle gaps and provides little spare space inside. Parts of the return yoke iron are laminated to allow the insertion of muon detectors into the gaps. Most components are located inside the magnetic field of the solenoid magnet.

The silicon microvertex detector (MVD) [7] closely abuts the beam pipe surrounding the target area. It provides secondary vertex sensitivity for particles with decay lengths on the order of 100 $\mu m$.

The central tracker features the straw tube tracker and several GEM tracking stations in the forward endcap part [5].

Two detectors of internally reflected Cherenkov light (DIRC) are to be located with the target spectrometer. The possibility of using thin radiators and placing the readout elements outside the acceptance favors the use of DIRC designs as Cherenkov imaging detectors for PID.

The lead tungstate (PWO) crystals of the electromagnetic calorimeters (EMC) [8] are read out with avalanche photo diodes (APD) or vacuum pentodes. Both, the crystal light output and the APD performance improve with lower temperature. Thus, plan is to operate the EMC at T = –25°C.

### 2.2.2 Forward Spectrometer

The Forward Spectrometer angular acceptance has an ellipsoidal form with a maximum polar angle of ±10 degrees horizontally and ±5 degrees vertically.

A dipole magnet is located at the most upstream part of the Forward Spectrometer to provide bending power with a vertical B-field. The majority of the detector systems (the exception being tracking sensors) are located downstream and outside the dipole magnet.

An aerogel RICH detector will be located between the dipole magnet and the Forward EMC. A Time-of-Flight wall covers slow particles below the Cherenkov light threshold.

In the Forward Spectrometer, a Shashlyk type electromagnetic calorimeter (scintillator fibers in a lead matrix) is followed by a scintillator-absorber-sandwich hadron calorimeter.

### 2.2.3 Data Acquisition

The data flow and processing is geometrically separated into the front-end electronics (FEE) located with the actual detector subsystems and comprises analog electronics, digitization, and (depending on subsystem) a first data concentrator stage.

After data transmission via optical fiber links the data acquisition (DAQ) is located in the Counting House, where the online analysis and data concentration (a factor 1000 according to the current paradigm) is handled before the data is sent to storage.

The overall data acquisition setup will have to cope with peak rates above the mean data rate of $\sim 200\,\text{GB/s}$. Bursts on different timescales are due to antiproton beam time structure, target density fluctuations (in case of pellet target) and luminosity decrease during the HESR operation cycle.

### 2.2.4 Infrastructure

The $\overline{\text{P}}$ANDA detector is located underground in an experimental hall, encased in smaller tunnel-like concrete structures, partially fixed, partially made of removable blocks. Most subsystems connect their FEE-components via cables and tubes placed in movable cable ducts to the installations in the counting house, where three levels are planned to accommodate cooling, gas supplies, power supplies and electronics. Electronic racks for only a few subcomponents will be placed directly next to the Target Spectrometer.

The EDD's FEE and digitization will be located inside the solenoid yoke of the target spectrometer. Cooling, low and high voltage, a nitrogen flow system and a laser calibration system will be supplied from the counting house.

## 2.3 The Particle Identification System

The charged particle identification (PID) will combine time-of-flight, tracking, dE/dx and calorimetry information with the output from imaging Cherenkov detectors.

The individual $\overline{\text{P}}$ANDA subsystems contributing to the PID and the combination into global PID information have been reviewed in the PID-TAG-report [9]. The contents relevant for this EDD-TDR is included below.

### 2.3.1 Tracking Detectors

Tracking information provides input to the DIRC analysis, as the Cherenkov angle is measured between the Cherenkov photon direction and the vertex of the radiating particle.

Tracking Detectors like the MVD and STT also provide information as the mean energy loss of charged particles per unit length depends on the velocity rather than the momentum. It is usually referred to as dE/dx and is described by the Bethe Bloch equation which depends on the velocity rather than the momentum of the charged particle. The dE/dx contributes to low energy PID at typically $p < 1\,\text{GeV/c}$ (see Figure 2.5).

Several GEM detector stations of the $\overline{\text{P}}$ANDA Target Spectrometer provide tracking information for the forward and endcap polar angles. The tracking information with an accuray of $\sigma \approx 1\,\text{mrad}$ enters into the analysis of the EDD.

### 2.3.2 EM Calorimeter

In the $\overline{\text{P}}$ANDA Target Spectrometer high precision electromagnetic calorimetry is required for a large energy range from a few MeV up to several GeV [8]. Lead-tungstate is chosen for the calorimeters in the target spectrometer due to its good energy resolution, fast response and high density, allowing a compact setup.

In addition to photon detection, the EMC is also the most powerful detector for the identification of electrons. The identification and measurement of this particle species will play an essential role for the physics program of $\overline{\text{P}}$ANDA.

The stopping power of an electromagnetic calorimeter is different for electrons, muons and hadrons. While muons and hadrons loose only a certain fraction of their kinetic energy by ionization processes, electrons deposit their complete energy in an electromagnetic shower. The ratio of the measured energy deposit in the calorimeter to the reconstructed track momentum (E/p) will be approximately one. Due to the fact that hadronic interactions within the crystals can take place, hadrons can have a higher E/p ratio than expected from ionization. Figure 2.6 shows the reconstructed E/p fraction for electrons and pions as a function of momentum.

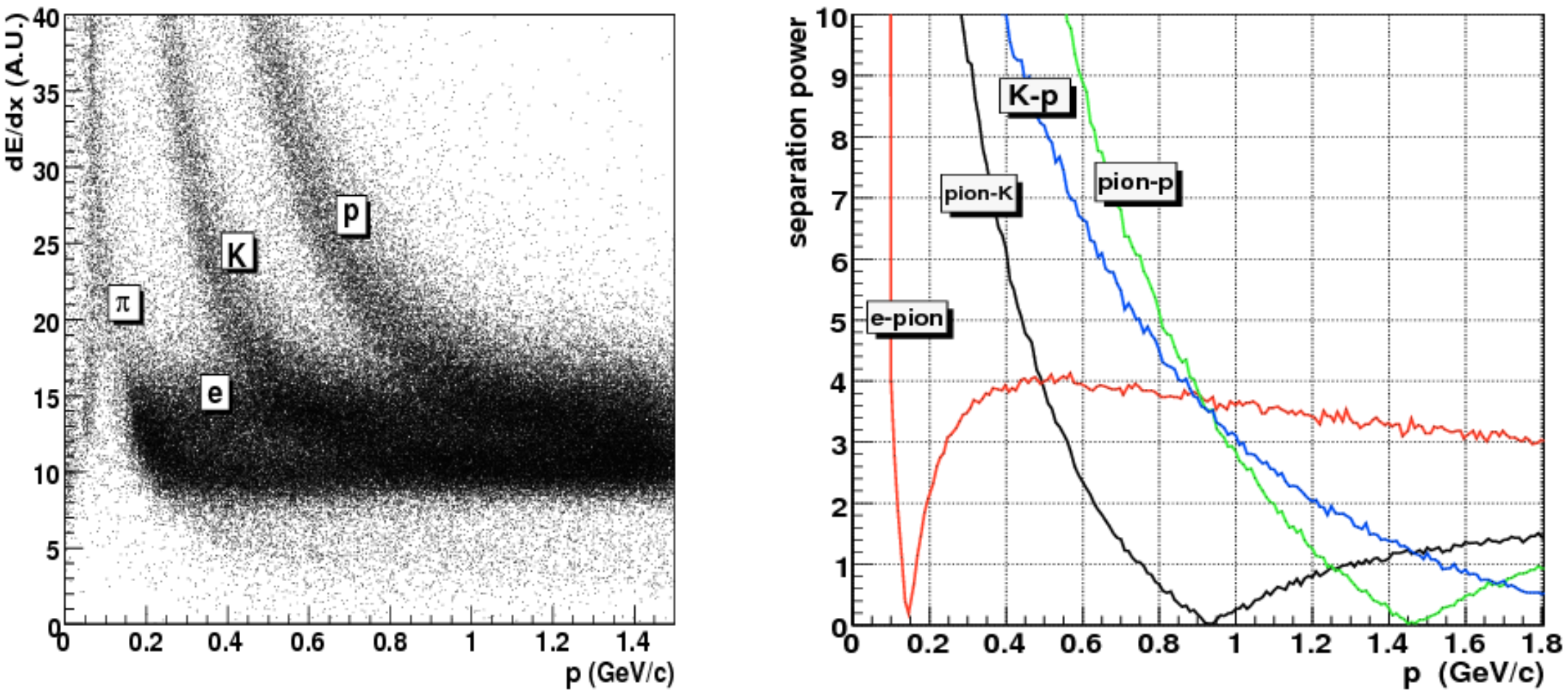

**Figure 2.5:** Energy loss by a 40% truncated mean algorithm for various particles versus the reconstructed momentum (left) and corresponding separation power (right) in the STT detector [9].

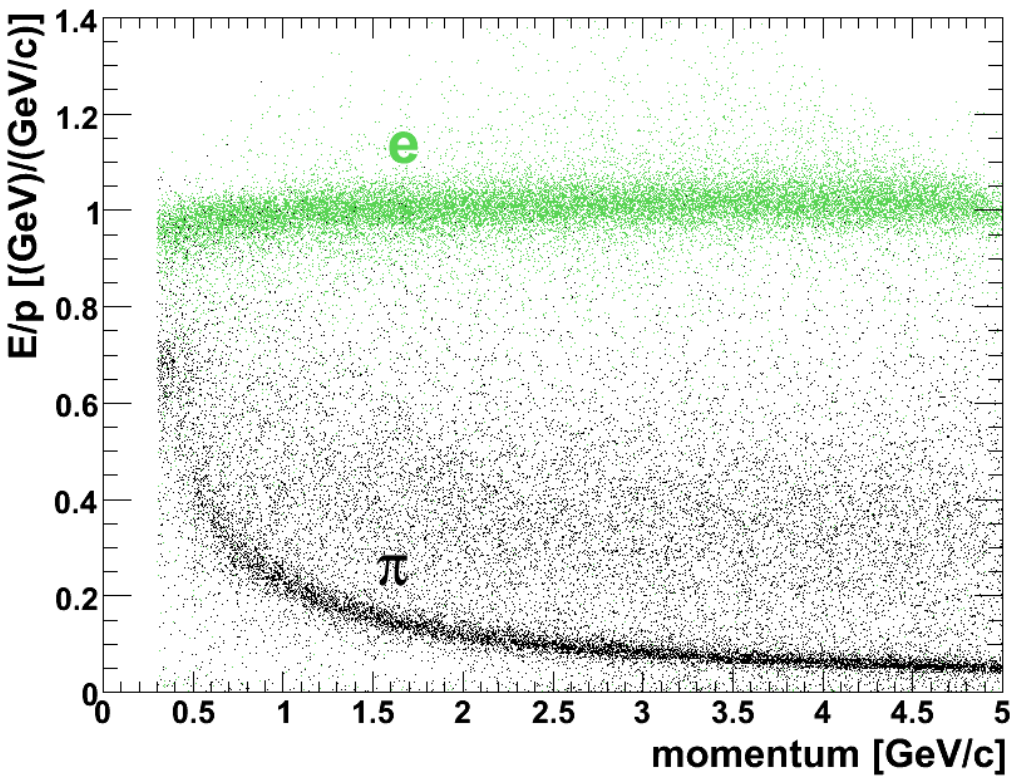

**Figure 2.6:** Simulated E/p versus track momentum for electrons (green) and pions (black) in the momentum range between 0.3 GeV/c and 5 GeV/c typical for Barrel and Endcap EMC [8, 9].

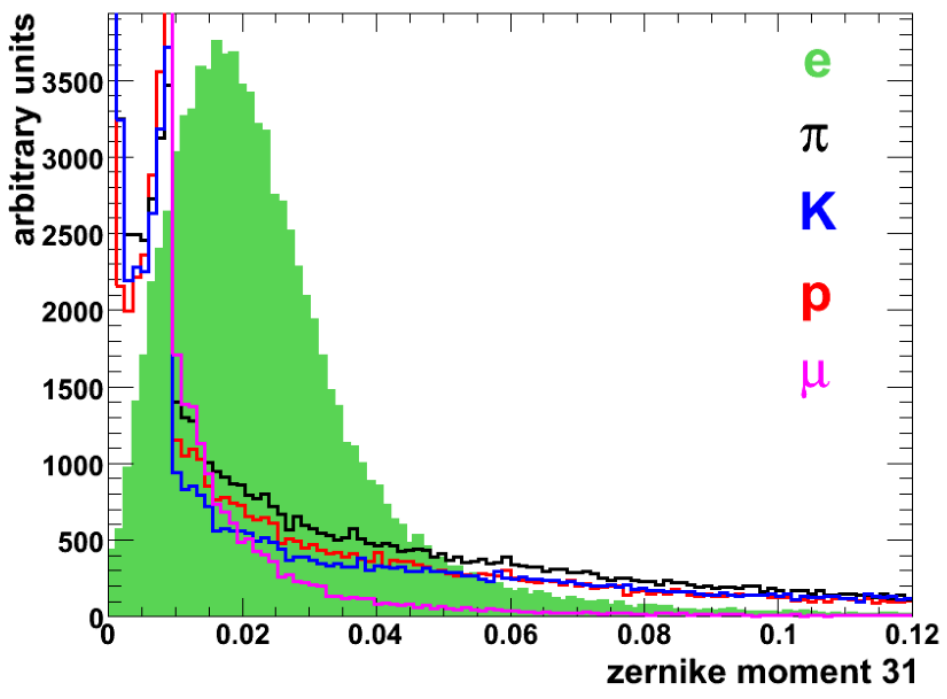

**Figure 2.7:** Zernike moment z31 for electrons, muons and hadrons [8].

Furthermore, the shower shape of a cluster is helpful to distinguish between electrons, muons and hadrons. Since the size of the crystals corresponds to the Molière radius of lead tungstate, the largest fraction of an electromagnetic shower originating from an electron is contained in a few crystals. A hadronic shower with a similar energy deposit has smaller lateral extension. The differences are reflected in the shower shape of the cluster, which can be characterized by single crystal to sum energy ratio, the lateral cluster moment and Zernike moments. Figure 2.7 shows the different distributions for the Zernike moment z31.

The crystal calorimeters of the Target Spectrometer are complemented in the forward direction with the shashlik type sampling calorimeter of the Forward Spectrometer consisting of 1404 modules of 55 × 55 $mm^2$ cell size covering 2.97 × 1.43 $m^2$.

### 2.3.3 Cherenkov Detectors

The endcap part of the Target Spectrometer covers forward angles up to $\vartheta = 22°$, excluding an inner elliptical acceptance of $\vartheta_x = 10°$ horizontal and $\vartheta_y = 5°$ vertical half-angles. The Endcap Disc DIRC (EDD) is will be located here.

A second DIRC, the Barrel DIRC covers the angular range from $\vartheta = 22°$ to 140°. The design [10] is based on the DIRC of the BaBar experiment [11].

An aerogel RICH detector will be located between the dipole magnet and the Forward EMC. Its design can be similar to the RICH detector in HERMES at DESY [12], possibly using four curved mirrors instead of two or a pure proximity focusing design using flat mirrors.

### 2.3.4 Time-Of-Flight Detectors

In the Forward Spectrometer a time-of-flight (TOF) measurement is planned that covers slow particles below the Cherenkov light threshold. For the target spectrometer, a scintillator based TOF system with a SiPM readout is foreseen.

### 2.3.5 Muon Detection

The return yoke for the solenoid magnet in the $\overline{\text{P}}$ANDA Target Spectrometer is laminated to accommodate layers of muon detectors. They form a range stack, with the inner muon layer being able to detect low energy muons. The cumulated iron layer thickness in front of the outer layers provide sufficient material to stop the high energy pions produced in $\overline{\text{P}}$ANDA.

### 2.3.6 Combined PID

The particle identification (PID) capability of several subsystem combinations (different tracker designs, with or without TOF, etc.) has been studied in the PID-TAG working group. The combined PID performance for particle pairs of (e, $\mu$, $\pi$, K, p) were presented in [9].

## References

[1] PANDA Collaboration. Technical Progress Report. 2005.

[2] PANDA Collaboration. Physics Performance Report for PANDA: Strong Interaction Studies with Antiprotons. 2009.

[3] Peter Spiller, Frank Becker, Oleksiy Dolinskyy, Lars Groening, Oliver Kester, Klaus Knie, Dieter Prasuhn, Hartmut Reich-Sprenger, Wolfgang Vinzenz, and Martin Winkler. The Accelerator Facility of the Facility for Antiproton and Ion Research. In *Proceedings, 6th International Particle Accelerator Conference (IPAC 2015): Richmond, Virginia, USA, May 3-8, 2015*, page TUBB2, 2015.

[4] PANDA Collaboration. Technical Design Report for the PANDA Internal Targets. 2012.

[5] PANDA Collaboration. Technical Design Report for the: PANDA Straw Tube Tracker. arXiv:1205.5441, 2012.

[6] PANDA Collaboration. Technical Design Report for the PANDA Solenoid and Dipole Spectrometer Magnets. arXiv:0907.0169, 2009.

[7] PANDA Collaboration. Technical Design Report for the: PANDA Micro Vertex Detector. arXiv:1207.6581, 2011.

[8] PANDA Collaboration. Technical Design Report for: PANDA Electromagnetic Calorimeter (EMC). arXiv:0810.1216, 2008.

[9] G. Schepers et al. Particle Identification at PANDA. *Report of the PID TAG*, March 2009.

[10] B. Singh et al. Technical design report for the PANDA Barrel DIRC detector. *Journal of Physics G: Nuclear and Particle Physics*, 46(4):045001, mar 2019.

[11] I. Adam et al. The dirc particle identification system for the babar experiment. *Nuclear Instruments and Methods in Physics Research Section A: Accelerators, Spectrometers, Detectors and Associated Equipment*, 538(1):281 – 357, 2005.

[12] N. Akopov et al. The HERMES dual-radiator ring imaging Cherenkov detector. *Nucl. Instr. Meth. A*, 479:511–530, 2002.

# 3 Design of the Disc DIRC

The Endcap Disc DIRC has been designed to provide charged hadron identification. It is positioned 194 cm downstream from the interaction point, directly in front of the endcap EMC [1] (Figure 3.1).

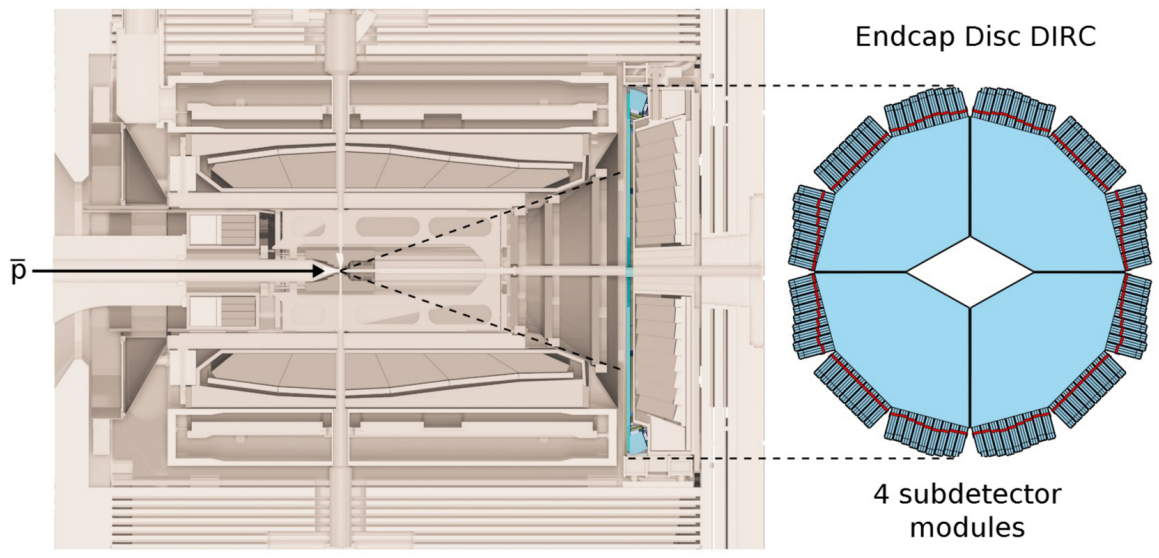

**Figure 3.1:** Position of the Endcap Disc DIRC of the $\overline{\text{P}}$ANDA target spectrometer [2]. The device will provide particle identification for charged hadrons with polar angles between $\theta < 22°$, and $\theta_x > 10°, \theta_y > 5°$.

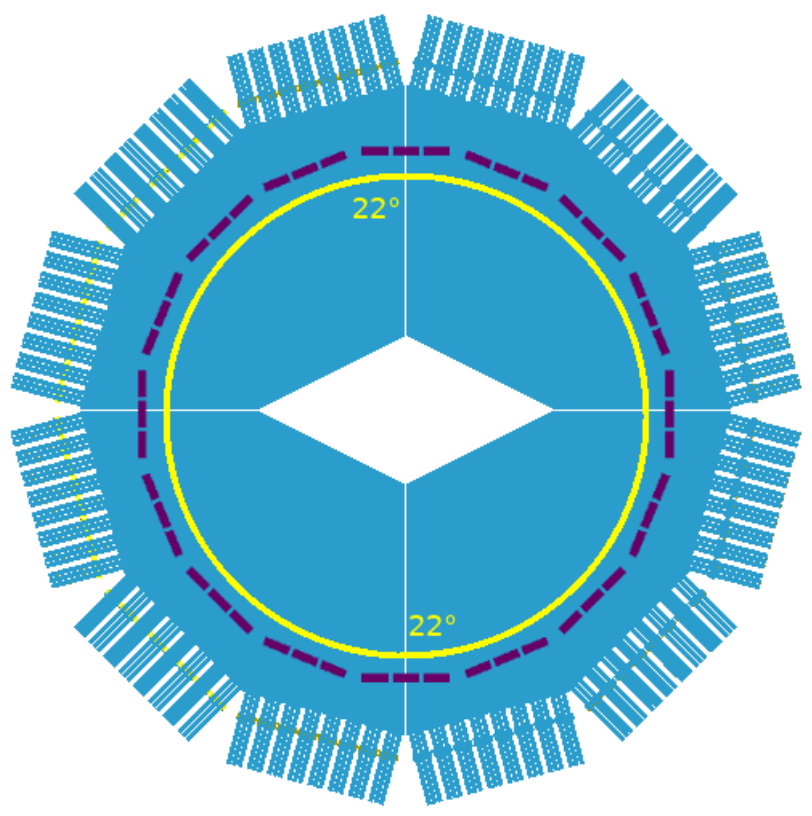

**Figure 3.2:** The angular acceptance for the Barrel DIRC (violet) and the Endcap Disc DIRC (yellow), calculated as projections of straight tracks originating in the interaction point.

The goal is to separate charged hadrons in the polar angle ranges between $\theta = 22°$ and $\theta_x > 10°$, $\theta_y > 5°$. Particles with smaller polar angles will be tracked and identified in the forward spectrometer. Larger polar angles between 22° and 140° will be covered by the Barrel DIRC. Because the acceptance of the Endcap Disc DIRC goes beyond 22°, there will be no gap in the covered phase space. Figure 3.2 illustrates the 2-dimensional angular acceptance for both detectors for straight tracks that origin in the interaction point.

## 3.1 Overview

While the principle idea of such a Disc DIRC has already been proposed by Kamae et al. [3] in 1996, a detector of this type has not yet been realized. During the development, several design approaches finally lead to a detector which is compatible with all $\overline{\text{P}}$ANDA requirements.

The $\overline{\text{P}}$ANDA magnetic field excluded the usage of MAPMTs (Multi-Anode PMTs) as photon sensors. The usage of mircochannel-plate photomultiplier tubes (MCP-PMTs) was considered, but prior to 2012 digital silicon photomultipliers (dSiPMs) were considered for the EDD because of their excellent efficiency, time resolution, and compactness. In addition, the lifetime of MCP-PMTs was too short to be used in the EDD.

A time-of-propagation (ToP) and a Focusing Disc DIRC (FDD) have been studied independently [4]. After showing that the ToP design has severe problems in dealing with the photon background in the measured patterns [5], a 3D Disc DIRC design has been proposed based on silicon photomultipliers [6].

This design combined the ToP and focusing techniques. The precise measurement of time and position allowed to reject the intrinsically large dark count rate of the dSiPM. A 3D-reconstruction method had been developed and tested with data samples from a time based simulation. The result showed that background rejection is indeed possible and sufficient resolution could be achieved.

However, irradiation studies in 2012 lead to the conclusion that the dSiPMs that were available at that time were not sufficiently radiation hard.

While the lifetime of MCP-PMTs improved significantly over the years, it was still not sufficient for the proposed detector designs. This lead to a rate limited design based on MCP-PMTs [6]. The idea was to increase the imaging resolution while decreasing the detected number of photons per incident particle to an acceptable rate for MCP-PMTs.

It was intended to reuse most of the methods already developed for the SiPM design and to use focusing optics similar to those developed for the FDD design. However, it turned out that further optimization of the optical system and the reconstruction method was mandatory to reach the re-quired detector performance with the lower number of detected photons [6]. While the optics for the FDD and dSiPM based designs had aspheric shapes, the new optical system makes use of a cylindrical shape which simplifies the manufacturing process. The resulting detector design is introduced in this chapter. Details on the individual components can be found in chapter 4.

## 3.2 Goals and Requirements

The primary task of the Disc DIRC is $\pi$/K-separation for tracks with polar angles between 5° and 22°. Figure 3.3 shows the superposition of kaon distributions from the channels

- $\bar{p}p \to D^0\bar{D}^0$
- $\bar{p}p \to D^0\bar{D}^0\gamma$
- $\bar{p}p \to D^{*0}\bar{D}^{*0}$
- $\bar{p}p \to D^{*0}\bar{D}^{*0}\gamma$
- $\bar{p}p \to \Lambda_c\bar{\Lambda}_c$
- $\bar{p}p \to D^+D^-$
- $\bar{p}p \to D^+D^-\gamma$
- $\bar{p}p \to D_s^+D_s^-$
- $\bar{p}p \to D_s^+D_s^-\gamma$
- $\bar{p}p \to D^{*+}D^{*-}\gamma$

at a beam momentum of 15 GeV/c. Especially in case of higher momenta, the Disc DIRC will have to provide kaon identification up to a momentum of 4 GeV/c or higher. Individual contributions have not been weighted by their total cross sections to attain an equal contribution to the phase space Figure [6]. In this range, the envisaged separation power for $\pi$/K is better than $3\,\sigma$. Figure 3.4 translates this requirement to the reconstructed Cherenkov resolution per track for different momenta.

The $\overline{\text{P}}$ANDA environment forms a major challenge for the design of this device as it has to be placed inside the target spectrometer. The whole device including its photon sensors will be exposed to a magnetic field of up to 2 T.

In front of the EMC, less than 42 mm in beam direction (z) are available for the radiator and housing of the EDD. At the perimeter of the radiator, where the optical system, sensors and electronics have to be placed, the available space is increased cross to about 148 mm × 210 mm (z×y) which runs along the inner edge of the EMC support frame. The approximate shape of this support frame is an octagon. One of the main challenges was to design a system which fits inside this predefined and asymmetric volume.

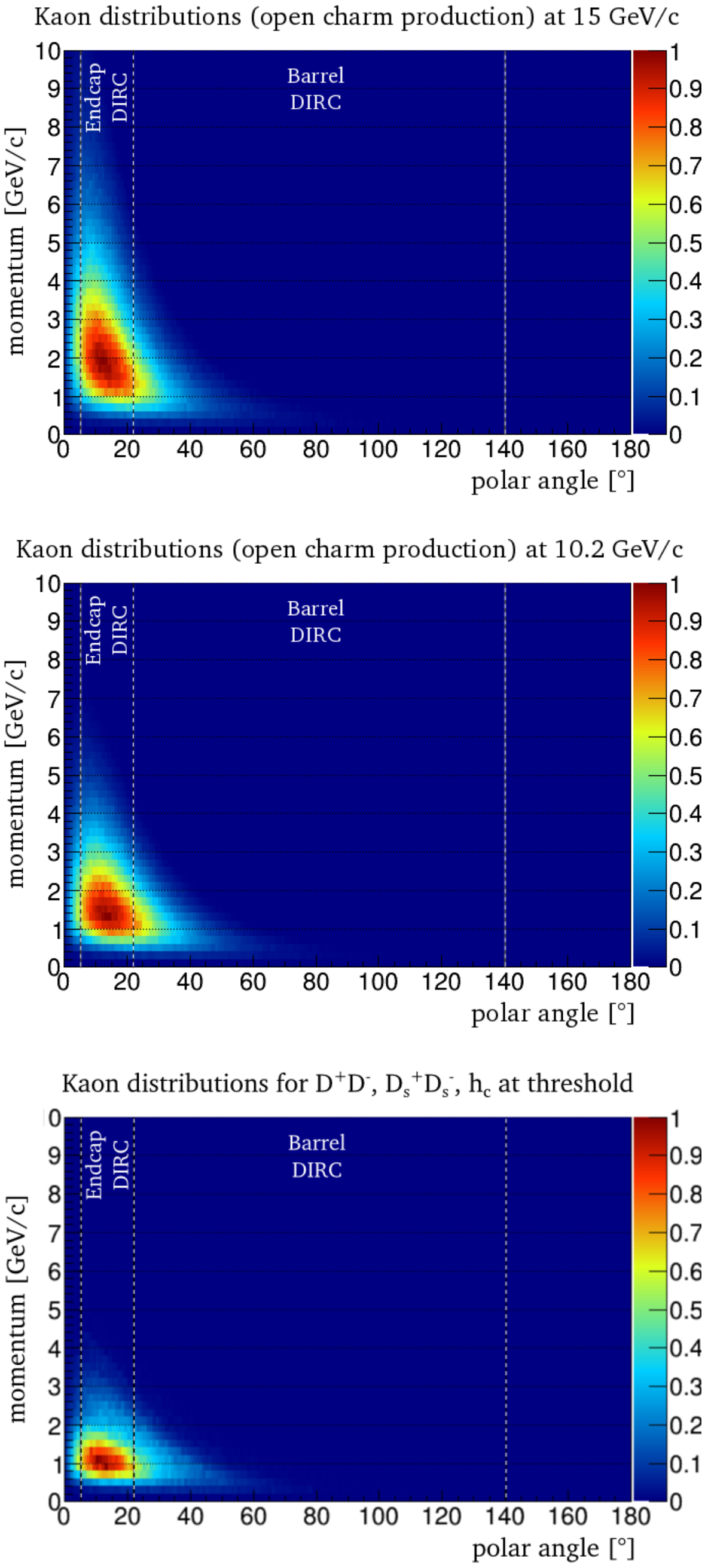

**Figure 3.3:** The upper two plots show the phase space distributions obtained by superimposing simulated kaon distributions of the selected benchmark channels at beam momenta of 15 GeV/c and 10.2 GeV/c. The lower plot shows the kaon distribution at the production threshold for the reactions $\bar{p}p \to D^+D^-$, $\bar{p}p \to D_s^+D_s^-$, $\bar{p}p \to h_c \to \eta_c\gamma$ [6]. The Endcap Disc DIRC covers the polar angles between 5° and 22°. The range from 22° to 140° is covered by the Barrel DIRC and polar angles smaller than the EDD acceptance are covered by a RICH detector in the forward spectrometer.

The Disc DIRC has to be able to cope with high interaction rates exceeding 20 MHz. A total ioniz-ing dose of up to 500 Gy is expected for the central parts of the radiator and about 10 Gy for the focusing optics which are shielded by the material of the barrel EMC. Due to the hadronic interaction, the neutron flux is a major contribution. The 1 MeV equivalent neutron flux is expected to exceed $10^{11}$ neutrons/cm$^2$ [6]

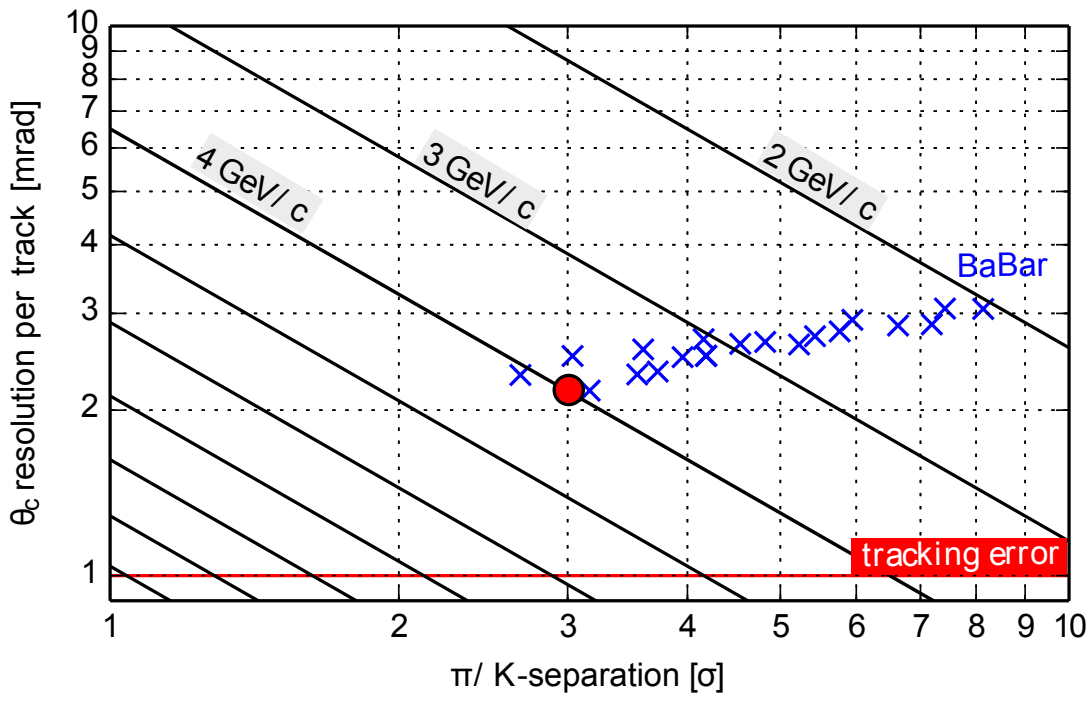

**Figure 3.4:** Correspondence of required Cherenkov angle resolution and $\pi/K$ separation in $\sigma$ at momenta from 1.5 to 10 GeV/c. The red circle indicates the targeted performance which translates to a track resolution of 2.1 mrad. For comparison, the BaBar DIRC separation power obtained from the channel $D^0 \to K^-\pi^+$ as shown in [7] has been added (blue crosses). A decrease of angle resolution towards lower momenta is an expected behavior as the effect of multiple scattering of particle tracks and energy loss increases with lower momenta.

## 3.3 Detector Design

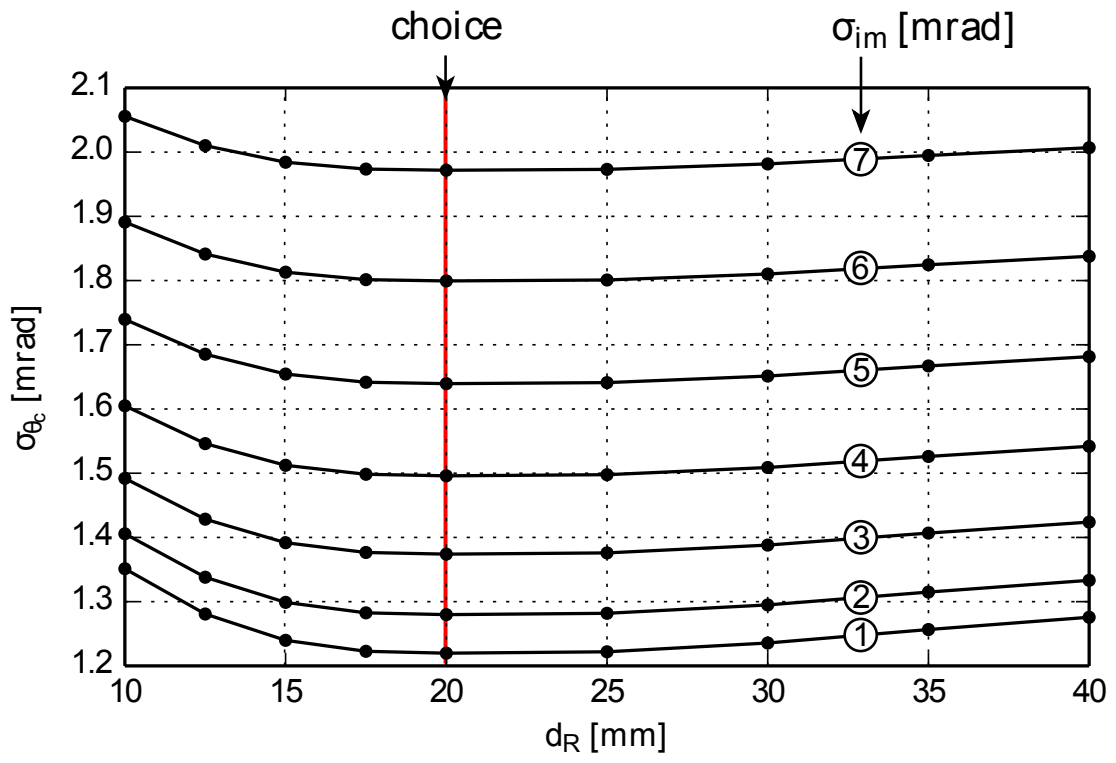

**Figure 3.5:** Error of measured Cherenkov angle as function of the radiator thickness $d_R$, forcing a fixed number of photons $N_{\text{ph.}} = 20$ by tuning the band width of the optical filter. The circled numbers correspond to an assumed image resolution. Details can be found in [6].

The area from 5° to 22° polar angle is covered by 2 cm thick radiator plates made from synthetic fused silica. This thickness was determined considering different error sources such as multiple Coulomb scattering (MCS), dispersion effects, and imaging errors (see also Figure 3.5). Synthetic fused silica is the preferred material as it combines a high radiation hardness with sufficient VIS/UV-transmission and low dispersion.

Charged tracks with

$$\beta \geq \frac{n_{\text{outside}}}{n_{\text{radiator}}} \tag{3.1}$$

emit Cherenkov photons inside the radiator. The photons with an angle

$$90° - \varphi > \arcsin \frac{n_{\text{air}}}{n_{\text{SiO}_2}} \tag{3.2}$$

relative to the surface normal are trapped due to total internal reflection. As indicated by the top right part of Figure 3.7, trapped photons are transported to the perimeter of the radiator by consecutive reflections at the down- and upstream surfaces. which have to be exactly parallel to conserve the angular information of the photon. In order to conserve the angular information of the photon the surfaces have to provide a high parallelism.

In total an area of about 2 m in diameter has to be covered with the precision polished radiator. Discussions with several manufacturers limited the maximum diameter of polished fused silica plates to roughly 1.5 m (60 inches). Therefore the proposed design divides the radiator into four identical quadrants which form four independent subdector modules as shown in Figures 3.1 and 3.7. Radiator edges do not need mirror coating as the majority of trapped photons will be internally reflected. This has been shown with simulation studies as presented in Figure 3.6 for three photon paths: direct, indirect on short side, and indirect on long side. Adding thin mirrors may introduce angular misalignments and reduce the reflectivity from 100% to values between 90% and 95%.

The radiator shape is a compromise between the $\overline{\text{P}}$ANDA geometry, the alignment of the sensors relative to the magnetic field lines and the number of edges to polish. The corners have to be convex and their number has to be minimized to reduce production costs.

The perimeter is equipped with readout modules (ROM, Figure 3.8) consisting of imaging optics and a position sensitive photon sensor. Both, hit time and hit position contain information about the Cherenkov angle. Dispersion errors are mitigated by using optical bandpass filters or photocathodes which limit the wavelength acceptance. In addition, the filters reduce the photon flux to enhance the lifetime of the sensors. The passband is chosen to minimize dispersion at the maximum photon flux that the sensors can handle. The width depends on the efficiency of the photon sensor. For standard MCP-PMTs with a lifetime of $5\,\mathrm{C/cm^2}$, the optimal band is 360 nm to 465 nm [6] which also has been used for the presented design studies. MCP-PMTs with improved photocathodes allow the selection of different bands which leads to an improved performance due to the lower chromatic dispersion error.

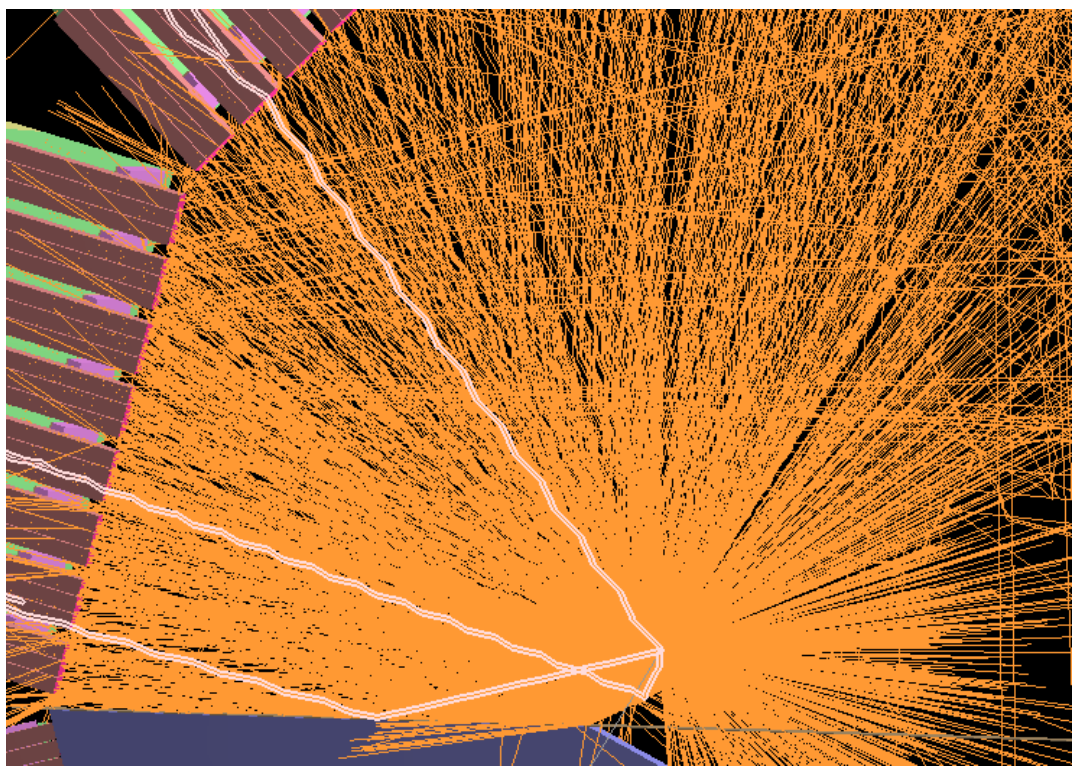

**Figure 3.6:** Reflections on two different positions at the outer rim without an additional mirror coating.

The readout module is the central component of the detector design, as it combines precision focusing optics and photon detection, signal digitization and optical link to the DAQ system in one unit. It consists of three 16 mm wide focusing elements (FELs) attached to fused silica bars, a 2 inch square MCP-PMT with custom fine granularity, an ASIC board for fast digitization of PMT signals and a front-end controller board (FCB) providing the connection to the DAQ system via an optical link. The individual ROMs are working independently featuring individual voltage supply, controller board and data link.

Photons enter the ROM at the bar-radiator interfaces and are transported by total internal reflection to the curved surface of the focusing element (FEL). The aluminum coated, convex, cylindrical mirror converts angular information into position information by imaging parallel light onto an area with a transversal spread of less than single pixel on the photocathode of the sensor. The high resolution MCP–PMT measures the position and arrival time of the photon. The choice of three FELs per ROM is a compromise between cost and performance, as the position resolution given by the width of the elements affects the separation power (see Figure 3.9).

The sensor will feature three columns of 100 strip-shaped pixels, one column per FEL, to match the resolution of the optical system (see also Figure 3.10). The magnetic field decreases the charge footprint on the PMT anode, so that a high resolution should be achievable with the availables.

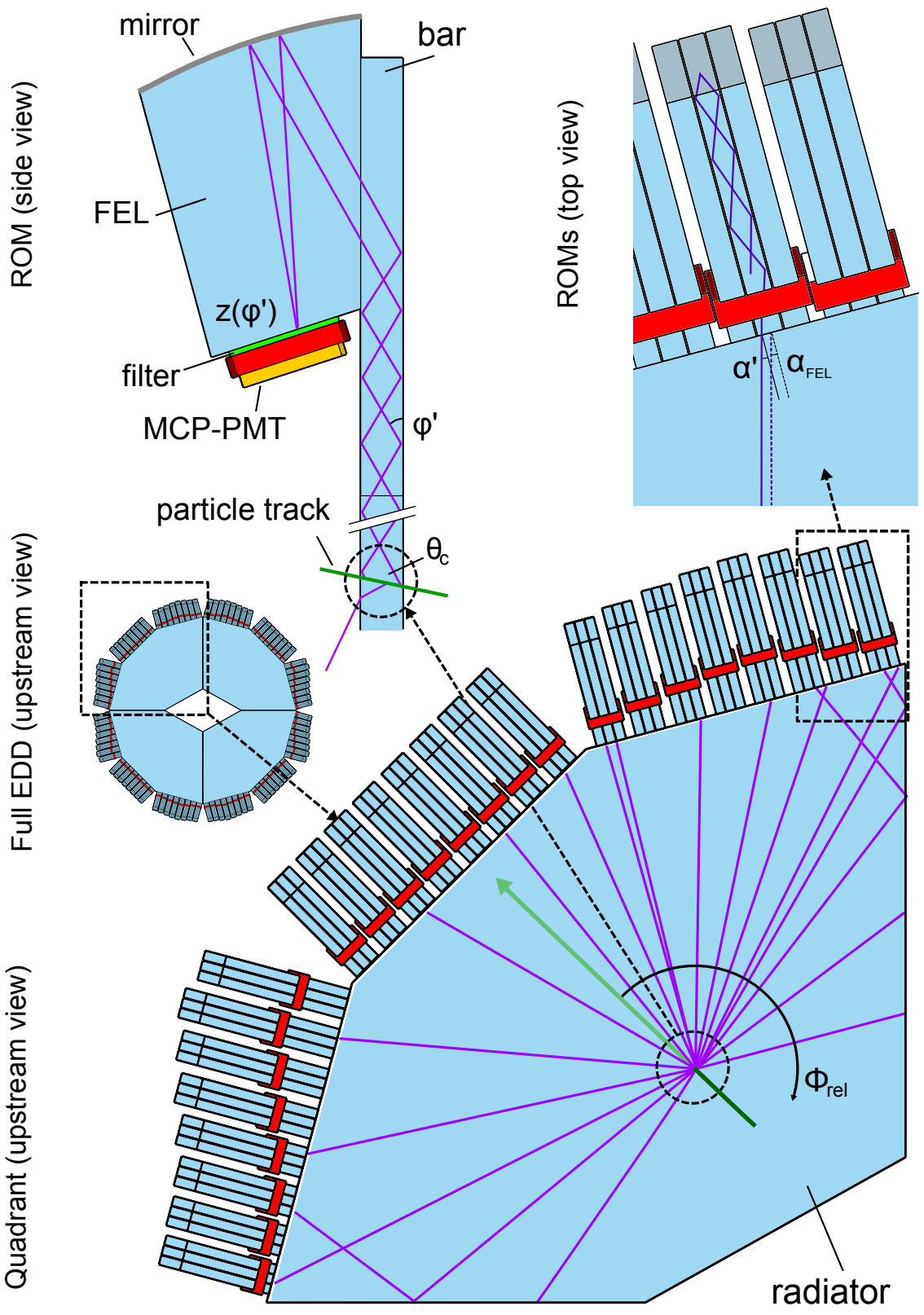

**Figure 3.7:** Schematic overview of one Disc DIRC quadrant and its working principle. Optical components are colored light blue, MCP-PMTs are colored red. The central figure shows a fully equipped quadrant with different photon paths (purple) originating from the same spot. Some of the photons are trapped by total internal reflection whereas the rest is lost by refraction on the surface. The top left part of the drawing illustrates the DIRC principle, where a charged particle (green line) passes through the radiator and Cherenkov photons are emitted along its path. The photons travel towards the edge of the radiator where they enter the bars and FELs where parallel entering photons are focused onto the MCP-PMT photocathode. The cylindrical-shaped surface is mirror-coated.

A very compact and fast readout will be available with the TOFPET-ASIC [8] which provides 64 channels per chip ($3.5 \times 7.1\,\mathrm{mm^2}$) with a maximum rate capability exceeding 100 kHz per channel and a built-in amplifier with a dynamic range of 300 pC. It will deliver a TDC resolution of 50 ps LSB, an energy measurement via time-over-threshold, self-triggered operation, and LVDS output. A PCB with 6 ASICs, attached to the back of the PMT,

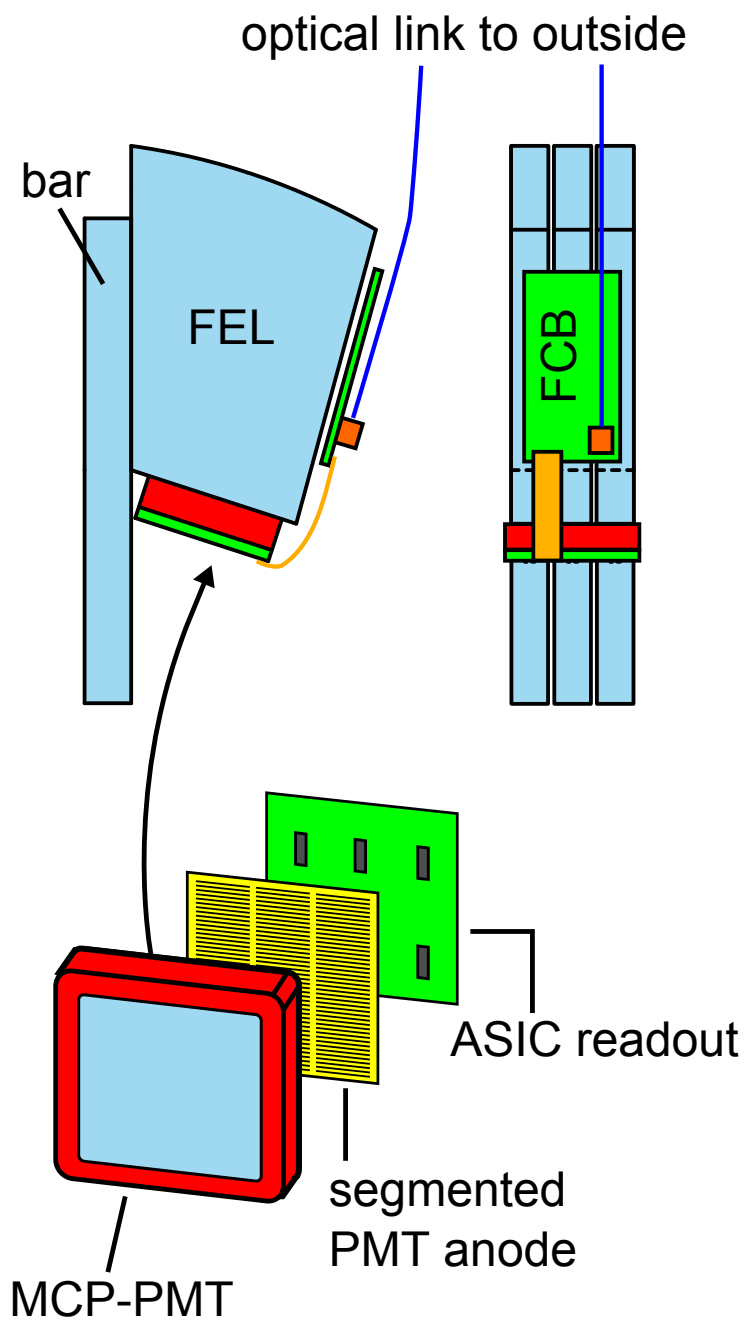

**Figure 3.8:** The drawing illustrates the setup and positioning of the MCP-PMT, the ASICs for readout and the front-end electronis board (FCB) together forming the readout chain.

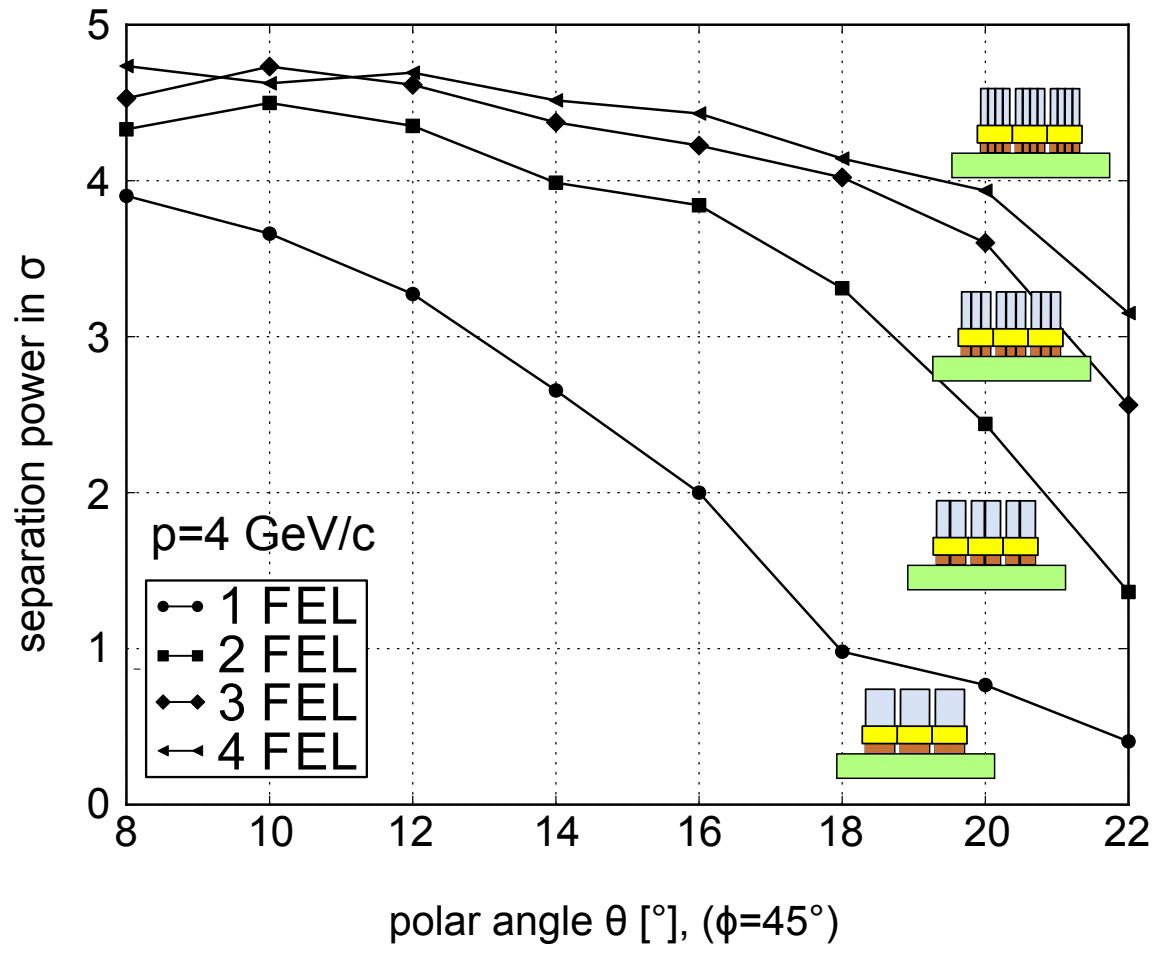

**Figure 3.9:** $\pi/K$ separation power computed from the distribution of reconstructed Cherenkov angles for a ROM consisting of a 2 inch-tube with 0.5 mm pixel width in focusing direction, 22 % peak PDE, 17 % reflecticity at the bialkali photocathode and 1 to 4 focusing elements [2]. The width of these optical elements has a significant impact on the 3D spot size on the image plane. The EDD design will use a setup with three focusing elements per ROM.

allows continuous sampling of data which is transmitted via LVDS to the front-end controller.

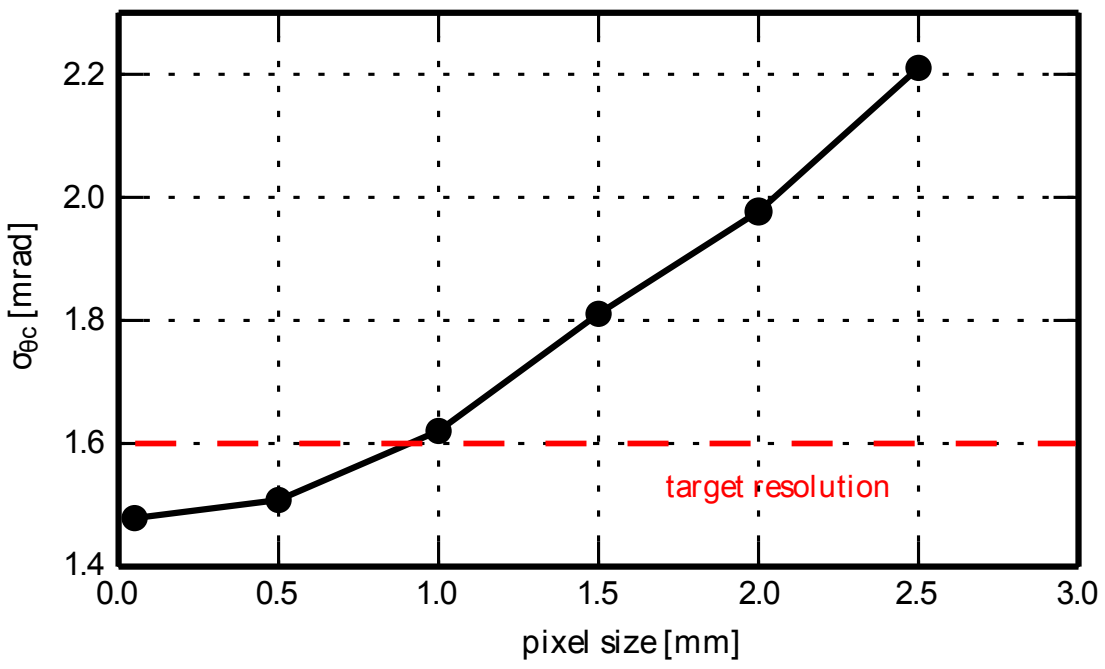

**Figure 3.10:** Dependence of the Cherenkov angle resolution on the pixel size for the design with bandpass filter. The points correspond to the reconstruction of 1000 $\pi$ at $(p, \theta_p, \phi_p)$=(4 GeV/c, 15°, 45°) [6].

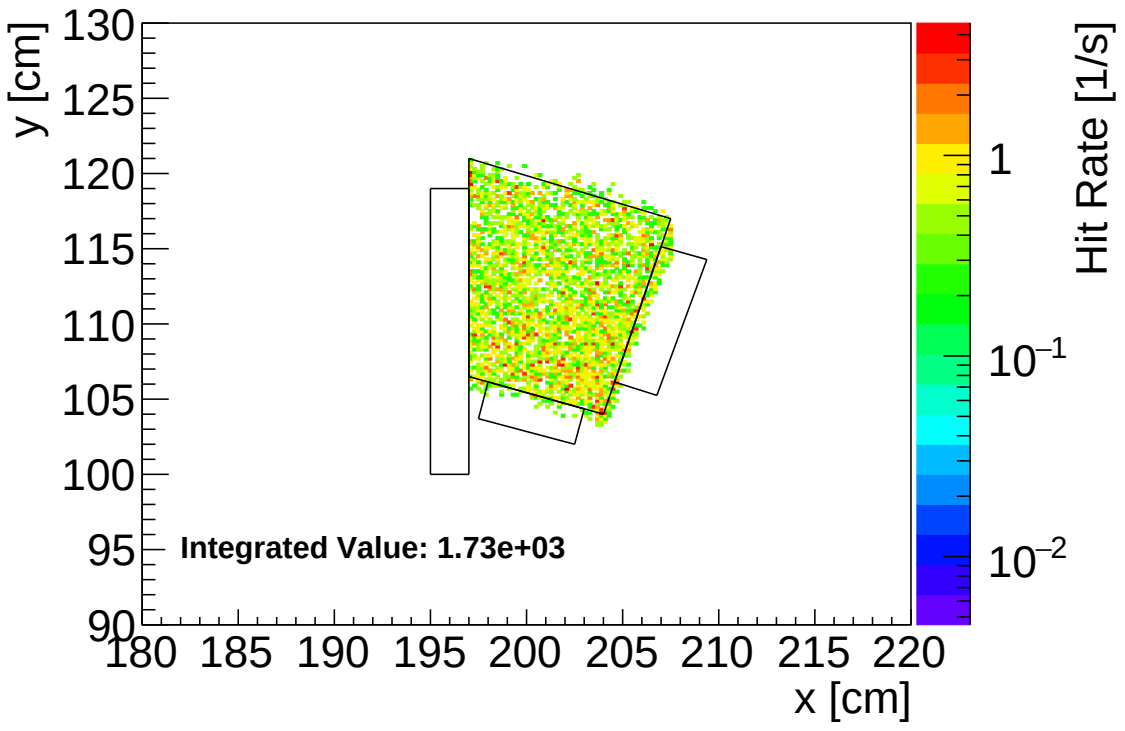

**Figure 3.11:** Distribution of charged tracks in the MC simulations using the Dual Parton Model particle generator of PandaROOT.

In order to estimate the radiation background, detailed Monte-Carlo studies have been persued. The FELs are shielded by the Barrel EMC, i.e. there are almost no direct charged tracks coming from the IP. In case a charged track passes an FEL, it produces additional light and "blinds" the MCP-PMT or makes it start "ringing". If this happens, the sensor will simply be deleted from the reconstruction for this event. According to the results shown in Figure 3.11, it is expected that the hit rate of charged tracks is 50 to 100 times smaller than that for the full radiator quadrant. It will be quite randomly distributed in the 3-dimensional phase space ($\varphi$, $z$, $\Delta t$) and will usually fail to pass the reconstruction criteria.

The presented design enables a direct calculation of the Cherenkov angle on a single photon level. Besides the information on the pixel and the time of propagation, the four-momentum $(E, \vec{p})$ of the charged particle has to be taken into account. The timing information is used to resolve ambiguities in connection with direct illumination of the ROMs or indirect illumination where the Cherenkov photons are reflected off a side surface. Depending on the azimuthal orientation $\phi_{\rm rel}$ between photon path and particle trajectory on the radiator plane, the angle $\varphi$, which is defined as the angle between the radiator surface and the photon path, can be calculated analytically by

$$\cos\varphi = \frac{A\cos\theta_c}{B} \pm \sqrt{\frac{\cos^2\theta_p - \cos^2\theta_c}{B} + \left(\frac{A\cos\theta_c}{B}\right)^2}, \quad (3.3)$$

where $\theta_p$ is the particle angle relative to the normal of the radiator plane [5]. The terms $A$ and $B$ are given by

$$A = \sin\theta_p \cos\phi_{\rm rel}, \quad B = A^2 + \cos^2\theta_p. \quad (3.4)$$

The FEL focusing allows to measure the transformed $\varphi'$ given by

$$\tan\varphi' = \frac{\tan\varphi}{\cos\alpha_{\rm FEL}} \quad (3.5)$$

with $\alpha_{\rm FEL}$ describing the angle between the photon path on the radiator plane and the orientation of the FEL (see also Figure 3.7). The angle $\varphi'$ is determined by the position $z$ which is given by the pixel that registers the signal. The Cherenkov angle can subsequently be determined separately by

$$\theta_c = \arccos\left(\sin\theta_p \cos\phi_{\rm rel} \cos\varphi + \cos\theta_p \sin\varphi\right) \quad (3.6)$$

for every hit. Averaging over a whole track with 16 to 25 individually detected photons, a resolution between 1.2 and 2.0 mrad can be achieved. This resolution fulfills the PID performance requirements of the EDD [6].

## References

[1] PANDA Collaboration. Technical Design Report for: PANDA Electromagnetic Calorimeter (EMC). arXiv:0810.1216, 2008.

[2] O. Merle et al. Development of an Endcap DIRC for PANDA. *Nucl. Instrum. Meth.*, A766:96–100, 2014.

[3] T. Kamae et al. Focussing dirc — a new compact cherenkov ring imaging device. *Nuclear Instruments and Methods in Physics Research Section A: Accelerators, Spectrometers, Detectors and Associated Equipment*, 382(3):430 – 440, 1996.

[4] Peter Schönmeier et al. Disc dirc endcap detector for panda@fair. *Nuclear Instruments and Methods in Physics Research Section A: Accelerators, Spectrometers, Detectors and Associated Equipment*, 595:108–111, 09 2008.

[5] O. Merle. Development of Reconstruction Methods and Algorithms for the PANDA Disc DIRC. Diploma thesis, Justus-Liebig-Universität Gießen, 2009.

[6] O. Merle. Development, design and optimization of a novel Endcap DIRC for PANDA. PhD thesis, Justus-Liebig-Universität Gießen, 2014.

[7] J. Schwiening et al. Construction and performance of the BABAR-DIRC. *J. Instrum.*, 2(10):P10004, 2009.

[8] M. D. Rolo et al. TOFPET ASIC for PET applications. *J. Instrum.*, 8(02):C02050, 2013.

# 4 Detector Components

## 4.1 Optical Components

The Cherenkov radiators are by far the largest non-mechanical parts of the P̄ANDA Endcap Disc DIRC. In comparison to a radiator in a RICH detector, a DIRC radiator also serves as a light guide, as the emitted Cherenkov light will be propagated inside the radiator towards its outer edges, where it enters focusing optics and/or photon sensors. Despite their large size, the radiator plates resemble precision optical components and have very limiting requirements regarding the choice of the material.

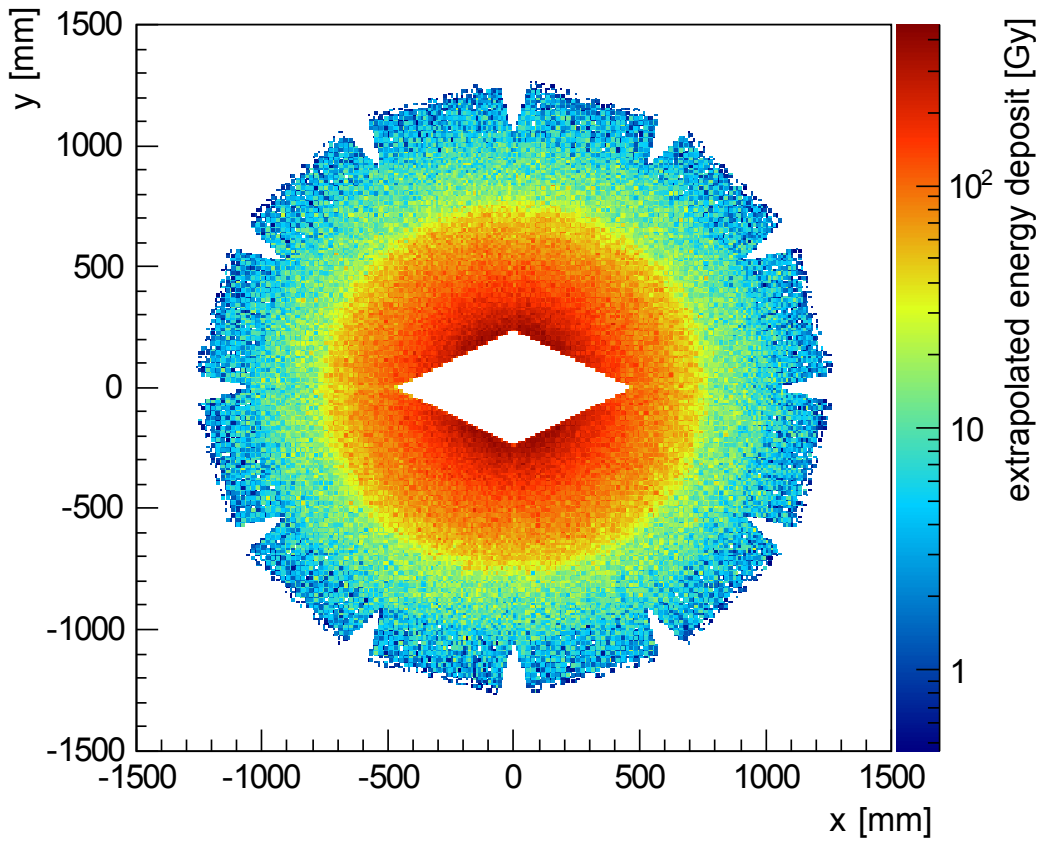

**Figure 4.1:** Energy deposit in the optical system extrapolated to 10 years at a 50% duty cycle and 10 MHz interaction rate [1]. The amount of deposited energy clearly drops beyond a radius of 75 cm around the center as the barrel calorimeter blocks this region.

### 4.1.1 Radiator

#### Choice of Material

The material for the optical components in the P̄ANDA Disc DIRC has to fulfill the following requirements:

- excellent optical properties such as transmission and small chromatic dispersion,
- radiation hardness,
- excellent machinability.

Quartz (chemically $SiO_2$), which basically meets the abovementioned criteria, exists in three different compositions. The crystalline form of quartz (natural quartz) is birefringent, contains a high level of impurities and hence cannot be used in the DIRC optics. An amorphous form of quartz (natural fused silica) is produced by crushing and melting natural quartz. Although the optical properties would fit the requirements, a considerably large amount of impurities remains, which depreciate radiation hardness.

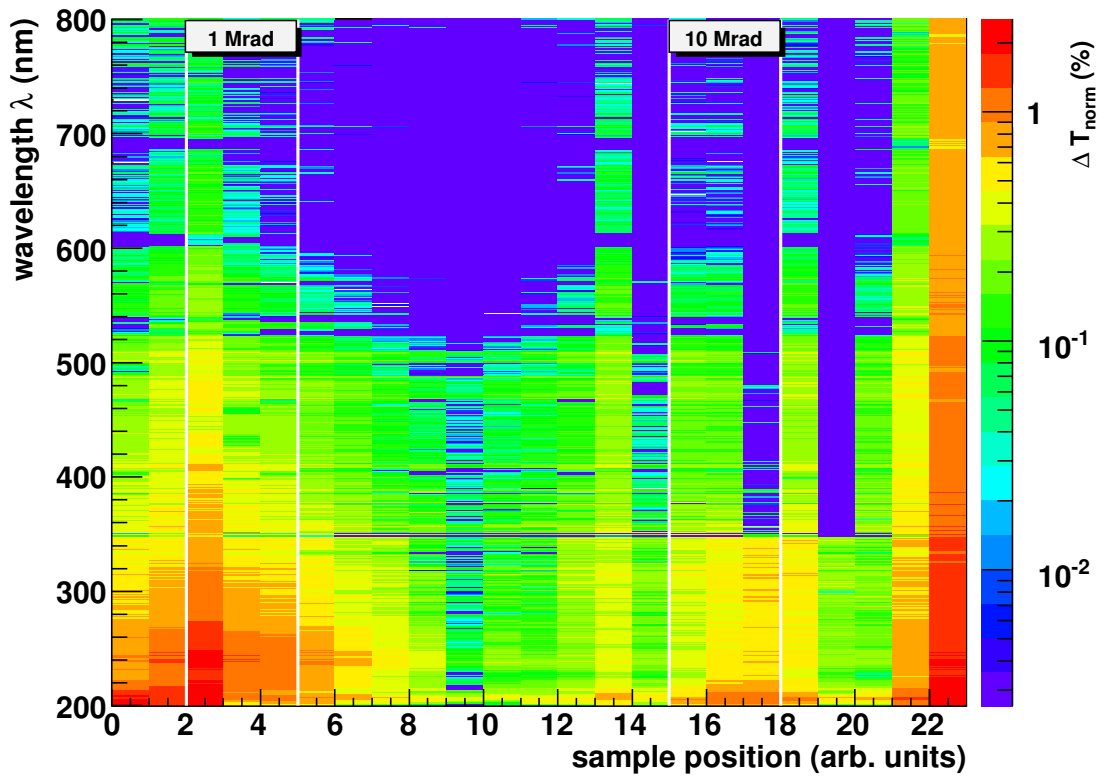

**Figure 4.2:** Normalized transmission difference $\Delta T_{norm}$ for Heraeus Suprasil 1 as a function of wavelength and position within the sample [2]. The vertical lines indicate the position of irradiation by a 150 MeV proton beam. No distinct features corresponding to the irradiation spots are observed within the obtained accuarcy. Large deviations at the corners are attributed to edge effects.

A third form of quartz (synthetic fused silica) is made of different feedstock such as silicon tetrachloride ($SiCl_4$). This feedstock is burned in an oxygen atmosphere at around 2000°C and forms a large ingot which can then be further processed [3]. This process results in a very pure material which is widely used in optical applications. Depending on the level of interstitial Hydrogen, the radiation hardness can be positively influenced. On the other hand very high Hydrogen levels can impair transmission properties [2]. Available materials (amongst others) are HPFS 7980 by Corning [4], Spectrosil 2000 and Suprasil 1 and 2 by Heraeus [5], NIFS by Nikon [6] and Lithosil Q0 by Schott [7].

#### Radiation Hardness

The optical properties of the radiator material are crucial for the overall performance of DIRC-type Cherenkov detectors. The generated Cherenkov photons travel a substantial pathlength inside the radiator material, unlike in traditional RICH counters, and undergo multiple reflections off the surfaces. Thus its optical properties must remain unchanged even after intense irradiation as encountered in $\overline{\text{P}}$ANDA. In case of the EDD the expected dose which has to be taken into account ranges from several hundred Gray near the beam to a few Gray at the outer rim (see Figure 4.1).

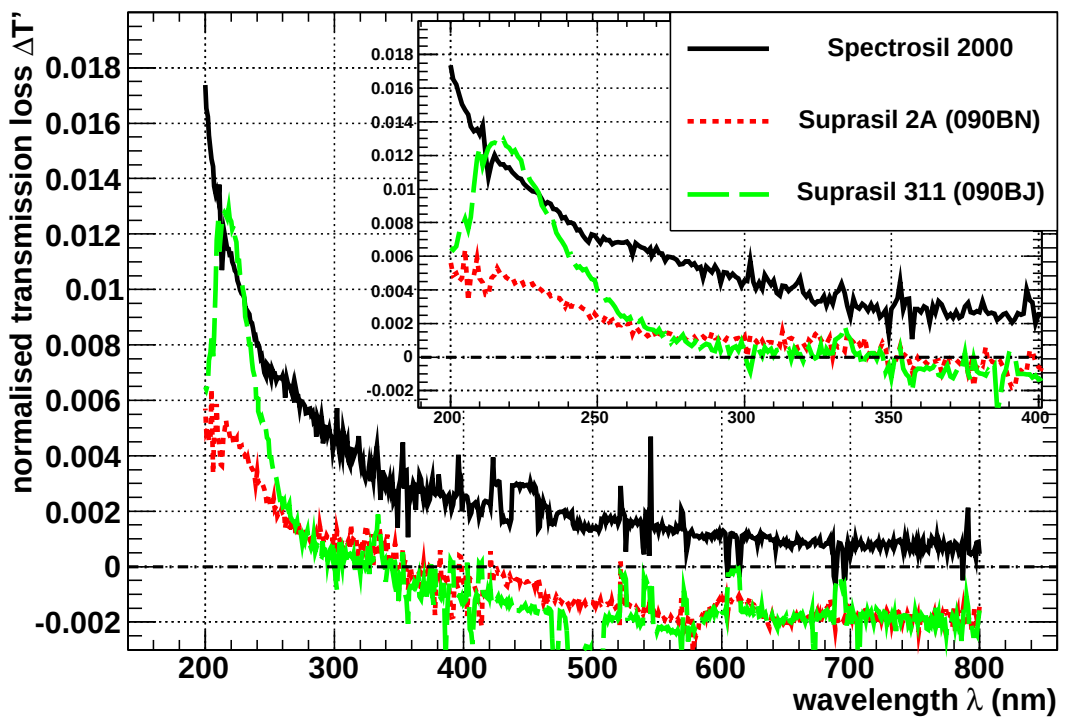

**Figure 4.3:** Comparison of normalized transmission loss $\Delta T'$ for Suprasil 2 A and 311 as well as Spectrosil 2000 [2]. Negative $\Delta T'$ values for both Suprasil samples above 350 nm are due to overall misalignment effects for the sets of measurements before and after the irradiation.

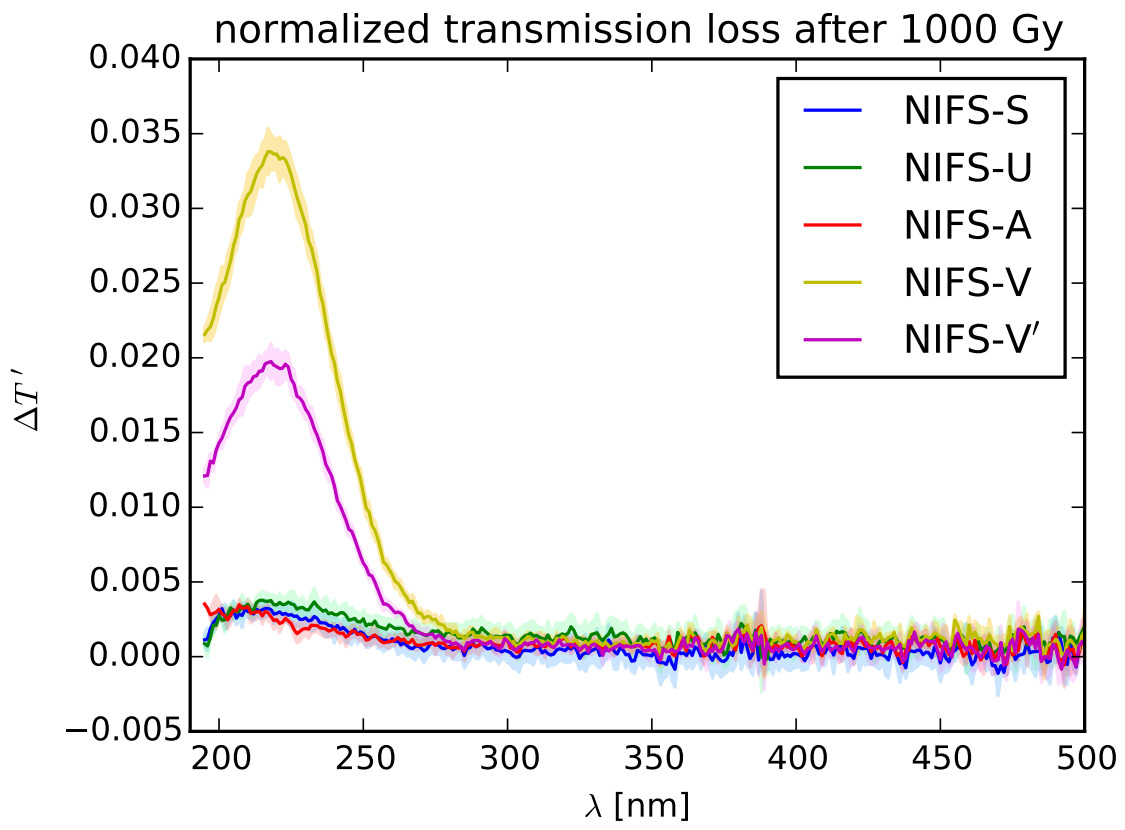

**Figure 4.4:** Normalized transmission loss for all NIFS-samples at wavelengths between 195 nm and 500 nm [8]. One sees a broad absorption band around 220 nm for the NIFS-V series. The other samples show a very small difference in the same region.

Fused silica has already been identified as the most suitable material for radiators in DIRC-type RICH detectors by the BaBar Barrel DIRC group [3]. The $\overline{\text{P}}$ANDA Cherenkov group investigated selected types of synthetic fused silica using proton beam and/or $\gamma$-ray irradiation studying the induced radiation damage and possible damage mechanisms [2, 9]. For all studies the transmission after irradiation was compared to the reference measurements prior to irradiation. The result is given as the normalized transmission loss

$$\Delta T' = \frac{T_{before} - T_{after}}{T_{before}} \tag{4.1}$$

which describes the change of transmission due to absorption effects inside the bulk material. Figure 4.2) shows a scan of a Heraeus Suprasil sample which was irradiated at two positions by 150 MeV protons with different dose levels (1 and 10 Mrad) at KVI Groningen. In the following these studies were extended using a $^{60}$Co source at the University of Giessen. Figures 4.3 and 4.4) depict the results for different types of Heraus Suprasil and Nikon Fused Silica (NIFS). The results corroborated previous findings by the BaBar group [3] and additionally established the crucial role that interstitial hydrogen plays in preventing radiation damage. A more elaborate summary of these studies can be found in the $\overline{\text{P}}$ANDA Barrel DIRC TDR [10].

**Table 4.1:** Specifications for the large radiator surfaces.

| category | specification |
|---|---|
| form error | $\sigma_{\text{slope}} \leq 5\,\mu$rad |
| TTV | $\leq 15\,\mu$m PV |
| wedge error | 2 arcsec |
| perpendicularity | 1 arcmin |
| roughness | 1.5 nm RMS |
| scratch/dig | 80/50 or better |
| chamfers | $< 0.5$ mm |

The expected integrated dose for the Disc DIRC over the $\overline{\text{P}}$ANDA lifetime is well below the dose applied in the irradiation tests mentioned above. Since at $\overline{\text{P}}$ANDA the wavelengths of the Cherenkov photons are cut off below about 300 nm due to the glue joints between the radiator and the bars, the irradiation dose induced reduction of the mean free path length in the UV range has no impact on the design of the $\overline{\text{P}}$ANDA Endcap Disc DIRC.

### Dimensions and Geometry

Early Disc DIRC designs proposed a single radiator, which covered the entire acceptance of the $\overline{\text{P}}$ANDA forward region [11],[12]. Due to limitations concerning the size of the available polishing machines worldwide, the radiator had to be divided into several smaller pieces, which than had to be glued together. In the final design a Disc DIRC with four

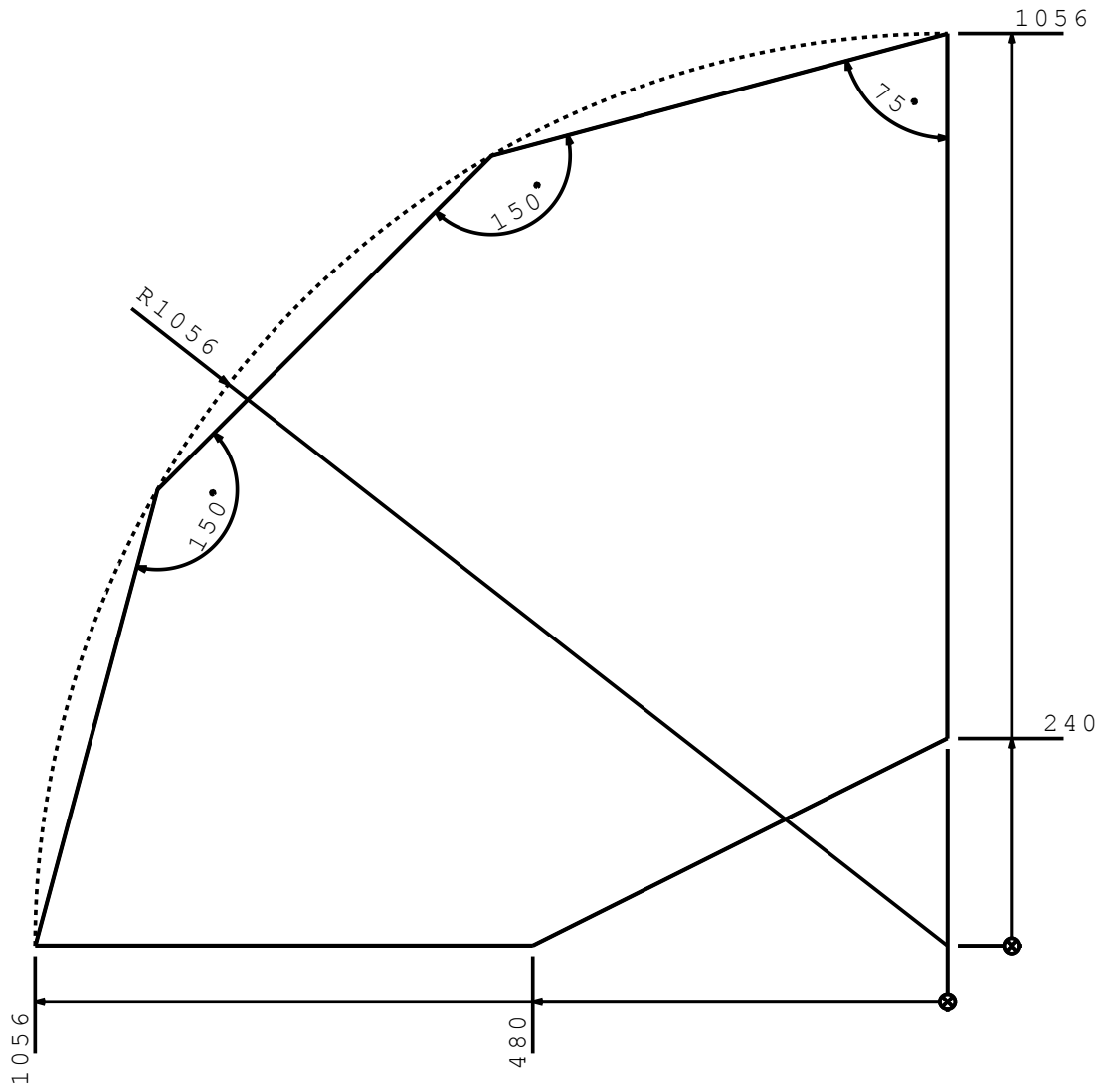

**Figure 4.5:** Technical drawing of a single radiator plate.

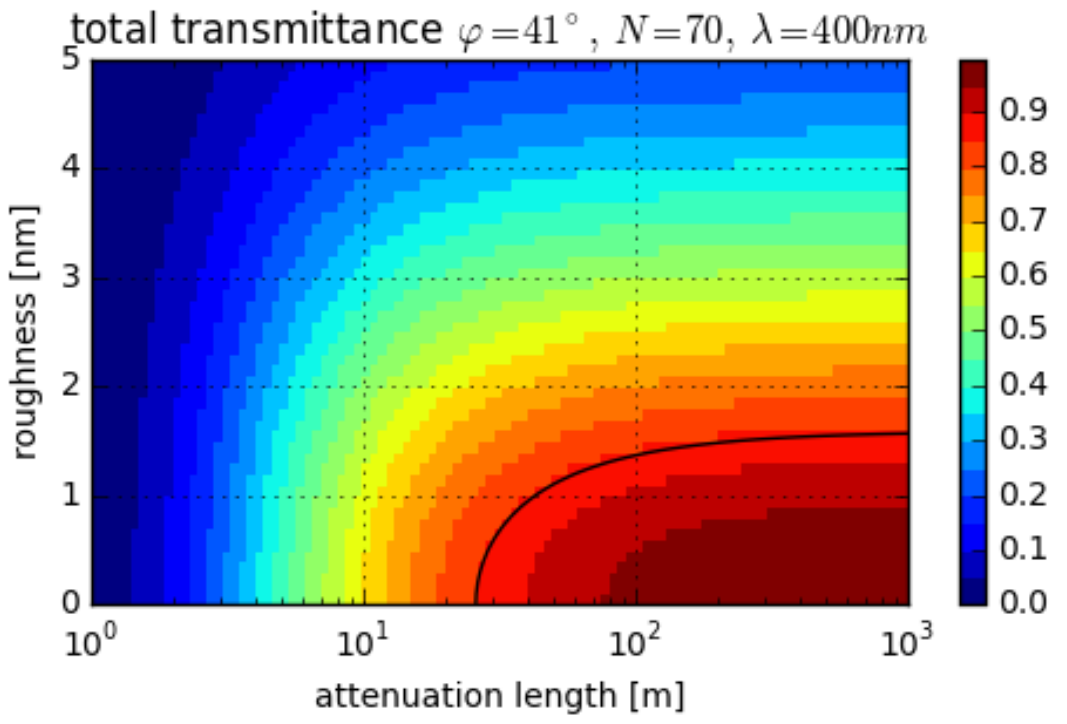

**Figure 4.6:** The total transmittance for different roughness values and attenuation lengths [8]. For attenuation lengths above 400 m the transmittance is completely dominated by the surface roughness. The black line indicates a total transmittance of 85%.

independent quadrants has been developed, where each quadrant houses one large radiator plate [13].

In order to reduce tooling costs, all radiator plates will have the same shape with a maximum corner-to-corner distance of 1493.4 mm and a thickness of 20 mm. All corners are convex. To avoid any risks during the polishing process and the mounting procedure, the sharpest edge will not fall below a value 75°. Four radiator plates will form a regular dodecagon with a radius of $1056 \pm 0.2$ mm. The length of the outer edges will be $546.6 \pm 0.5$ mm. Eight ROMs will be connected to each out side. On the opposite side a triangle with a height of $240 \pm 0.2$ mm and a width of $480 \pm 0.2$ mm will be cut out to leave an acceptance hole for the forward spectrometer (see Figure 4.5).

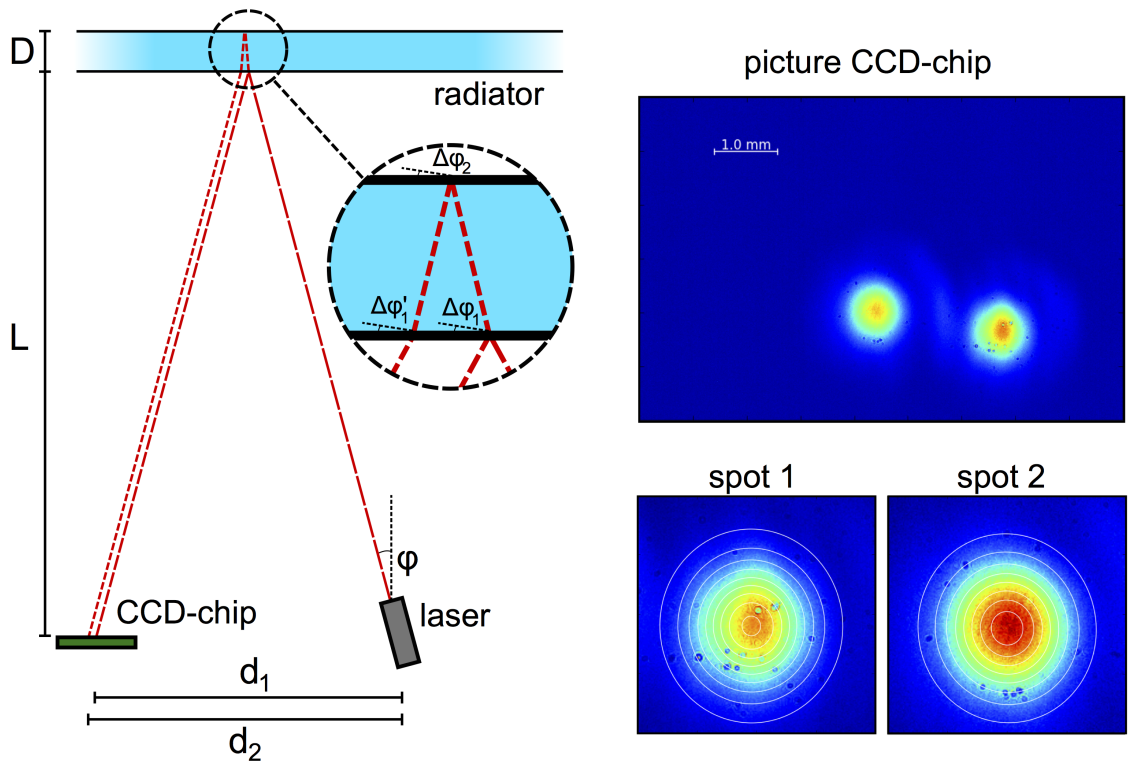

**Figure 4.7:** The left part schematically shows the top view of setup for the surface profile measurement. The top right picture is a colored example picture taken by the CCD chip which shows the direct (indirect) reflection off the (back-) surface of the radiator and the corresponding 2-dimensional Gaussian fits.

## Specifications

In order to conserve the angle of the propagating photon and to avoid light loss, high demands are made on the squareness of the side to face angles and parallelism as well as on the surface specifications of the radiator plates. However, one also has to be careful not to over-specify the price-enhancing tolerances. In case of the BaBar DIRC high effort was put into finding the appropriate surface specifications for their fused silica bars [3]. These very valuable methods have been adapted to the requirements and geometry of the disc radiators and own simulations resulted in a new set of specifications tailored to the needs of the $\overline{\text{P}}$ANDA Disc DIRC.

In comparison to a Barrel DIRC, photons have relatively few internal reflection inside the radiator (up to one hundred instead of several hundred). Scalar scattering theory for smooth surfaces predicts that the light loss due to surface scattering is proportional to the square of the surface roughness. This leads to a reduced surface roughness specification of 1.5 nm RMS for the large surfaces and 5 nm RMS for the small surfaces, where in good approximation only up to one reflection is expected (see also Figure 4.6). Therefore also the squareness can be reduced to 1 mrad for side-to-face angles. In order to reduce angle smearing the parallelism of the large face sides has to be within 2 arcsec and the slope of the form error must be below 15 $\mu$m peak-to-valley (see also table 4.1).

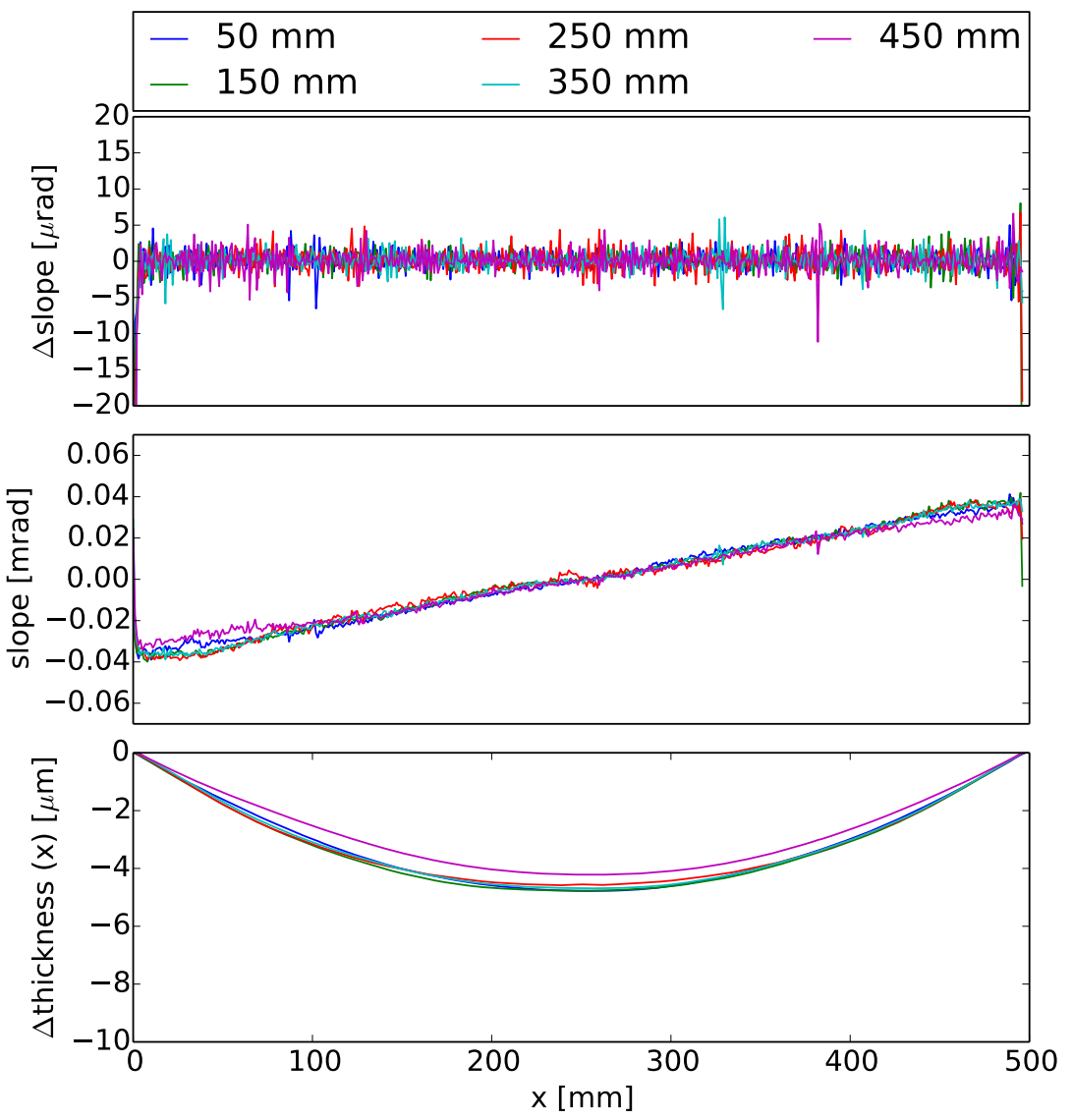

**Figure 4.8:** Fine horizontal scan of the $50 \times 50\,\mathrm{cm}^2$ Nikon prototype radiator at different vertical positions as indicated in the legend [8].

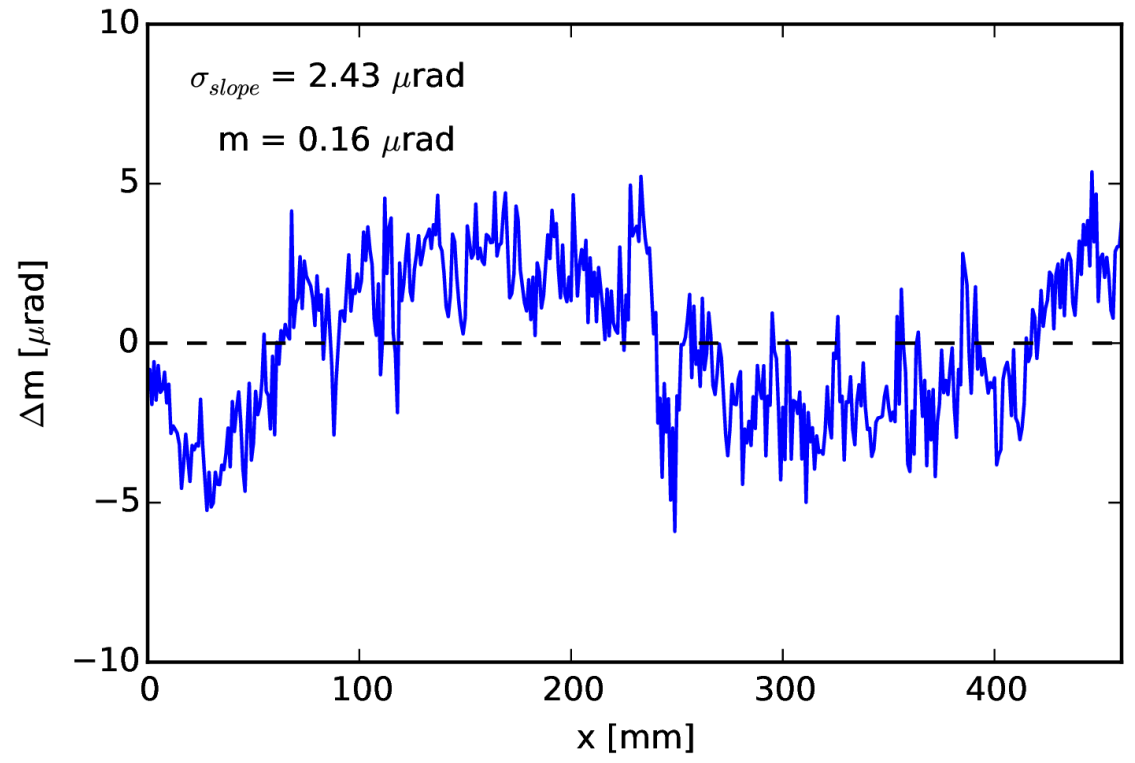

**Figure 4.9:** Deviation of the slope from a perfectly linear progression of the thickness variation [8]. $m$ represents the slope of the fit of the slope at $y = 250\,\mathrm{mm}$ in Figure 4.8 and $\sigma_{\mathrm{slope}}$ the RMS error.

### Optical Quality Insurance

The optical and mechanical quality of the DIRC radiators is of critical importance for the PID performance of the EDD. Imperfections can influence the photon yield as well as the single photon resolution of the Cherenkov angle. Setups for the measurement of the internal transmission and surface roughness of different types of radiators have been developed and used by nearly all groups working on DIRC-type detectors [3, 14, 10].

In order to complement one another the EDD group focused on the evaluation of the slope or thickness variation of the manufactured radiators which is crucial regarding the angle conservation of the Cherenkov photons. The utilized measurement setup is shown in Figure 4.7. It uses the relative position change between a direct reflection spot from the front surface facing the laser and CCD-chip and an indirect reflection spot from the back surface. As with interference fringes, this method does not determine the absolute thickness, but the relative thickness can be determined to sub-$\mu$m accuracy [8]. Figure 4.8 depicts the relative slope, absolute slope and thickness variation for a fine horizontal scan with a $50 \times 50\,\mathrm{cm}^2$ prototype radiator by Nikon. An important observation is that the relative slope is mostly constant and shows no large deviation at both ends of the radiator. The slope of the evaluated prototype radiator can be approximated by a first order polynomial and the mean deviation from this form could be measured to be only a few $\mu$rad as shown in Figure 4.9). In Figure 4.10 the thickness profile of a two-dimensional scan is shown which reveals a concave shape.

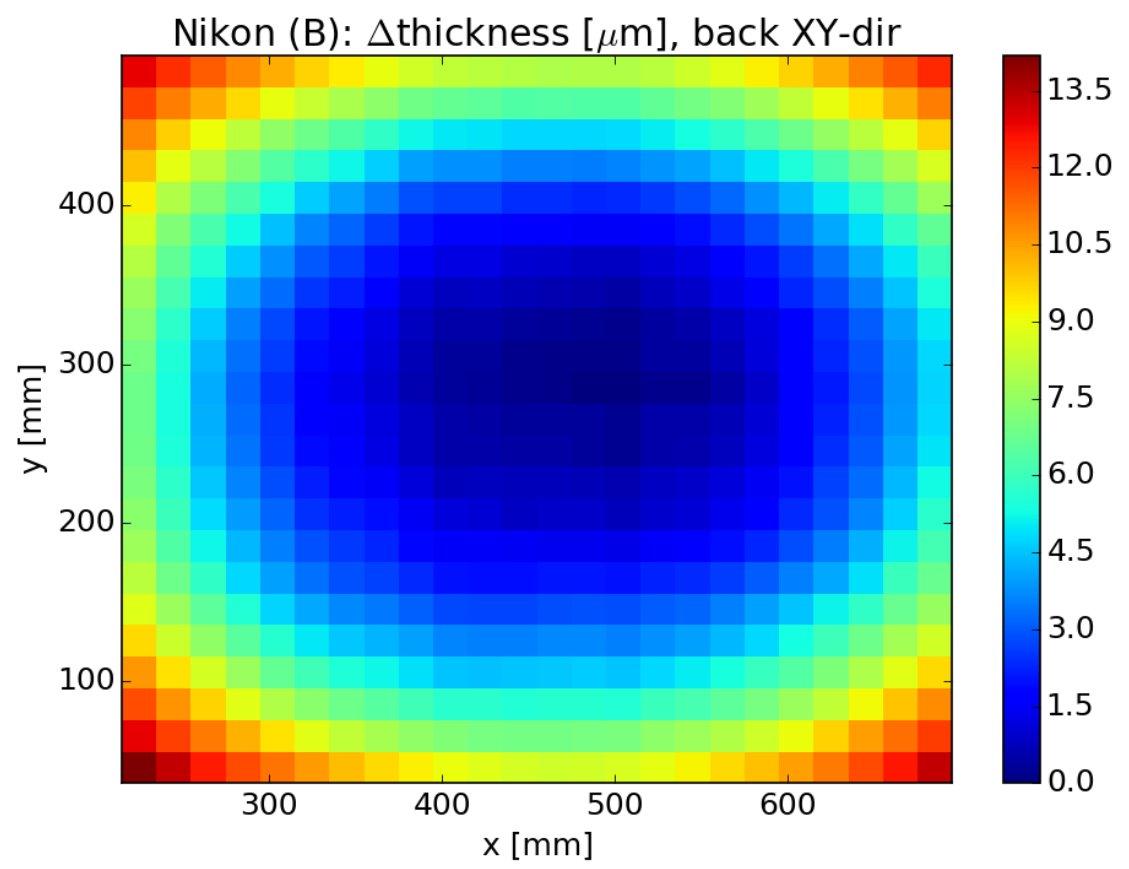

**Figure 4.10:** Thickness variation in $\mu$m for the $50 \times 50\,\mathrm{cm}^2$ Nikon prototype radiator obtained by numerical integration [8].

Measuring the relative radiator thickness is not only important to qualify delivered components. Knowing the exact shape opens the possibility to estimate the systematic errors and even to correct for them in the reconstruction software where the original Cherenkov angle is calculated.

## 4.1.2 Imaging Optics

The proposed Disc DIRC design is supposed to work with a limited number of photons to be compatible with the aging properties of current MCP-PMTs. This is achieved by limiting the detected wavelength range. The lower number of photons is compensated by the reduced dispersion errors in the lim-ited spectral range and the precise determination of the photon angles becomes of prime importance. The imaging optics developed for the EDD have been optimized to provide the maximum resolution while staying compatible with the geometrical con-straints and the alignment of the MCP-PMT axis to the magnetic field [1]. An example for the expected sensor field along the tube is shown in Figure 4.14. As a positive side effect, the new optical system does not make use of aspheric or acylindric shapes, which are complicated to produce as they require non-standard equipment for both production and metrology.

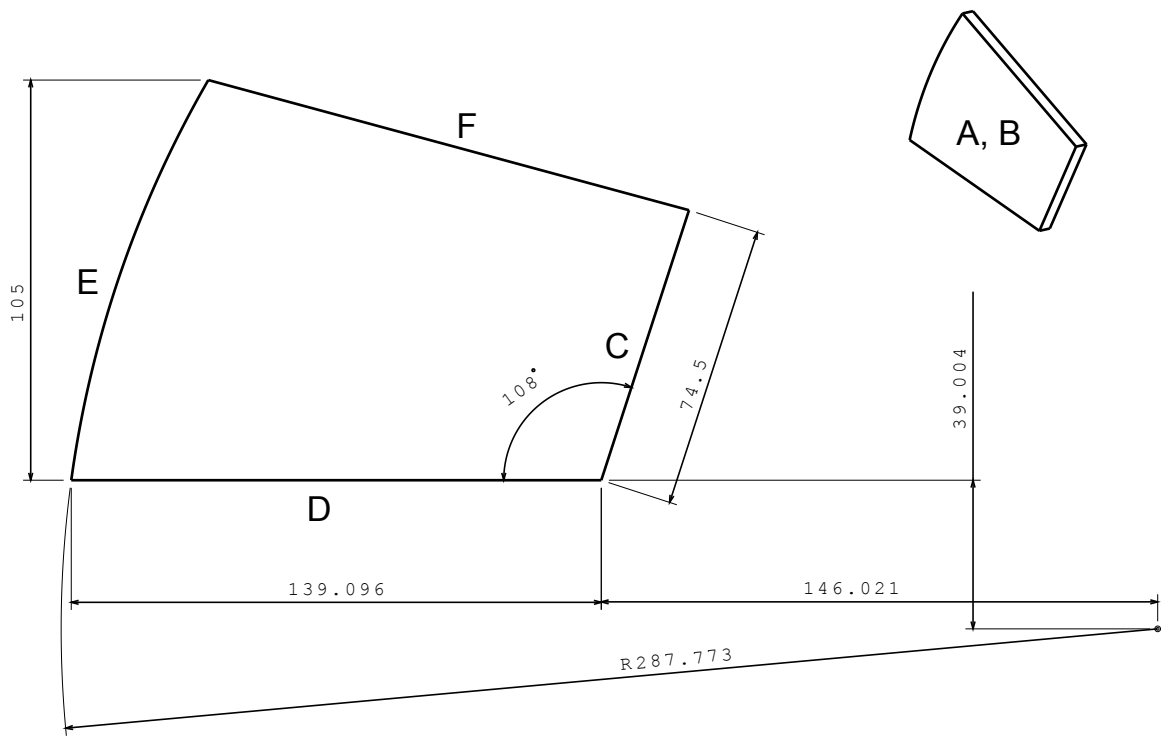

**Figure 4.11:** Technical drawing of a focusing light guide.

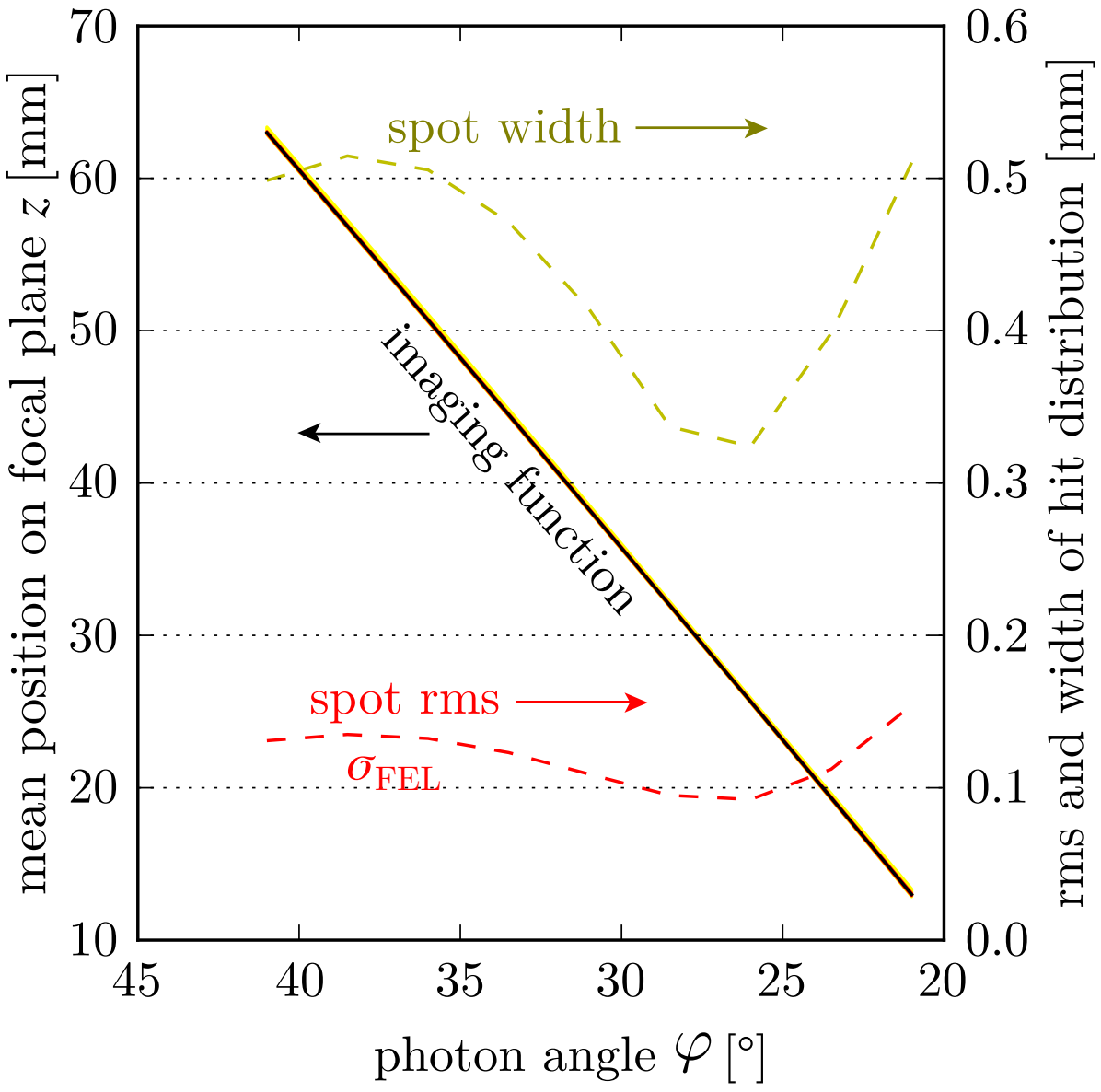

**Figure 4.12:** Imaging performance (2D) of the focussing optics [13]. The photon angle $\phi$ is almost a linear function of $z$.

**Table 4.2:** Specifications for the FEL surfaces.

| category | side | specification |
|---|---|---|
| form error | E | $\lambda/2$ PV |
| | A-D | $2\lambda$ PV |
| roughness | A-E | 2 nm |
| coating | A | Al-coating |
| chamfers | A-F | < 0.2 mm |

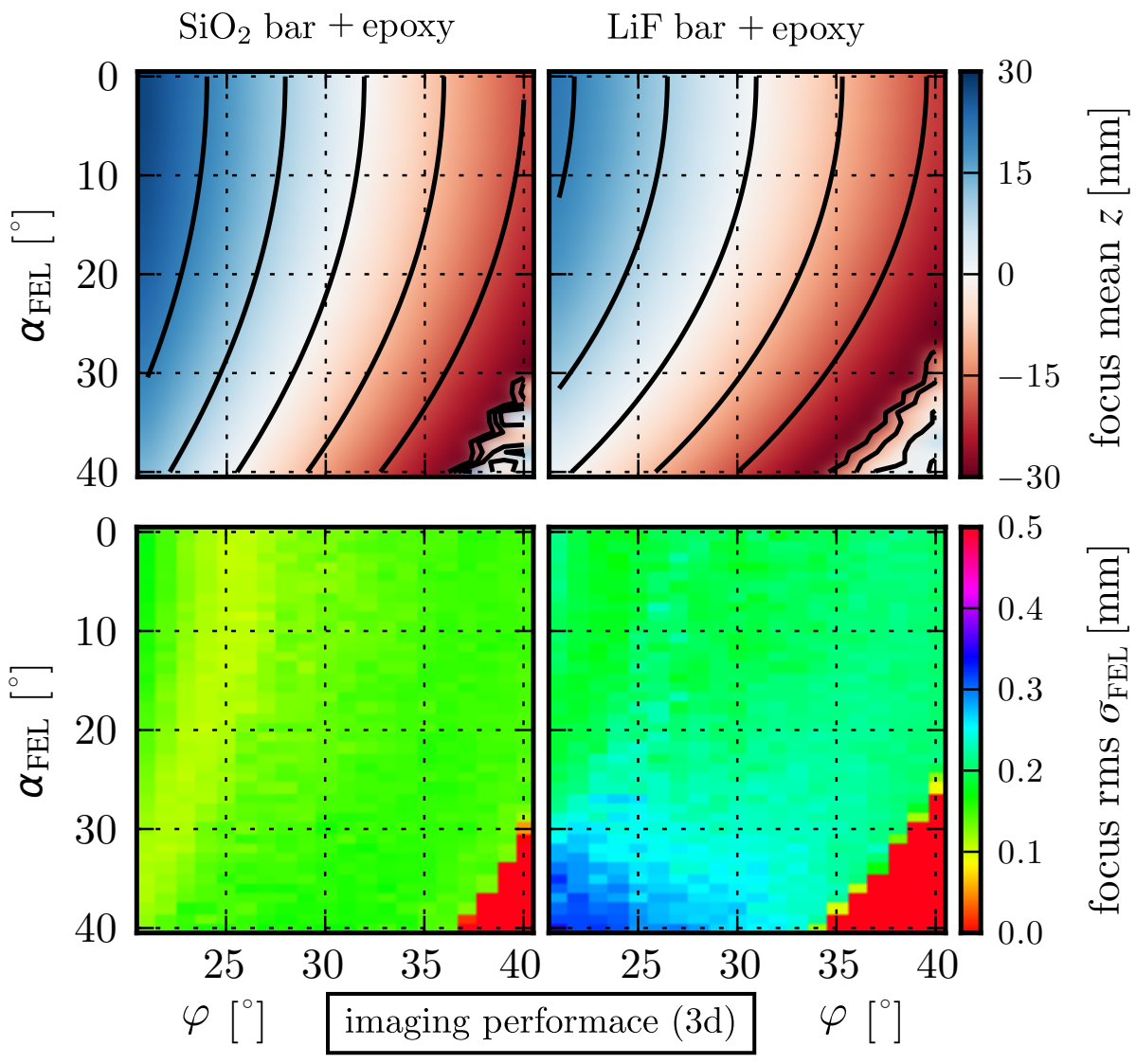

**Figure 4.13:** The 3D FEL imaging performance varies differently for the LiF bar option with $\alpha_{FEL}$ [13].

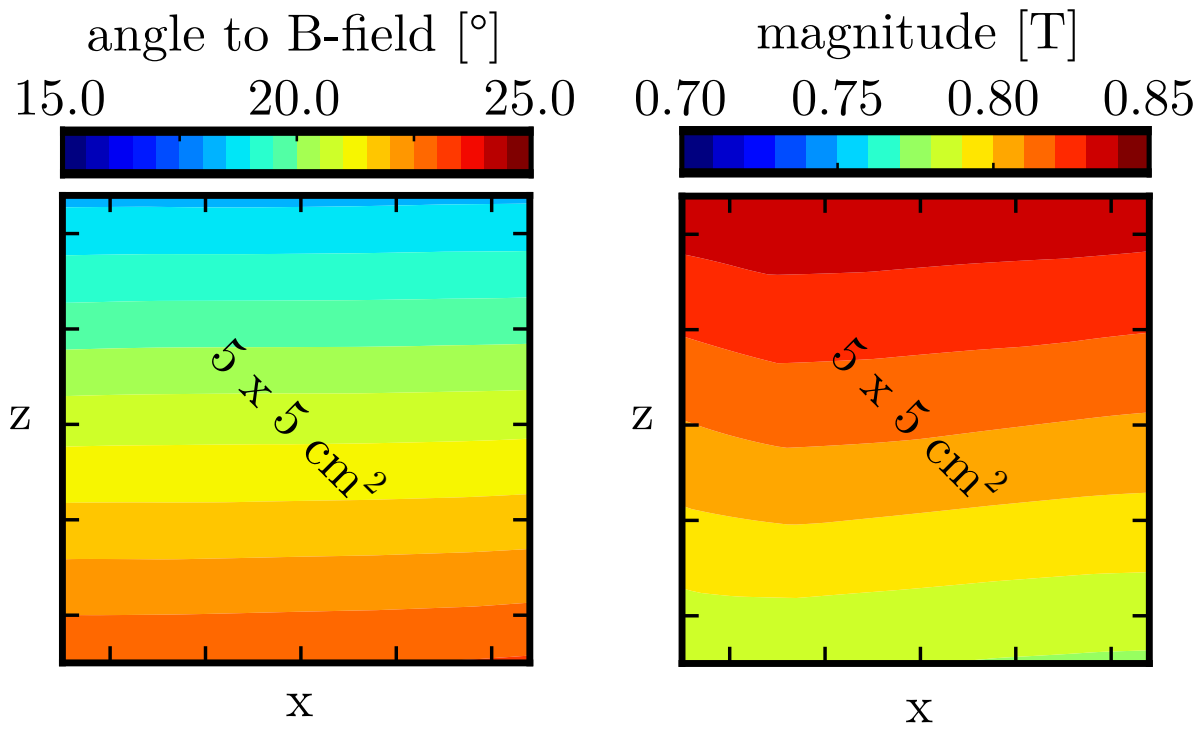

**Figure 4.14:** Magnetic field at one sensor (exemplarily, as fields vary with position) [13].

The complicated polynomial shapes of the optics proposed for the SiPM and FDD designs have been replaced by a simple cylindrical mirror. This goal can only be reached by accepting a trade-off between acceptance and imaging quality. The corresponding specifications for a FEL can be found in Table 4.2 and Figure 4.11. The imaging performance is presented in Figure 4.12 (2D ray-tracing) and in the left panel of Figure 4.13 (3D ray-tracing).

A drawback of this solution is that the focusing ele-ment cannot be bonded directly to the radiator. As shown in Figure 3.7, additional quartz bars have to be used as a connection. The reason for this is the constrained size of the radiator combined with the need of alignment of the tubes with the magnetic field. However, the combination of a focusing element and a bar does not lead to increased cost if compared to an aspherical layout.

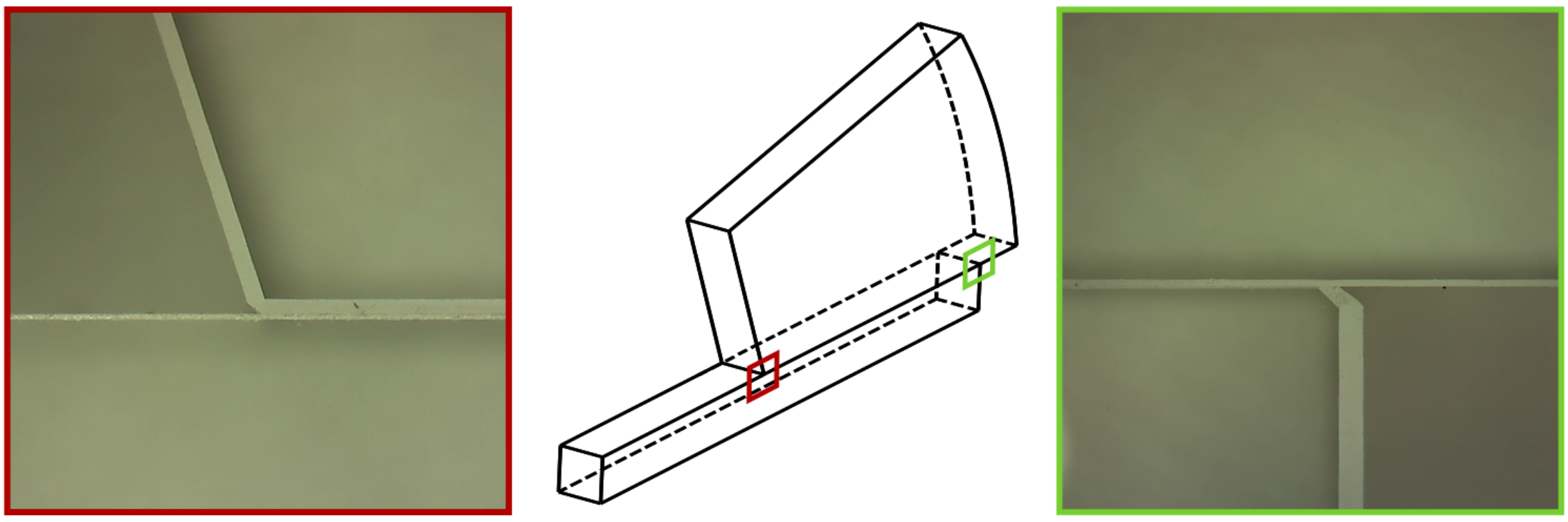

**Figure 4.15:** FEL and bar contacted via optical contact bonding. The pictures correspon to the front and rear part of the connection as indicated in the drawing.

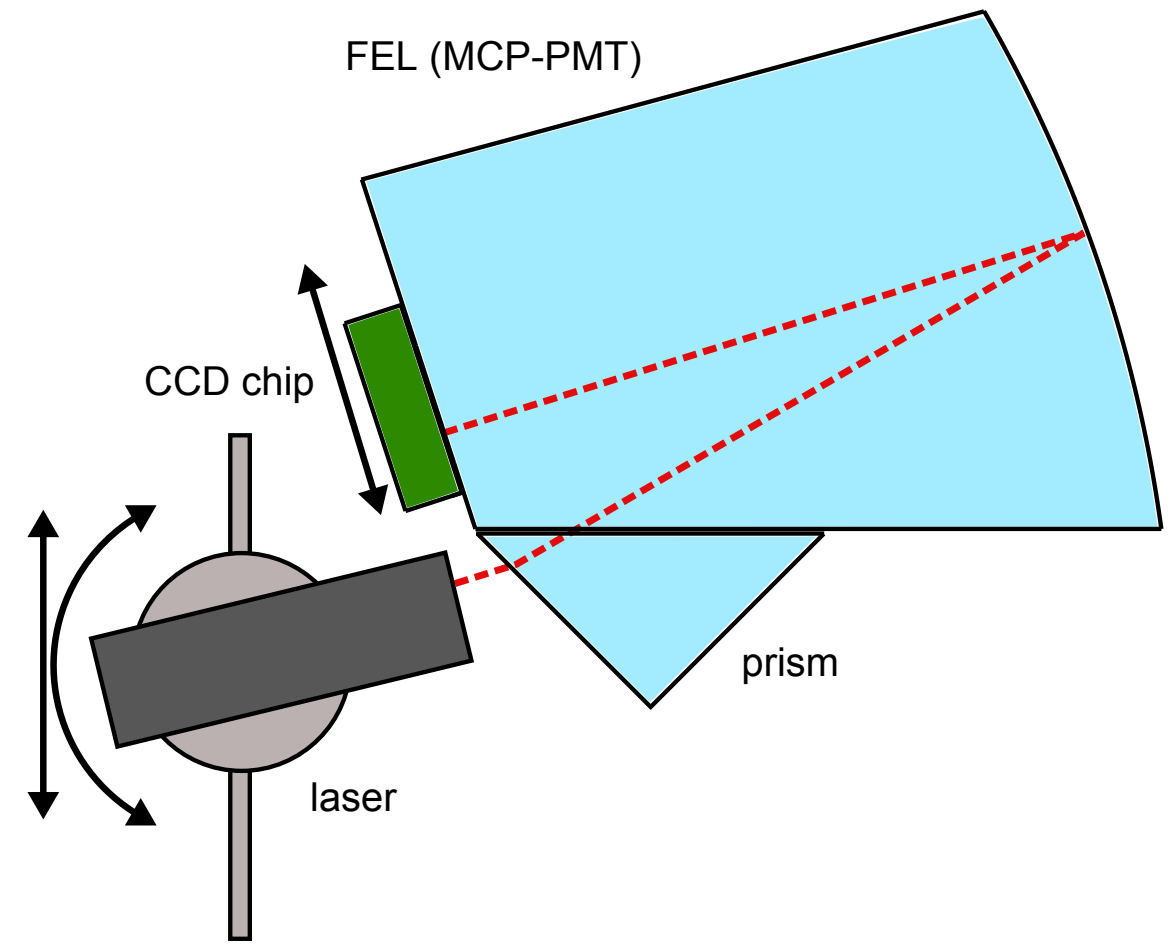

**Figure 4.16:** Schematic drawing of the quality measurement for the FEL [8]. The position of the CCD-chip can be varied to cover the full image width of the FEL.

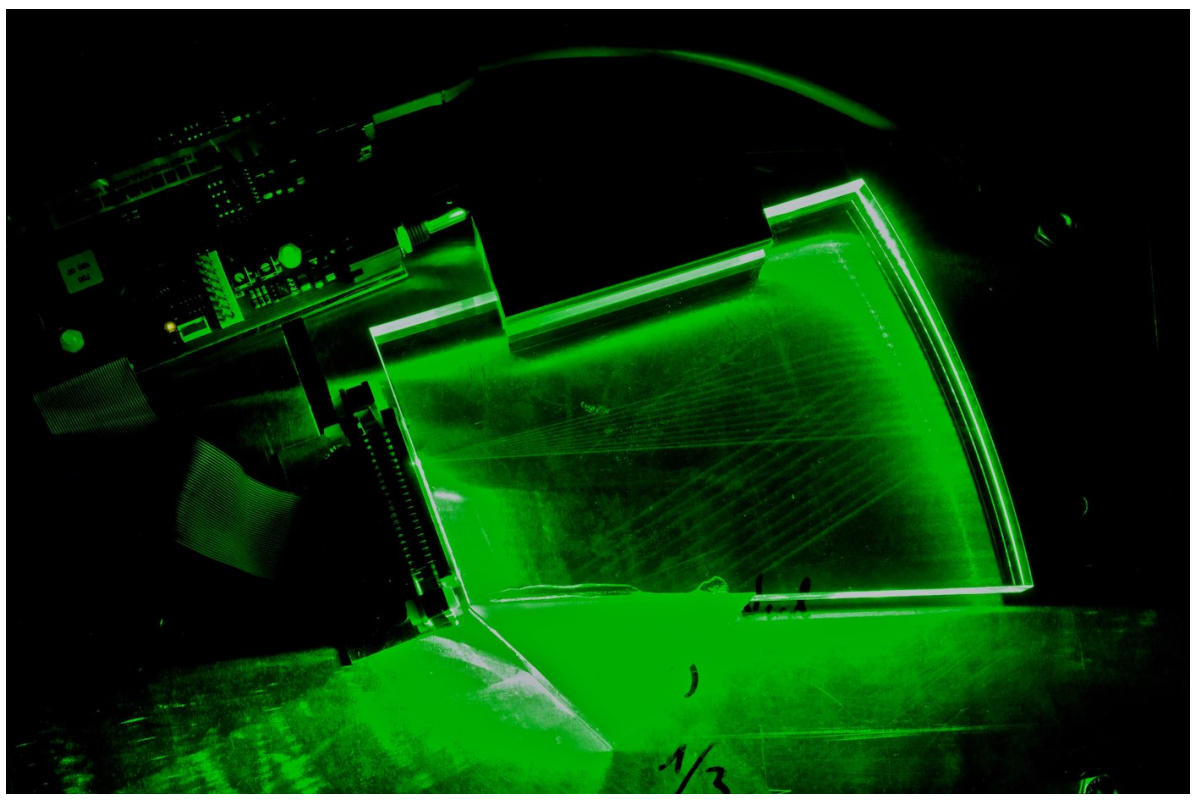

**Figure 4.17:** A picture of the FEL focusing taken with long time exposure for different laser positions [8].

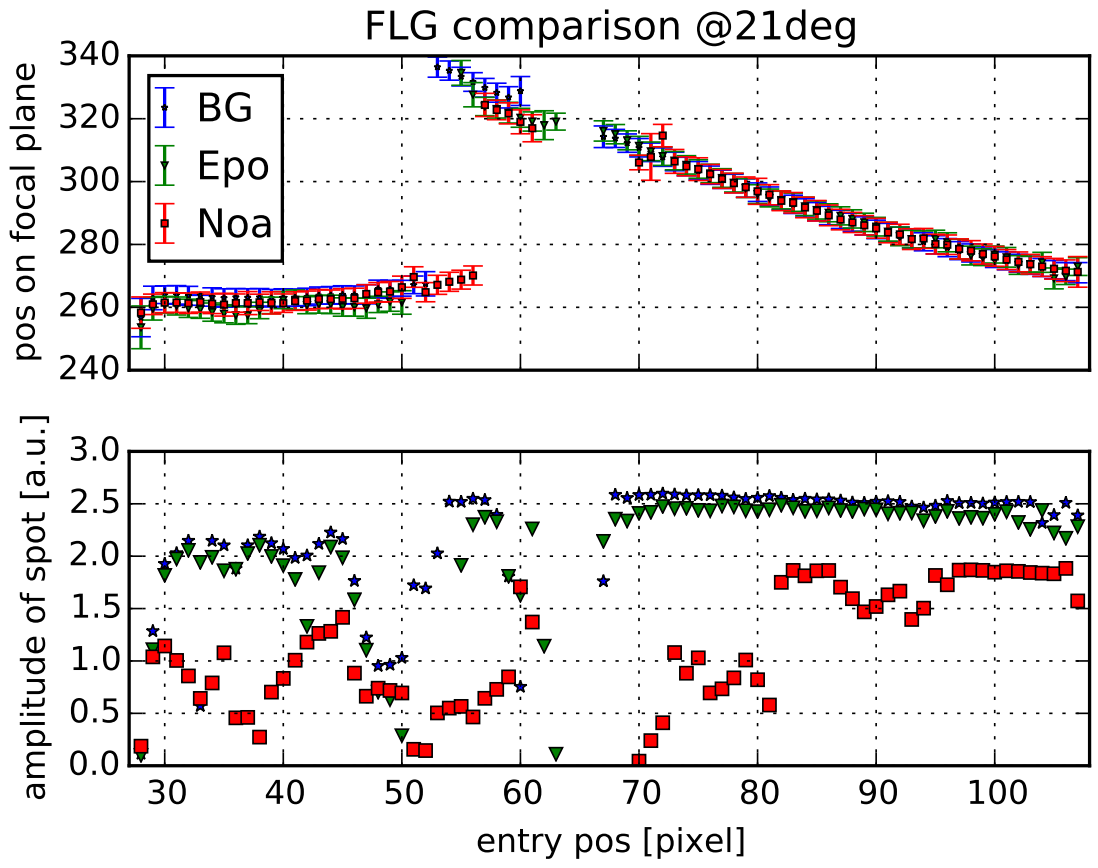

**Figure 4.18:** Scan of three FELs connected to bars at $\phi = 21°$ [8]. S2 was non-adhesively contacted (blue stars), S6 (green triangles) and S8 (red squares) were glued using APM Epicol 348990 [15] and NOA-61 [16]. The glue joint for S8 turned out to be deficient after it developed a large net of bubbles which leads to a crucial absorption especially for smaller angles and low entry positions. Bad data points have been removed from the plot. One pixel corresponds to $9\mu$m.

The position on the sensor is a function of the internal reflection angle $\varphi$ and the angle $\alpha_{\mathrm{FEL}}$ between the photon propagation plane and the normal of the focusing element (Figure 3.7). Therefore the 2D-imaging performance can be misleading, as it ignores important distortions. The 3D imaging function $z(\varphi, \alpha_{\mathrm{FEL}})$ is shown in Figure 4.13, with the z-position encoded as color. For larger $\alpha_{\text{FEL}}$ and a given $z$, an error $\delta\alpha_{\text{FEL}}$ translates to an error in $\varphi(z, \alpha_{\text{FEL}})$. In the experiment, $\alpha_{\text{FEL}}$ is roughly determined by the location of the emitting particle track and the rim position where the photon entered the bar. As the tracking errors are considerably smaller, $\delta\alpha$ is primarily defined by the width of the focusing element and its distance to the track. Tracks with larger polar angles are closer to the rim and $\delta\alpha$ increases. The influence on the detector performance can be seen in Figure 3.9 on page 15. The chosen number of 3 FELs per ROM is a compromise between imaging performance and production cost. A design with a single focusing element per MCP-PMT cannot be used.

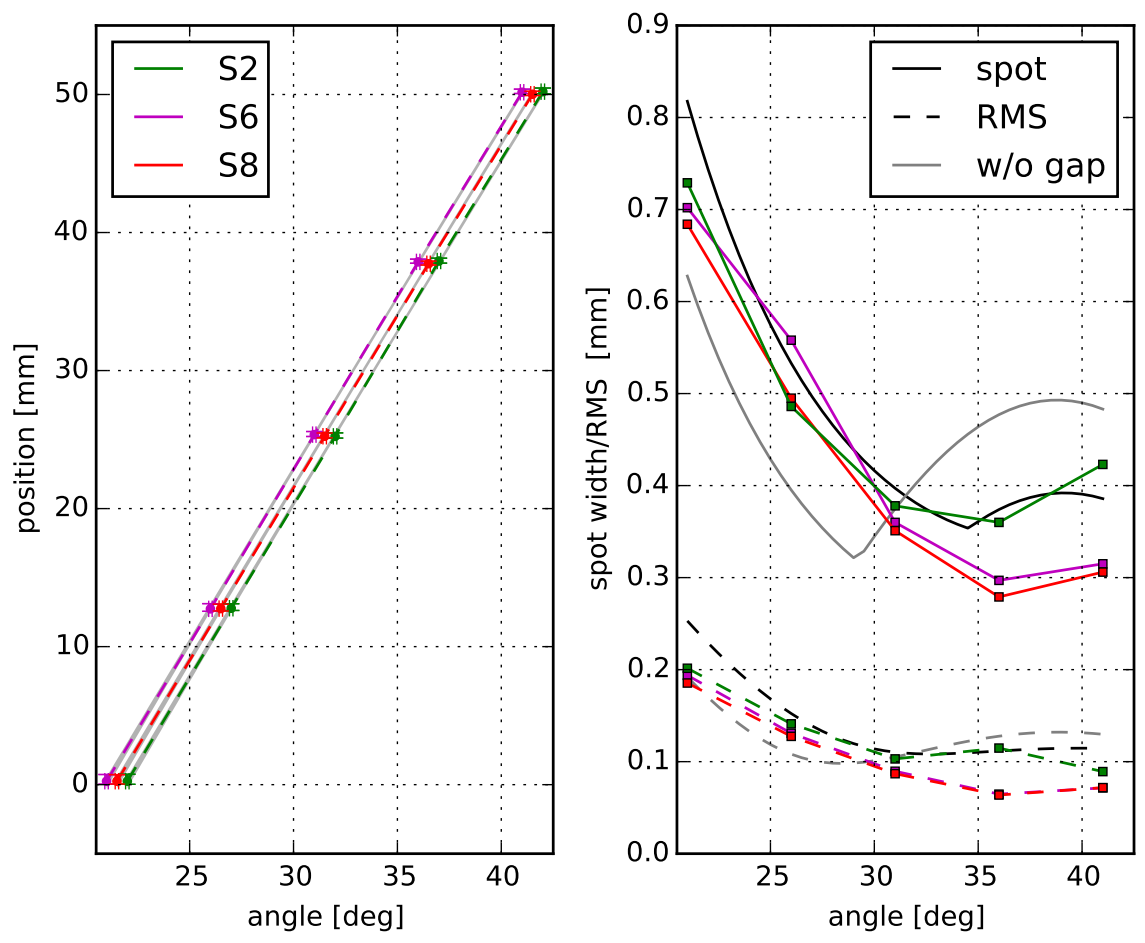

**Figure 4.19:** Left: Measured imaging function for all MCP-PMT FELs (S2 and S8 have been shifted to the right by 0.5° and 1° for a better display). Right: Spot width and RMS for the FELs [8]. The black lines represent data obtained by simulation. The gray lines show the original spot widths and RMS values with a directly coupled 2 mm entry window as shown in Figure 4.12.

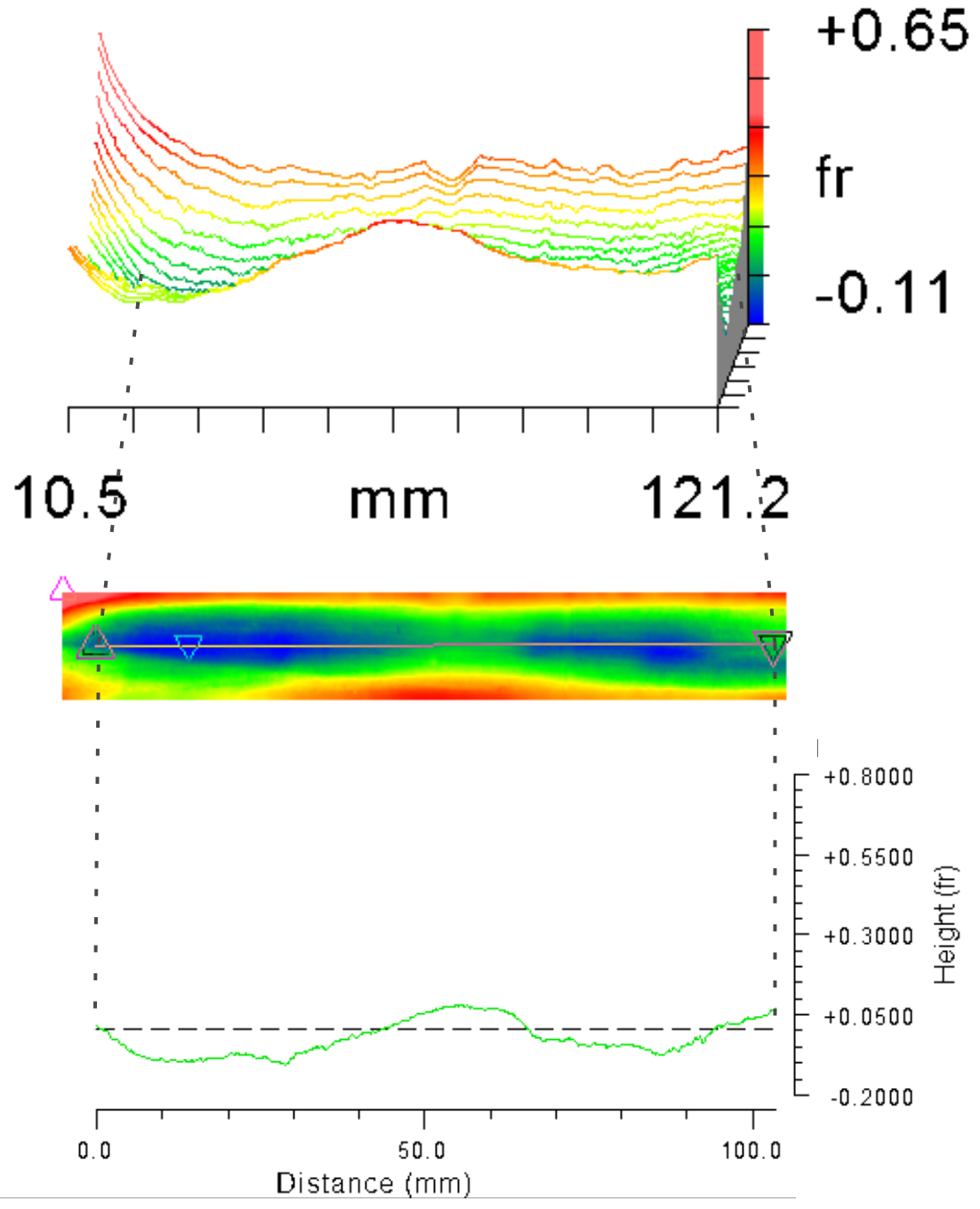

**Figure 4.20:** Measurement of the surface error of the curved side of the FEL [17]. It has a radius of 287.77 mm and form error of $\lambda/2$ PV is required.

The idea to build an apochromat based on a LiF bar, as proposed by Morosov [18] and later Föhl [19], has also been investigated. In this case the bar is realized as LiF crystal. The corresponding cylindrical FEL is of similar size and shape, while providing a slightly better alignment of the sensor to the field. The RMS spot size is roughly 40% larger than for the $SiO_2$ version but the consequent error will be outweighed by the reduced chromatic aberration. Practically, there are several arguments against such a design. First of all, LiF is a single-crystal and therefore expensive. In addition its coefficient of thermal expansion (CTE) is 67 times higher than the CTE of fused silica which makes glue joints difficult. The different index of refraction causes reflections at the bar-FEL interface which reduce the imaging quality (Figure 4.13, right). Also the image blur due to $\delta\alpha_{\text{FEL}}$ is larger due to refraction. These complications lead to the decision, that the LiF based dispersion correction is not a safe and prefered option.

In order to evaluate and qualify first three FEL prototypes a test setup was developed which is schematically shown in Figure 4.16. A laser is coupled into the FEL via a prism and the resulting spot is measured using a CCD-chip for different angles of incidence (see Figure 4.17). One FEL prototype was coupled to the bar via optical contact bonding. This technique demands ultra-fine polished surfaces but at the same time provides the perfect optical transmission properties. Figure 4.15 shows the coupling which was done by Berliner Glas [17]. The two remaining prototypes were glued to the bars after a first inspection with an in-house setup using APM Epicol 348990 [15] which is a two-component epoxy and NOA-61 [16]. In case of NOA-61 an initially small inclusion led to a large defect structure which severely degraded the transmission between the bar and the FEL. The defect was probably pandered by insufficient UV-curing. Figure 4.18 shows a measurement at $\phi = 21°$ for all three FEL-bar pairs. The focusing performance is similar for all three samples. However, optical contact bonding shows the best result regarding the image quality as well as the relative transmission. The transmission of the NOA-61 sample is clearly worse due to the described defect structure that shows up as bubbles in the glue joint.

Comparing the results for different angles of incidence to the simulation a good agreement for the imaging function as well as the maximum spot width of the distribution and its RMS value can be obtained as shown in Figure 4.19. Because the produced FELs were optimized to work with an MCP-PMT entry window thickness of 2 mm the simulation was adapted to the prerequisites of the test setup with a slightly shifted geometry.

### 4.1.3 Optical Joints

In addition to the optical components made of synthetic fused silica different optical joints have to be taken into account to correctly estimate the transmission of the Cherenkov photons. Besides the internal transmission a matching of the refractive index is advantageous to reduce Fresnel losses at the boundaries. Although the joints will be shielded by the EMC a radiation dose of roughly 10 Gy is expected that might affect the glue.

The presented design requires a glue joint between the bars and the radiator. Typical candidates for this purpose are Epotek 301-2 [20] which is a two-component epoxy or NOA-61 [16] which cures by being exposed to UV light. In connection with earlier DIRC developments (BaBar and Belle II) both options have been evaluated. As Epotek 301-2 provides the strongest bonding and at the same time high transmission and radiation hardness this option is favored.

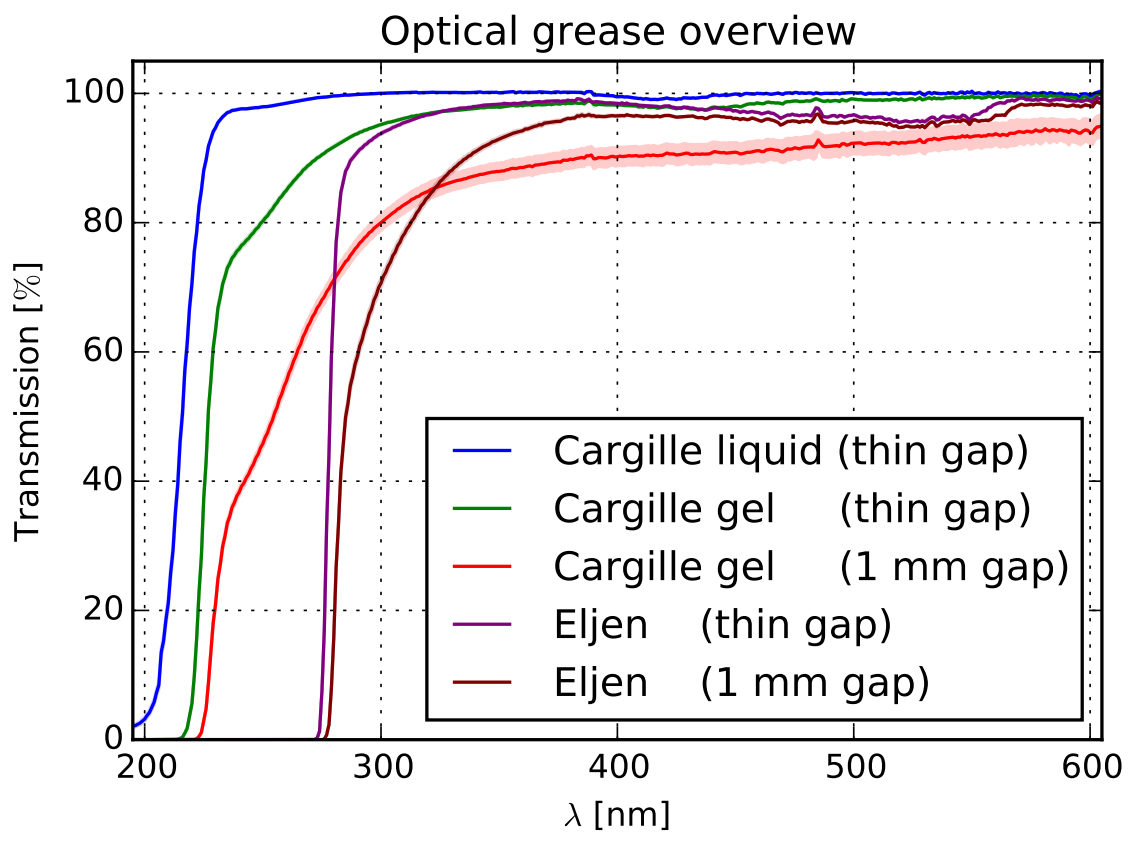

**Figure 4.21:** Overview of three different optical grease options with different thicknesses [8].

In case of the intersection between the bars and the FEL optical contact bonding is foreseen (see also section 4.1.2 and Figure 4.15 on page 22). This option enables a high transmission even for steep angles which otherwise could be lost by additional reflections.

In case of the optical joint between the FEL and the MCP-PMT entry window different options are available. Because the sensors have to be exchangeable a removable solution is required. For prototype tests different optical greases have been used (BC-630 by Saint Gobain [21], EJ-550 by Eljen Technology [22], Fused Silica Matching Liquid Code 06350 and Optical Gel Code 0607 by Cargille [23]. Figure 4.21 gives an overview for different greases and thicknesses to be used in MC simulations. However, this option is not recommended for an experiment which runs several years as the grease might dissolve or float out of the contact region.

The GlueX FDIRC prototype used an optical cookie from Shin-Etsu with the name RTV SES 403, in order to create a removable glue joint between FBLOCK and Wedge of the bar box. The chosen RTV gap was 1 mm. According to the requirements, it was possible to cut the glue with a thin razor in case of an unplanned removal necessity. However, the photon detectors were coupled to quartz with an air gap.

## 4.2 Photon Sensors

### 4.2.1 Requirements

The reconstruction of the Cherenkov image requires the reconstruction of two spatial coordinates. Additional measurements can be used to over-constrain the system and thus improve the detector performance and suppress background. Designs using two spatial coordinates plus a timing measurement in their reconstruction are known as 3D DIRC systems. The Endcap Disc DIRC (EDD) relies mainly on the reconstruction of two spatial coordinates. A reasonably precise time information will be employed in aid of the reconstruction of the Cherenkov pattern and for background suppression.

The envisaged detector system is subject to severe constructional constraints requiring the readout to be placed inside the solenoid of the $\overline{\text{P}}$ANDA target spectrometer (TS). The expected field in this region ranges up to 1.5 T. As magnetic shielding would decrease the geometrical coverage and thus hamper the performance of the detector, a highly efficient photon detection system insensitive to mag-netic fields of that order of magnitude has to be identified and constructed.

Since the image planes will be rather compact and because the antiproton-proton average annihilation rate inside $\overline{\text{P}}$ANDA will be $2 \cdot 10^7$, the single photon density at the sensor surface can reach 10 MHz/tube. This requires a very high rate capability and a long lifetime of the counters in terms of integrated anode charge. By measuring the time-of-propagation of the Cherenkov photons from their creation point to the sensor surface the $\pi/K$-separation can be optimized. Especially direct photons which go without additional reflections at the side surfaces of the radiator can be distinguished from those who do. However, to fully exploit this information a time resolution better than 100 ps ($\sigma$) is necessary. Finally, each particle traversing the radiators produces only a limited number of Cherenkov photons. As a consequence, the photon sensors have to be of very high quality in terms of quantum, collection and geometrical efficiency and require a low dark count rate.

The identification of the particle type relies on the reconstruction of the Cherenkov cone. This requires a position sensitive readout plane with a high ratio of active to passive surface. The best available options are improved multi-anode micro-channel plate PMTs. The design of the EDD foresees MCP-PMTs using a micro-channel plate of 10 $\mu$m pore diameter and a segmentation of at least $3 \times 100$ pixels.

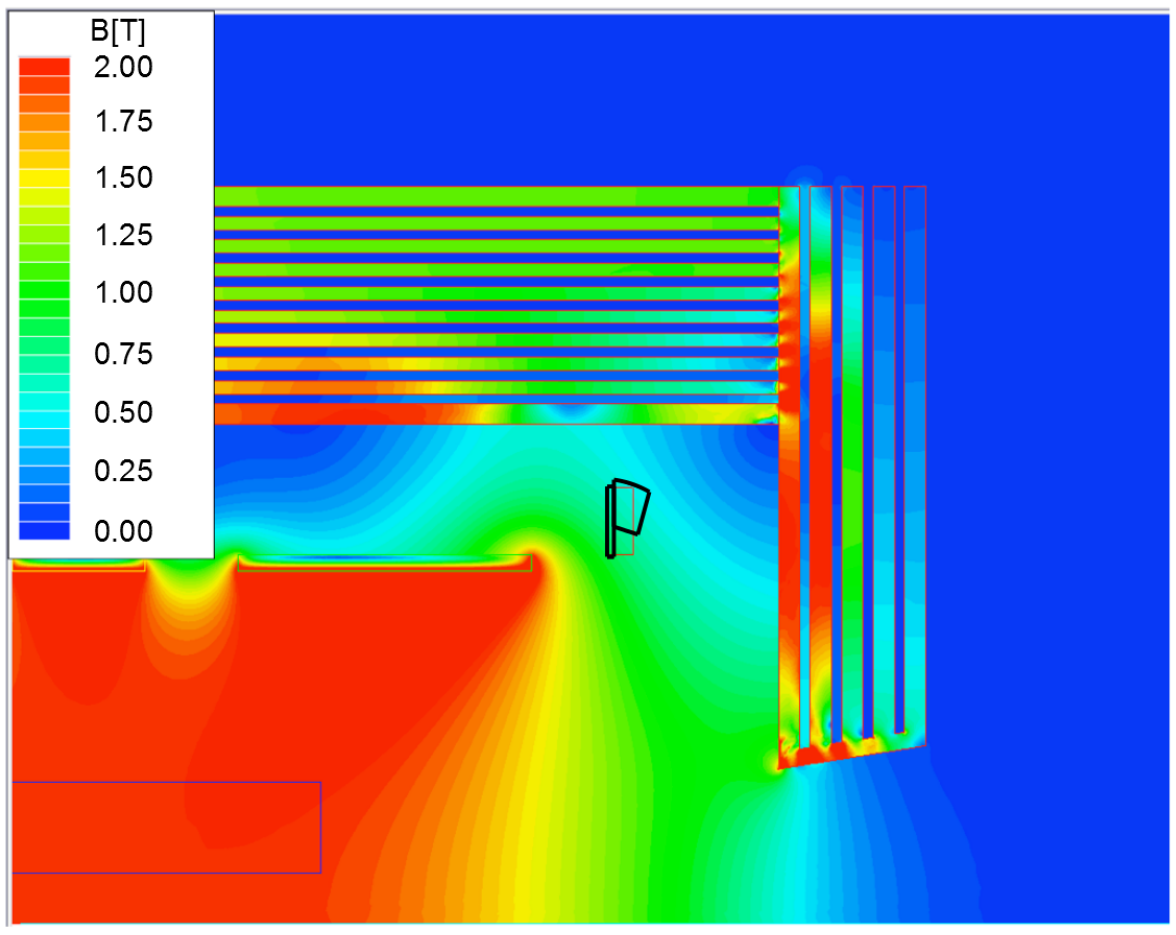

**Figure 4.22:** Magnetic field maps showing the strength of the solenoidal field in the EDD readout region

### Magnetic Field

The compact design of the $\overline{\text{P}}$ANDA target spectrometer requires the photon detection system and initial digitization stages to be located inside the return yoke of the target solenoid. The photon detection system of the EDD is exposed to a magnetic field between 0.70 and 0.85 T (see Figure 4.22). The available construction space allows a moderate optimization of the PMT orientation relative to the magnetic field lines. The compact design and high geometrical filling factor necessary do not allow the installation of magnetic shielding. A suitable photon detection system should therefore work inside a magnetic field of up to 1 T (allowing for a safety margin of 20% in the prediction of the magnetic flux).

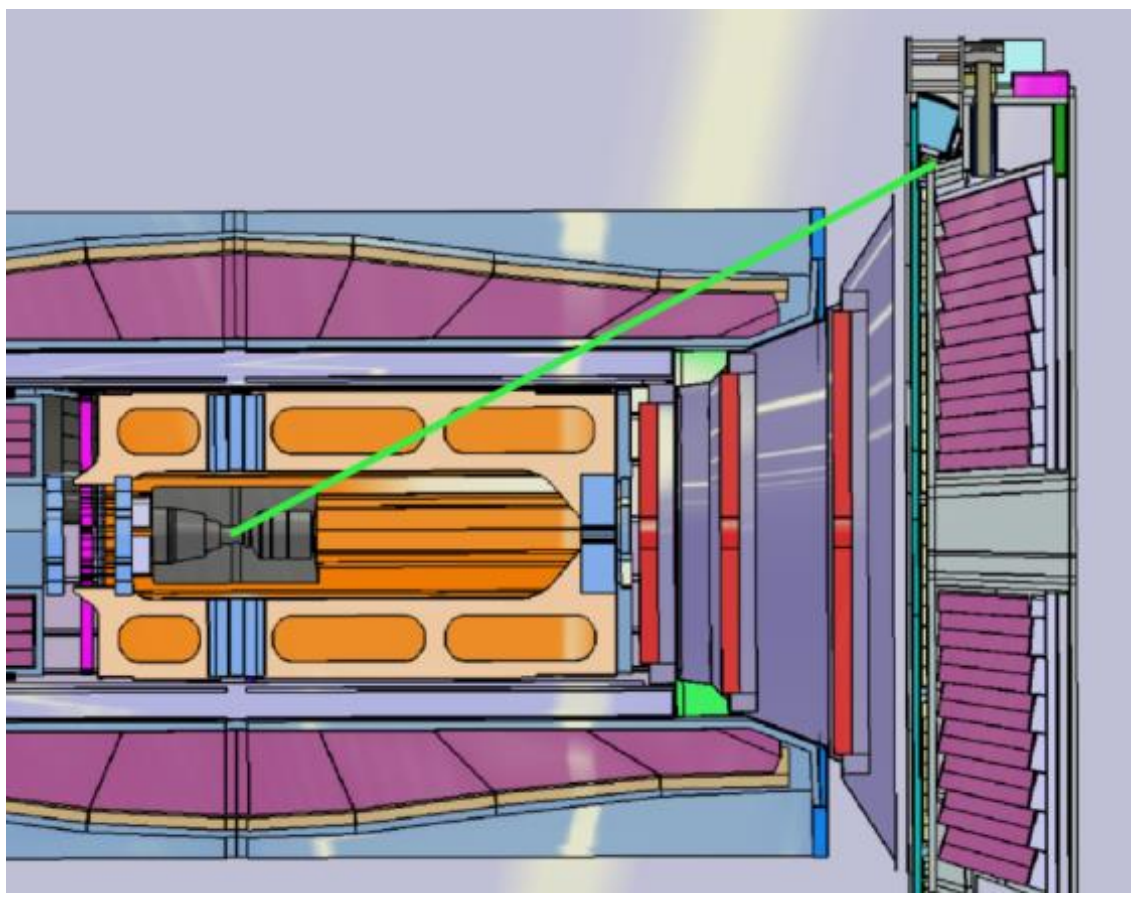

**Figure 4.23:** The sensors and the readout of the EDD are in the shadow of the EMC as shown by the green line.

### Radiation Hardness

The $\overline{\text{P}}$ANDA experiment with hadronic interactions at high luminosity will produce a large integrated radiation dose that the detectors have to withstand. It consists mainly of neutrons and some electromagnetic background. An integrated radiation dose corresponding to $10^{11}$ neutrons/cm$^2$ is expected in this area over the lifetime of $\overline{\text{P}}$ANDA. This dose is not an issue for PMTs, but digital readout might be affected. As shown in Figure 4.23 the readout region of the EDD is in the shadow of the EMC and therefore less affected than e.g., parts in the MVD.

### Area and Pixel Size

In the design of the focusing elements (FELs) a focal plane area of $50 \times 50\,\text{mm}^2$ was assumed to match the photon detection candidates. To achieve the required PID power, the photon detection system of the $\overline{\text{P}}$ANDA EDD requires approximately 100 strips of spatial resolution. For a focal plane of about 5 cm, this yields a strip pitch of 0.5 mm. A reduction of the strip size will not enhance the performance much further due to the limitations of the 3D optical design (see also Figure 3.10 on page 15).

### Time Resolution

The design of the EDD requires time resolutions on the order of 100 ps. It is not suggested to use the photon time of propagation to reconstruct the Cherenkov angle or to mitigate dispersive effects. Time-resolution will however help in the event correlation and more important in the image reconstruction and background suppression. Furthermore, a time-walk correction and charge calibration are not necessary, since the obtained time resolution is sufficient to fulfill the requirements.

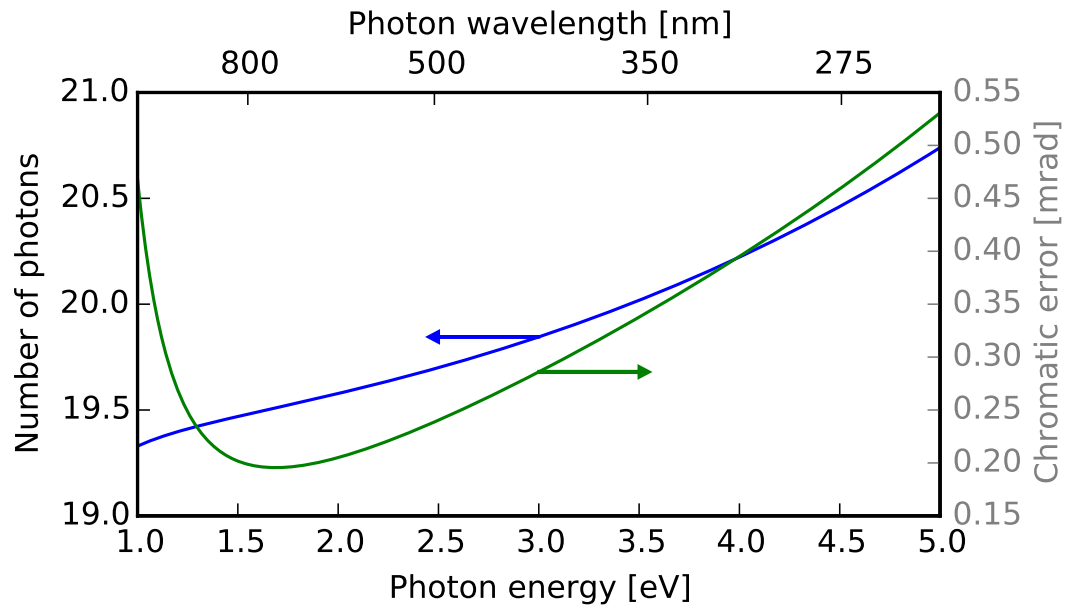

**Figure 4.24:** The larger the wavelength range is, the higher is the number of detected photons, but for a fixed photon energy interval, the statistics is almost independent of the wavelength [24]. The black line shows the number of detected photons (left scale) for an arbitrary detection efficiency of 10% and a band-pass filter of 0.5 eV width. The gray line shows the resulting chromatic error of the reconstructed Cherenkov angle (right scale).

### Spectral Range

Cherenkov photons are produced in a "blue" spectrum as a function of the wavelength. The probability of emission decreases with the square of the wavelength, but the small wavelengths have the disadvantage of large dispersion effects. Studies have shown that for an application like the EDD an optimum for a fixed energy interval of Cherenkov photons can be found at red visible spectrum (see also Figure 4.24).

The radiator material is transparent for visible and ultraviolet light and is not restricting the spectral sensitivity of candidate photon detection systems. The spectral range will be restricted to $\lambda > 300$ nm by optical glue and by an optical filter or a matching photocathode to increase the MCP-PMT lifetime and reduce the chromatic error of the image.

### Rate

The average interaction rates at $\overline{\text{P}}$ANDA are estimated to be about 10 MHz at full luminosity. During short timescales bursts of interactions can drive the rate up to 26 MHz. Each proton–antiproton interaction averages about three charged tracks that create Cherenkov light in the radiator of the EDD. Each charged track in the radiator generates about a few hundred photons. About 18 of those that are focused and registered in the MCP-PMTs. Assuming that these are homogeneously distributed over all available channels one can derive the hit rates for the individual readout pixels as follows:

$$= \frac{2.6\times10^{7}\frac{\text{events}}{\text{s}}\times 3\frac{\text{tracks}}{\text{events}}\times 22\frac{\text{photons}}{\text{tracks}}}{96\text{ modules}\times 300\frac{\text{channels}}{\text{module}}} \quad (4.2)$$

$$= 60\text{ kHz} \quad (4.3)$$

By adding a safety margin of 25% because of possible charge sharing and cross-talk between pixels, a total count rate of about 75 kHz/channel (90 kHz/channel in case of a photocathode without a filter which is sensitive in the green spectrum and registers about 26 Cherenkov photons per track on average) is achieved, keeping in mind that this value corresponds to extreme conditions where short bursts temporarily create very high photon rates. Therefore, the maximum rate capability of 600 kHz, as promised by PETSYS Electronics, will fully satisfy our requirements.

Furthermore, the high luminosity runs correspond to an average event frequency of 1/(5 ns) which compares to the 40 ns deadtime of each channel that is defined by the front end electronics. The probability to get more than 2 photons in the same readout channel is roughly given by

$$p(2\gamma) \approx \frac{22}{7200} \approx 0.3\% \quad (4.4)$$

if 7200 channels are taken into account.

### Lifetime

Simulations show that the expected photon rates at the readout plane of the EDD will be up to 2 MHz/cm$^2$ for the highest luminosity. However, as far as lifetime calculations are concerned, it is important to average the interaction rates over the adequate lapse of time, namely one machine cycle of the HESR (compare chapter 2). Assuming a gain of $10^6$ for the MCP-PMTs and 50% duty cycle for $\overline{\mathsf{P}}$ANDA simulations predict that the integrated anode charge over 10 years can accumulate to more than 7 C/cm$^2$ (8 C/cm$^2$ in case of a "green" photocathode) for the Endcap Disc DIRC.

### 4.2.2 Photon Sensor Options

Summarizing the requirements outlined above, the EDD needs a pixelated photon detection system capable to work in a magnetic field up to 1 T and with a rate of 75 kHz per channel. Possible solutions were investigated to find the best option for the EDD:

- Silicon Photomultiplier arrays (SiPMs)
- Multianode dynode PMTs (MAPMTs)
- Microchannel-plate PMTs (MCP-PMTs)

MAPMTs had to be ruled out in the early stages of the detector development mainly due to their sensitivity with respect to a magnetic field which leads to a rapid gain drop even in small to moderate magnetic fields.

SiPMs are resistant to magnetic fields and capable of low light detection why an EDD design based so-called digital Silicon Photomultipliers (dSiPMs) had been developed. However, irradiation with ionizing particles lead to strong increase in the dark count rate which would make proper event reconstruction impossible inside the $\overline{\mathsf{P}}$ANDA experiment.

The last option, MCP-PMTs, are the ideal sensors for applications where a low noise and sub 100 ps single photon detection is required inside a high magnetic field if properly aligned. They are available as multi-anode devices and provide a good active area ratio while still being rather compact in size. However, until recently the major drawback of MCP-PMTs have been serious aging issues. Ions in the residual gas produced by the electron avalanche are accelerated towards the photo cathode (PC) which gets damaged from this permanent bombardment. As a consequence the quantum efficiency (QE) drops while the integrated anode charge increases. Until recently the rate conditions in $\overline{\mathsf{P}}$ANDA were far beyond the reach of any commercially available MCP-PMT where the QE had dropped by more than half after typically <200 mC/cm$^2$ [25, 26], while for the EDD at least 7 C/cm$^2$ are needed to allow a detection of enough photons for the reconstruction of the Cherenkov angle.

### 4.2.3 Evaluation of MCP-PMTs

This section gives an overview of the various tests for MCP-PMTs which have been performed. The work has been shared by the groups at Friedrich-Alexander-Universität Erlangen-Nürnberg and Justus-Liebig Universität Gießen. The group at Erlangen has gained a lot of experience over the past years and evaluated different characteristics such as lifetime, dark count, gain stability, cross talk and also the behavior of the sensors inside a magnetic field for different sensor types.

The work at Gießen was focused on the evaluation of MCP-PMTs with highly-segmented anodes and their performance inside a magnetic field. This is important because the relatively fine resolution of the anode strips (pitch of 0.5 mm) is a critical issue, as the electron cloud that is emitted from the MCP plates has a certain transverse size. The charge cloud leaving the micro channels is a result of the amplification of the primary photoelectron, and has a finite width when it strikes the readout plane. As a consequence the charges are possibly distributed over various readout anodes. In principle this is not a big problem because by weighting the charges collected in the anodes, the initial origin of the charge cloud can be derived. However, the rate requirements of the readout electronics scale directly with the average number of anodes involved in a meaningful charge weighting process for a single Cherenkov photon. The ideal case for the EDD would be if the charge clouds are small enough to be completely caught by a maximum of two anodes, hence reducing the rate requirements. The operation of the MCPs in a high magnetic field will reduce the lateral diffusion of the electrons due to the small gyro-radius of the electrons in the approximately aligned magnetic field of the order of 1 T. To further reduce the smearing effect, the longitudinal distance of the MCP plates to the electrodes is reduced. It is essential to verify the cross-talk and resolution of the new MCP-PMTs in a strong magnetic field.

Starting early on, many types of MCP-PMTs have been investigated: circular-shaped single anode tubes from the Budker Institute of Nuclear Physics (BINP) in Novosibirsk, various square-shaped 2×2 inch$^2$ 8×8 pixel Planacon MCP-PMTs with different layouts from PHOTONIS, and several of the 1×1 inch$^2$ array R10754X with four strips or 16 pads from Hamamatsu. Very recently, Hamamatsu has developed a larger square-shaped 2×2 inch$^2$ prototype MCP-PMT R13266 with 8×8 and 128×6 pixels, which is currently under investigation for possible usage in the $\overline{\mathsf{P}}$ANDA DIRC detectors. The technical characteristics of some of the investigated sensors are listed in Tab. 4.3.

**Table 4.3:** Characteristics of the investigated lifetime-enhanced MCP-PMTs.

| Manufacturer | BINP | PHOTONIS | | Hamamatsu | |
|---|---|---|---|---|---|
| Type | | XP85112 | R10754X-M16 | R10754X-M16M | R13266-M64 |
| Counter ID | #1359/#3548 | 1223/1332/1393 | JT0117 | KT0001/KT0002 | JS0022 |
| Pore diameter ($\mu$m) | 7 | 10 | 10 | 10 | 10 |
| Number of anodes | 1 | 8×8 | 4×4 | 4×4 | 8×8 |
| Active area (mm$^2$) | $9^2$ $\pi$ | 53×53 | 22×22 | 22×22 | 53×53 |
| Total area (mm$^2$) | $15.5^2$ $\pi$ | 59×59 | 27.5×27.5 | 27.5×27.5 | 61×61 |
| Geom. efficiency (%) | 36 | 81 | 61 | 61 | 75 |
| Comments | better vacuum; e-scrubbing; new PC | better vacuum; 1-/1-/2-layer ALD surfaces | film between $1^{st}$&$2^{nd}$ MCP | ALD surfaces; film between $1^{st}$&$2^{nd}$ MCP | ALD surfaces; film in front of $1^{st}$ MCP |

The sensors were illuminated with a PiLas [27] laser which produces fast light pulses of 14 ps width ($\sigma$) at a wavelength of 372 nm; its maximum repetition rate is 1 MHz. The light is guided through glass fibers, attenuated to the single photon level by neutral density filters and then focused onto the surface of the MCP-PMT with a system of micro lenses, which allows light spots from a few tens of $\mu$m to several cm in diameter. With the smaller spot sizes and an XY-scanner the gain and crosstalk behavior of the multi-pixel MCP-PMTs were investigated as a function of the surface position in steps of about 0.5 mm. For measurements of the rate capability typically a large laser spot was used.

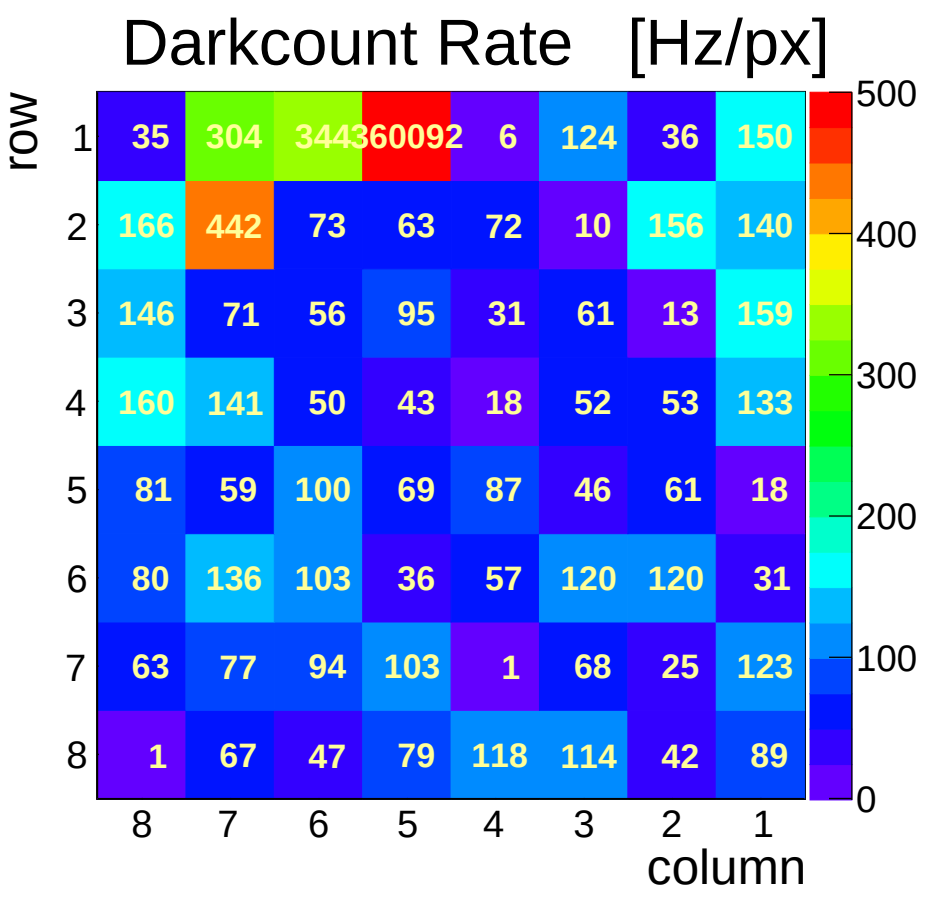

**Figure 4.25:** Dark count map of the new Hamamatsu R13266-07-M64 (JS0035) MCP-PMT. The white numbers reflect the dark count rates per pixel.

Measurements of gain and time resolution as a function of the magnitude and the direction of a magnetic field were performed at a dipole magnet at the Forschungszentrum Jülich in Germany, which delivers a homogeneous field of up to 2.2 T over a pole shoe gap of 6 cm height. Usually the MCP-PMT signals were passively split after a 200-fold amplifier (Ortec FTA820A, 350 MHz bandwidth). One signal was directly fed into an ADC, while the other was discriminated (Philips Scientific 705) to determine the time delay between the MCP-PMT anode signal and the reference signal of the laser control unit. CAMAC and VME data acquisition systems were used to record the anode charge and the time delay for the signals of each pixel.

The most precise time resolution measurements were made with a LeCroy WavePro7300A with 3 GHz bandwidth and 20 GS/s sampling rate. This oscilloscope allows the determination of time resolutions at the few pico-second level.

#### 4.2.3.1 Dark Count Rate

Each charged track will create several hundred Cherenkov photons. After many reflections and other losses along the radiators and taking into account the Quantum Efficiency (QE) of the photon sensors only a few tens of these photons will actually be detected. Therefore, it is important to use sensors with a moderately low dark count rate. From our measurements we find that at a gain of $10^6$ and a threshold of 0.5 photo electrons the typical dark count rate for most of the tested MCP-PMTs is below 1 kHz/cm$^2$. These numbers are sufficient for the Endcap Disc DIRC. Only the new BINP MCP-PMT with a modified photo cathode shows a dark count rate of more than 100 kHz/cm$^2$, while the Hamamatsu R10754 and R10754X show a significantly lower rate of ~100 Hz/cm$^2$. We also observed that often the main fraction of the dark count rate comes from rather localized spots in the MCP-PMT indicating that most anode pixels have a very low dark count rate of only a few ten Hz (see Figure 4.25).

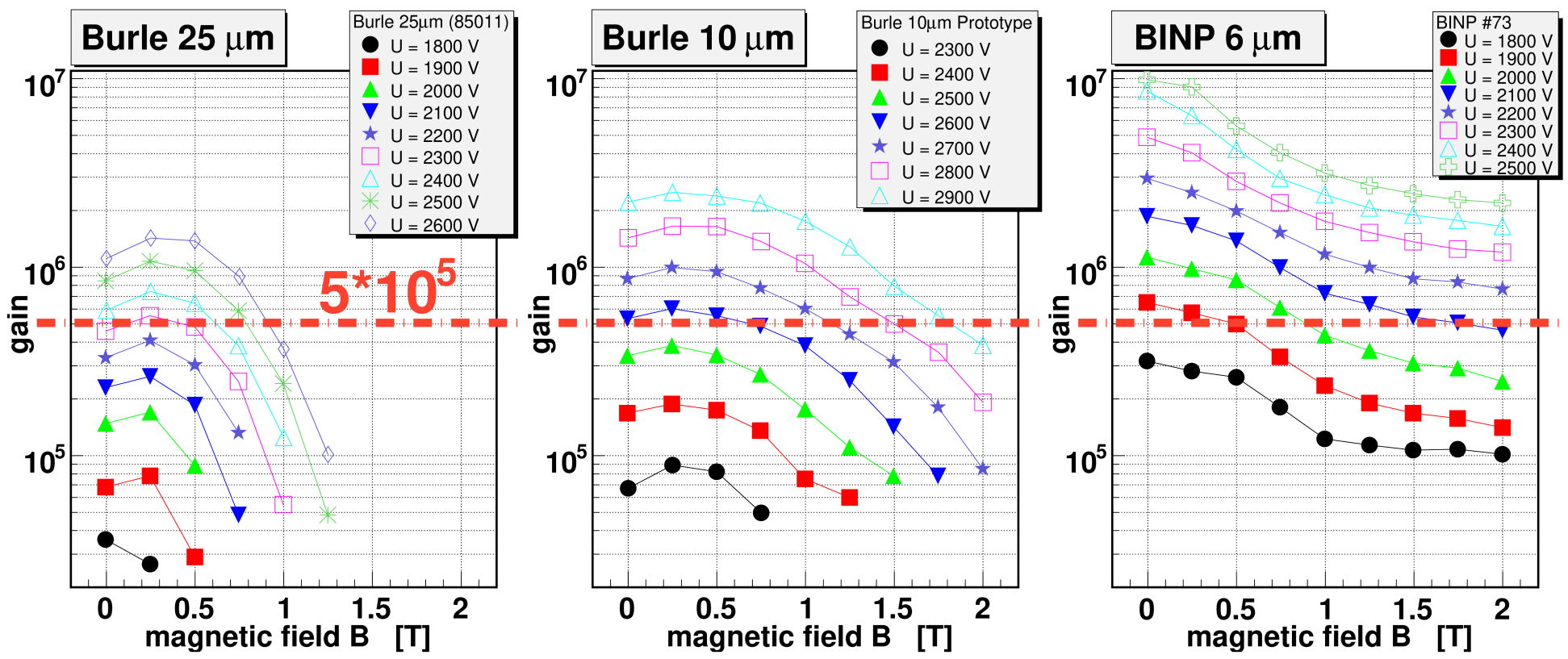

**Figure 4.26:** Gain as a function of the magnetic field for different high-voltage settings. Compared are MCP-PMTs of Burle-PHOTONIS with 25 $\mu m$ pore diameter (left), a PHOTONIS prototype with 10 $\mu m$ (middle) and a BINP device with 6 $\mu m$ pore diameter (right). The minimum gain of 5 x $10^5$ for an efficient single photon detection is indicated by the dash-dotted line.

#### 4.2.3.2 Gain inside Magnetic Field

The behavior of the gain as a function of the magnetic field is shown in Figure 4.26 for different high-voltage settings of three MCP-PMTs with different pore sizes. Clearly, the maximum gain reachable with the MCP-PMT depends on the pore diameter. The 25 $\mu m$ device reaches just above $10^6$ while with the MCP-PMT with 6 $\mu m$ pore size a gain of almost $10^7$ is possible. These results are compatible with earlier measurements [25].

The dash-dotted line indicates the minimum gain of about $5 \cdot 10^5$, which is still acceptable for an efficient single photon detection. From the plots it is obvious that the gain of the 25 $\mu m$ version of the PHOTONIS Planacon XP85012 reaches this limit only at large high-voltage settings. Since the gain collapses completely just above 1 T this device does not meet the requirements for the EDD. The PHOTONIS Planacon XP85112 with a smaller pore diameter of 10 $\mu m$ exhibits a larger gain and in principle it is still operable in the full 2 T field of the $\overline{\text{P}}$ANDA solenoid. Efficient single photon detection appears possible up to at least 1.75 T, a high voltage setting close to the recommended maximum for this device is needed though. The best gain performance in a high magnetic field is observed for the BINP MCP-PMT with 6 $\mu m$ pore diameter. The $\overline{\text{P}}$ANDA gain limit for single photon detection is reached at moderate operation voltages even in a 2 T field.

Usually the gain reaches a maximum at ~0.5 T and drops at higher fields. At a pore size of 25 $\mu$m the gain totally collapses just above 1 T, which can be attributed to the Larmor radius of the avalanche electrons at this field. Therefore, to efficiently detect single photons up to 1.2 T, as required in the EDD, a pore size of ≤10 $\mu$m is needed [28].

For the BINP MCP-PMT (see [28]), the PHOTONIS XP85012, and the Hamamatsu R10754-00-L4 measurements of the gain dependence on the orientation of the PMT axis with respect to the field direction were also performed. The results for the two latter devices are displayed in Figure 4.27. In the upper row the gain dependence on the tilt angle $\phi$ between the PMT axis and the field direction is shown: this demonstrates that up to $\phi \approx 20°$ no significant gain change is observed, while at larger angles the gain at higher field values starts to drop rapidly. Still, even at moderate tilt angles MCP-PMTs can be used for an efficient single photon detection in high magnetic fields. This is important for the PMT orientation in the EDD and an enormous advantage compared to standard dynode-based PMTs.

In the lower row of Figure 4.27 the gain behavior at different rotation angles $\theta$ of the PMT around the field axis and at a tilt angle $\phi \approx 15°$ is shown: there is a significantly different slope at $\theta = 180°$, when the capillaries of one of the two MCP layers point almost exactly along the field direction. At all other measured rotation angles the gain follows roughly the same slope.

It is important to note here that the gain behav-ior of ALD-coated MCP-PMTs inside a magnetic field are similar to the tubes described above. However, with ALD MCP-PMTs we observed a trend towards a faster dropping gain at higher magnetic fields, also the maximum gain is shifted to lower B-fields. This is probably caused by saturation effects inside the pores because the secondary electron yield (SEY) is significantly higher at the ALD coated surfaces compared to lead glass. Nevertheless, the ALD-coated MCP-PMTs behave well inside magnetic fields and can be used for the $\overline{\text{P}}$ANDA DIRCs.

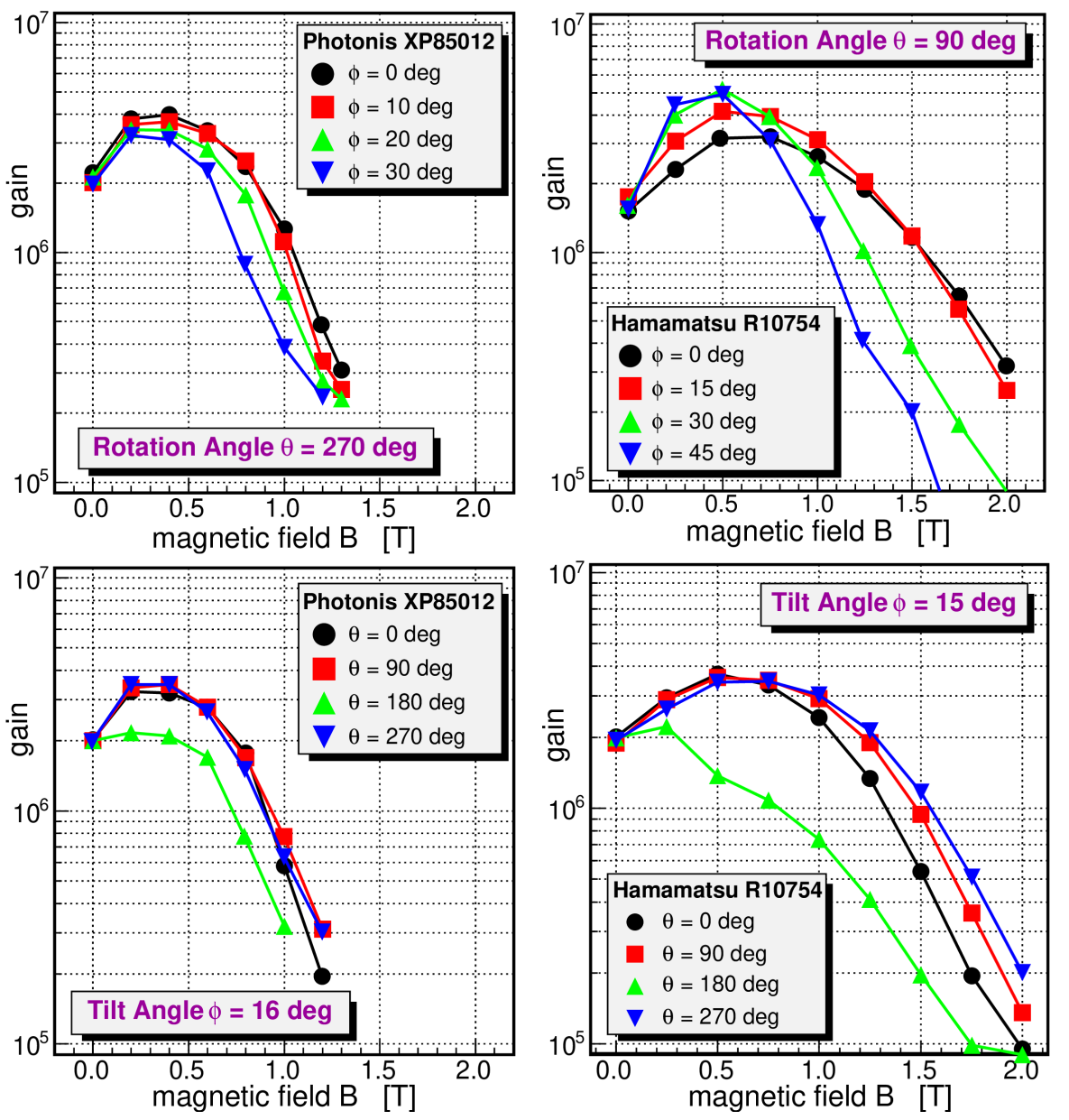

**Figure 4.27:** Gain as a function of the magnetic field direction for the PHOTONIS XP85012 (left column) and the Hamamatsu R10754-00-L4 (right column). In the upper row the dependence on the tilt angle $\phi$ is shown, in the lower row that on the rotation angle $\theta$.

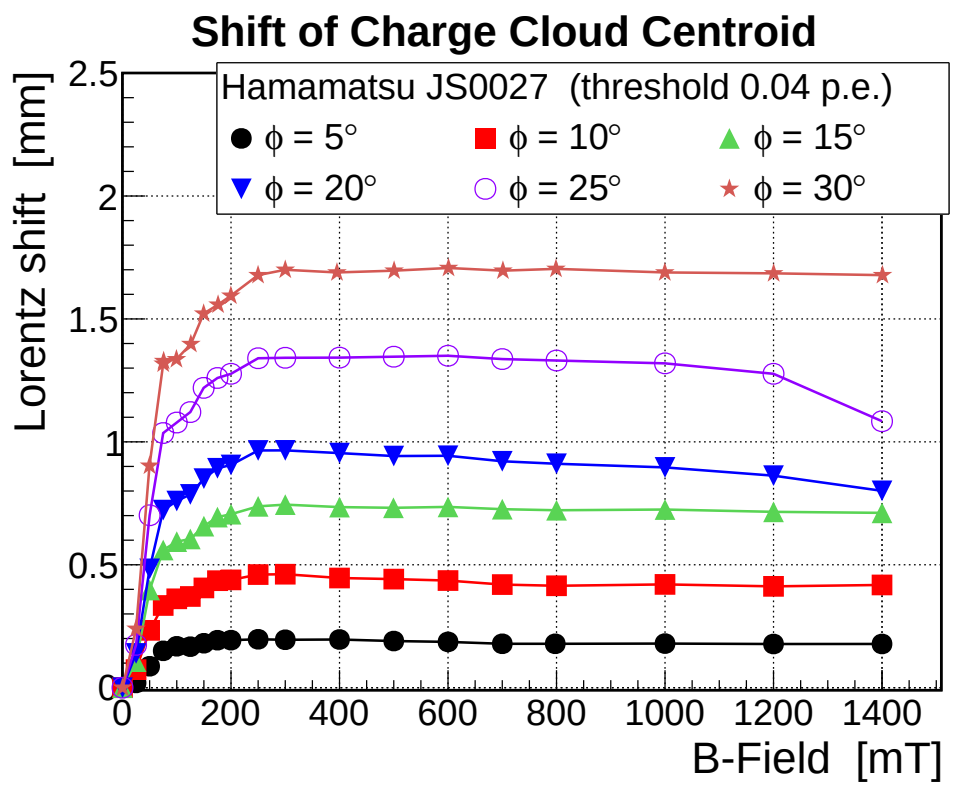

**Figure 4.28:** Lorentz shift for different tilt angles $\phi$ measured inside the new 2-inch 128×6 pixel Hamamatsu R13266-07-M768 (JS0035) MCP-PMT.

Another important measurement is the determination of the Lorentz shift of the charge cloud in the MCP-PMT when operated inside a B-field. In our measurement a focused laser spot was always pointed at the same pixel and 7 adjacent pixels were read out. When the PMT axis is tilted versus the direction of the magnetic field at an angle $\phi$ around a rotation axis (x) perpendicular to the high resolution coordinate (y), the centroid of the charge cloud at the anode plane is shifted along the direction of the high pixelation (y). By calculating the charge weighted positions along the y-axis one can determine the Lorentz shift which is specific to every MCP-PMT type dependent on the inner structure of the tube. In the case of the Hamamatsu R13266-07-M768 MCP-PMT we observe a shift of 1.7 mm at $\phi = 30°$, as shown in Figure 4.28. This would correspond to a position shift of 4 pixels with respect to the photon interaction point at the PC and has to be taken into account in the later analysis of the experimental data. In the final experiment, centroiding will not be required. Because of the strong magnetic field of the PANDA solenoid magnet, the electron cloud becomes very small, and only one single pixel will be hit as shown during lab tests.

#### 4.2.3.3 Time Resolution

In Figure 4.29 the time resolutions of the PHOTONIS MCP-PMT XP85012 with 25 $\mu m$ pores, the BINP MCP-PMT with 6 $\mu m$ pores, and the new 2-inch Hamamatsu R13266 prototype MCP-PMT with 10 $\mu m$ pores are compared. For the latter two tubes resolutions of 27 ps and 22 ps, respectively, were measured. These results still contain contributions from the finite time resolution of the electronics devices, the input channels of the oscilloscope, and in particular of the laser pulses. These resolutions were measured independently to be about 5-6 ps/channel for the oscilloscope channels and the same for the readout electronics devices used. The PiLas laser contributes 14 ps. Unfolding these contributions results in a net transit time resolution for single photons of $\sigma_t \approx 20$ ps for the BINP MCP-PMT and well below that for the Hamamatsu R13266 MCP-PMT.

The distribution of the measured time resolutions [28, 29] usually consists of a narrow peak ($\sigma_t$) and a tail to one side which originates mainly from photo electrons backscattered at the MCP entrance. This behavior was seen for all investigated MCP-PMTs, though with different fractions. As listed in Table 4.4, the width of the peak was always $\leq$50 ps, with the best resolutions of 27 ps and 22 ps (at $10^6$ gain and after ×200 amplification of the MCP-PMT anode signal) for the BINP MCP-PMT with 6 $\mu$m pore diameter and for the Hamamatsu prototype R13266 with 10 $\mu$m pore diameter, respectively. All listed time resolutions are without any correction for the resolutions of the used electronics modules and the laser pulse width.

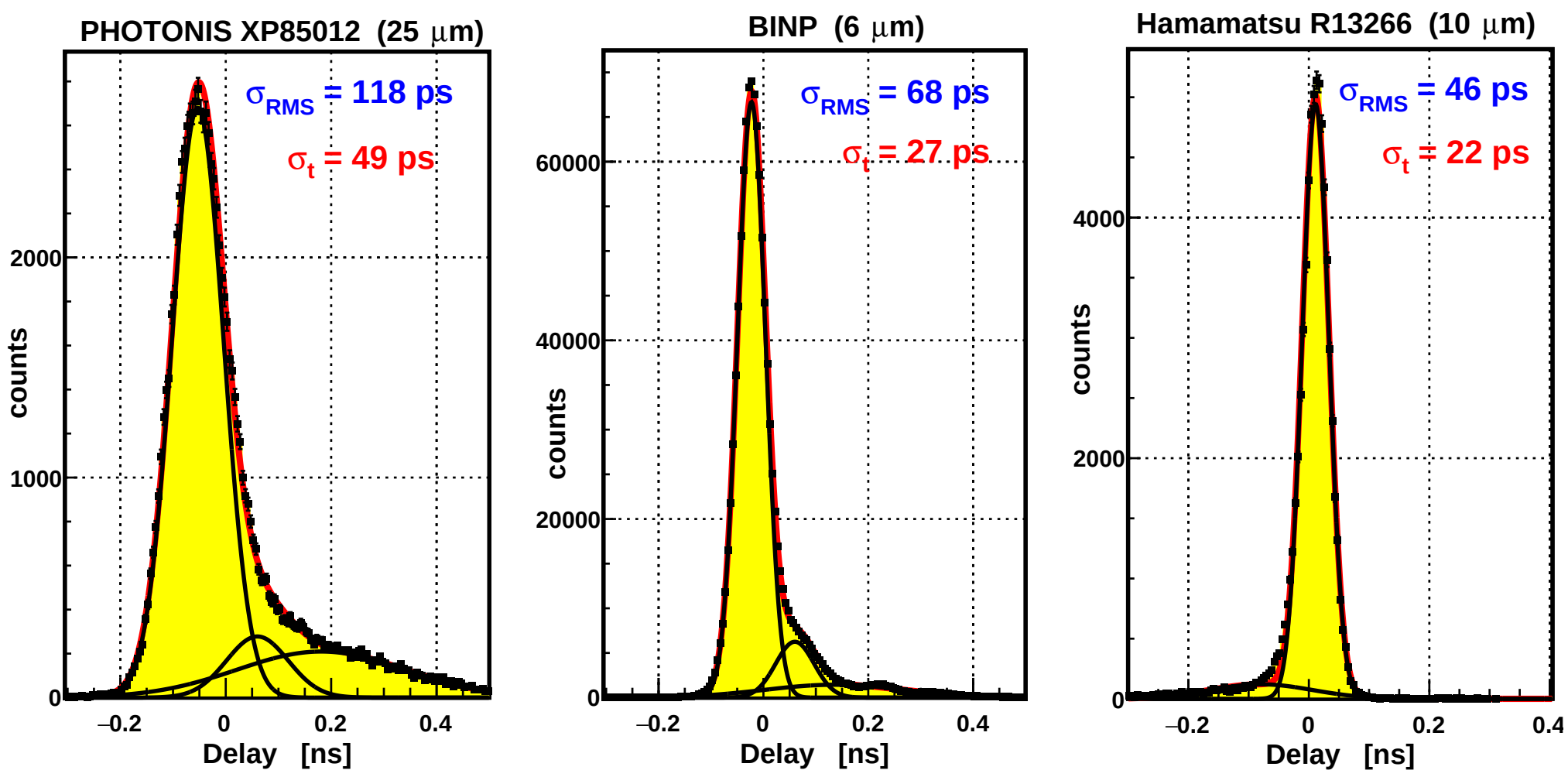

**Figure 4.29:** Single photon time resolution for a PHOTONIS MCP-PMT with 25 $\mu$m (left), a BINP device with 6 $\mu$m (middle), and the new Hamamatsu 2"×2" R13266 prototype MCP-PMT with 10 $\mu$m (right) pore diameter measured with a 3 GHz / 20 Gs oscilloscope. The tails are caused by photo electrons recoiling from the MCP entrance; left and right tails simply reflect different start and stop times in the delay measurements. A LeCroy 821 leading edge discriminator and an Ortec VT120A amplifier were used.

The RMS width of the time distribution depends strongly on the height and extension of the recoil tail. This can be partly controlled by building MCP-PMTs with a narrow gap between the PC and the first MCP, which reduces the amount of backscattered photo electrons reaching an MCP pore. In general it appears that all MCP-PMTs had an RMS time resolution in the 100 ps region and lower, while the new Hamamatsu R13266 even reaches <50 ps.

The time resolutions were also measured as a function of the magnitude of the magnetic field, with no significant deterioration at higher fields being observed.

**Table 4.4:** Single photon time resolutions of many investigated types of MCP-PMTs

| Manufacturer | Type | Pores [$\mu m$] | $\sigma_t$ [ps] |
|---|---|---|---|
| BINP | #73 | 6 | 27 |
| PHOTONIS | XP85112 | 10 | 41 |
| | XP85011 | 25 | 49 |
| | XP85013 | 25 | 51 |
| | XP85012 | 25 | 37 |
| Hamamatsu | R10754 | 10 | 32 |
| | R10754X | 10 | 33 |
| | R13266 | 10 | 22 |

#### 4.2.3.4 Gain Homogeneity and Crosstalk

The response of the multi-anode MCP-PMTs was investigated with XY-scans across the active surface. The gain of the different pixels in a device can vary by a factor 3 to 5 as measured in some Hamamatsu and PHOTONIS tubes [29]. The new high-QE XP85012 MCP-PMTs with 25 $\mu$m pores of the latter manufacturer show typical gain variations of about a factor 2 across the 64 pixels, as plotted in Figure 4.30 (lower row). The lowest gains are usually observed for the edge pixels and especially at the corners. In contrast, the new 2-inch Hamamatsu R13266 still shows a significant gain inhomogeneity (Figure 4.30, upper row), with measured fluctuations of almost a factor 10. This clearly has to be improved to be suitable for the $\overline{\text{P}}$ANDA Endcap DIRC.

A lower gain may cause a reduced detection efficiency of the pixel. In Figure 4.31 the number of counts of each pixel in a row is shown, when the active surface of the MCP-PMT was illuminated in 0.5 mm steps along the x-coordinate (or col-umn) while the y-position (or row) was kept constant. A fair count rate homogeneity is observed. These plots also show the crosstalk among the anode pixels. For the PHOTONIS XP85012 crosstalk is mainly visible at the transition to the adjacent pixels, most likely caused by charge sharing at the edge of the anode pixels and by backscattered photo electrons at the MCP entrance, while pixels further away are hardly affected. While this was different for the Hamamatsu R10754-00-L4 (a significant response of all other pixels was observed when a certain pad was illuminated; even pixels far from the light spot were firing), the crosstalk behavior of the new Hamamatsu R13266-07-M64 is very similar or even better than that of the PHOTONIS XP85012. The crosstalk of both MCP-PMT types appears acceptable for the $\overline{\text{P}}$ANDA DIRCs.

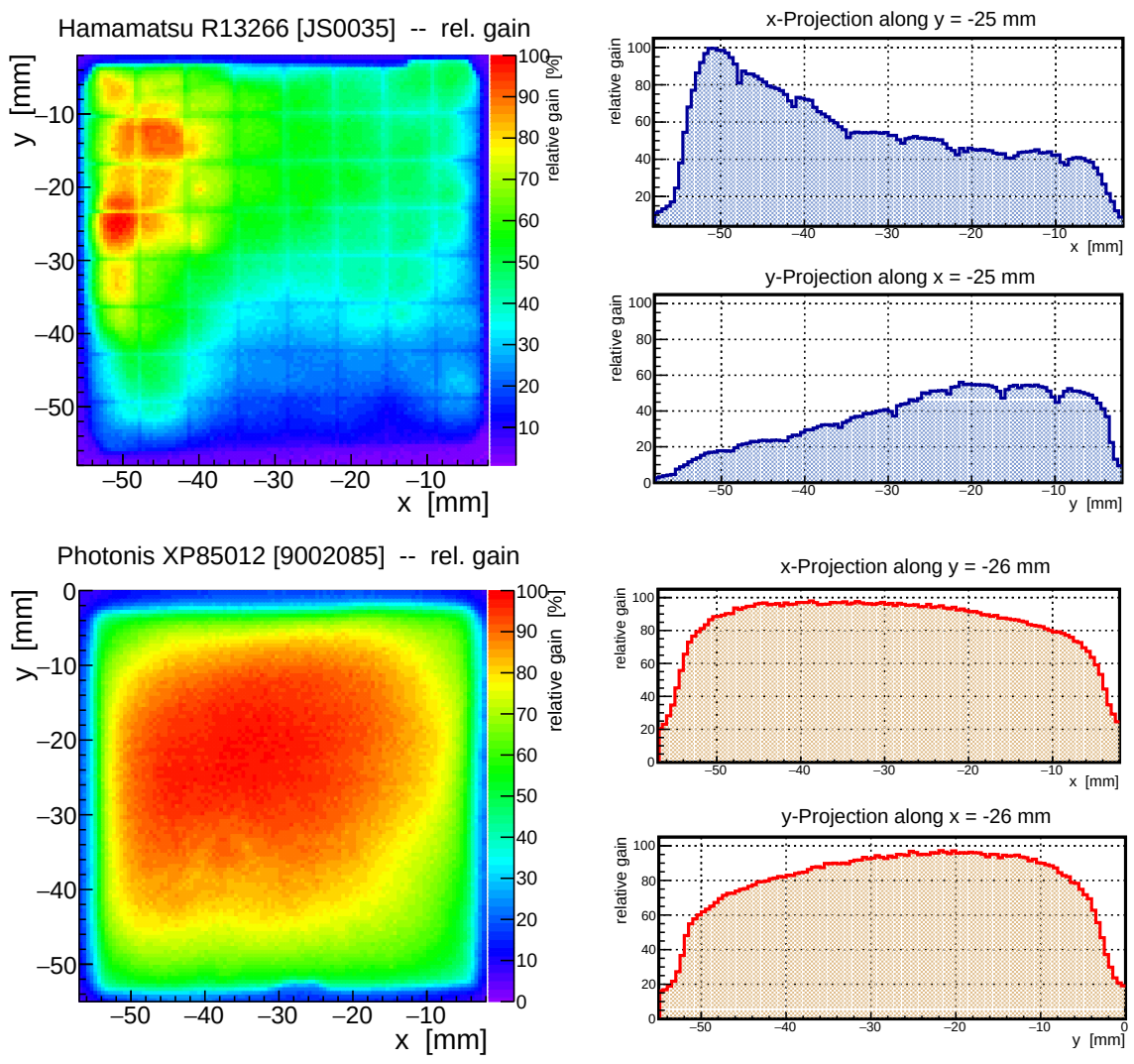

**Figure 4.30:** Relative gain (measured with the anode current) across the active area for the new PHOTONIS high-QE XP85012 MCP-PMT (lower left) and the new 2-inch Hamamatsu R13266 MCP-PMT (upper left). 100% relative gain corresponds to an absolute gain of $\sim 10^6$. Also shown are projections of the 2D plot along the x- and the y-axis (right column).

## Gain Stability at High Rates

The rate capability of MCP-PMTs is one of the most critical issues in high rate experiments like $\overline{\text{P}}$ANDA. The expected photon density (after QE) at the EDD readout (anode) plane is $\sim$1 MHz/cm$^2$. At these photon rates the current in the high resistive material of the MCP capillaries may not flow off fast enough, which causes charge saturation effects. The result of this is a rapidly decreasing gain as seen in Figure 4.32 where the normalized gain is plotted versus the anode current. Assuming a certain gain of the tube (e.g., $10^6$ in the figure) this current can be translated into a single photon density which is given at the upper axis.

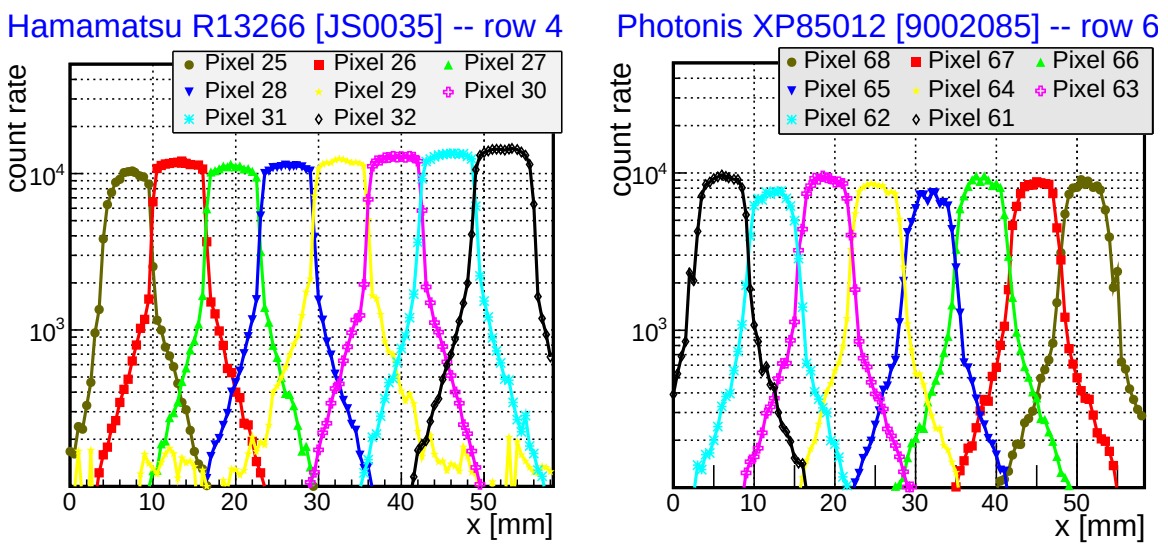

**Figure 4.31:** Crosstalk behavior along one row of pixels for the new high-QE PHOTONIS XP85012 MCP-PMT (right) and the new 2-inch Hamamatsu R13266 MCP-PMT (left).

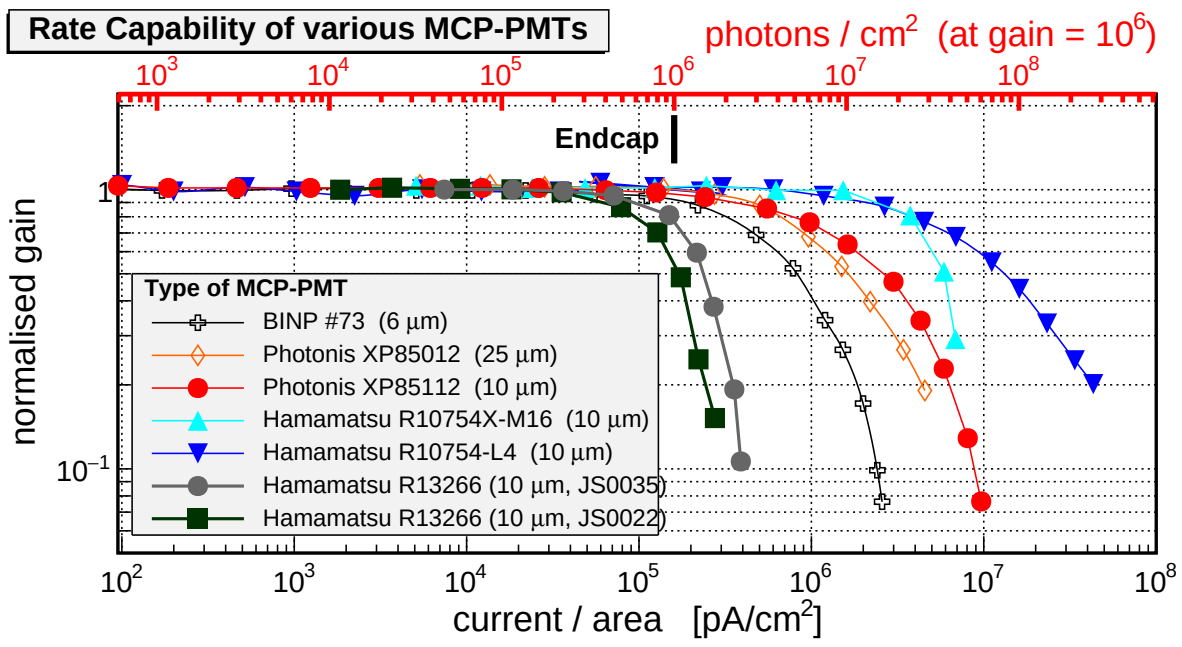

**Figure 4.32:** Rate capability of various models of MCP-PMTs: the relative gain is plotted versus the anode current. At the upper axis the translation into a rate for single photons assuming a constant gain of $10^6$ is given. The expected rate of detected photons for the Endcap Disc DIRC is also indicated.

The gain of most of the more recent MCP-PMTs starts dropping at photon densities just at or well beyond 1 MHz/cm$^2$ (e.g. PHOTONIS XP85012, XP85112 and Hamamatsu R10754 in Figure 4.32). However, the new 2-inch Hamamatsu R13266 prototype MCP-PMTs R13266 shows a significantly lower rate capability and would not quite qualify for the EDD at the moment.

### 4.2.3.5 Aging and Lifetime

Ageing of an MCP-PMT usually manifests itself in a reduction of its gain, its dark count rate and in particular its quantum efficiency (QE) when the integrated anode charge (IAC) accumulates. While a lower dark count rate is desirable and the reduced gain can to some extend be compensated by a higher PMT voltage, the diminishing QE harms the experiment. The main cause of the QE drop appears to be feedback ions from the residual gas, especially heavy atoms like lead, which impinge on the photo cathode (PC) and damage it. It has also been speculated that neutral rest gas molecules like oxygen and carbon dioxide may pollute the PC surface and change its work function [30].

### Methods for Lifetime Improvement

An obvious way of reducing the amount of residual gas in the tube is to bake the microchannel plates at high temperatures to outgas the glass material and desorb the surfaces. Additionally, the vacuum inside the MCP-PMT is improved and the manufacturers often apply electron scrubbing to clean and polish the MCP surfaces. Besides these approaches the three main manufacturers of MCP-PMTs apply the following techniques to extend the QE lifetime:

- In their latest MCP-PMT models the Budker Institute of Nuclear Physics (BINP) in Novosibirsk applies a special treatment to the bialkali PC which is baked in a vapor of caesium and antimony. This seems to increase the PC's hardness against feedback ions, but significantly increases the dark count rate of the tube [31].
- A new and innovative approach is pursued by PHOTONIS. The surfaces and pores of the MCPs are coated with a very thin layer of secondary electron emissive material by applying an atomic layer deposition (ALD) technique [32, 33, 34]. This layer is expected to significantly reduce the outgassing of the MCP material.
- Hamamatsu first tried to eliminate the ion back flow from the anode side of the MCP-PMT by putting a thin protection layer of aluminum (film) between the two MCPs. In addition, potential gaps between the MCPs and the metal walls of the tube's frame were sealed with ceramic elements to hinder neutral atoms and molecules from the rear part of the MCP-PMT in reaching the PC [30]. In their most recent MCP-PMTs Hamamatsu also applies the ALD technique, often combined with a film in front of or between the MCPs.

In the recent years we have measured the lifetime of several MCP-PMTs of the three manufacturers mentioned above. The first tubes from BINP (#82) and PHOTONIS (XP85012-9000298 and XP85112-9000897) were still without the above-listed improvements (see Figure 4.33). A list of the characteristics of the lifetime-enhanced MCP-PMTs discussed in this report are given in Tab. 4.3.

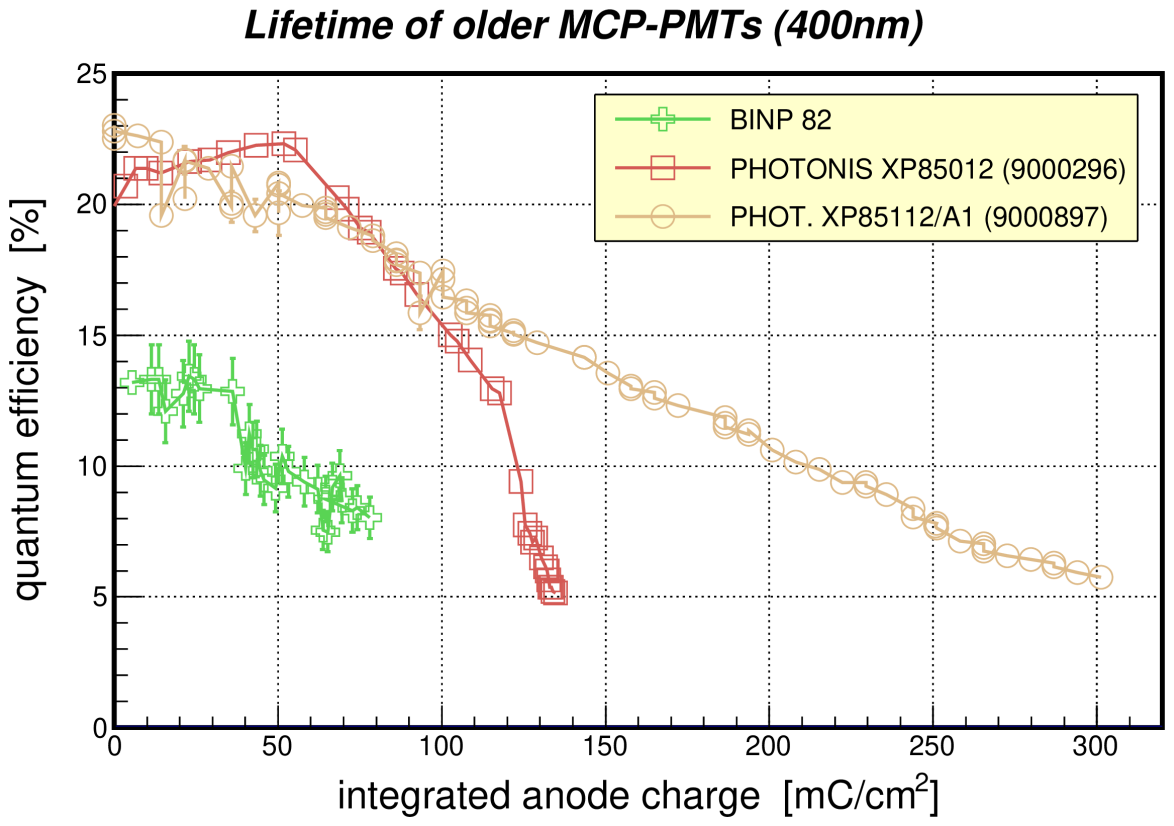

**Figure 4.33:** Comparison of our aging measurements (status 2011) with not yet lifetime-enhanced MCP-PMTs: quantum efficiency as a function of the IAC at 400 nm.

### Setup of Lifetime Measurements

Until recently only few quantitative results on the lifetime of MCP-PMTs were available [25, 35]. Moreover, these were obtained in very different environments and therefore difficult to compare. The standard way of measuring the lifetime of an MCP-PMT is to determine the gain and especially the QE as a function of the IAC. If the QE has dropped by a certain percentage (e.g. 50%) of its original value the sensor is presumed unusable. The $\overline{\text{P}}$ANDA experiment is expected to run for at least 10 years at a 50% duty cycle. Assuming the average antiproton-proton annihilation rate of 20 MHz and a sensor gain of $10^6$, simulations show an IAC of at least 7 C/cm$^2$ expected for the MCP-PMTs of the Endcap Disc DIRC.

The lifetimes shown in Figure 4.33, which we determined for the first MCP-PMTs of BINP (#82) and PHOTONIS (XP85012-9000298 and XP85112-9000897), were by far not sufficient for $\overline{\text{P}}$ANDA. The QE had dropped by >50% after only ≈200 mC/cm$^2$ IAC [26, 36].

The setup of our lifetime measurements is described in the publications [26, 37]. The MCP-PMTs were permanently and simultaneously illuminated with a blue (460 nm) LED at a rate comparable to that expected in $\overline{\text{P}}$ANDA. The entire photo cathode of the MCP-PMT was homogeneously illuminated with near-parallel light. At the entrance window the light was attenuated to a level of ∼1 photon/cm$^2$ per pulse; at a gain of $10^6$ this corre-sponds to an IAC of ~14 mC/cm$^2$/day at the EDD photon rate of 1 MHz. The stability of the LED was controlled by measuring the current of a photodiode placed close to the MCP-PMTs. The MCP-PMTs' responses were continuously monitored by recording the pulse heights with a DAQ system at a highly prescaled rate. In irregular time intervals (a few days at the beginning, a few weeks later) the Q.E. of the photo cathode of each illuminated MCP-PMT was determined over a 300-800 nm wavelength band. The setup for the Q.E. measurements [38] consisted of a stable xenon lamp, a monochromator with 1 nm resolution and a calibrated reference diode (Hamamatsu S6337-01).

For each MCP-PMT, and in intervals of a few months, the photo current across the whole PC surface was measured in small steps of 0.5 mm at a wavelength of 372 nm to identify the regions where the QE degradation possibly starts.

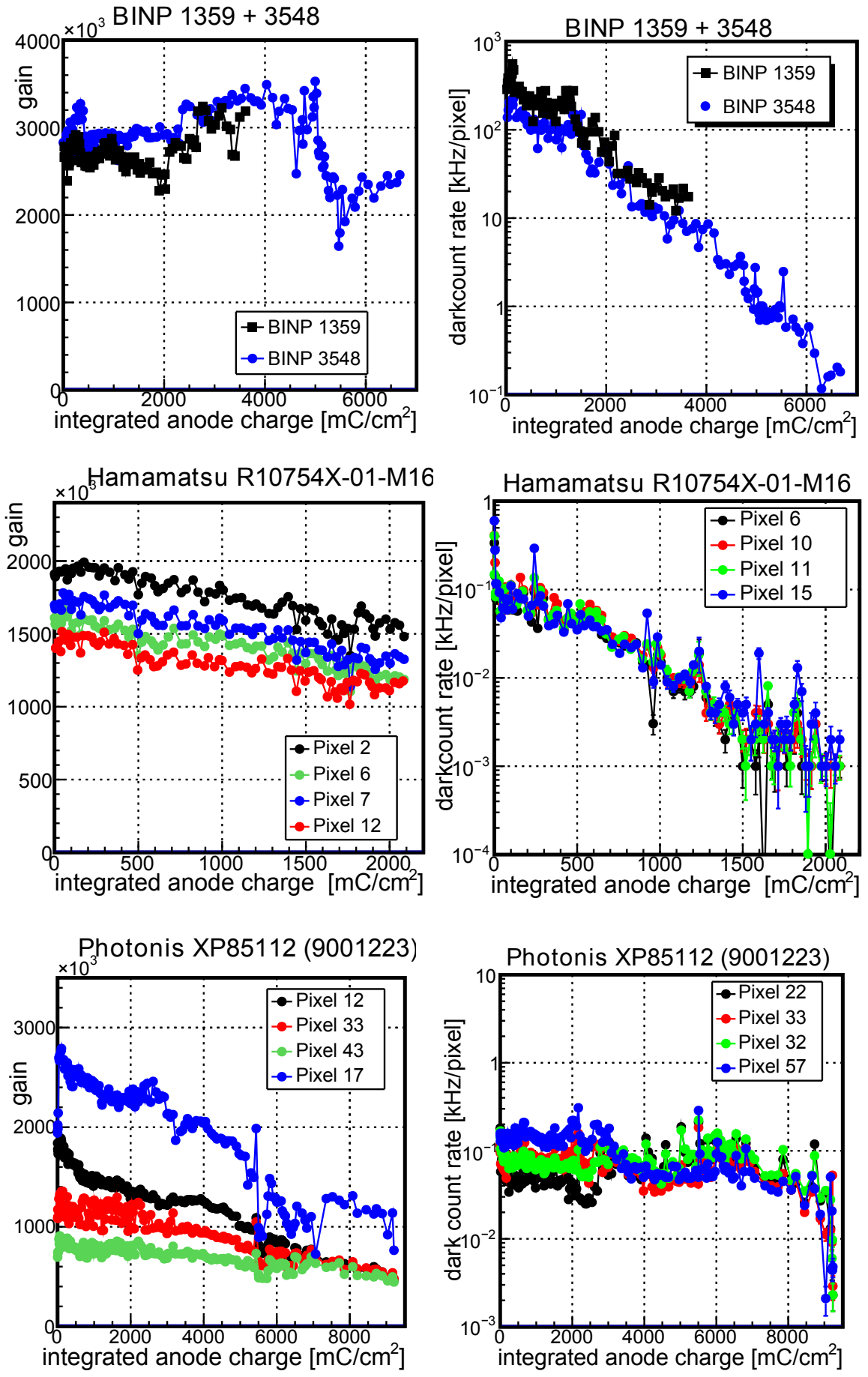

**Figure 4.34:** Gain (upper row) and dark count rate (lower row) as a function of the IAC for selected MCP-PMTs: BINP with a modified PC (left); Hamamatsu with a film but without ALD coating yet (middle); PHOTONIS with ALD-coating (right).

### Results of Lifetime Measurements

Important quantities for Cherenkov detectors are the gain and dark count rate of the sensors. The gain has to be high enough for an efficient single photon detection and the dark count rate should be low since the photon yield per track is usually rather moderate. These quantities were measured as a function of the IAC as shown in Figure 4.34. We observe that the gain changes are only moderate for most of the pixels of the displayed sensors and can easily be compensated for by increasing the tube's voltage. On the other hand the dark count rate has dropped by more than two orders of magnitude for the BINP and Hamamatsu MCP-PMTs. This finding might indicate a change of the PC's work function during the illumination of the sensor. The PHOTONIS XP85112 does not show these massive changes in the dark count rate.

It was reported earlier [25, 26] that the QE degrades faster for red than for blue light. To study the observed wavelength dependence we have measured the spectral QE as a function of the IAC for all investigated new MCP-PMTs. The results for different wavelengths are displayed in Figure 4.35 for representative samples of MCP-PMTs treated with different techniques to reduce aging: modified PC from BINP (left column); film at the MCP entrance, but no ALD-coating yet, from Hamamatsu (middle column); first ALD-coating from PHOTONIS (right column). It is obvious from the plots that the MCP-PMTs of the three manufacturers behave differently. While the QE of the Hamamatsu R10754X with a film as ion barrier starts dropping significantly beyond ~1 C/cm$^2$ the QE of the BINP #3548 with its modified PC shows a constant decline up to almost 7 C/cm$^2$. The PHOTONIS XP85112 (9001223) shows practically no QE degradation up to 6 C/cm$^2$ and deteriorates fast beyond that integrated anode charge. Once the QE starts dropping a clear spectral dependence of the QE is visible in the R10754X and the XP85112 (lower row in Figure 4.35). This could also indicate a change in the work function of the PC, possibly due to residual gas atoms and molecules adsorbed at the PC surface. The displayed BINP 3548 and PHOTONIS XP85112 (9001223) MCP-PMT up to 6 C/cm$^2$ IAC do not exhibit a clear QE dependence upon the wavelength.

In Figure 4.36 the gain, dark count rate (DCR), absolute and relative QE (normalized to 350 nm) are compared for the most recent lifetime-enhanced MCP-PMTs with ALD-coating from PHOTONIS

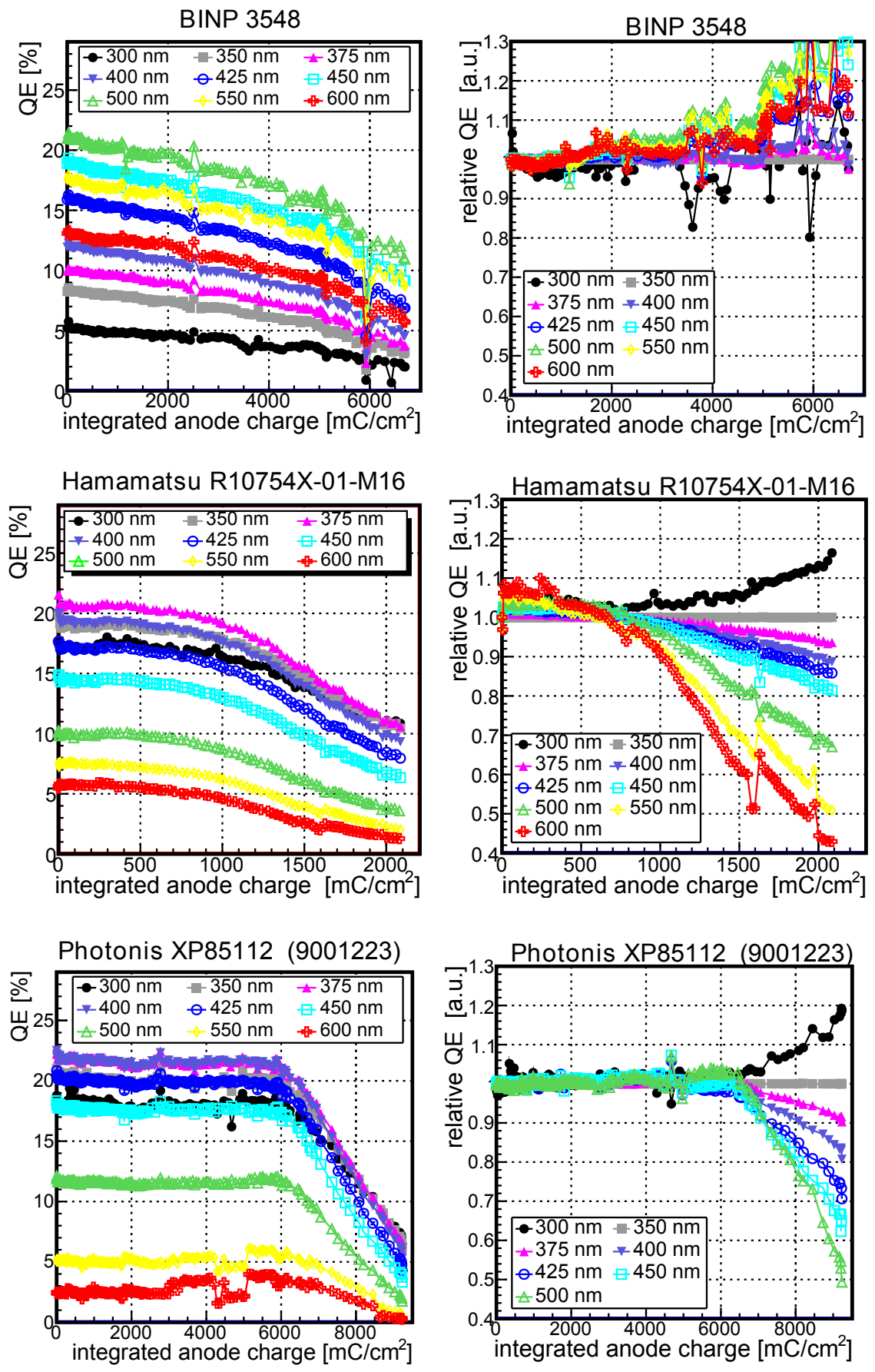

**Figure 4.35:** QE (absolute and relative to 350 nm) as a function of the IAC and for different wavelengths.

XP85112 9001393 (2 ALD layers, upper row), Hamamatsu R10754X-M16M KT0001 (1-inch, 1 ALD layer + film between first and second MCP, middle row), and Hamamatsu R13266 JS0022 (2-inch, 1 ALD layer + film in front of the first MCP, lower row). In comparison to the MCP-PMTs shown in Figures 4.34 and 4.35 the gains and the DCRs remain more stable when the IACs increases. However, both Hamamatsu MCP-PMTs show a wavelength-dependent QE degradation already starting at moderate IACs. In contrast, the PHOTONIS XP85112 with 2 ALD layers shows no sign of aging up to 12 C/cm$^2$.

The results of the QE scans across the PC surface are displayed in Figure 4.37 to Figure 4.42 for six MCP-PMTs with different properties. The upper left plot always shows a QE chart of the full PC surface with the IAC accumulated at the time of writing this document. For a better judgment of the magnitude of the observed QE changes three projections along the x-axis at different positions of y are plotted for each MCP-PMT. The histograms in these plots correspond to different anode charges, from the beginning of the illumination or the time when no QE degradation was observed yet to the highest accumulated anode charge.

It is obvious from the plots that all MCP-PMTs without ALD-coating (BINP #3548 in Figure 4.37 and Hamamatsu R10754X in Figure 4.38) show clear QE damages already after moderate anode charges of 1-2 C/cm$^2$. From the QE chart and its projections it appears that the QE degradation starts at the corners (R10754X) or at the rim (#3548) of the sensor. With progressing illumination the QE drop extends more and more to the inner regions of the PC. After an anode charge of 5025 mC/cm$^2$ and 1765 mC/cm$^2$ for the BINP and Hamamatsu MCP-PMT, respectively, the QE has dropped by more than 50% of its original value in certain regions. Also the new 2-inch Hamamatsu prototype MCP-PMT R13266 (JS0022, with ALD) in Figure 4.41 shows significant QE damage at the right PC half ($x > 0$ mm) already after $\sim$1.3 C/cm$^2$. The left half ($x < 0$ mm) was covered during the illumination and did not see photons. However, we expect that the lifetime of this MCP-PMT type will increase with newer and improved prototypes.

The situation is somewhat different for the ALD-coated sensors PHOTONIS XP85112 9001332 (Figure 4.39) and Hamamatsu R10754X-M16M KT0001 (Figure 4.40), where the QE degradation either evolves very slowly (KT0001) or does not start before >10 C/cm$^2$ (9001332). Beyond this anode charge the sensor 9001332 shows the development of some QE damage at the left side, when a clear step emerges around $x = 0$ mm. This stems from the fact that the right half of the PC ($x > 0$ mm) of the sensor was covered during the illumination process. In the KT0001 a 50% QE drop is reached at $\sim$11 C/cm$^2$, in the 9001332 this value is $\sim$14 C/cm$^2$. This would be enough charge for the Endcap Disc DIRC. As in the other parameters gain and DCR (see Figure 4.36) the PHOTONIS XP85112 9001393 with two ALD-layers (Figure 4.42) shows no sign of QE damage up to 12 C/cm$^2$. Up to now this is the by far best performing MCP-PMT.

Finally, in Figure 4.43 the QE at 400 nm is compared for all investigated MCP-PMTs. Clearly, the older MCP-PMTs (labeled "no countermeasures", open symbols at the very left, see also Figure 4.33) show a fast declining QE which drops below 50% after <200 mC/cm$^2$. The situation is very different for the new lifetime-enhanced tubes. The QE of the Hamamatsu R10754X-01-M16 with a protection film is exhausted at <2 C/cm$^2$, while the QE of the two BINP MCP-PMTs (#1359 and #3548, labeled "new PC") is continuously dimin-ishing up to ∼3.5 C/cm² and ∼7 C/cm², respectively. This is much better for the ALD-coated de-vices (lower plot of Figure 4.43). Except for the four new 2-inch Hamamatsu R13266 prototype MCP-PMTs all other ALD-coated MCP-PMTs show a stable or only slowly degrading QE up to at least 6 C/cm$^2$. The two 1-inch Hamamatsu R10754X MCP-PMTs (KT0001 and KT0002) show a constantly degrading QE up to >10 C/cm$^2$ IAC. While for the PHOTONIS 9001223 the QE starts dropping beyond 6 C/cm$^2$ the QE of the identically constructed 9001332 is basically unaffected up to ∼10 C/cm$^2$ IAC. The PHOTONIS 9001393 with its different design with two ALD layers is stable up to at least 12 C/cm$^2$.

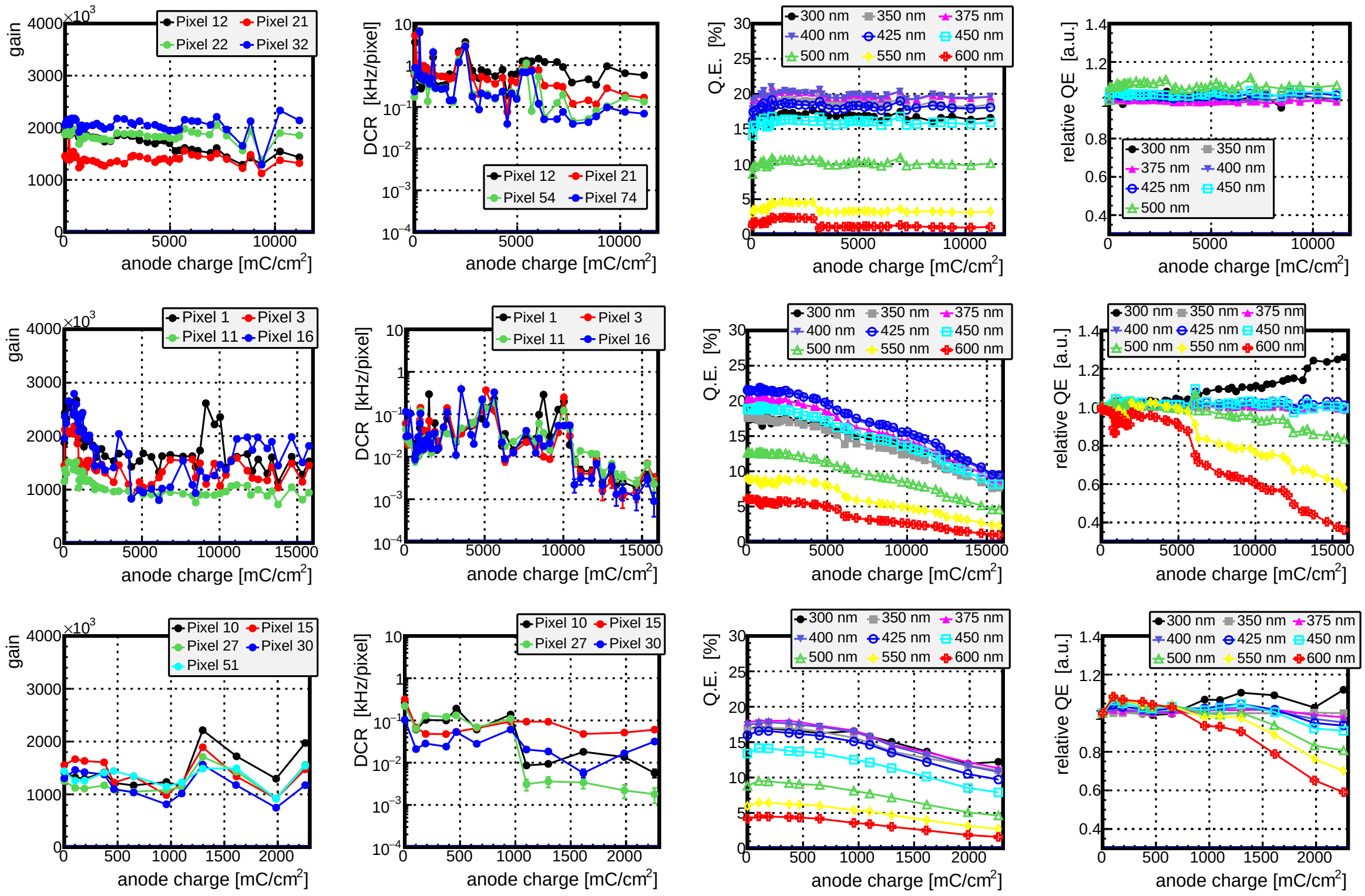

**Figure 4.36:** Gain, dark count rate (DCR), and QE for different wavelengths (absolute and relative to 350 nm) as a function of the IAC. In the upper row a PHOTONIS MCP-PMT XP85112 (9001393) with 2 ALD layers is displayed. The middle and lower rows show ALD-coated Hamamatsu MCP-PMTs with one- (R10754X-M16M KT0001) and two-inch (R13266 JS0022) active area, respectively.

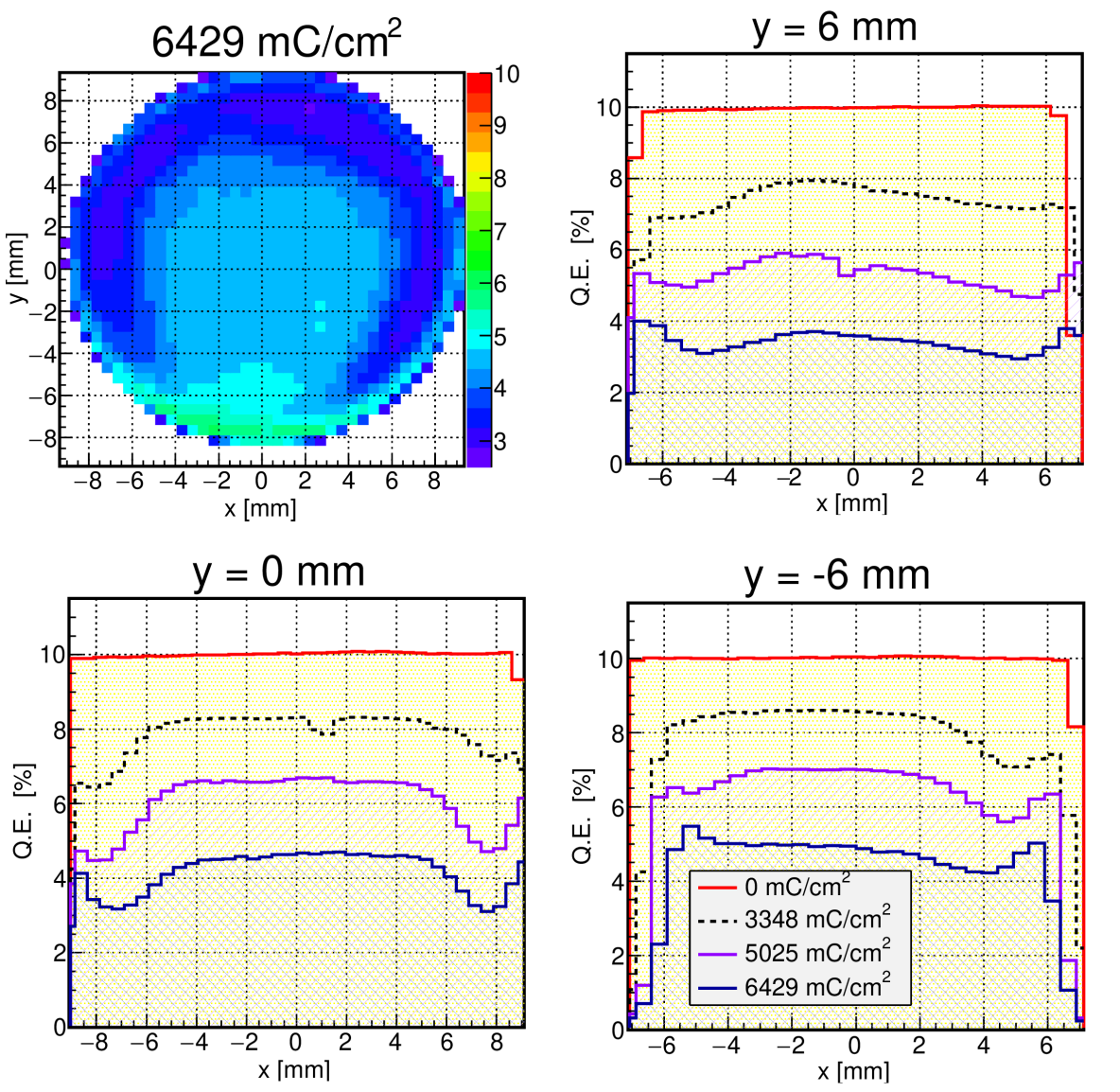

**Figure 4.37:** QE at 372 nm as a function of the PC surface for the BINP #3548 MCP-PMT with an active area of 18 mm diameter. Upper left: two-dimensional QE chart (in % [color level]); other plots: QE x-projections at different y-positions and anode charges.

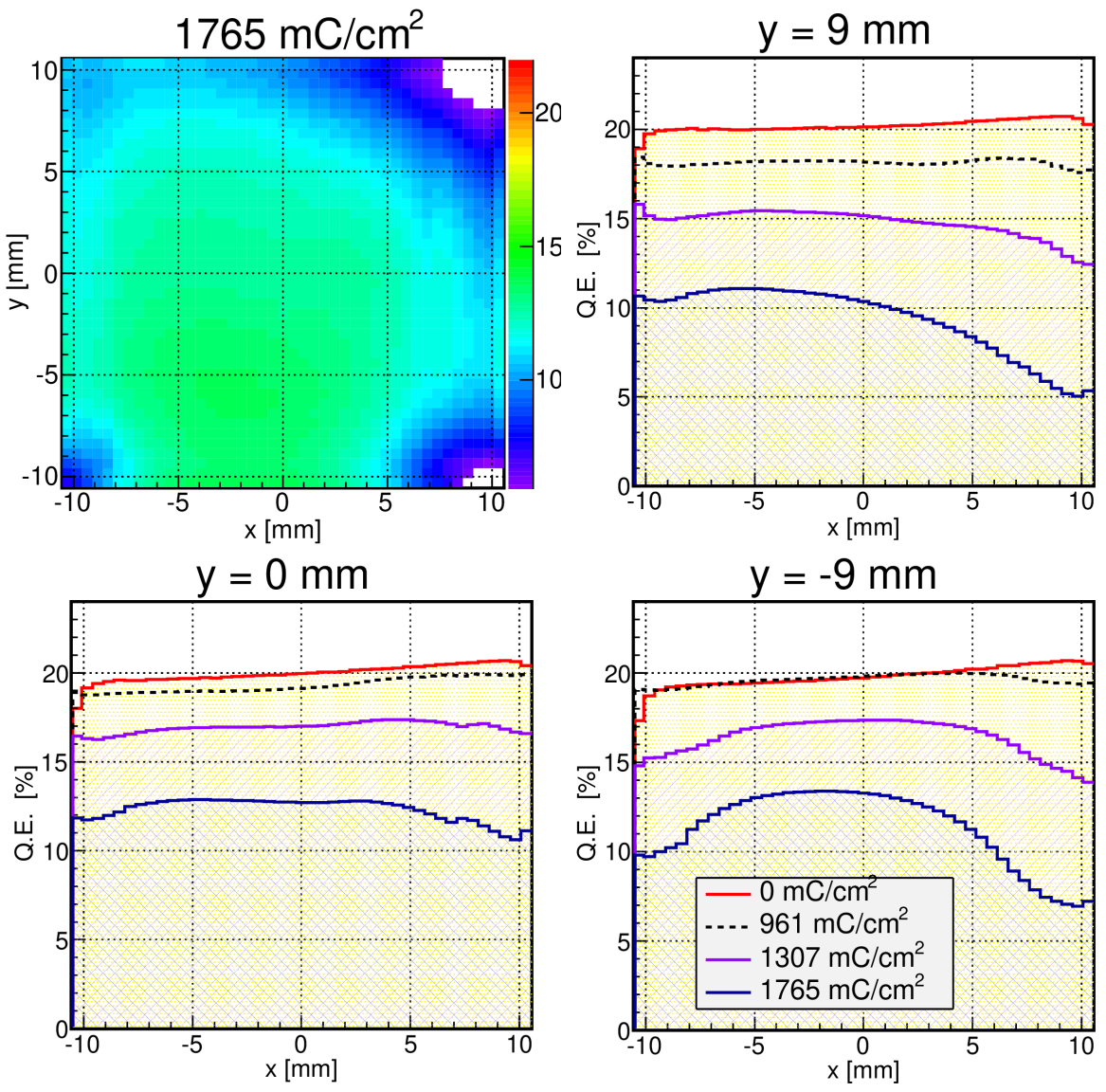

**Figure 4.38:** QE at 372 nm as a function of the PC surface for the Hamamatsu R10754X-01-M16 (JT0117) MCP-PMT with an active area of 22×22 mm². The four plots display the same properties as in Figure 4.37.

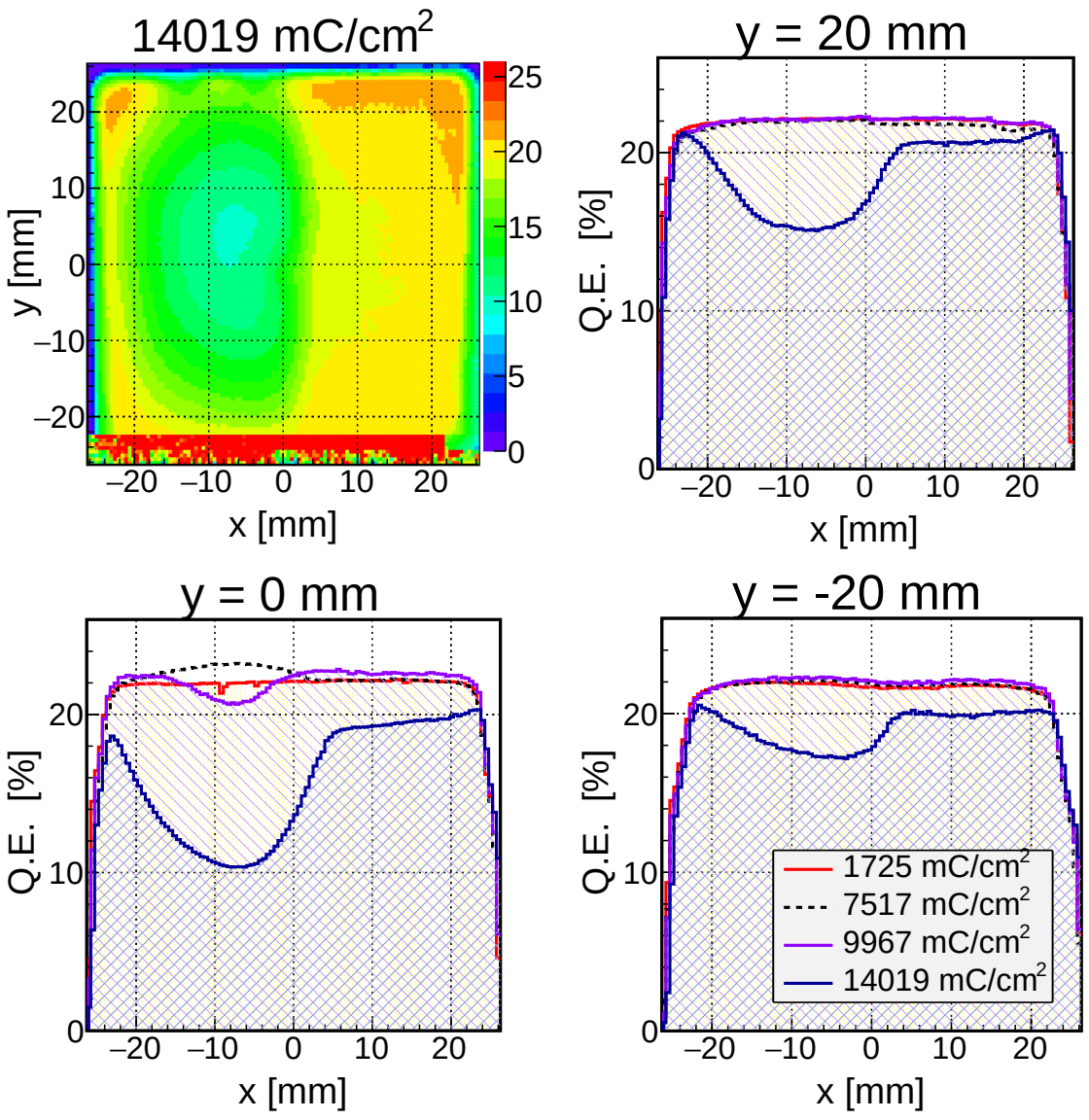

**Figure 4.39:** QE at 372 nm as a function of the PC surface for the PHOTONIS XP85112 (9001332) MCP-PMT with an active area of 53×53 mm$^2$. The four plots display the same properties as in Figure 4.37.

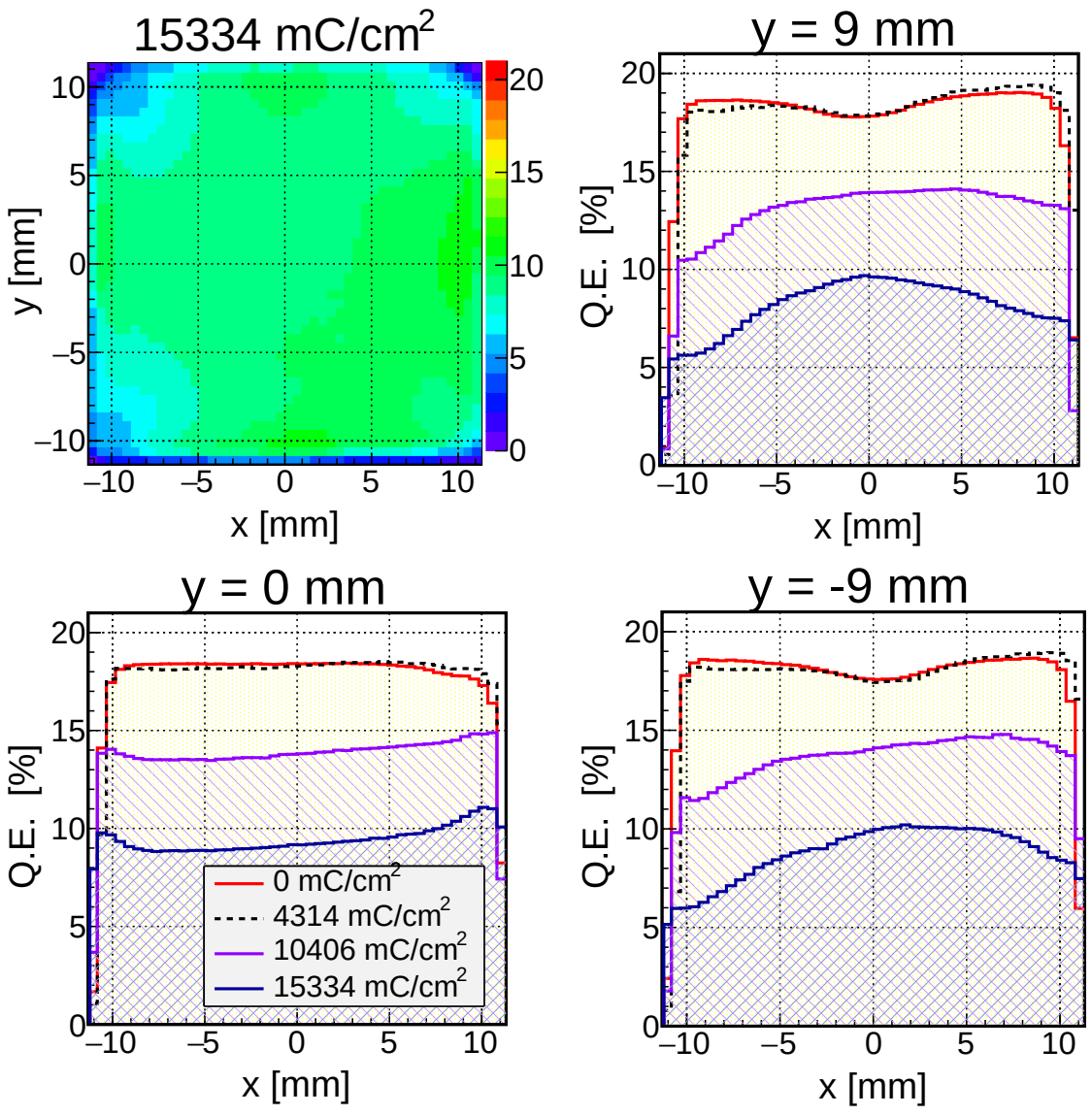

**Figure 4.40:** QE at 372 nm as a function of the PC surface for the 1-inch Hamamatsu R10754X-07-M16M (KT0001) MCP-PMT with an active area of 22×22 mm$^2$. The four plots display the same properties as in Figure 4.37.

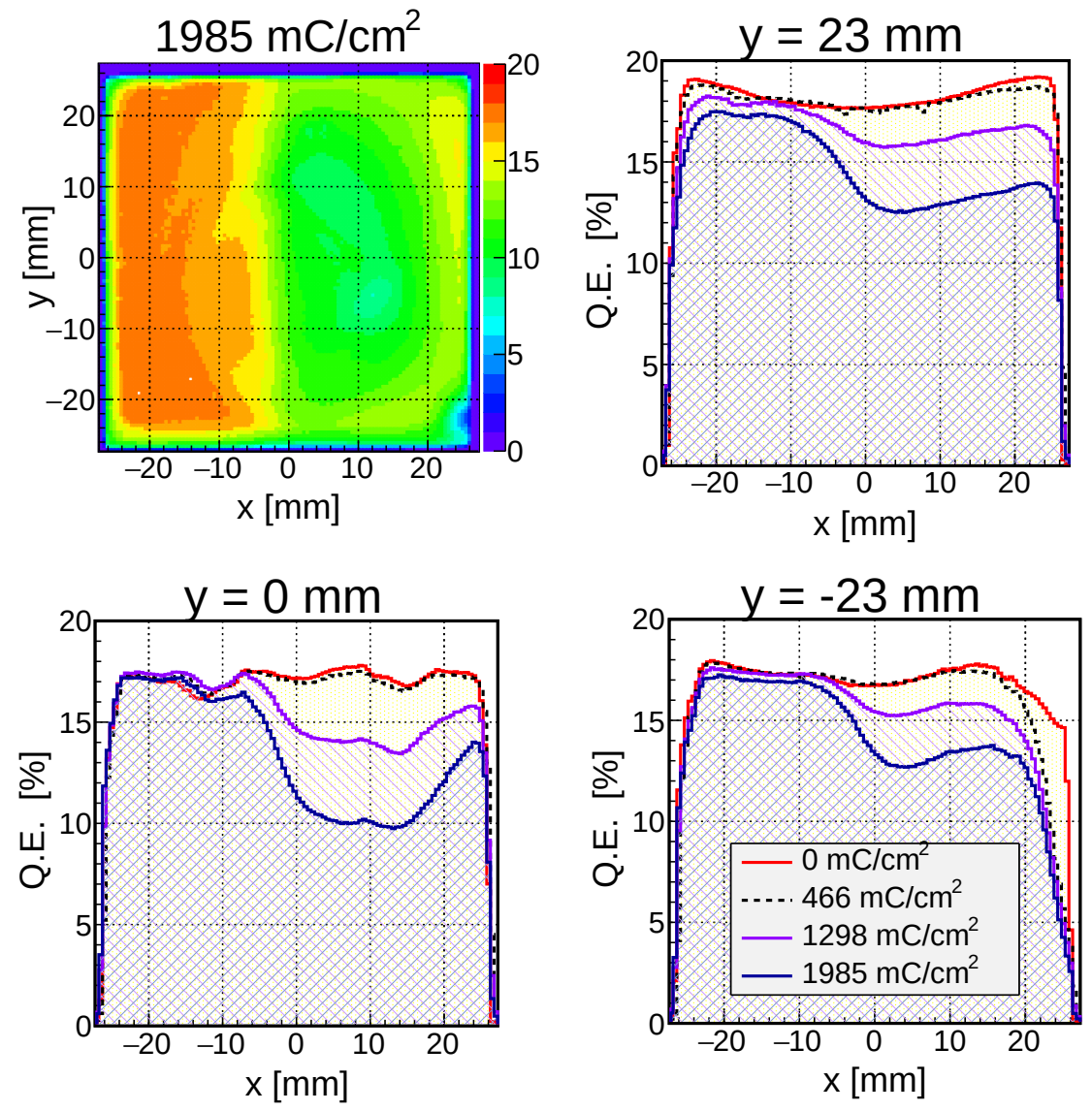

**Figure 4.41:** QE at 372 nm as a function of the PC surface for the 2-inch Hamamatsu R13266 (JS0022) prototype MCP-PMT with an active area of 53×53 mm$^2$. The four plots display the same properties as in Figure 4.37.

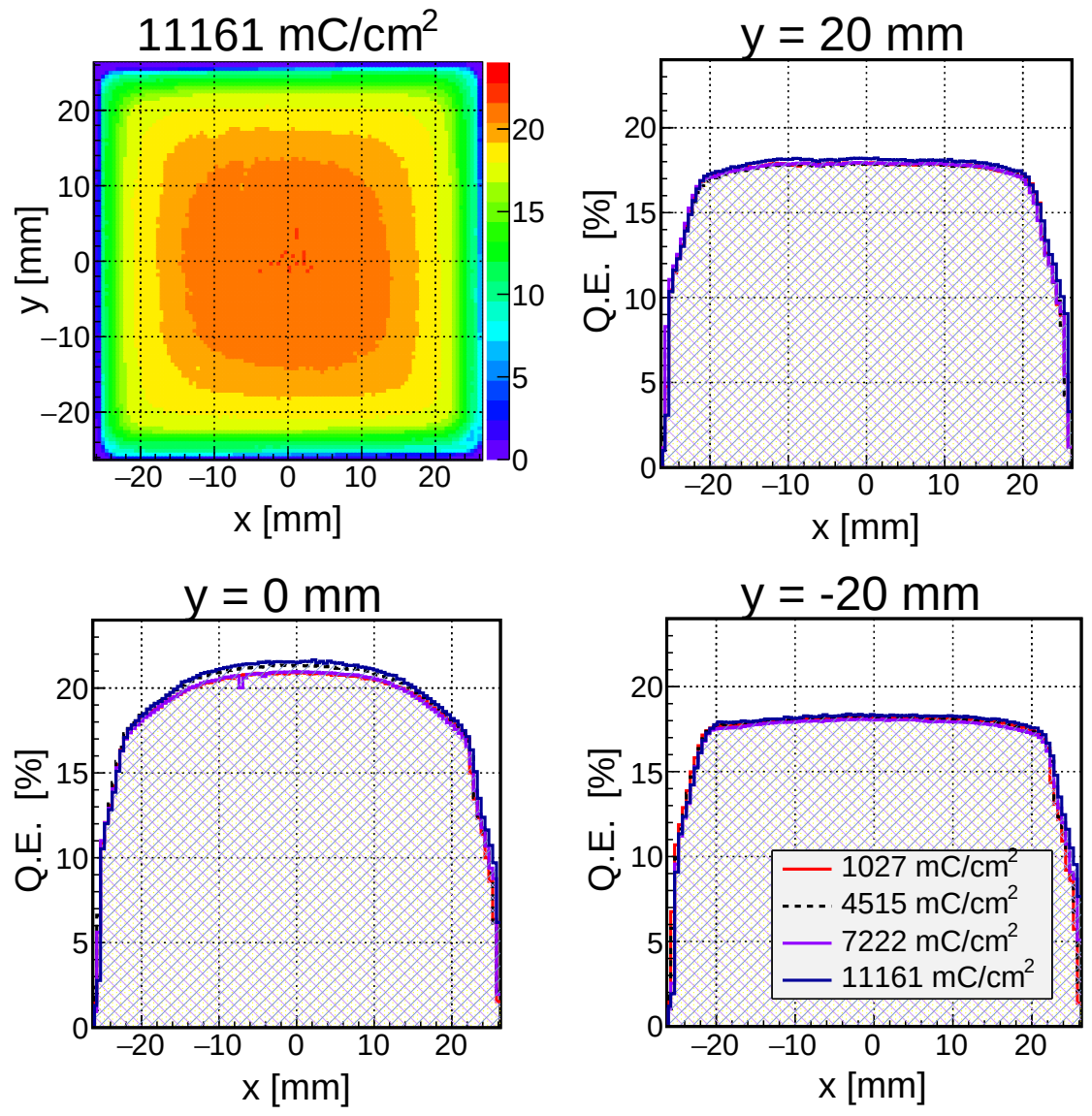

**Figure 4.42:** QE at 372 nm as a function of the PC surface for the PHOTONIS XP85112 (9001393) MCP-PMT with 2 ALD layers and an active area of 53×53 mm$^2$. The four plots display the same properties as in Figure 4.37.

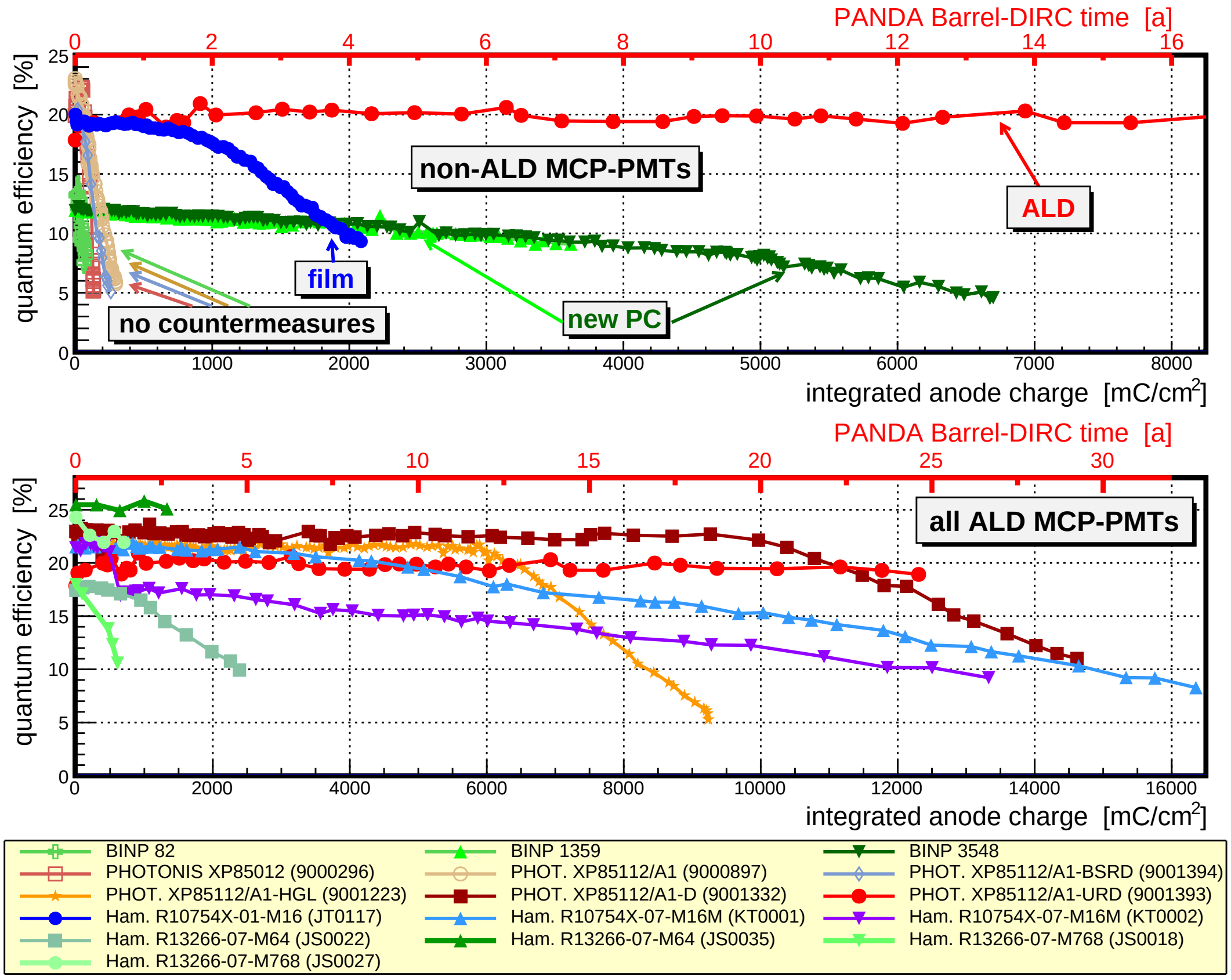

**Figure 4.43:** Comparison of our MCP-PMT measurements: quantum efficiency as a function of the IAC at 400 nm.

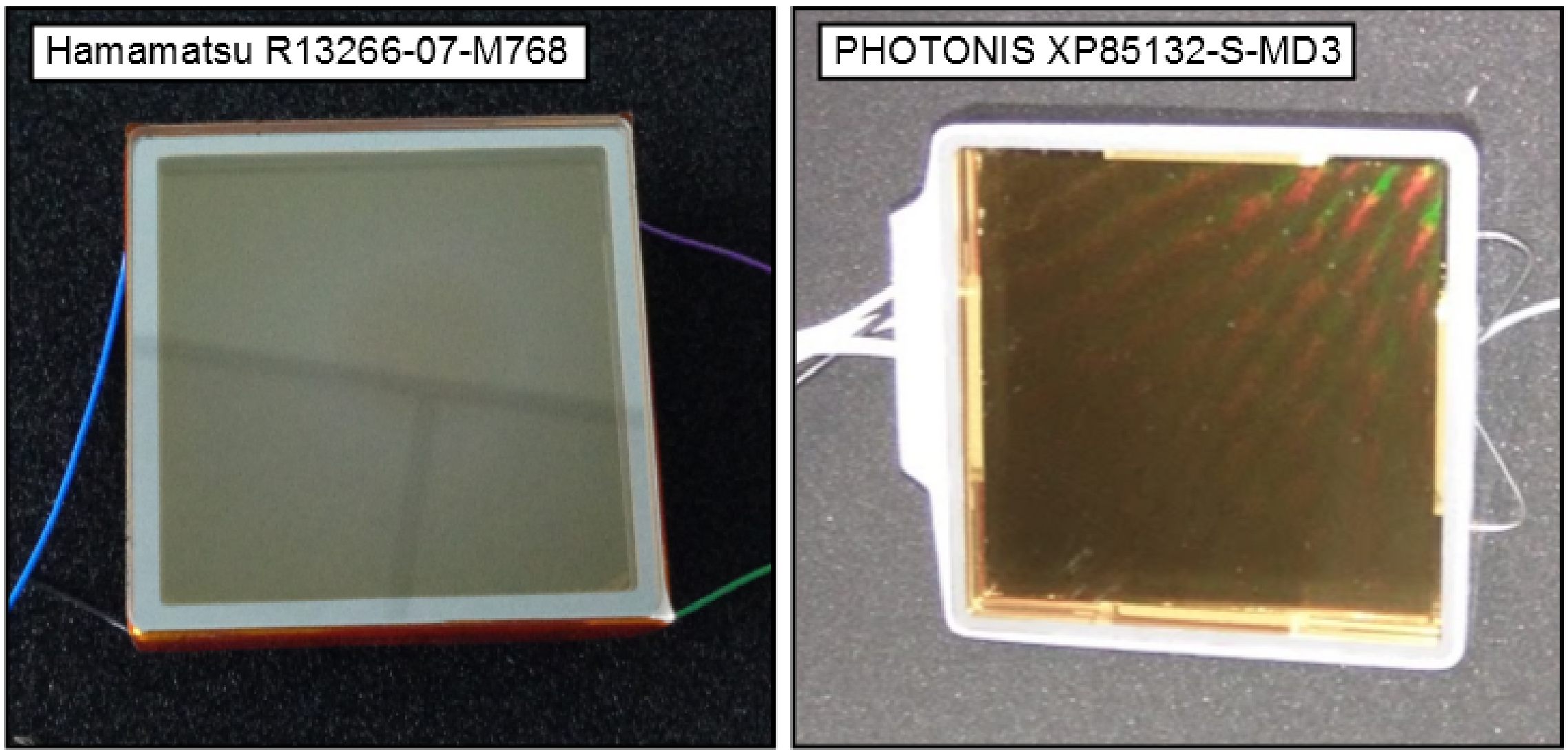

**Figure 4.44:** Photographs of prototypes of the main sensor candidates for the EDD.

#### 4.2.3.6 Highly Segmented MCP-PMTs

In comparison to the MCP-PMTs which will be used for the $\overline{\text{P}}$ANDA Barrel DIRC the EDD requires highly segmented anodes with a pixel pitch of not more than 0.5 mm. One crucial requirement of the MCP-PMTs is their position resolution. The PID performance of the entire detector heavily depends on a reliable and precise reconstruction of the points of impact of the Cherenkov photons in the focal plane of the FELs. There are different processes af-fecting the position resolution of the MCP-PMTs. In order to evaluate the resolution especially in the context of charge sharing and its dependence on an external magnetic field various measurements have been performed at the University of Gießen. Two sensors which are promising to meet the criteria for the final detector were evaluated: a Hamamatsu R13266-07-M768 with a multialkali photocathode, a 3 mm thick entry window, $6 \times 128$ pixels with a pitch of 0.4 mm and a PHOTONIS XP85132-S-MD3 with bialkali photocathode, a 2 mm thick entry window, $3 \times 100$ pixels with a pitch of 0.5 mm. Both MCP-PMTs have an active area of $53 \times 53\,\mathrm{mm}^2$ and contain MCPs of the chevron type with a channel diameter of $10\,\mu$m (see also Figure 4.44).

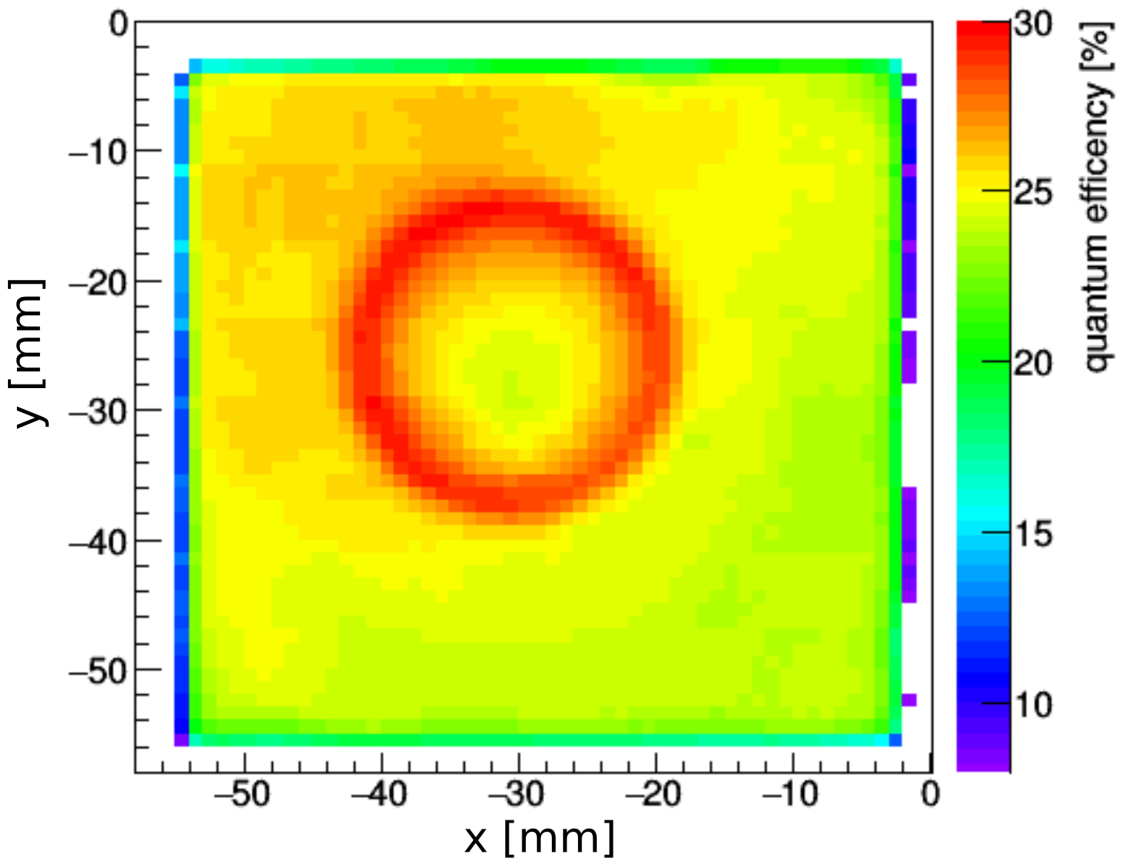

**Figure 4.45:** Two-dimensional QE chart in % at 372 nm as a function of the PC surface for the Hamamatsu R13266-07-M768 [39].

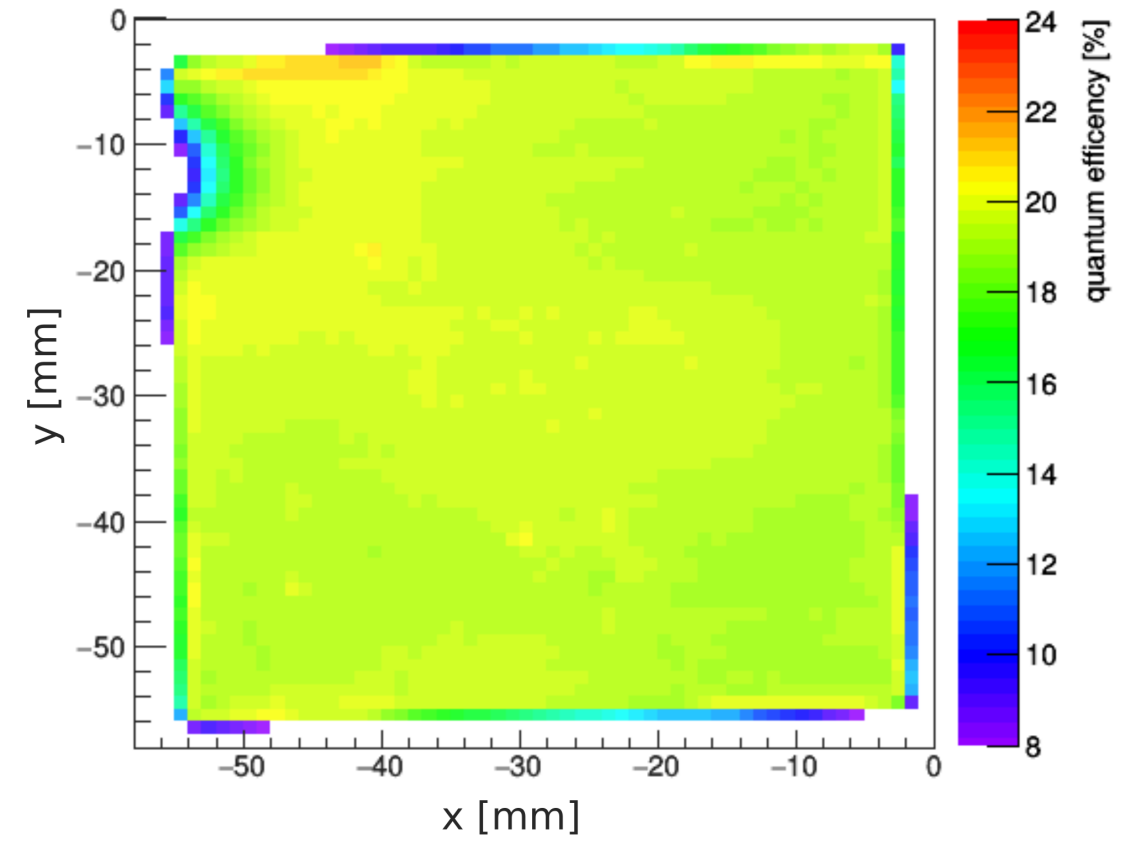

**Figure 4.46:** Two-dimensional QE chart in % at 372 nm as a function of the PC surface for the PHOTONIS XP85132-S-MD3 [39].

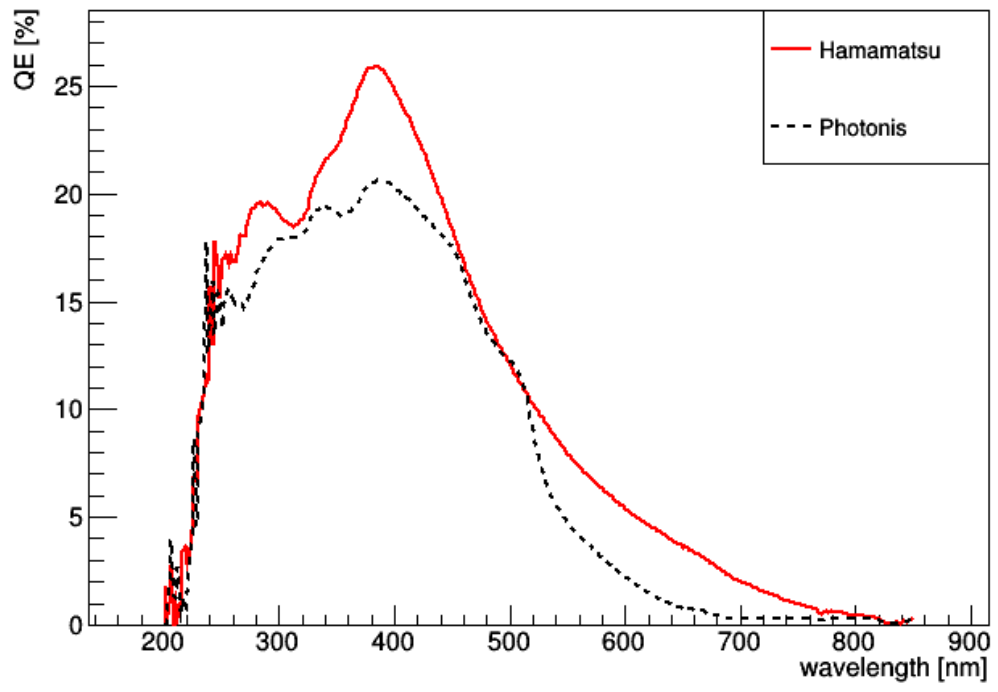

**Figure 4.47:** Spectral sensitivity of both MCP-PMT prototypes [39].

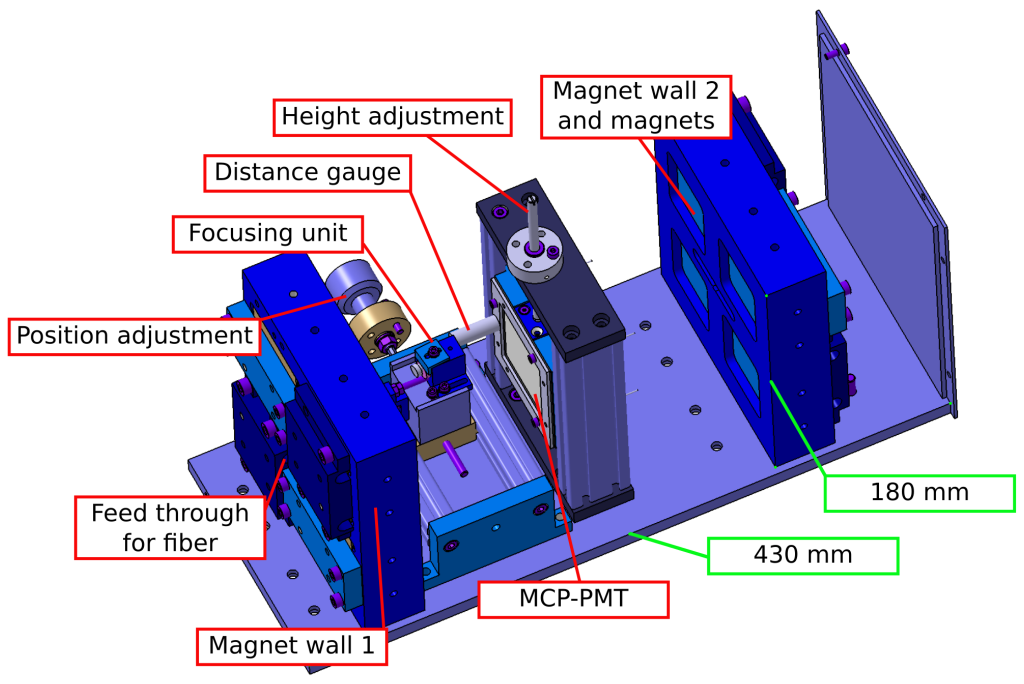

**Figure 4.48:** Setup for testing the MCP-PMTs with permanent magnets [40]. Two walls hold magnets with surface fields above 500 mT

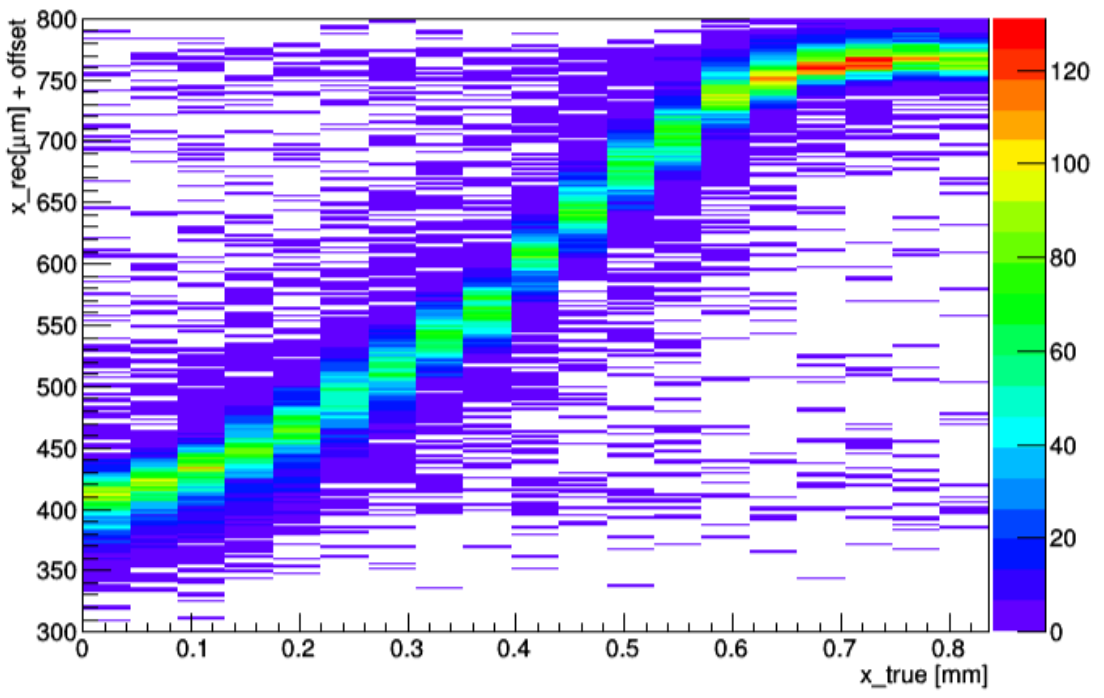

**Figure 4.49:** Reconstructed position with the Hamamatsu MCP-PMT prototype at 65 mT [39]. The high resolution is achieved by centroiding.

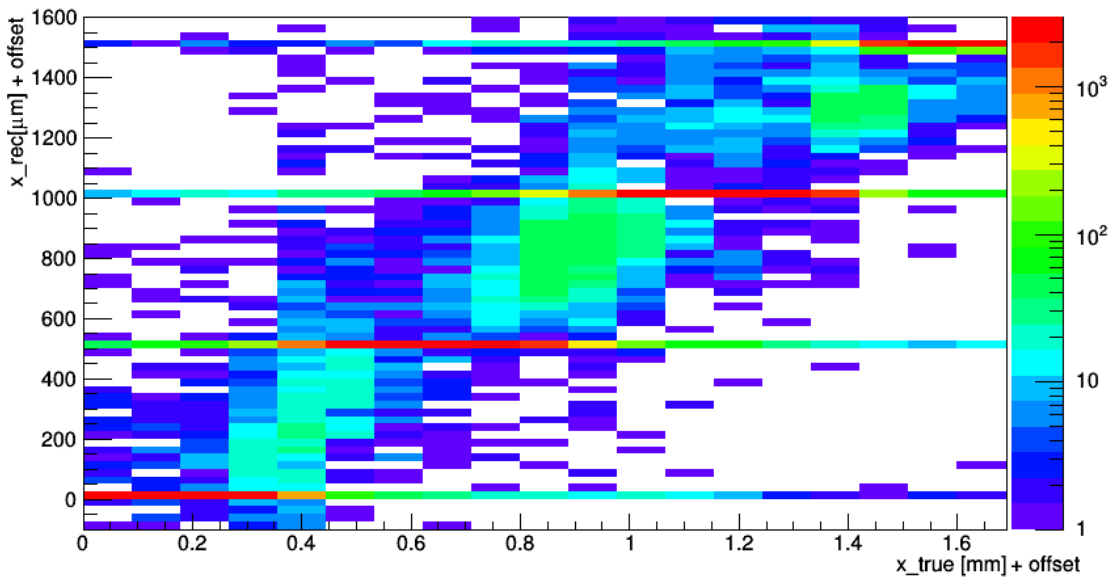

**Figure 4.50:** Reconstructed position with the PHOTONIS MCP-PMT prototype at 65 mT [39]. Most of the signals appear in the red bands.

Figures 4.45 and 4.46 show the quantum efficiencies for both sensors. The PHOTONIS MCP-PMT prototype has a better homogeneity and varies around 20% over the full active area. The QE decreases around the region where the HV supply cables have been welded. In case of the Hamamatsu MCP-PMT prototype the overall QE is higher with values between 20 and 25%. However, the measurement also reveals a circular region with an increased QE up to 30%. In addition to the spatial dependence at 372 nm the spectral sensitivity was measured as shown in Figure 4.47.

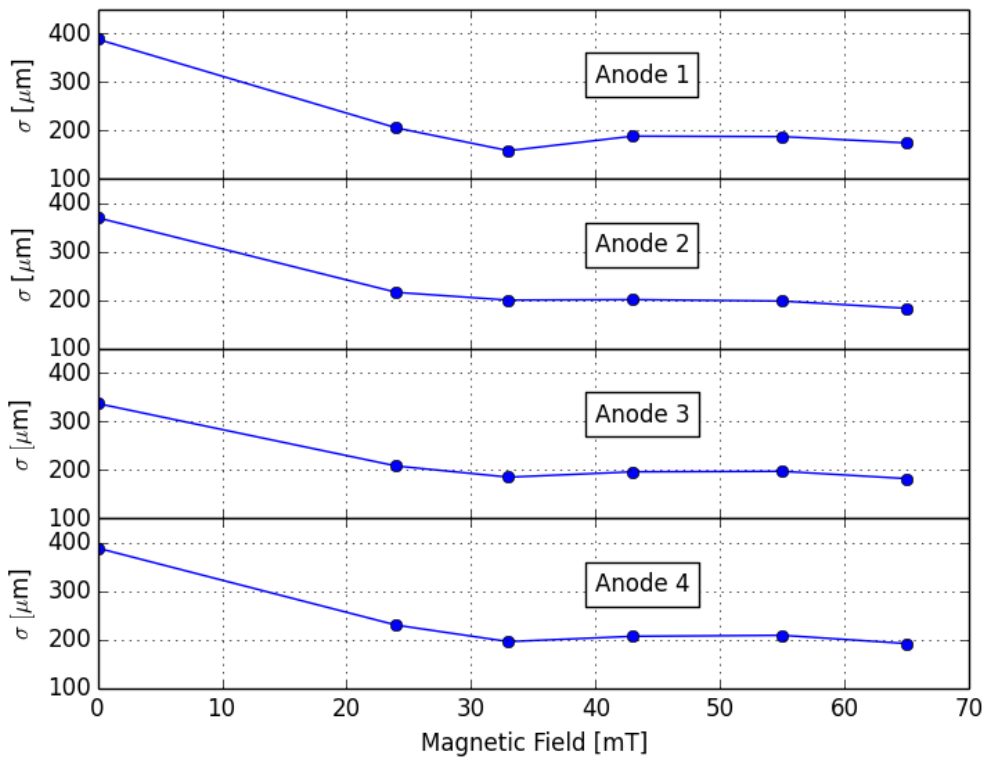

**Figure 4.51:** Width of the count rate curves as a function of the magnetic field for four neighboring anodes of the PHOTONIS MCP-PMT prototype [39].

In order to evaluate the sensors in a magnetic field a setup with permanent magnets has been developed [39]. A CAD image of the setup is shown in Figure 4.48. It includes two walls which are equipped with four permanent magnets each. The surface field strength of the magnets is about 500 mT. However, at the position of the MCP-PMT inside the setup a maximum field strength of 65 mT can be reached. One of the walls contains a feed through for a laser fiber wich can be focused on the MCP-PMT photocathode taking into account the thickness of the MCP-PMT's entry window.

Figures 4.49 and 4.50 show the results of a position scan between two (four) anodes in case of the Hamamatsu (PHOTONIS) MCP-PMT. If a threshold of 150 mV was exceeded for at least one anode of the Hamamatsu MCP-PMT the position of impact was reconstructed by linear weighting of the amount of charge collected. In case of the Hamamatsu MCP-PMT this centroiding allows to obtain a position resolution of around 20 $\mu$m. Nevertheless an increased number of active pixels translates into a higher data traffic. For the PHOTONIS MCP-PMT a threshold of 15 mV was required. The reason for the factor 10 difference between both photo sensors is related to the fact that the Hamamatsu MCP-PMT was a prototype and labeled by the company. It has been confirmed that the standard Hamamatsu and Photonis sensors have comparable gains if operated at their nominal voltages.

As shown in the corresponding figure most of the hits are recorded by a single active anode which leads to the red bands which have a width corresponding to the size of the pixel. The width of the count rate curve for the PHOTONIS MCP-PMT prototype with respect to the magnetic field inside the test setup is shown in Figure 4.51. According to this measurement the width reaches its minimum already between 30 and 40 mT. As expected the magnetic field improves the position resolution and collimates the charge cloud such that charge sharing only appears when a position between two anodes has been struck.

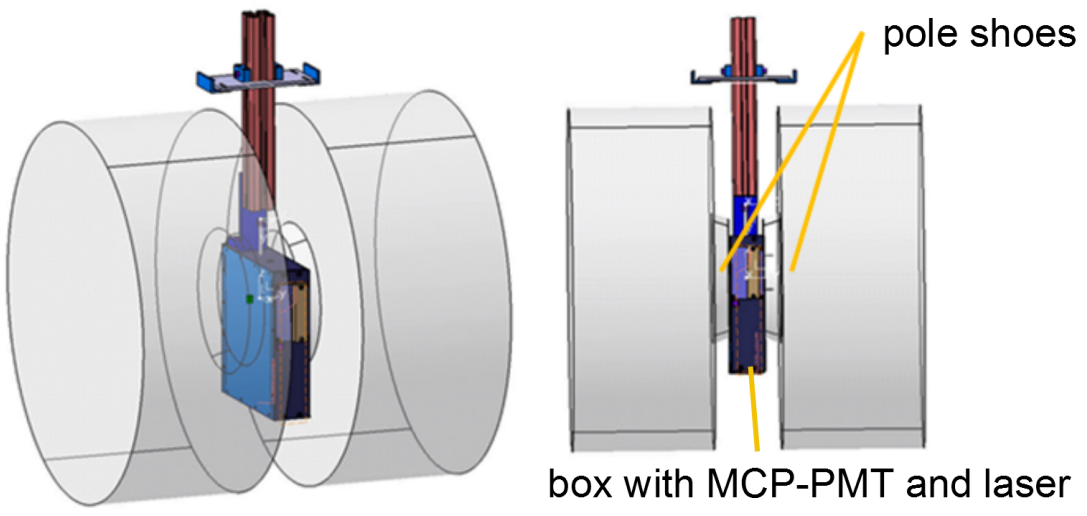

**Figure 4.52:** Setup for measuring the MCP-PMT performance inside a dipole magnet at Gießen.

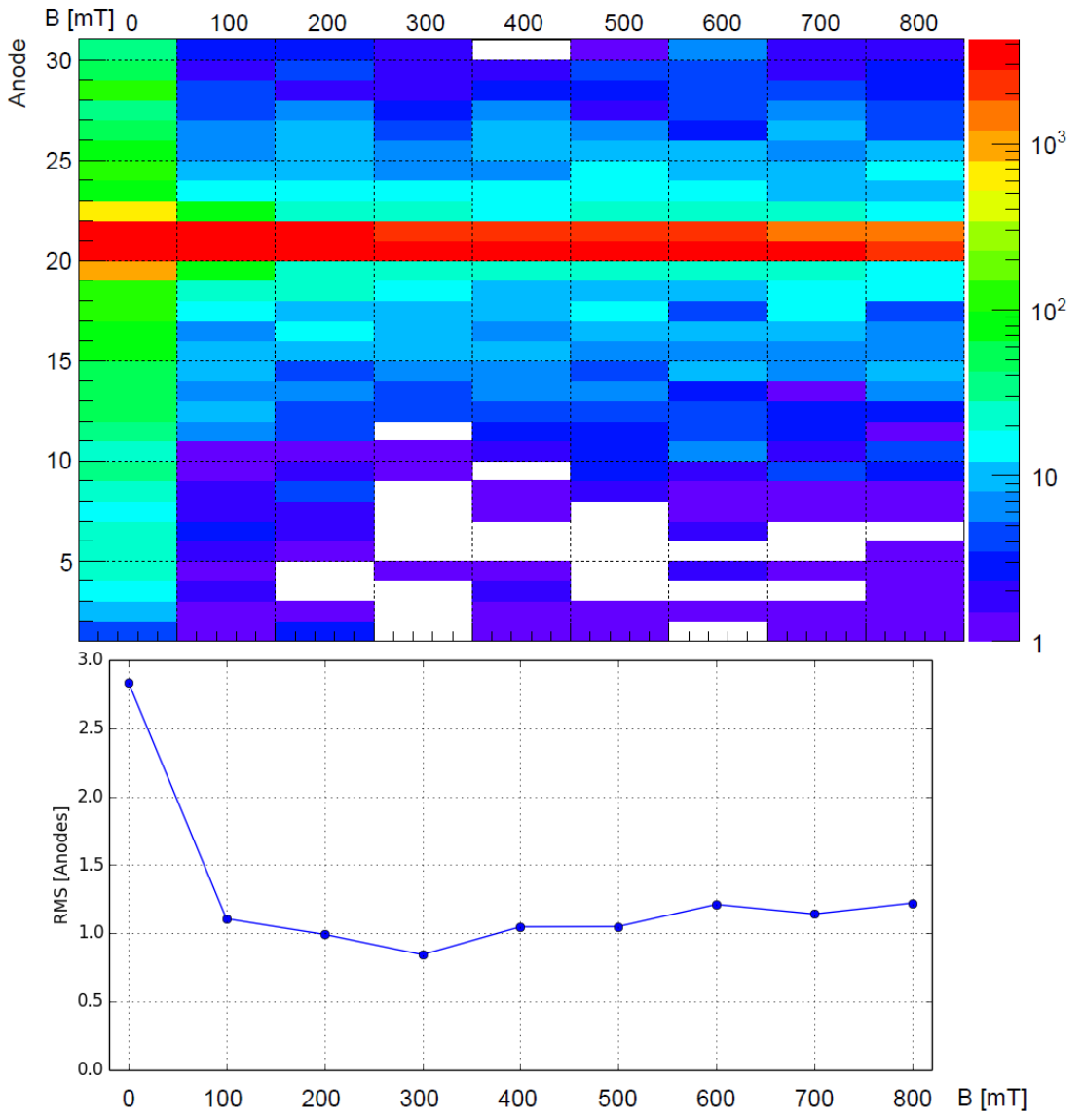

**Figure 4.53:** Scan of the magnetic field inside the dipole magnet at Gießen for the PHOTONIS MCP-PMT prototype [39]. The laser spot was placed between two anodes to maximize the effect of charge sharing.

This result was confirmed by an additional measurement using a dipole magnet at the Justus-Liebig

Universität Gießen which reaches field strengths up to 800 mT. For this measurement a special box was constructed to fit the MCP-PMT and the laser, which is now illuminating the photocathode via a mirror, between pole shoes of the magnet (see Figure 4.52) [39]. In contrast to the EDD inside $\overline{\text{P}}$ANDA field lines in this setup are perpendicular to the MCP-PMT surface. The readout was done with a TOFPET Evaluation Kit.

Figure 4.53 shows a magnetic field scan ranging from 0 to 800 mT. In agreement with the findings using the permanent magnet setup the resolution is not further reduced after moving beyond 100 mT. The laser spot was positioned between two anodes to maximize the effect of charge sharing and search for the effect of undersampling. As expected from the studies done at Erlangen and by other groups the total amount of registered hits drops slightly with an increasing field.

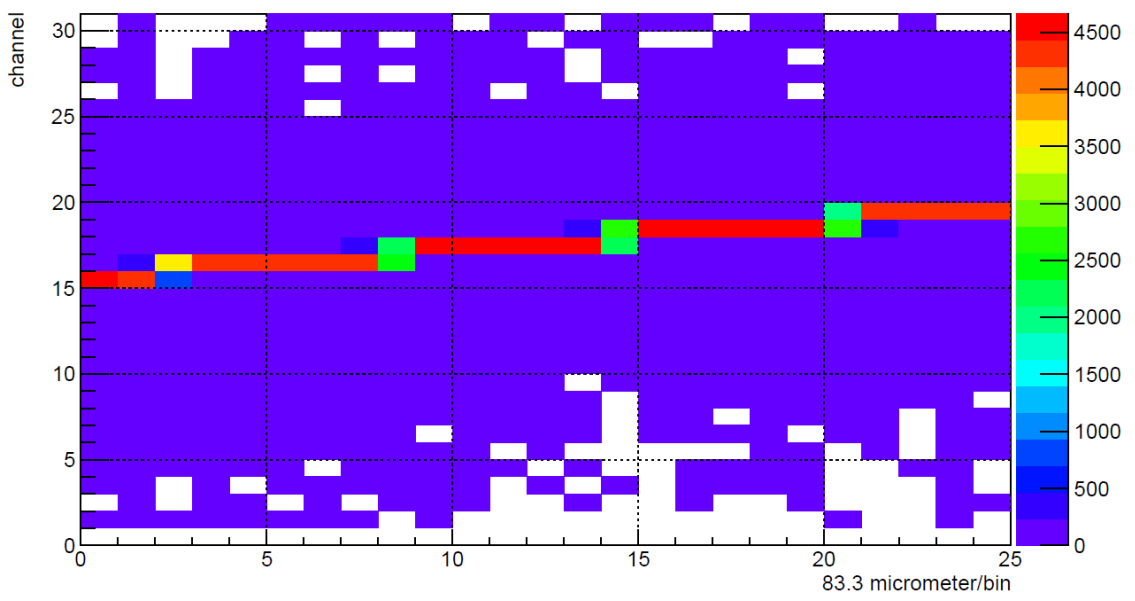

**Figure 4.54:** Position scan at 800 mT using the PHOTONIS MCP-PMT prototype [39].

Finally a position scan at 800 mT over multiple pixels was performed. The result is depicted in Figure 4.54. As previously shown the magnetic field collimates the charge cloud such that only one anode will be struck in the majority of the events. Only in the transition area between two anodes charge sharing can be observed.

### 4.2.4 Conclusions

The intensive search for suitable photon sensors for the EDD DIRC leads to the conclusion that MCP-PMTs are the most appropriate candidates. The tubes with 10 $\mu m$ pore size fulfill the requirements in terms of magnetic field immunity, time resolution, dark count rate, and gain stability at high photon rates. The recently developed techniques to prevent the photo cathodes from aging led to a "quantum jump" in the lifetime of these devices. Especially by coating of the MCPs with an ALD technique the lifetime of MCP-PMTs can be extended to values which even exceed the anticipated 10 years running of the EDD DIRC at the highest $\overline{\text{P}}$ANDA luminosity.

All important parameters (quantum efficiency, gain, etc.) of the possible MCP-PMTs are measured in the Erlangen group. In addition, the described lifetime studies of each individual MCP-PMT are ongoing together with an investigation of the occurring cross-talk and charge sharing effects. In order to monitor the behavior of the MCP-PMTs during run-time, a waveform digitizer can be integrated to read out certain channels, depending on the spatial constraints.

## 4.3 Frontend Electronics

### 4.3.1 Requirements

The frontend electronics (FEE) of the EDD includes amplification, digitization and data concentration of the MCP-PMT signals. It has to cope with a large number of channels in an environment with large magnetic field of roughly 1 T and high radiation dose from ionizing particles and neutrons fluxes. The heart of the system is an ASIC with high precision timing and capable to digitize large rates. As derived in section 4.2.1 the worst case rate requirements are in the order of 75 kHz per channel. These rate requirements do not only challenge the digitization stage of the analog signals from the MCP-PMTs but also the data transmission technology which has to handle and merge many of these channels.

A special difficulty arises from the fact that $\overline{\text{P}}$ANDA will operate in trigger-less mode. This means that the entire read-out has to run continuously, be buffered, pre-processed and hit patterns have to be partially reconstructed to be able to provide PID information for online event filtering, if required. Very important for a common event building and event reconstruction is the synchronization of the clocks of all subdetectors using the SODANET framework [41].

The FEE has to operate calibration, monitoring and slow control functionality, like setting thresholds and monitoring temperatures. Last, but not least, it has to cope with the tight volume constraints and low power consumption.

### 4.3.2 Fast Digitization

During the prototype development an FPGA based fast digitization board called TRB3 time-to-digital converter readout board (TRB3) [42, 43] was used. It is an an advanced version of the trigger readout boardX (TRB2) [44] that was originally developed for the High-Acceptance Dielectron Spectrometer (HADES) experiment. The time resolution was determined to be less than 20 ps root mean square for all the channels. The TDCs are capable to detect multiple hits, with a maximal hit rate of 50 MHz. However, using the TRB3 for the EDD has two major drawbacks: its space requirement and limited radiation hardness.

In contrast ASICs can be realized in very compact dimensions. It is even feasible to abut a PCB with ASICs directly to the back of the MCP-PMT where the analog signals are immediately digitized when they enter the ASIC. This is advantageous as no discrete analog electronics is needed, no long analog signal cables are needed, and a cross-talk free optical fibre brings the information to the outside world instead of 300 individual analog cables per MCP. Moreover, the use of compact devices reduces the material budget of the EDD and the space requirements for the FEE.

One of the biggest advantages of the ASICs is their radiation hardness. Implementing the same functionality on FPGAs would be easier to accomplish but non of the commercially available FPGA is guaranteed to work under the conditions present in $\overline{\text{P}}$ANDA where single event upsets (SEU) or permanent latch-ups compromise the functioning of the FPGA and make it error prone. These errors might result in compromised data packages or even in the failure of the entire FPGA.

Low power consumption and heat generation of the front-end electronics is another important requirement for the EDD. Even though there is enough space left to provide proper cooling to the electronics it is desirable to reduce this requirement to a minimum. In the tight volume where the EDD will be installed, temperature stability is important to minimize thermal effects on the optics. The EDD fused silica radiator plate is temperature controlled by a nitrogen gas flow system. The thermal energy that is produced by the front-end electronics has to be brought out of the tight area which has the thermal shield of the cooled endcap EMC on one side and the solenoid on the other side. ASICs can generally be designed to operate at lower levels of power consumption compared to equivalent circuits implemented in FPGAs.

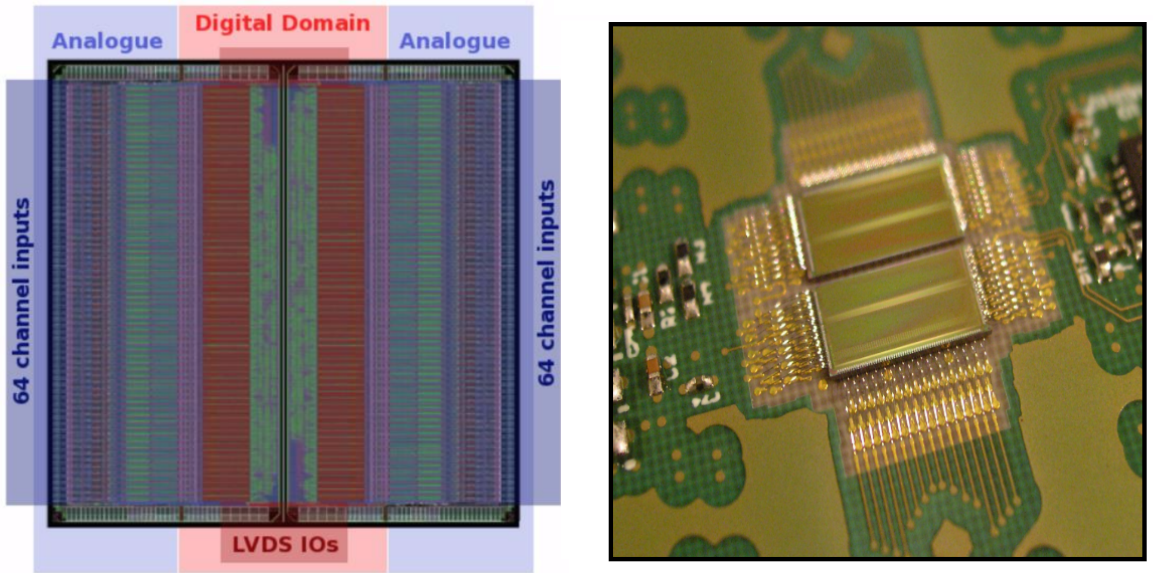

**Figure 4.55:** [45] Two abutted TOFPET ASICs that create a compact readout unit with 128 channels.

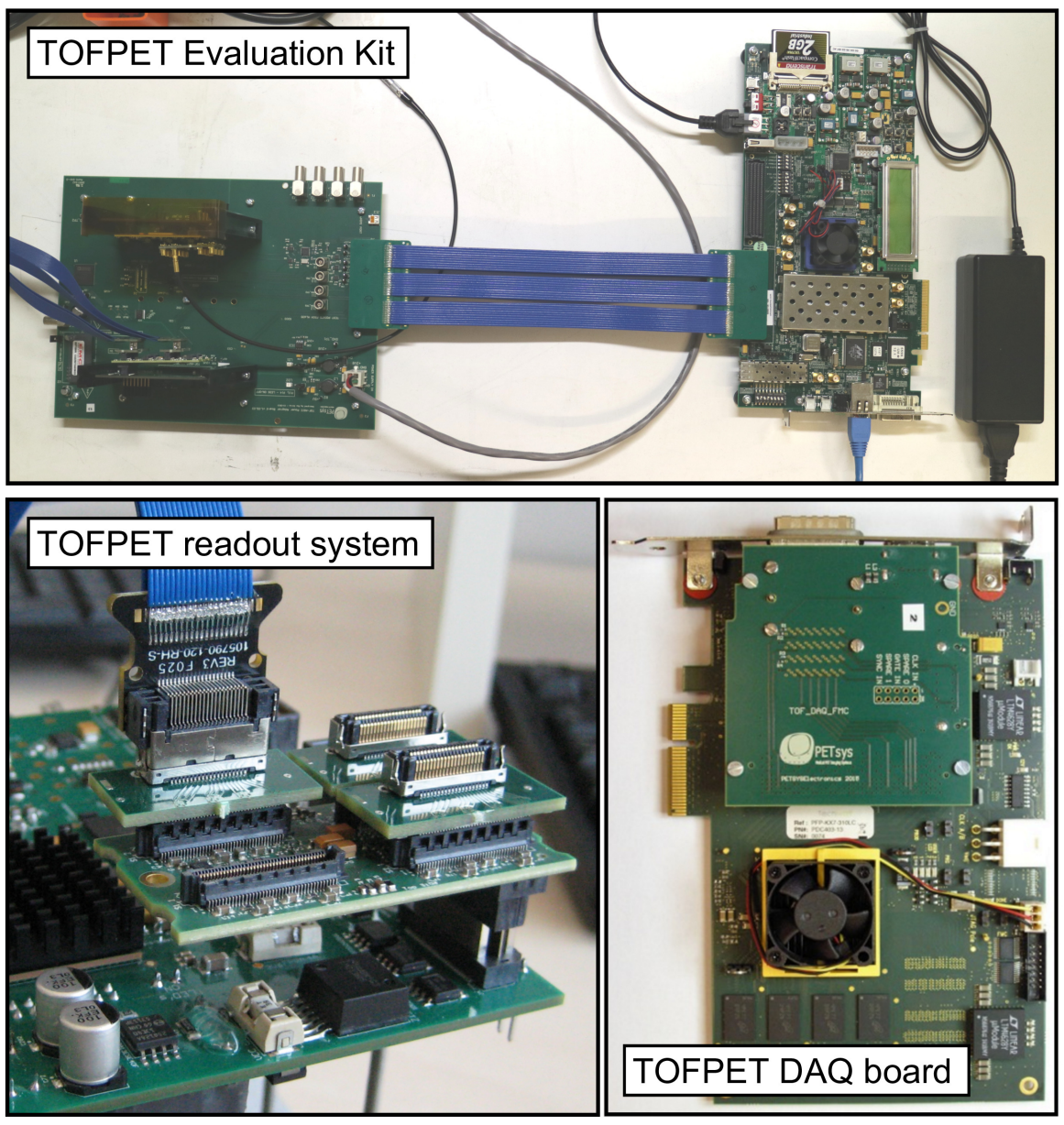

**Figure 4.56:** A TOFPET Evaluation Kit with 256 channels (top). A TOFPET readout with different frontend boards which contain the TOFPET ASICs and FPGAs to further process the data (lower left). The system communicates with a special DAQ board which is build into a PC (lower right).

#### 4.3.2.1 TOFPET ASIC

The proposed TOFPET ASIC (shown in Figure 4.55), a commercially available product distributed by PETSYS Electronis [46], has been evaluated as the most promising candidate. It has the advantage that some of the developers of the chip are part of the $\overline{\text{P}}$ANDA collaboration. Other detector subgoups of $\overline{\text{P}}$ANDA like the Barrel Time-of-Flight and MVD will also use the TOFPET or a modified versions of the TOFPET for their FEE [47, 48]. An integration into the EDD FEE should be doable with moderate design efforts.

The intrisic time resolution of the chip corresponds to 25 ps RMS and allows a charge measurement via time-over-theshold. Each ASIC features 64 input channels with built-in amplifiers that provide a dynamic range of 300 pC for the total input charge but is also able to do photon counting. It operates at clock frequencies between 80 and 160 MHz and uses LVDS for digital IO. A maximum channel hit rate of 160 kHz can be reached with a maximum data output rate of 160 kHz. The power consumption stays comparably low with 8 − 11 mW per channel. Although there is enough space left to provide proper cooling to the electronics it is desirable to reduce this requirement to a minimum. It is planned to cool the electronics with a dedicated liquid or gas cooling system close to the ASICs and the FPGA. A prototype cooling setup, based on alcohol cooling, has already been successfully tested. It allows to set the temperature of the electronics in a 10°C range on the ASICs surface which is suitable for our design.

In the tight volume where the EDD will be installed, temperature stability is important to minimize the thermal effects on the optics. The EDD fused silica radiator plate is temperature controlled by a nitrogen gas flow system. The thermal energy that is produced by the front-end electronics has to be brought out of the tight area which has the thermal shield of the cooled endcap EMC on one side and the solenoid on the other side. Currently, a second improved version of the chip is developed which includes a reduced power consumption between 5 and 8 mW per channel and an increased event rate up to 600 kHz per channel. Until now, there is no reliable estimation about the heat flow from EMC to the EDD volume. Therefore, the required cooling power and coolant flow has to be adjusted according to the final design of PANDA.

Originally, the TOFPET ASIC was developed for the readout of SiPMs in medical PET applications. Tests with single photon signals from SiPMs have been carried out successfully and show that the ASIC is well suited for digitizing these signals. In order to verify their compatibility with MCP-PMT photo sensors two setups were tested with available sensors: a TOFPET Evaluation Kit providing four ASICs with a total of 256 channels and a full TOFPET system consisting of a DAQ card and several frontend boards to read out a larger amount of channels (see Figure 4.56).

Both systems were tested with the available PHOTONIS MCP-PMT prototype. Printed circuit board (PCBs) were designed to provide a reliable electrical connection for all pins on the backside of the sensor as well as to reroute the channels according to the needs of the TOFPET readout. Because of the high channel density fine pitch SMD components were used. The analog signals were transmitted via shielded low pitch coaxial cables. All cables and PCBs were designed with a transmission line impedance of 50 Ω to reduce signal reflections on electrical interfaces.

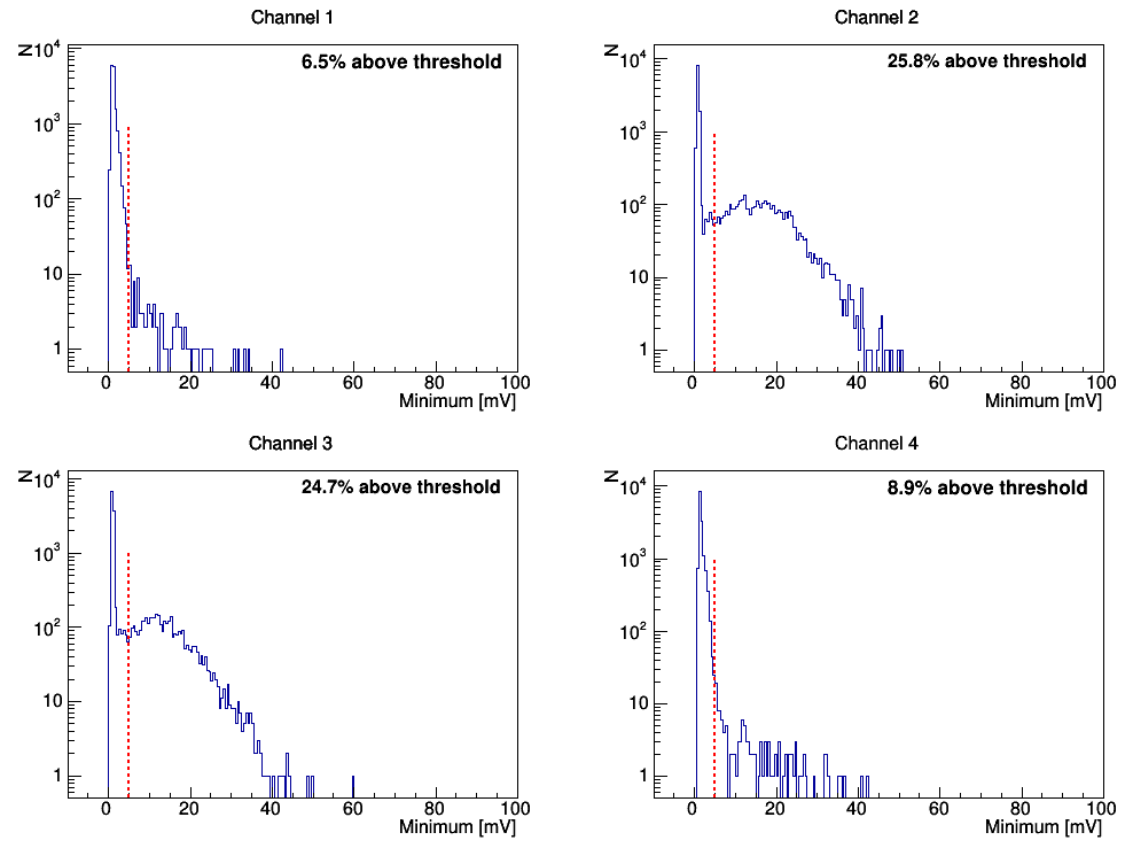

**Figure 4.57:** Amplitude distributions of the PHOTONIS MCP-PMT prototype under single photon illumination measured with a scope [39] . The focused light spot is placed between channel 2 and 3 to test the case with maximum charge sharing.

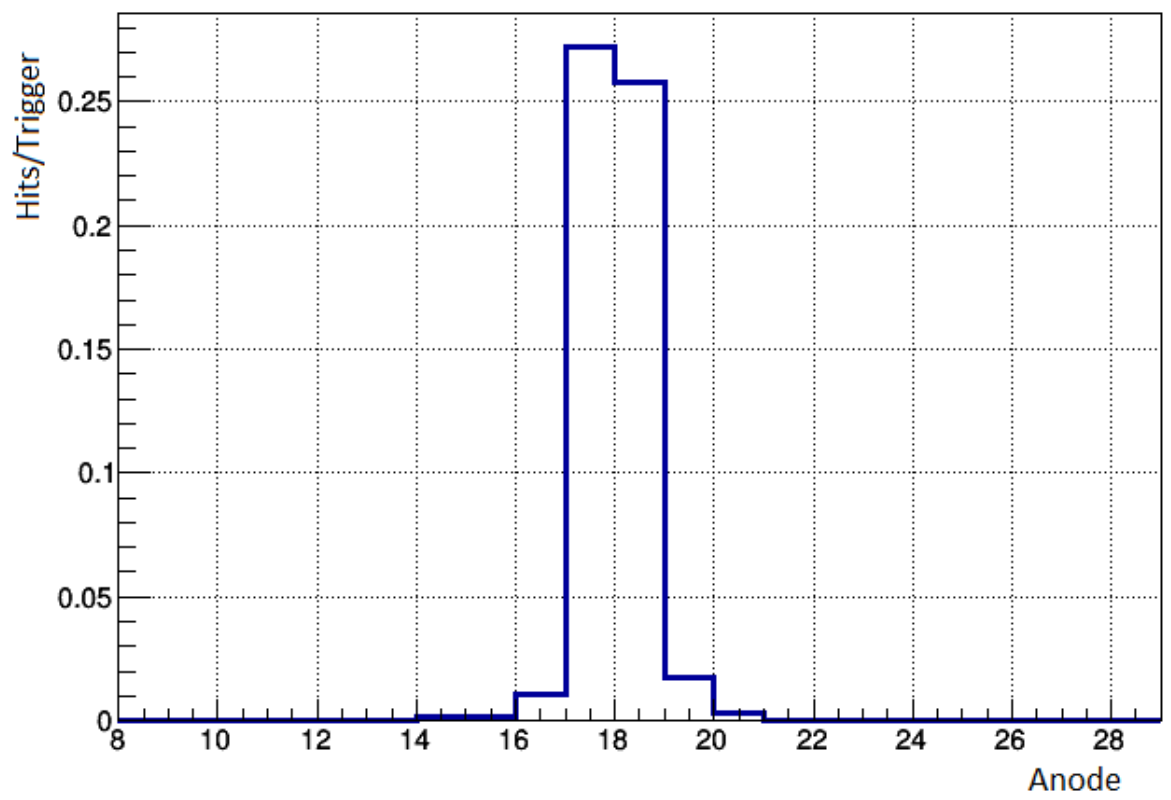

**Figure 4.58:** Spatial distribution of the PHOTONIS MCP-PMT prototype under single photon illumination measured with the TOFPET ASIC [39]. All other parameters of this measurement are as shown in Figure 4.57.

Small pulses with amplitudes between 1 and 2 mV were transmitted without distortions of the pulse shapeAn efficiency scan showed that at least 99.44% of the pulses that were created by a pulse generator and were fed into the ASIC through a capacitor could be registered. To evaluate the TOFPET performance for single photon signals coming from MCP-PMTs the setup shown in Figure 4.48 with a magnetic field strength of 65 mT was used. In total four neighboring channels were read out. The laser spot was placed between the central two channels to maximize the effect of charge sharing. The intensity of the laser was set to a single-photon level (with a fraction of detected multi-photon events below 3.4%). Figure 4.57 shows the result of the first measurement which was done with a scope. As soon as a pulse exceeds a threshold of 5 mV it was counted. The two central pixels registered a photon in about 25% of the pulses and the outer channels counted photons in less than 9%.

After this measurement the TOFPET ASIC was connected to the test setup leaving all remaining parameters unchanged. With the TOFPET a large part of the column (64 channels) could be read out. After applying proper time-cuts to elimate dark counts Figure 4.58 was obtained. Similar to the measurement that was done with the scope the central columns registered a photon in about 25% of the pulses. This measurement shows that the TOFPET ASIC is in fact well suited for efficient single photon detection from MCP-PMTs [39].

All studies will be repeated with the new TOFPET ASIC version 2 and a dedicated readout board designed PETSYS Electronics. A first test of this new version showed an unstable behavior in the amplification and threshold settings. This bug was found during a testbeam campaign at CERN in 2018 and is going to be fixed in the next hardware revision.

### 4.3.3 Data Concentration and Computing

The data concentrator has an impact on and control of the data rate, data loss and cooling infrastructure of the detector. Its task is to provide a network link to DAQ and slow control. The DAQ requires time tagging of individual hits based on the common $\overline{\text{P}}$ANDA reference time distributed by the SODA (synchronization of data acquisition) system.

Currently two options are shortlisted to be used for the EDD: an FPGA, e.g. the Altera Arria GX which has been found to be compatible with the environment or a dedicated ASICs [1]. Such an ASIC has been developed by the CERN community called gigabit transceiver (GBT) [49] which comprises a feature-rich architecture for data transmission, a radiation hard optical link (called Versatile Link), a serializer-deserializer ASIC (SERDES) and an optical slow control interface ASIC (GBT-SCA). The user rate provided by the GBT chipset goes up to 3.28 Gb/s.

An important feature in connection with $\overline{\text{P}}$ANDA and SODA is the possibility to provide an external reference and hence synchronization of the GBTX clock. The TOFPET ASIC is completely compatible with the GBTX and up to nine TOFPET ASICs can be read out at 320 MBit/s. Due to the simplicity of the solution and the minimal development effort in comparison with the implementation of such features on FPGA the favourite solution is to combine these two ASICs (see also Figure 4.59).

The recorded data will be transfered to the outside of the dector via an optical link with up to 6.5 Gb/s. There it will be further processed by compute nodes which can be a simple computer, multi-processor system or an FPGA board, like the multipurpose, ATCA (advanced telecommunications computing architecture) compliant FPGA board which has been developed for the $\overline{\text{P}}$ANDA DAQ System [50]. Such actions can be on any kind ranging from cluster finding to online particle identification. Due to the high effort of implementing certain algorithms onto an FPGA by using a hardware description language it is advisable to consider FPGA solutions only in cases where software solutions on conventional compute nodes are limited in performance or cost [1].

## 4.4 Detector Control, Monitoring and Calibration

### 4.4.1 Detector Control System

The detector control system (DCS) for $\overline{\text{P}}$ANDA will be based on EPICS which is a set of Open Source software tools, libraries and applications developed collaboratively and used worldwide to create distributed soft real-time control systems for scientific instruments. Details of implementation will be done at a later stage to adapt common $\overline{\text{P}}$ANDA standards in the graphical user interfaces (GUI).

The following list of parameters has to be monitored for the EDD:

- high voltages (HV) and currents of the MCP-PMTs,
- supply voltages (LV) and currents of the read-out electronics,
- temperatures at the PCBs and the optical elements,
- gas flow rate and temperatures of the dry nitrogen that flushes the radiator plate and the read-out electronics.

Figure 4.60 shows an overview of the EDD DCS. The system enables the shift crew and the detector experts to control the detector operation, to archive the detector parameters, to modify detector parameters, and to have tools for visualization, calibration and simple analyses of the device. Automatic error detection, alarm and shutdown operations should be included.

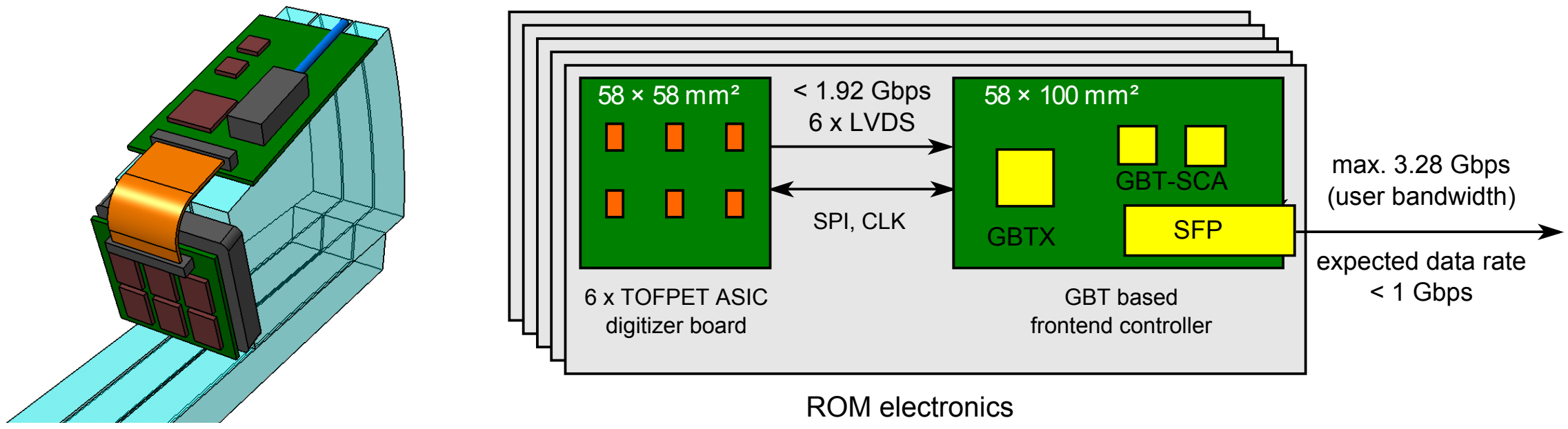

**Figure 4.59:** Left: Position of the ASICs and FEE close to the MCP-PMT on the FEL which together form the readout chain for one ROM (the ROM cover has been removed). Right: Schematic for the FEE.

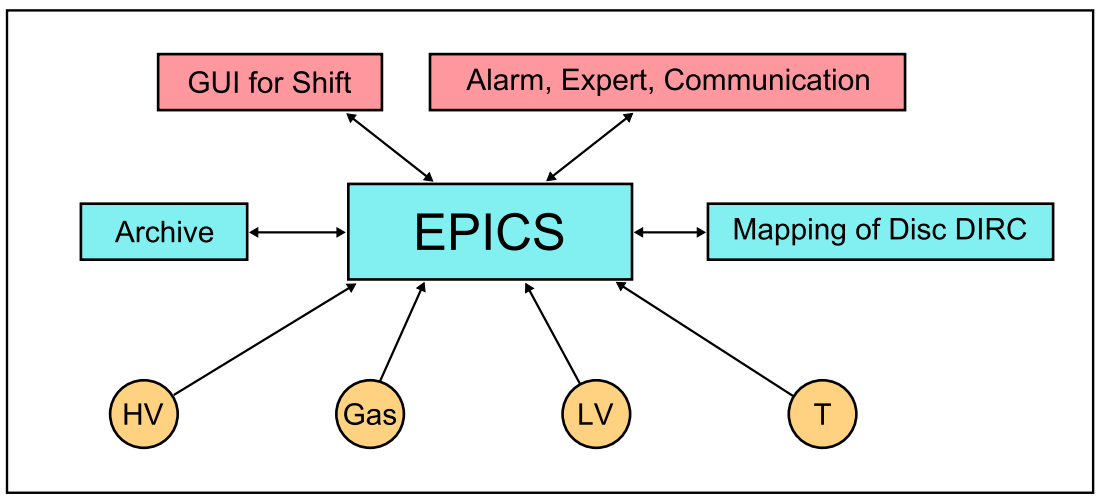

**Figure 4.60:** Disc DIRC detector control scheme (DCS). The software should use Control System Studio (CSS) and communication between EPICS and GUI.

## 4.4.2 Laser Monitoring System

As described above, the EDD slow control parameters like HV, LV, temperatures etc. will be monitored and will activate alarm signals on failure.

Beyond that, a monitoring system that controls the stability of the optical system, the functionality of the photon sensors and of the read-out system would be of great help for early failure and error detection of the whole EDD system. One of the most complex detector monitoring system is the one build for the ATLAS Tile-Calorimeter, where three types of controls were used: a charge injection system (CIS) to check the FEE, a Laser Pulse Injection System to monitor the photon detector response, and a radioactive source for a long-term gain control [51]. As the EDD works in a single photon mode, the absolute single photon calibration of the photon sensors can be done offline. For our application, the most important control system is a laser monitoring system (LMS), as it will allow to perform a fast check before data taking, where we can verify that all pixels are operating and that the optical transmission and reflection properties are stable.

As shown in Figure 4.61 a pulsed laser beam is split into 5 fractional beams. One of them is fed to a calibrated photon detector to monitor the power of the laser beam pulses. The other 4 beams of equal intensity are fed into an EDD quadrant each and illuminate the radiator plate in an approximate homogeneous way. Each pixel will register photons at a well defined rate. The system will monitor any degradation of the optical transport system and of individual pixels or whole MCP-PMTs. Timing and gain can be calibrated with this system.

The laser, beam splitter and the calibrated photon detector are situated outside the $\overline{\text{P}}$ANDA detector, preferably in a temperature controlled, dry and light-tight box. Optical fibers will transport the light inside the $\overline{\text{P}}$ANDA detector. The fibers are placed along the stabilizing cross between the radiator plates and will inject the light at the diamond shaped inner rim of the quadrants isotropically into the radiator plates. For radiation hardness, the optical fibers should be quartz fibers. Inside the plates the light will be transported by internal total reflection to the MCP-PMTs.

A simplified version of the LMS has been tested successfully at our DESY/CERN prototype DIRC. As photon source we used a PiLas laser emitting a pulse of about 400 mW with a very small pulse jitter ($\sim$ 40 ps). Figure 4.62 shows the functionality of 16 channels of the PMT number 7, used in the test experiment and tested by the laser pulse system. These measurements were done within a few minutes in between test runs and were archived for the offline analysis to document efficiencies and relative timing.

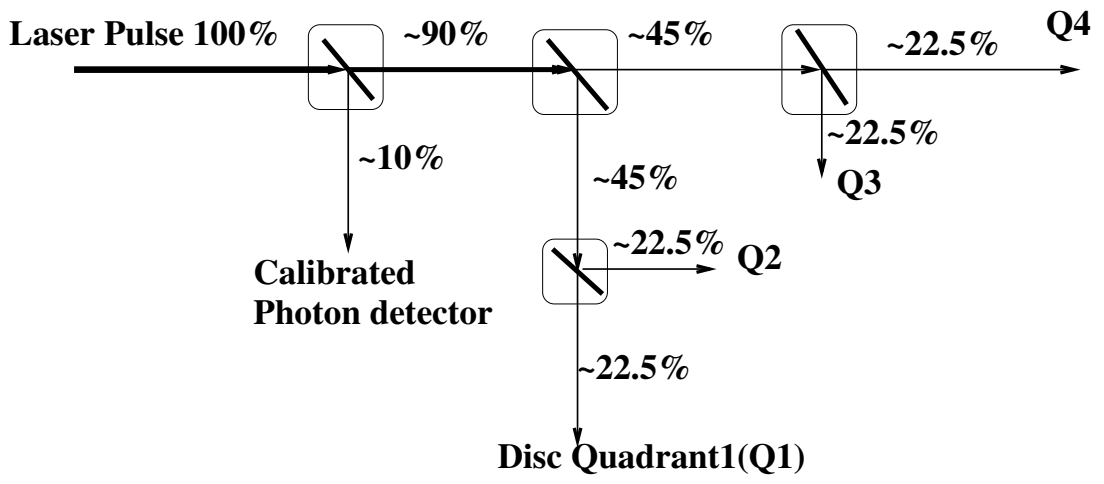

**Figure 4.61:** Laser beam injection system: A pulsed laser beam is split in to four beams that enter the four EDD quadrants, and in to a 5th beam that is monitored by a calibrated photon detector and is used to control the stability of the laser.

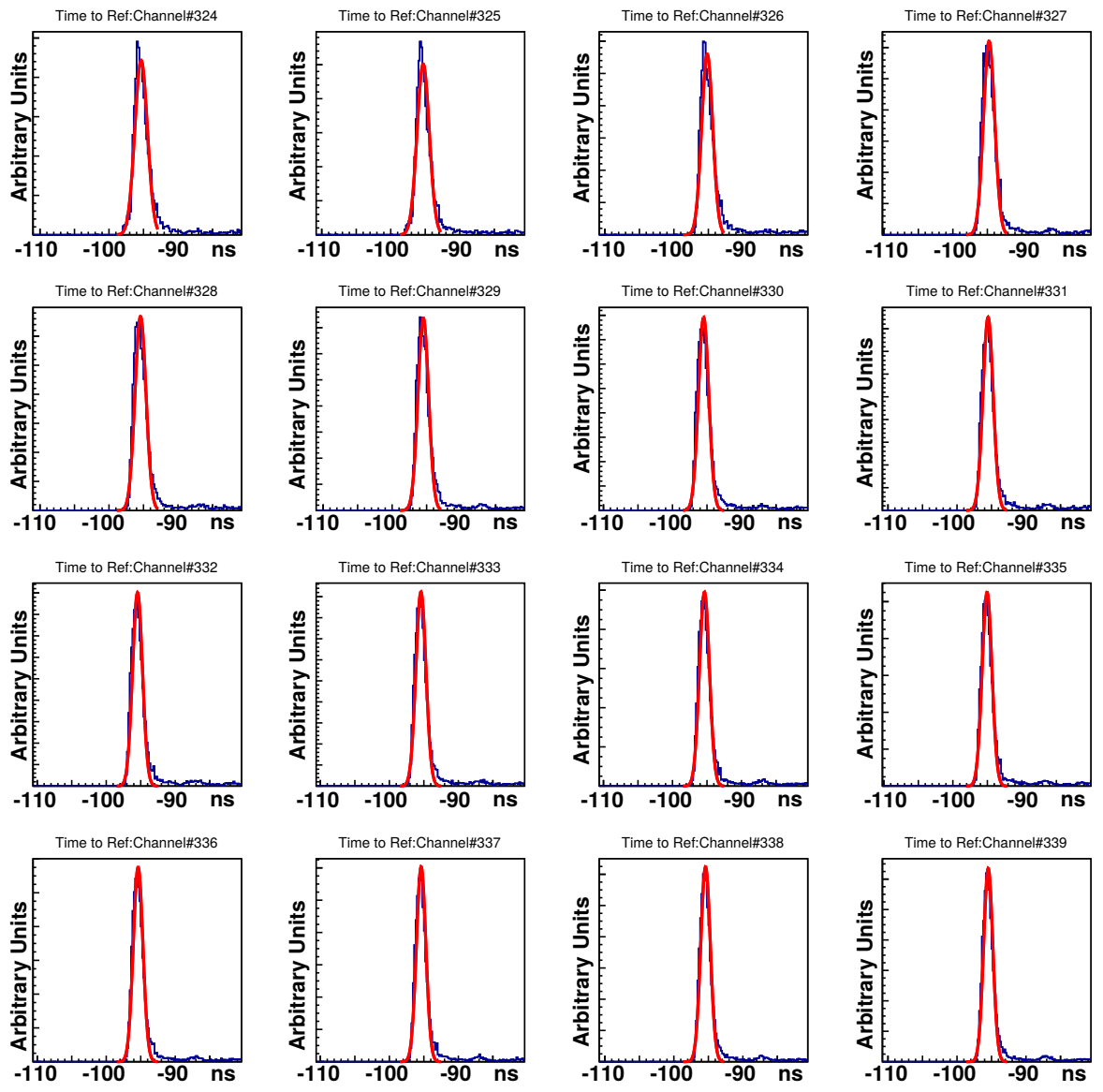

**Figure 4.62:** A simplified laser monitoring system has been tested at the DESY and CERN prototypes. This figure shows result of one run, done in $\approx$ 2 minutes, for each of the 16 channels of the Hamamatsu PMT (PMT number 7). Plotted is a histogram of the count rate as function of the time difference between trigger signal (trigger output from laser) and the single channel PMT responce. Using this scheme one can monitor the timing, the time spread, and the integral of detected photons. This information will be collected in the EDD calibration database continously. For all 27 functional PMT we have similar spectra.

### 4.4.3 Optical Calibration and Alignment

If a particle traverses the radiator plate and produces Cherenkov light, the complicated Cherenkov pattern is registered by certain pixels of the numerous MCP-PMTs. The reconstruction software calculates back the Cherenkov angle from the known pattern. To be able to do that, it needs the exact position and orientation of all optical elements and of the photon sensors and their pixels. The optical measurements with a laser beam during the production and assembly of the quadrants will determine all these parameters. Fitting and likelihood methods can check the measured photon patterns during operation to monitor charges of the optical system or even correct for imperfections of the optics.

### 4.4.4 In-beam Calibration and Alignment

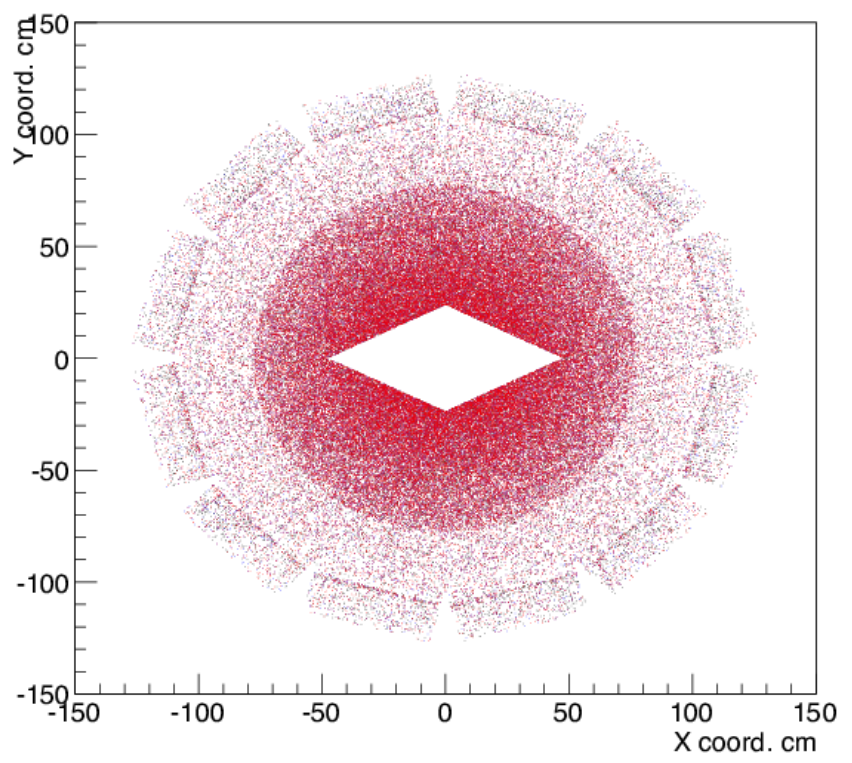

**Figure 4.63:** The hit map of the EDD radiator plate is well filled with pions (black) and protons (red) from the process $\bar{p}p \to \Lambda\bar{\Lambda}$. Charged kaons (blue) come from the process $\bar{p}p \to \Phi\Phi$.

It is highly recommended to verify the functionality of the quadrant in a real particle beam before they are installed in $\overline{\text{P}}$ANDA. As a first check, they will be measured in horizontal position on the laser table in Giessen, using the cosmic test stand. It is also possible to measure two or more quadrants on top of each other to compare the results for individual cosmic muons. Especially the comparison of two or more photons from the same cosmic muon in a single radiator gives useful information on the single photon resolutions.

Nevertheless, if the FAIR schedule allows for it, the quadrants will be tested in an external beam faciity with good spatial, angular and time resolution and either different particle species or different velocities of the same particle will be used, to crosscheck the measured Cherenkov patterns with the predicted ones. As the resolution of the MCP-PMTs improves in a magnetic field, a test of the final performance of the EDD will show up only after installation in $\overline{\text{P}}$ANDA, as a magnet that fits the required strength and direction of the $\overline{\text{P}}$ANDA magnet will not be available at test beam facilities.

After installation in $\overline{\text{P}}$ANDA, the large, dominating pion background can be used to do an online calibration and alignment using the measured pion tracks.

A more sofisticated verification of the PID capabilities of the EDD uses pions, muons, kaons and protons in $\overline{\text{P}}$ANDA which are identified independently of the EDD by kinematic fits, vertex fits or identification from other detectors. The muons can be identified by the muon chambers and give a clean source of $\beta = 1$ particles. Decays from pair production of $\Lambda\overline{\Lambda}$, $\Phi\Phi$ or $K_s K_s$ can be used to get well defined, clean samples of pion, kaon and proton according to the decays

$$\overline{p}p \rightarrow \Phi\Phi \rightarrow K^+K^-K^+K^- \quad (4.5)$$
$$\overline{p}p \rightarrow \Lambda\overline{\Lambda} \rightarrow p, \overline{p}, \pi^-, \pi^+ \quad (4.6)$$
$$\overline{p}p \rightarrow K_s K_s \rightarrow 4\pi^\pm. \quad (4.7)$$

Figure 4.63 shows the distribution of hit patterns of certain channels on the EDD radiator plate.

## References

[1] O. Merle. Development, design and optimization of a novel Endcap DIRC for PANDA. PhD thesis, Justus-Liebig-Universität Gießen, 2014.

[2] M. Hoek et al. Tailoring the radiation hardness of fused silica. *Nucl. Instr. Meth. A*, 639:227–230, 2011.

[3] J. Cohen-Tanugi et al. Optical properties of the dirc fused silica cherenkov radiator. *Nucl. Instr. Meth. A*, 515:680–700, 2003.

[4] Corning Incorporated (USA). One Riverfront Plaza, Corning, NY 14831, USA.

[5] Heraeus Holding GmbH. Heraeusstraße 12-14, 63450 Hanau, Germany.

[6] Nikon Corporation Glass Business Unit. 10-1, Asamizodai 1-chome, Minami-ku, Sagamihara, Kanagawa 252-0328, Japan.

[7] Schott North America, Inc. 555 Taxter Road Elmsford, NY 10523, USA.

[8] E. Etzelmüller. Developments towards the Technical Design and Prototype Evaluation of the PANDA Endcap Disc DIRC. PhD thesis, Justus-Liebig-Universität Gießen, 2017.

[9] E. Etzelmüller et al. Tests and developments of the PANDA Endcap Disc DIRC. *J. Instrum.*, 11:C04014, 2016.

[10] PANDA Collaboration. Technical Design Report for the PANDA Barrel DIRC Detector. arXiv:1803.10642, 2017.

[11] PANDA Collaboration. Technical Progress Report. 2005.

[12] S. Spanier B. N. Ratcliff. DIRC dreams: research directions for the next generation of internally reflected imaging counters. *Nucl. Instr. Meth. A*, 433:456–463, 1999.

[13] O. Merle et al. Development of an Endcap DIRC for PANDA. *Nucl. Instrum. Meth.*, A766:96–100, 2014.

[14] K. Inami et al. TOP counter for particle identification at the Belle II experiment. *Nucl. Instr. Meth. A*, 766:5–8, 2014.

[15] APM Technica AG. Max-Schmidheiny-Str. 201, 9435 Heerbrugg, Switzerland.

[16] Norland Products Inc. 2540 Route 130, Suite 100, Cranbury, NJ 08512, USA.

[17] Berliner Glas KGaA Herbert Kubatz GmbH & Co. Waldkraiburger Straße 5, 12347 Berlin, Germany.

[18] B. Morosov et al. Proposal of a disc DIRC. *GSI Document Server*, DOC-2004-Dec-68, 2004.

[19] K. Föhl. A Focussing Disc DIRC design for particle identification in PANDA. *Hyperfine Interact.*, 194:171–175, 2009.

[20] Epoxy Technology, Inc. 14 Fortune Drive, Billerica, MA 01821, USA.

[21] Compagnie de Saint-Gobain, Zweigniederlassung Deutschland. Krefelder Straße 195 , 52070 Aachen, Germany.

[22] Eljen Technology. 1300 W. Broadway, Sweetwater, TX 79556, USA.

[23] Cargille Laboratories. 55 Commerce Rd., Cedar Grove, NJ 07009, USA.

[24] M. Düren et al. The Endcap Disc DIRC of PANDA. *Nucl. Instr. Meth. A*, in press, 2017.

[25] N. Kishimoto et al. Lifetime of MCP–PMT. *Nucl. Instr. Meth. A*, 564:204–211, 2006.

[26] A. Britting et al. Lifetime-issues of MCP-PMTs. *J. Instrum.*, 6:C10001, 2011.

[27] Advanced Laser Diode Systems A.L.S. GmbH. Schwarzschildstraße 6, 12489 Berlin, Germany.

[28] A. Lehmann et al. Performance studies of microchannel plate PMTs in high magnetic fields. *Nucl. Instr. Meth. A*, 595:173–176, 2008.

[29] A. Lehmann et al. Position sensitive photon detectors for nuclear physics, particle physics and healthcare applications. *J. Instrum.*, 4:P11024, 2009.

[30] T. Jinno et al. Lifetime-extended MCP-PMT. *Nucl. Instr. Meth. A*, 629:111–117, 2011.

[31] M.Yu. Barnyakov and A.V. Mironov. Photocathode aging in MCP PMT. *J. Instrum.*, 6:C12026, 2011.

[32] D.R. Beaulieu et al. Nano-engineered ultra-high-gain microchannel plates. *Nucl. Instr. Meth. A*, 607:81–84, 2009.

[33] M. Wetstein et al. Development of sub-nanosecond, high gain structures for time-of-flight ring imaging in large area detectors. *Nucl. Instr. Meth. A*, 639:148–150, 2011.

[34] O.H.W. Siegmund et al. Atomic layer deposited borosilicate glass microchannel plates for large area event counting detectors. *Nucl. Instr. Meth. A*, 695:168–171, 2012.

[35] A.Yu. Barnyakov et al. R&D of microchannel plate phototubes. *Nucl. Instr. Meth. A*, 567:17–20, 2006.

[36] F. Uhlig et al. Performance studies of microchannel plate PMTs. *Nucl. Instr. Meth. A*, 695:68–70, 2012.

[37] A. Lehmann et al. Significantly improved lifetime of micro-channel plate PMTs. *Nucl. Instr. Meth. A*, 718:535–540, 2013.

[38] Herold B. et al. Study of 40K-induced rates for a KM3NeT design option with multi-PMT optical modules. *Nucl. Instr. Meth. A*, 626-627:S234–S236, 2011.

[39] J. Rieke. Design of a compact photon detection system for the PANDA Disc DIRC prototype. PhD thesis, Justus-Liebig-Universität Gießen, in preparation.

[40] J. Rieke et al. Resolution changes of MCP-PMTs in magnetic fields. *J. Instrum.*, 11:C05002, 2016.

[41] I. Konorov et al. Soda: Time distribution system for the panda experiment. In *Conference Record of the IEEE Nuclear Science Symposium 2009 (NSS08)*, page 1863, 2009.

[42] A. Neiser et al. TRB3: a 264 channel high precision TDC platform and its applications. *J. Instrum.*, 8:C12043, 2013.

[43] C. Ugur et al. A 16 channel high resolution (smaller 11 ps RMS) Time-to-Digital Converter in a Field Programmable Gate Array. *J. Instrum.*, 7:C02004, 2012.

[44] I. Frohlich et al. A General Purpose Trigger and Readout Board for HADES and FAIR-Experiments. *IEEE Transactions on Nuclear Science*, 55(1):59–66, 2008.

[45] Online documentation of the TOF-PET ASIC by PETsys. `http://www.petsyselectronics.com/web/products/1`.

[46] PETsys Electronics, SA. Taguspark, Edificio Tecnologia I, 26, 2740-122 Oeiras, Portugal.

[47] PANDA Collaboration. Technical Design Report for the PANDA Barrel Time-of-Flight. 2017.

[48] PANDA Collaboration. Technical Design Report for the: PANDA Micro Vertex Detector. arXiv:1207.6581, 2011.

[49] Online documentation of the The GBT Project.

[50] H. Xu et al. An ATCA-based High Performance Compute Node for Trigger and Data Acquisition in Large Experiments. *Phys Procedia*, 37:1849–1854, 2012.

[51] C. Mayer et al. The ATLAS Tile Calorimeter Calibration and Performance. *EPJ Web of Conferences*, 60:20051, 2013.

# 5 Performance of the Detector System

## 5.1 Prototype Tests

Many test beam campaigns have been done by the group over the last years where simple and more complex Cherenkov detectors have been built and tested in particle beams. Not all of the results are relevant for the final EDD design, but it has always been an important lesson for new group members to work in a real experimental condition and to experience that there is a big difference between a condition where all pieces of a detector work at home in the lab and getting a complete detector system working at a given time.

As the initially available MCP-PMTs were not able to provide the necessary lifetime for using them in the $\overline{\text{P}}$ANDA experiment [1] early prototypes were based on an initial design using SiPMs for photon detection [2, 3, 4, 5].

### 5.1.1 A fast Cherenkov Counter Prototype using dSiPMs in 2010

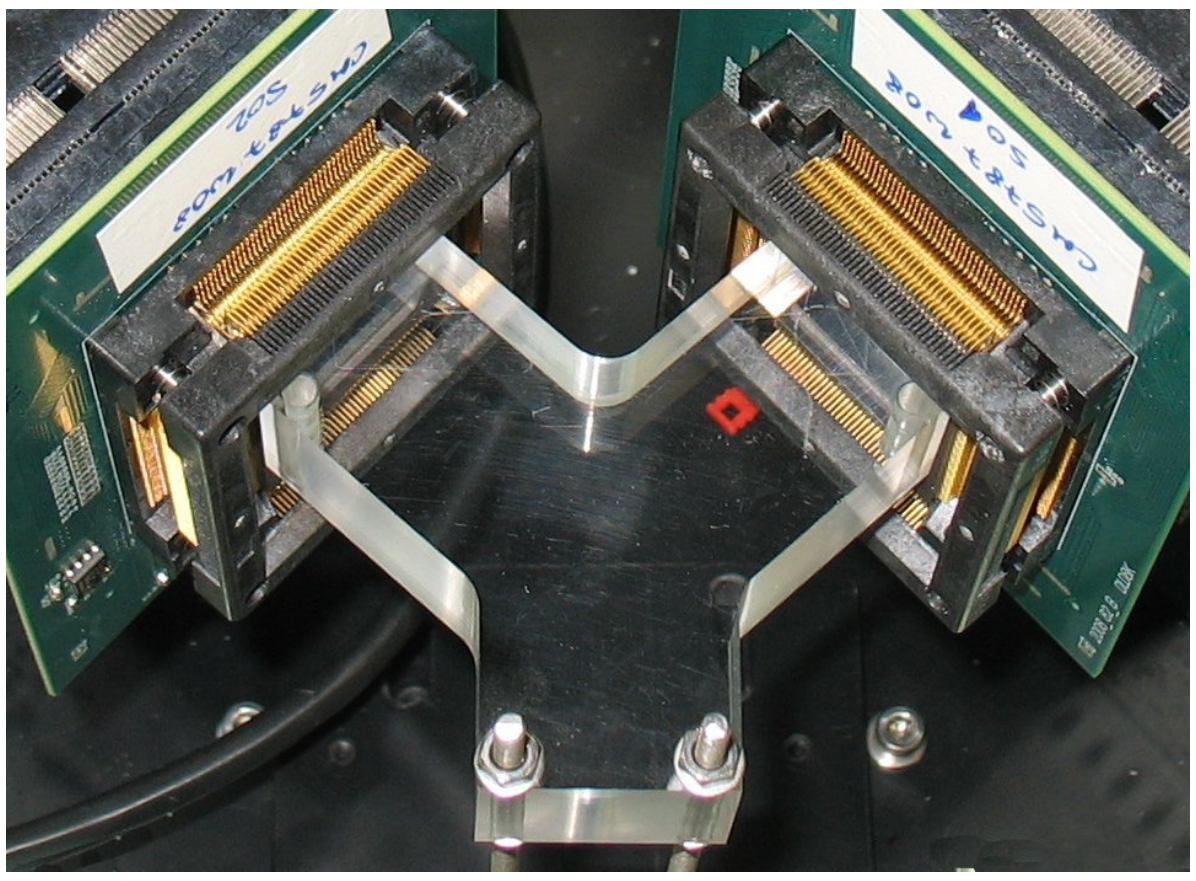

**Figure 5.1:** Y-shaped acrylic glass Cherenkov radiator attaching to two Philips digital SiPMs.

The Philips dSiPM Prototype together with a Y-shaped acrylic glass radiator has been tested at CERN in a 2010 beam time (Figure 5.1). With a refractive index of $n \approx 1.49$ Cherenkov light is emitted at $\vartheta_C = 48.2°$. The timing performance of this set-up has been evaluated by measuring the time difference between the two legs. The resolution depends on the number of photons registered in the detectors as shown in Figure 5.2. A single photon timing of $\sigma = 60$ ps has been achieved, mainly limited by geometrical constraints. The set-up was built and tested in cooperation between Philips and JLU Giessen.

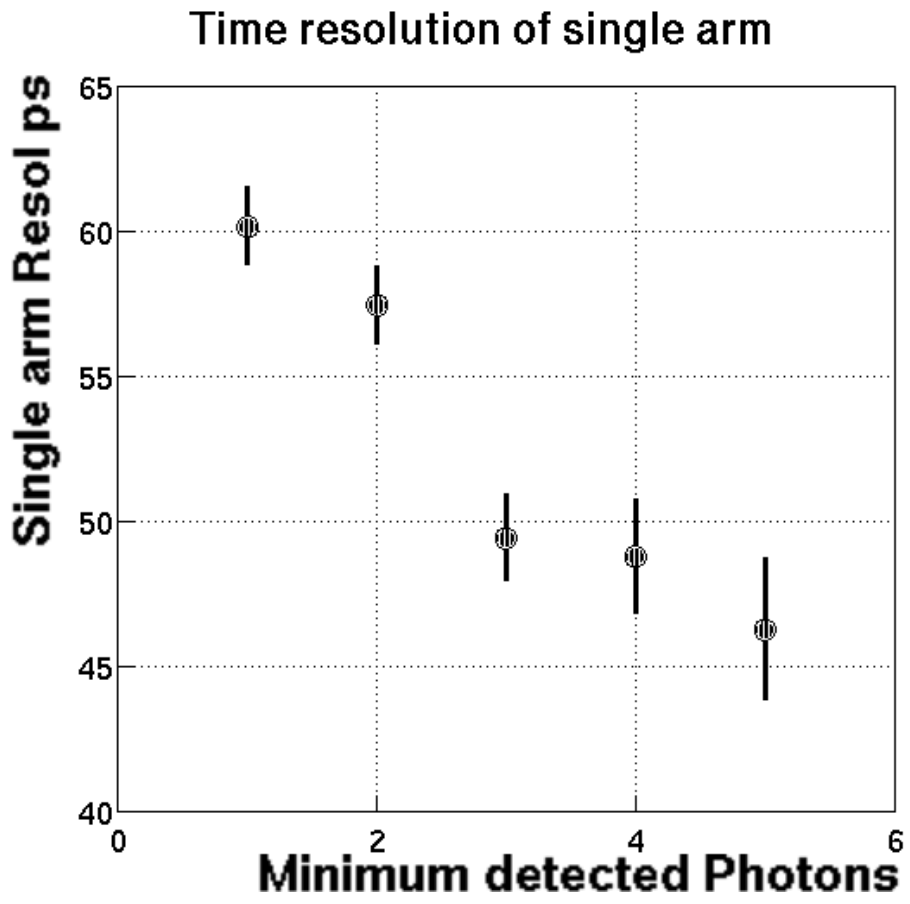

**Figure 5.2:** Detector resolution versus photon multiplicity.

### 5.1.2 Measurement of DIRC Cherenkov Patterns at DESY in 2011 and 2012

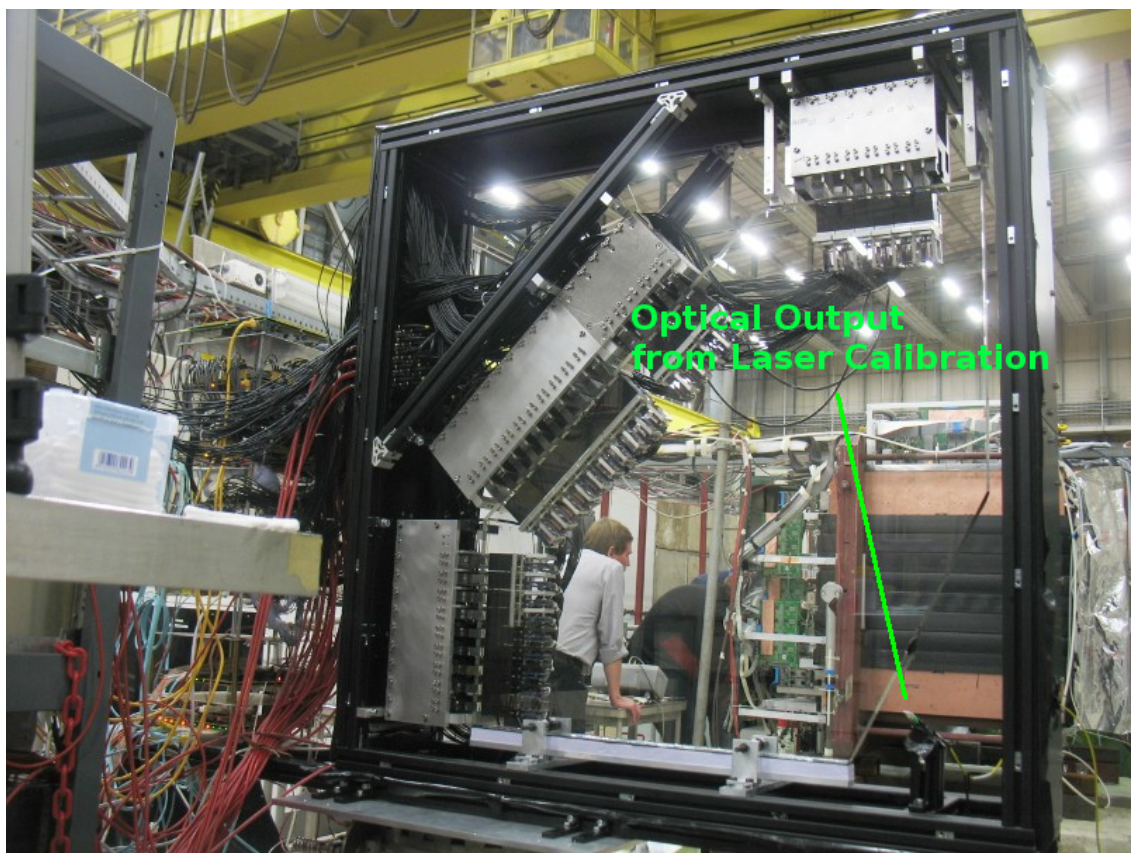

**Figure 5.3:** Fully equipped borofloat radiator DIRC prototype in 2012.

A demonstrator Disc DIRC prototype for the SiPM design, one quarter segment scaled down to 80 percent of the $\overline{\text{P}}$ANDA geometry, was constructed at Giessen university. For the test in an electron beam of 3 GeV/c at the T22 beam-line at DESY it was equipped with 5 (15) MaPMTs for test the beam campaign in 2011 (2012). The two large surfaces of the radiator plate made of borofloat glass have a typical surface roughness of 2 nm to 3 nm RMS. Figure 5.3 shows the setup with the light-tight covers still removed. The inner three faces not equipped with sensors were prepared differently: the bottom face was coupled with BC630 optical grease to a metal mirror, the diagonal face was covered with black masking tape, and at the vertical face the fine-milled surface was left unchanged.

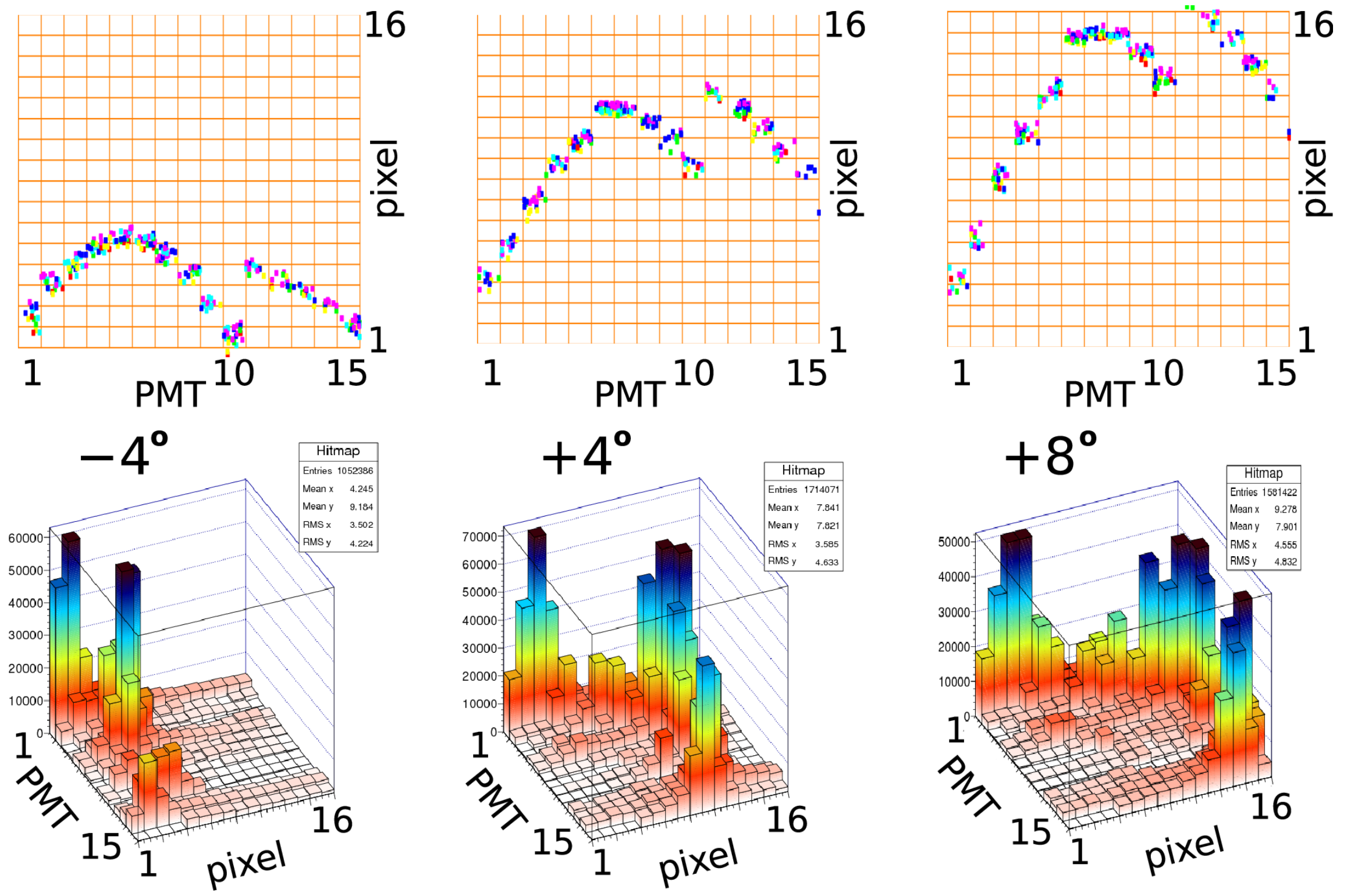

**Figure 5.4:** Results from the DESY 2012 test beam, showing cumulative electron hit patterns (pixel position on PMT versus ordinal number of the PMT) [6] . As predicted (top panels), the experimental pattern (bottom panels) moves as the beam-incidence angle of the electron beam is changed with respect to the normal of the radiator plate.

For availability and cost reasons, this prototype was instrumented with Hamamatsu H10515B100 multi-anode PMTs with 16 pixels of 16 mm width and a pitch of 1 mm. The MaPMT readout surface of 16x16 mm$^2$ limits the angular acceptance for the Cherenkov photons in the prototype design. The FELs were made of acrylic glass. Their shape – originally designed for a larger sensor – was kept unchanged, as – combined with the 0.8 mm pixel width – one still obtains a useful position resolution (see Figure 5.5). In comparison to the FELs which were optimized for the final detector design using MCP-PMTs these FELs did not have a cylindrical or spherical but multi-polynomial focusing surface.

A data acquisition system was used, that was originally developed for the HADES experiment at GSI. It consisted of a time-of-flight board (TOF add-on) for digitization and of a trigger readout board (TRBv2) for the time measurement [7]. The TOF add-on was equipped with 32 NINO ASICs [8] for pre-amplification and discrimination. The board included a charge-to-output pulse width function. The four HPTDCs [9] on the TRB board measured the arrival times of these pulses and their pulse widths.

Each board had 128 channels. The whole setup consisted of five TOF add-on and five TRB boards. The data streams were accumulated in a hub and sent to the DAQ computer via optical links. In addition there was a trigger control board (CTS, central trigger system) in order to create a coincidence trigger logic from the two scintillators in front of the prototype and behind.

Figure 5.4 shows predicted and measured patterns of Cherenkov photons for three settings of the angle scan which clearly shows for DIRC detectors typical distorted Cherenkov ring-like pattern that moves as the angle is changing.

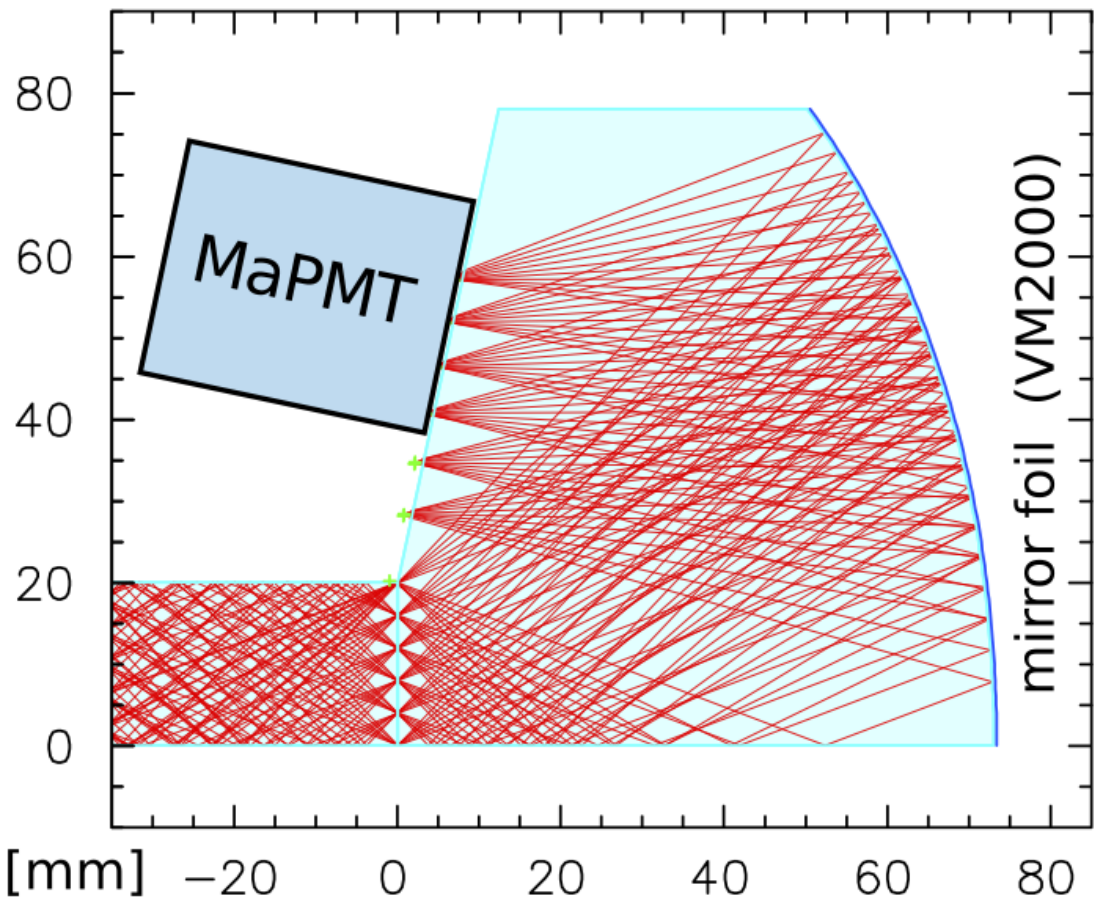

**Figure 5.5:** Shape of the focusing element for the SiPM design along with different photon paths [6]. The position of the MaPMT sensor is indicated. For reflection on the curved face, a VM2000 foil from 3M was attached. As the spatial acceptance of the MaPMT is smaller than for the dSiPM only part of the image plane was read out.

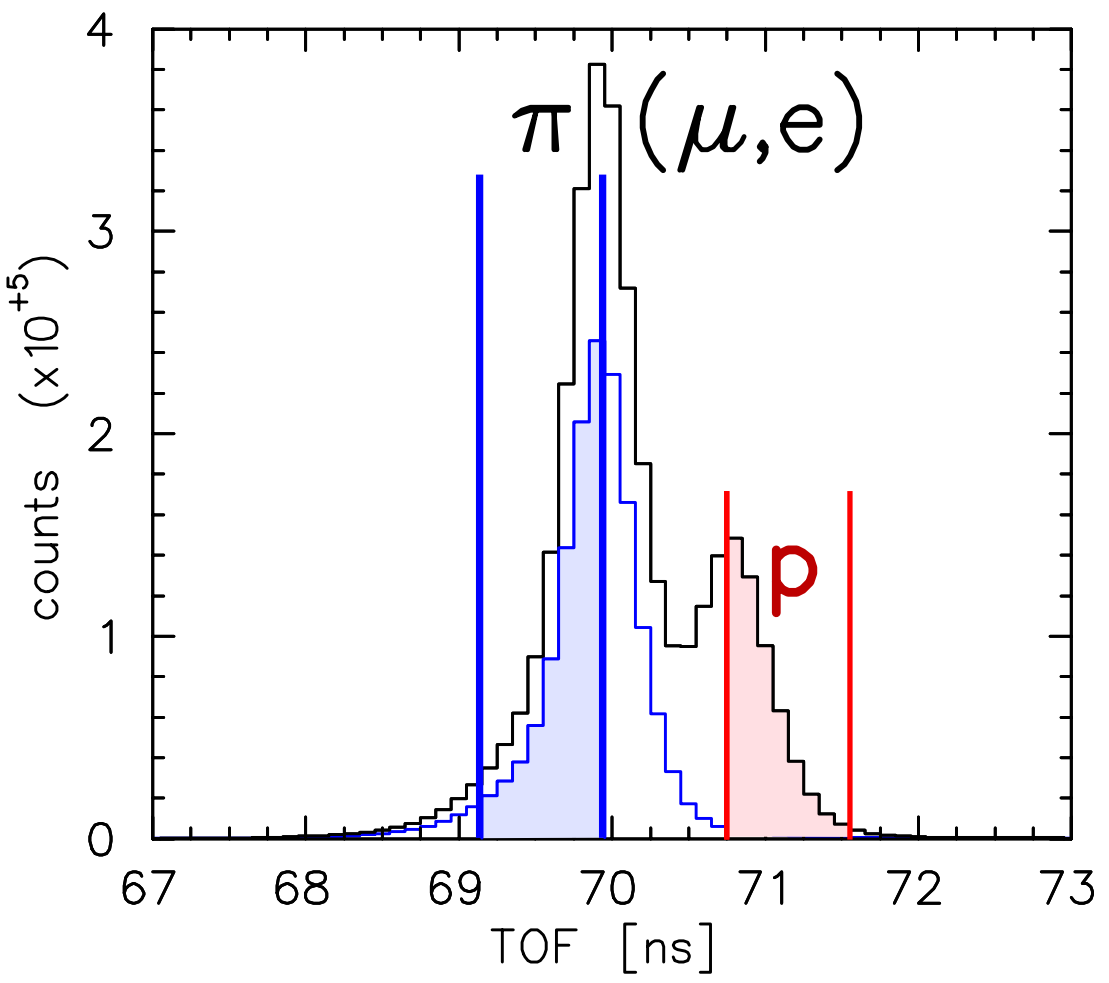

**Figure 5.6:** Time-of-Flight histogram measured with two MCP sensors at positions TOF1 and TOF2; The smaller histogram underneath the pion peak shows the threshold-Cherenkov tagged TOF, the marked time intervals are used for selecting pions and protons (darker parts).

### 5.1.3 Particle Identification with a DIRC Prototype at CERN in 2012

For the test beam campaign the demonstrator prototype from the previous test beams at DESY had been fully equipped with the nominal number of 30 optical readout elements and photon sensors, 480 sensor pixels in total. Particles of several GeV/c in a secondary hadron-rich mixed beam delivered by the PS/T9 test beam-line at CERN were measured in late 2012. First results were reported in [6] and further analyzed in [10].

The chosen compact FEL shape (Figure 5.5) is inscribed in a cuboid of $73\times78\times15\,\mathrm{mm}^3$ and images an angular range of 25.7° onto a focal plane length of 32 mm. This FEL shape requires a mirror coating on the curved surface. It was originally designed to fit candidate Philips modules of position-sensitive dSiPMs with $4\times4$ chip arrays covering a 32 mm×32 mm square area.

The full setup is shown in Figure 5.7. The coincidence of scintillators S1 (15 cm diameter) and S2 (5 cm diameter) gave the trigger signal. A TOF measurement was taken between the positions TOF1 and TOF2 about 7 m apart where two PHOTONIS XP85112 MCP sensors were placed. These detected Cherenkov light produced in their entry windows. At 3.5 GeV/c protons could be separated from light particles, but the time resolution was not sufficient to separate kaons.

The beam-line magnets were set to parallel optics. With the external particle identification procedure using TOF and threshold Cherenkov data, particle samples could be prepared and then compared to the Disc-DIRC prototype data.

The TOF spectrum for $p = 3.5$ GeV/c is shown in Figure 5.6, showing reasonable but overlapping distributions for pions (with additional light particles) and protons. Tight time cuts have been chosen with selection intervals from each peak maximum to $3\,\sigma$ outwards. In addition, the signal from the gas threshold Cherenkov counter with 5 meters of $CO_2$ radiator gas, pressurized to 2.5 bar (just above the Cherenkov threshold for pions), provides additional sample cleanup and cross-check.

Figure 5.8 shows cumulative hit patterns for $p = 3.5$ GeV/c and $\vartheta = 3°$ incidence angle. The patterns for pions (which also include muons and electrons) and protons are located at different heights. Qualitatively, the measured Cherenkov patterns reproduce the predicted shape and the relative shift of protons compared to pions.

In Figure 5.9 the misidentification of the Disc DIRC prototype is illustrated as a function of the number of detected photons. Due to the limited number of PMTs, most of the events have a small number of photons, but those that have a large number of registered photons show a good proton-pion separation. The misidentification depends mainly on the momentum, angular resolution, and the relative flux factors of the protons and pions. In contrast to that, the efficiency of the detector as a function of the number of photons can only be determined by Monte-Carlo simulations because it depends on the fraction of tracks that can be reconstructed.

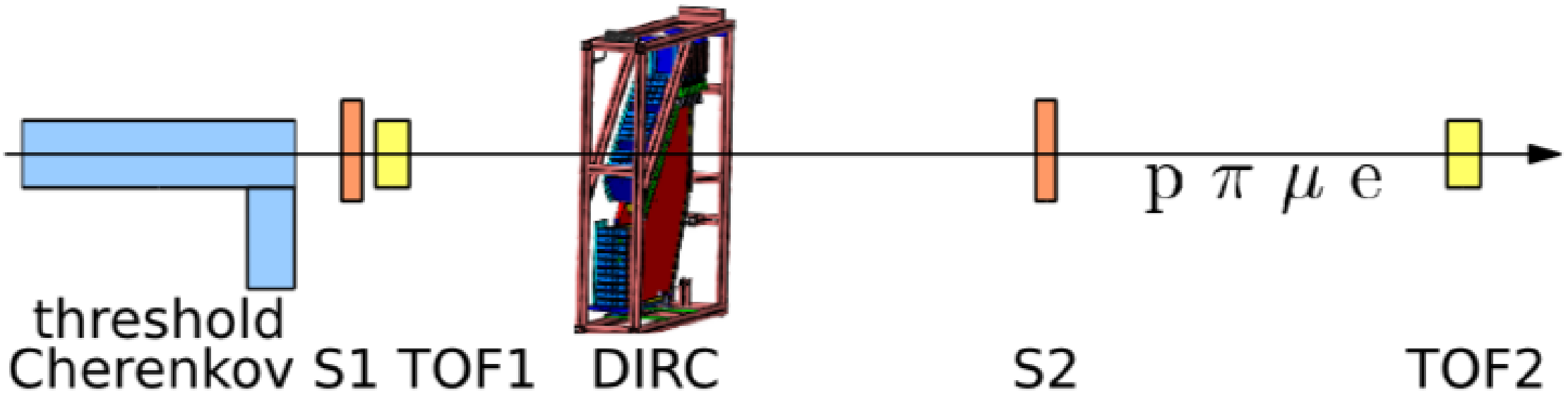

**Figure 5.7:** Experimental setup at CERN. The coincidence of the scintillators S1 upstream and S2 downstream of the DIRC provided the trigger signal. Time-of-Flight was measured between the TOF1 and TOF2 positions. The threshold Cherenkov detector provided a tag for pions and lighter particles.

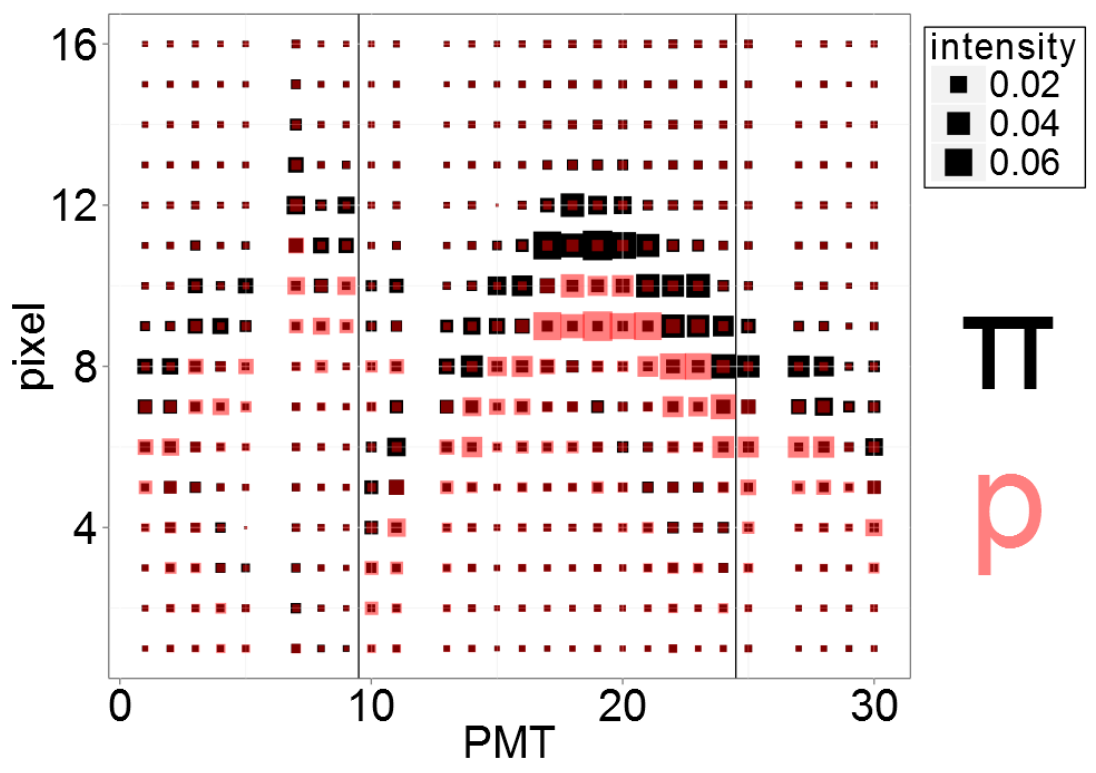

**Figure 5.8:** Cumulative hit patterns for light particles ($\pi$, $\mu$, e, black squares) and for protons (red squares). Each vertical column shows the 16 pixels of one MaPMT.

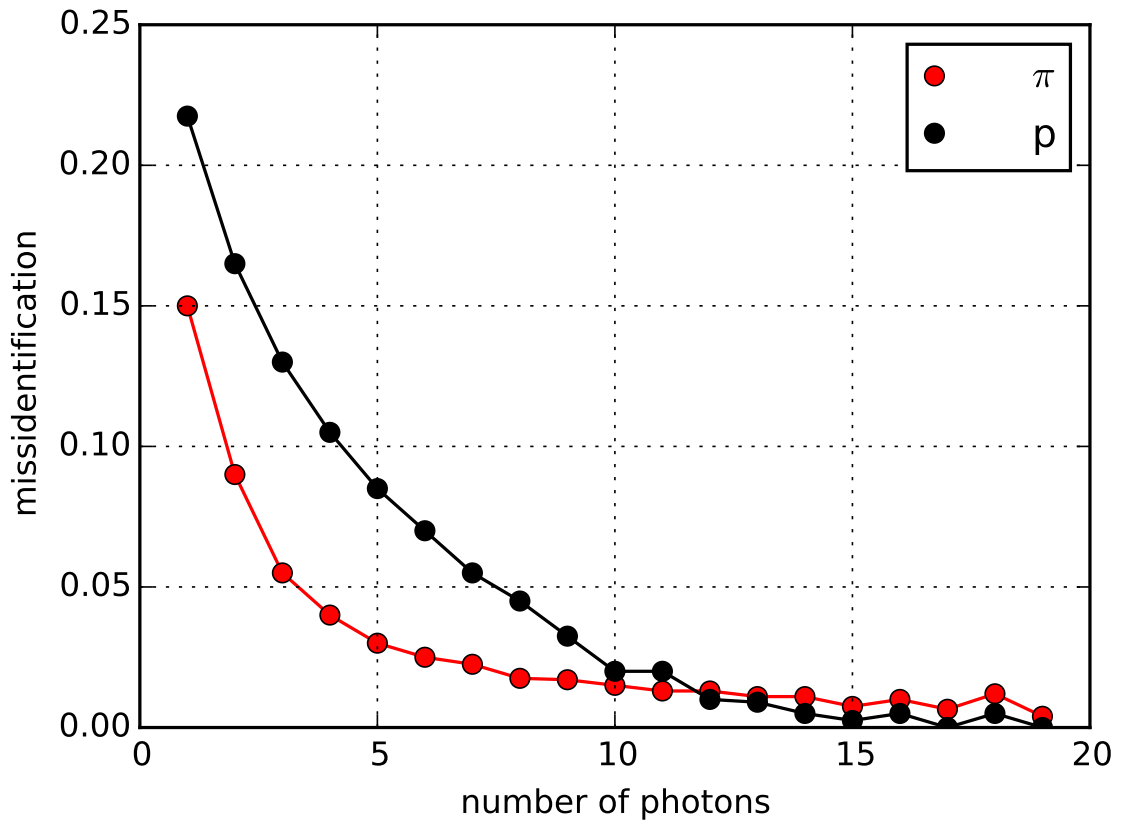

**Figure 5.9:** Misidentification fractions for pions (together with other light particles) and protons for a specific setting of the likelihood cuts in the analysis. Misidentification is defined here as the relative fraction of all pions (protons) that are identified as protons (pions).

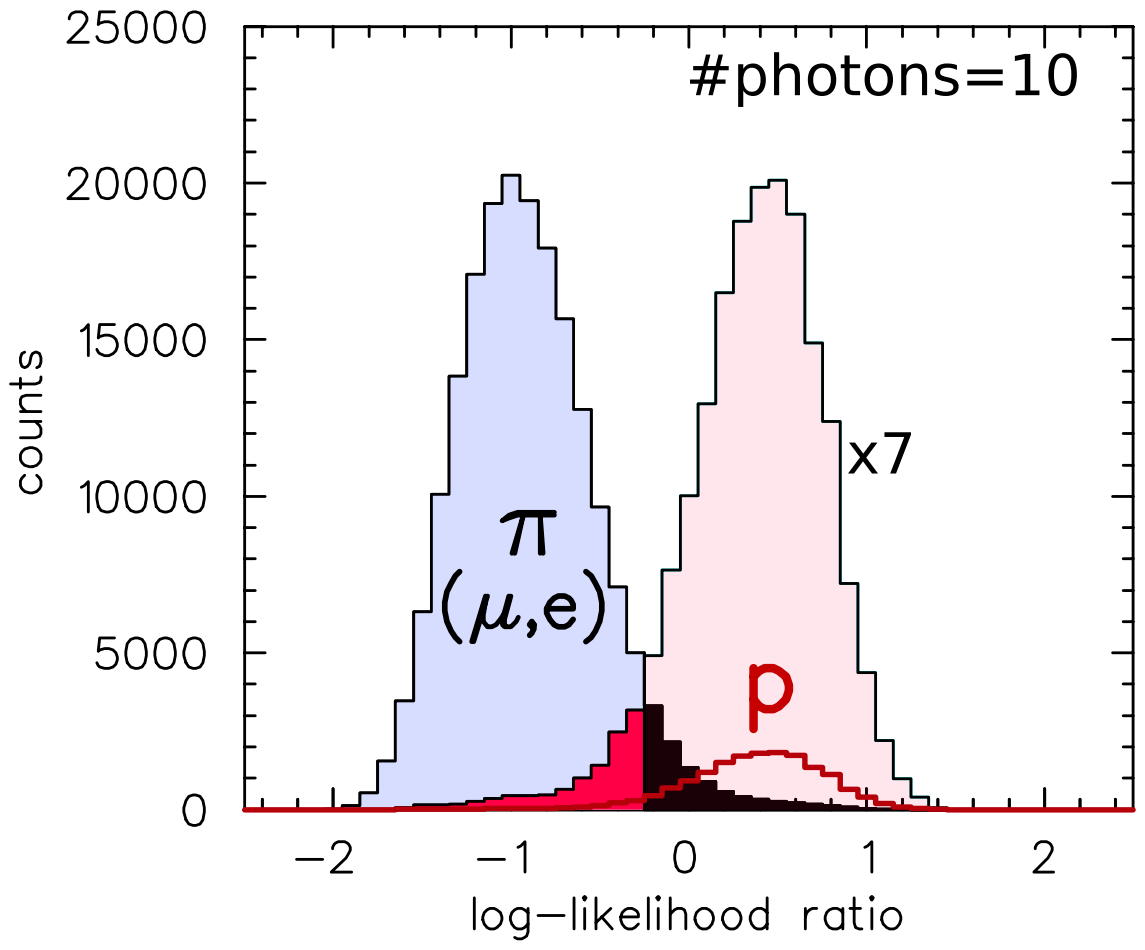

**Figure 5.10:** Histogram of the summed log-likelihood values for N = 10 photons. The two event sets are selected with the TOF and threshold Cherenkov information. The proton distribution is scaled up by a factor 7 to have both distributions at equal height. Dark black histogram (right of selection boundary): pions misidentified as protons; red histogram (left of selection boundary): protons misidentified as pions.

Figure 5.10 shows the log-likelihood ratio distribution for the case of 10 detected Cherenkov photons. The pion-proton identification works reasonable at 3.5 GeV/c with a small overlap of misidentified particles.

The main limitations of this prototype compared to the anticipated $\overline{\text{P}}$ANDA EDD were the simple borofloat radiator, the acrylic glass light-guides with limited optical performance [11] and the PMTs with a pixel pitch of 1 mm. However, with protons and pions of a 3.5 GeV/c particle beam, the particle identification capability of a Disc DIRC has experimentally been demonstrated for the first time.

### 5.1.4 Prototype Test with MCP-PMT Sensors at CERN in 2015

Because the radiation hardness of dSiPMs was insufficient (see also section 4.2.2 on page 27), and it became clear that the development of radiation hard dSiPMs was outside our time window, crucial design changes of the EDD were necessary, to replace the dSiPMs by MCP-PMTs. The transition took about two years as new components had to be specified, ordered and tested before being assembled to a full system for a testbeam campaign at CERN in 2015.

Here the 2015 EDD prototype and the $\overline{\text{P}}$ANDA Barrel DIRC prototype were tested in parallel [12]. In addition an MCP-TOF system and silicon fiber tracker (SFT) by the University of Erlangen-Nuremberg[13] and a segmented highly precise start counter (FLASH) from the University of Mainz were installed. The full setup is schematically shown in Figure 5.11. The PS/T9 beam line provides a secondary beam of mixed hadrons with momenta up to 15 GeV/c.

In comparison to the previously used prototypes the 2015 EDD prototype consisted of relatively few but high-quality components. The radiator was made of synthetic fused silica and was provided by Nikon[14] (see also section 4.1.1 on page 20). It has a square shape with dimensions $500 \times 500 \times 20\,\text{mm}^2$, the remaining specifications being chosen according to the requirements of the final radiator design. A surface roughness of less than 0.7 nm RMS was reached.

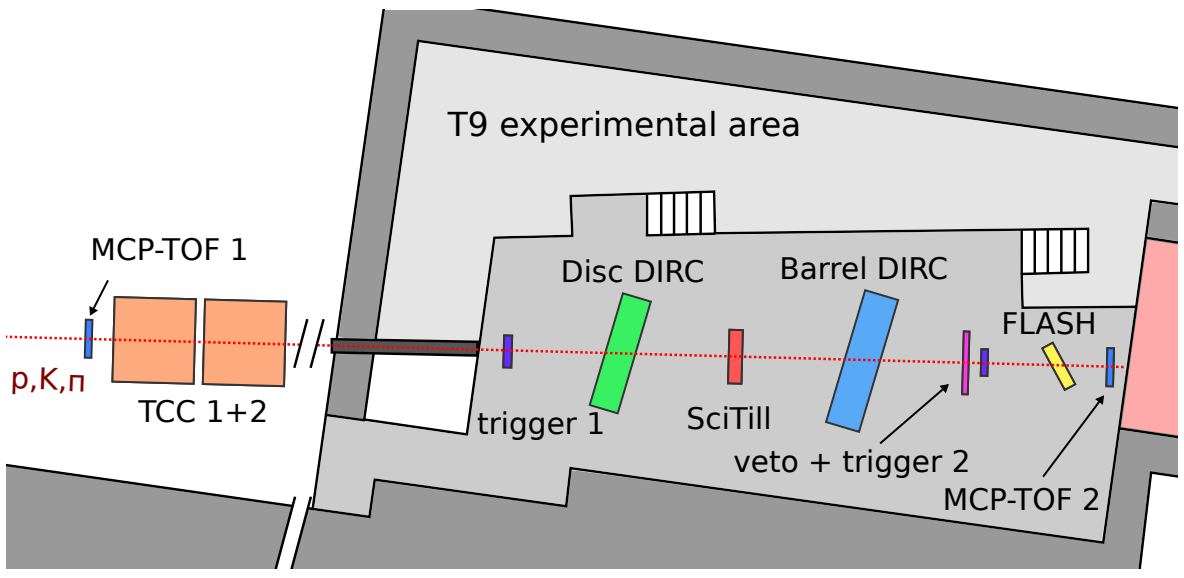

**Figure 5.11:** The experimental setup inside the T9 area at CERN during the 2015 testbeam. The sizes of the prototypes are not to scale.

The optical system was completed by three FEL-bar pairs which were manufactured by Berliner Glas [15] according to the final specifications for an MCP-PMT based EDD design (see also section 4.1.2 on page 4.1.2). Hence, for the first time, the optical system consisted of fused silica components only.

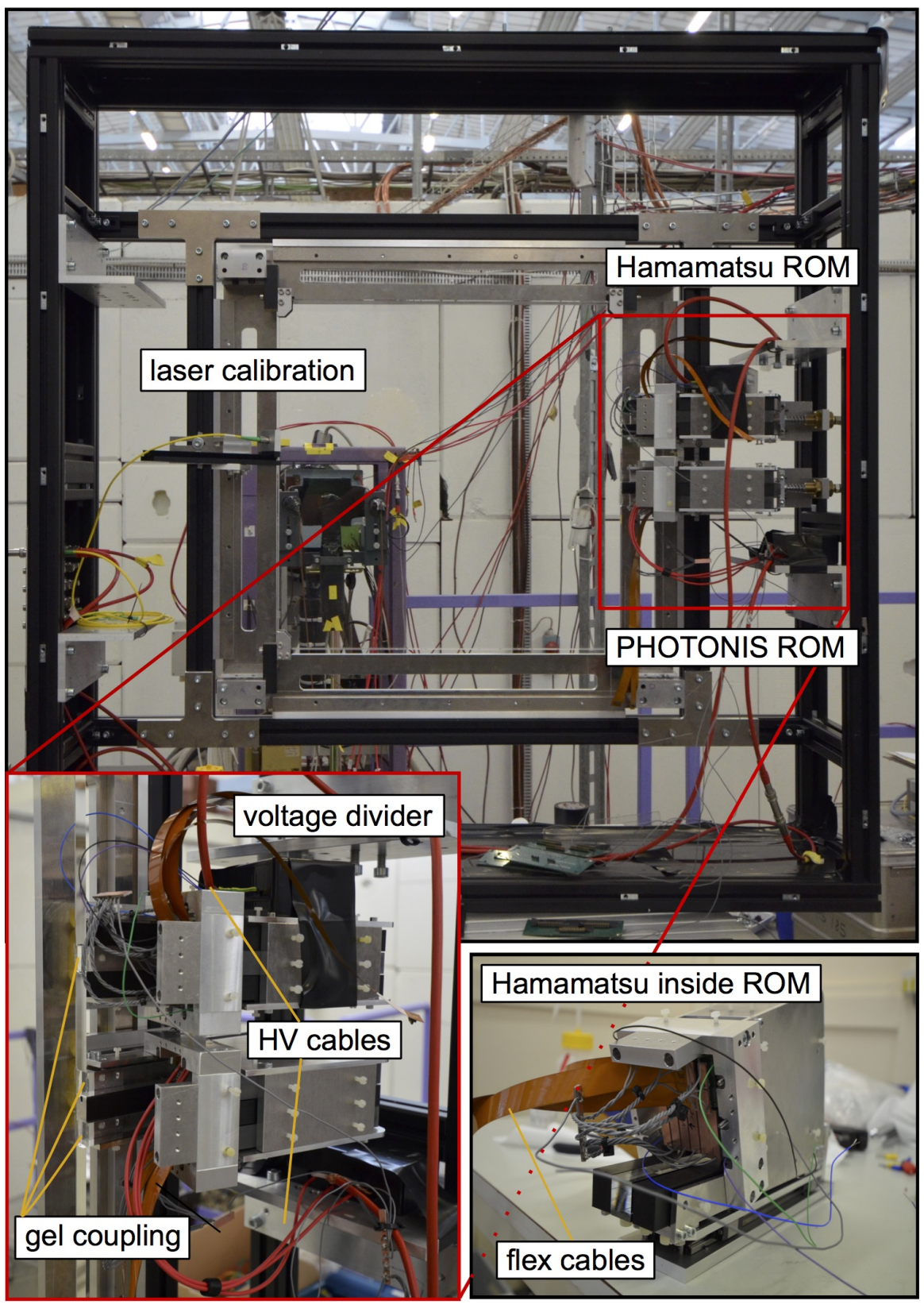

**Figure 5.12:** The 2015 EDD prototype at CERN.

For the photo detection two MCP-PMTs were used: a 2 inch square PHOTONIS MCP-PMT with $3 \times 100$ pixels and a 2 inch square Hamamatsu MCP-PMT with $6 \times 128$ pixels. Each MCP-PMT was assembled on first ROM prototypes which housed one (in case of the Hamamatsu MCP-PMT) and two FELs (in case of the PHOTONIS MCP-PMT). The setup is shown in Figure 5.12. The coupling of optical components including the MCP-PMT entry windows was done using optical gel code 0607 by Cargille [16].

As the TOFPET readout was not yet available the readout and data acquisition was performed by the TRBv3 system which was jointly used by all groups. All TRBv3 boards inside the T9 beam area were synchronized by a central trigger system (CTS). Due to firmware problems the data acquisition was impeded resulting in data loss and insufficient timing. For this reason the MCP-TOF system could only be used to get a reference start time for the event selection but not for a reliable particle iden-tification.

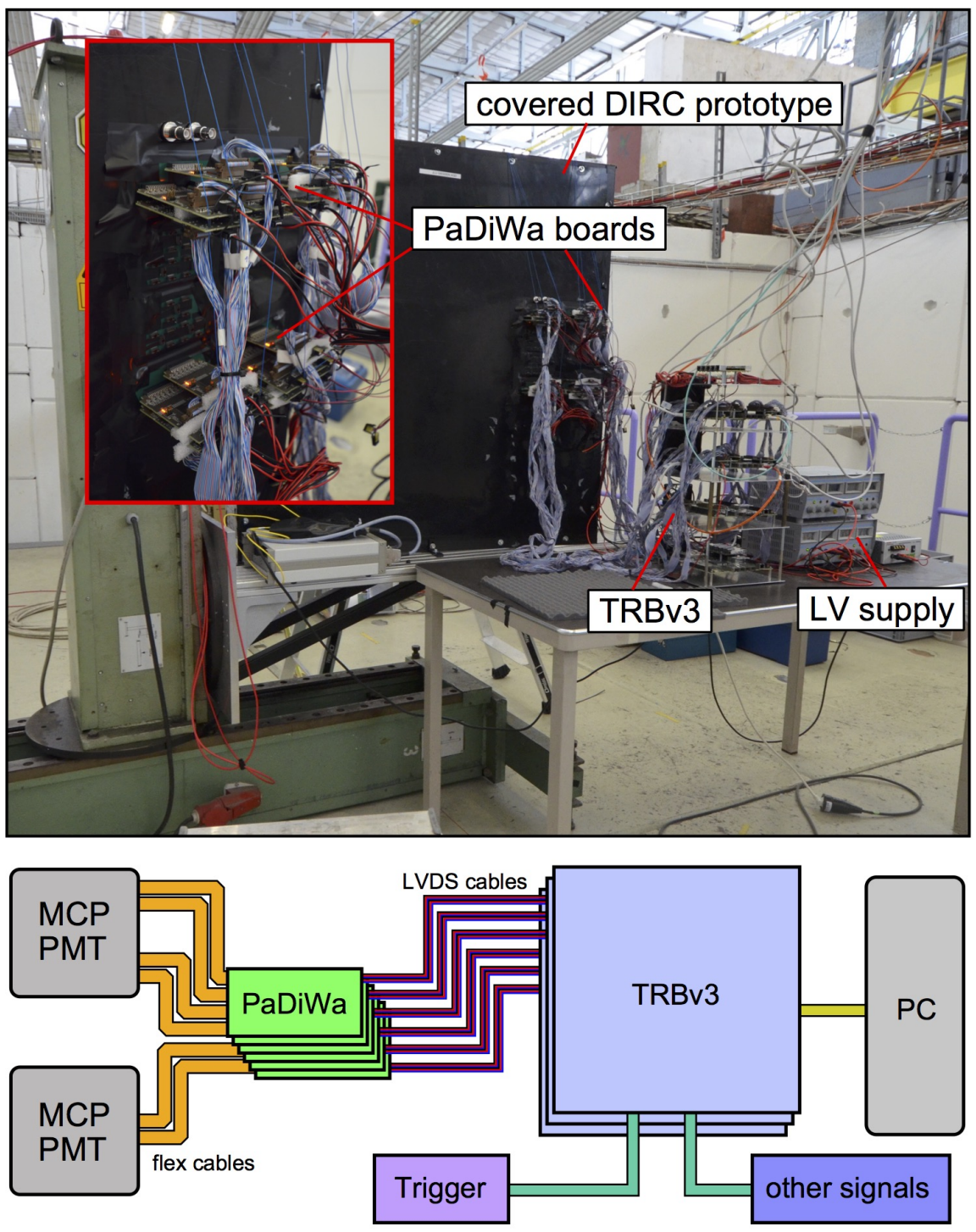

**Figure 5.13:** The top panel shows an upstream view of the detector on the translation stage during the 2015 testbeam at CERN. The PaDiWa boards are plugged to the outside of the prototype box. The bottom sketch gives a schematic picture of the DAQ scheme.

As depicted in Figure 5.13 bottom the analogue signals from the MCP-PMTs were forwarded by printed flex cables that were connected to PCBs which were integrated into the light-tight cover of the prototype box. Front-end boards (PaDiWa) that amplify and discriminate the signals were plugged to the outside of these PCBs. The now digitized signals were forwarded to the TRBv3 by LVDS cables.

The main goal of the 2015 testbeam campaign was to measure the single photon resolution for the available setup and to compare it with predictions by Monte Carlo simulations. The single photon resolution of this setup was limited by the large size of the beam spot, as the position of individual beam particles could not be measured. Also the accepted wavelength interval of the sensors without filters or dedicated photocathodes reduced the single photon resolution compared to the one of the final EDD.

### Cherenkov Resolution Measurements at 10 GeV/c

A large amount of statistics was collected during a 10 GeV/c angle scan. At this momentum the Cherenkov angles for pions and protons differ by less than 4 mrad which corresponds to 1.14 (1.46) pixels on the PHOTONIS (Hamamatsu) MCP-PMT.

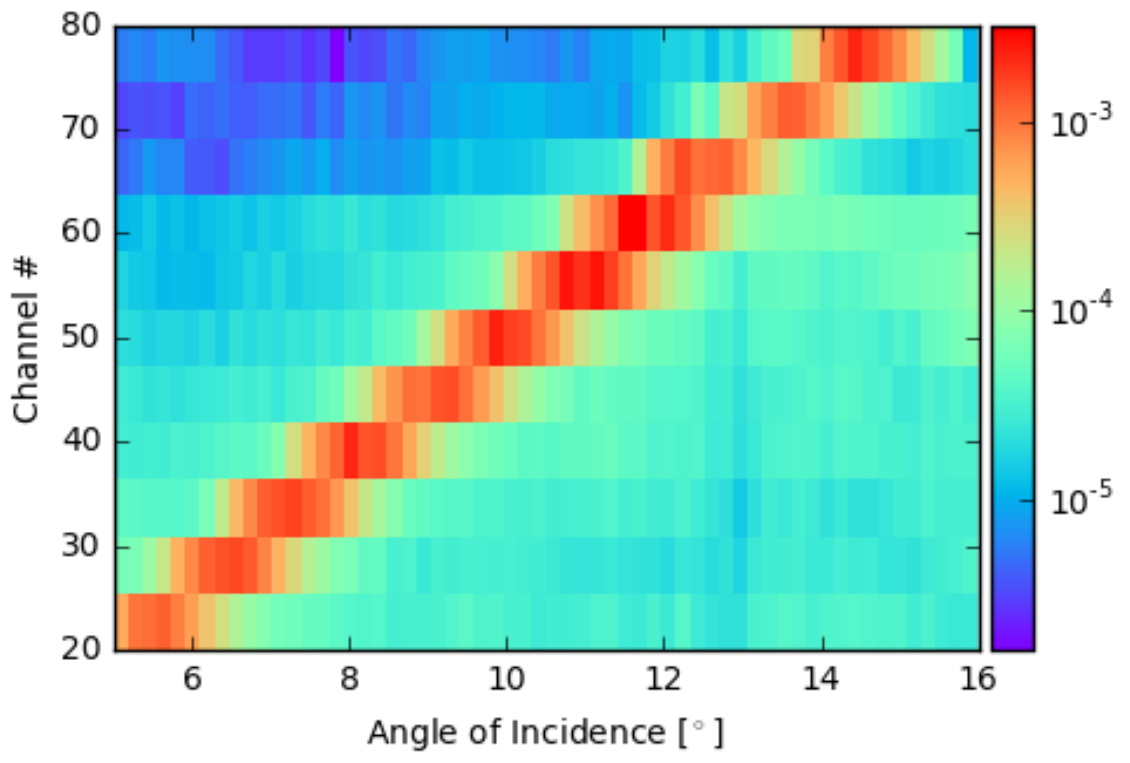

**Figure 5.14:** Hitmap for a 10 GeV/c angle scan. The linear progression for different angles can be nicely observed.

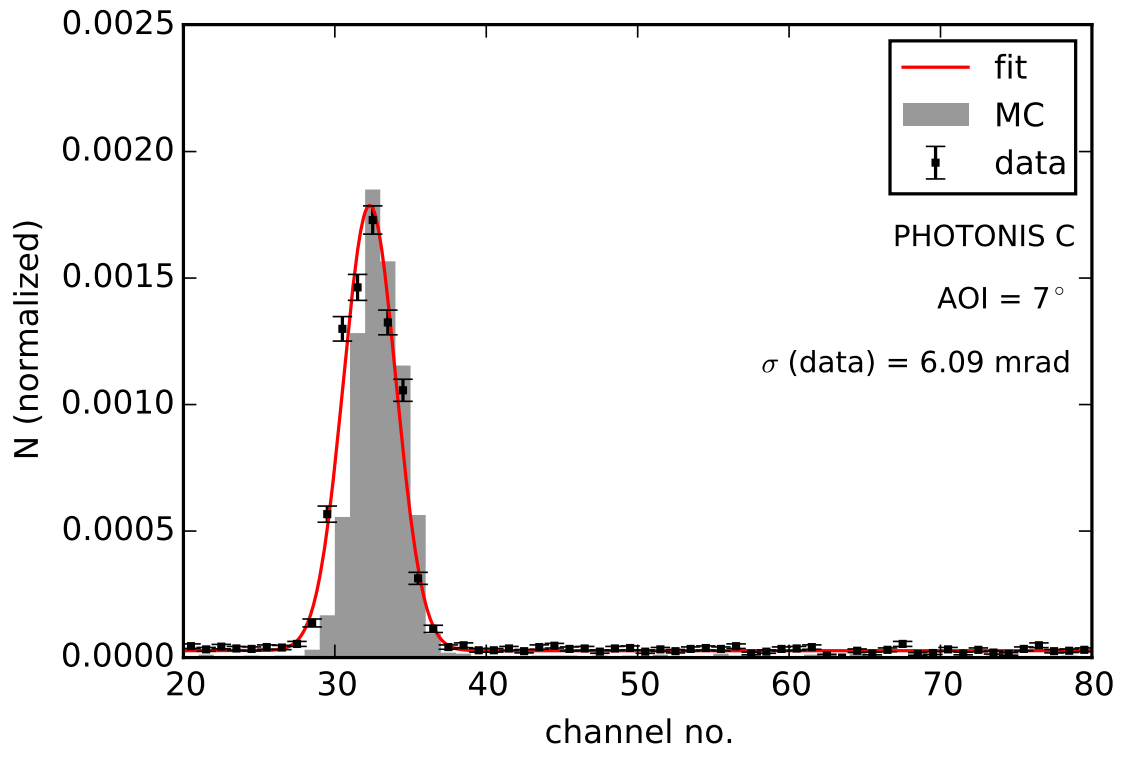

**Figure 5.15:** Single photon resolution on the PHOTONIS MCP-PMT at an angle of incidence of 7° during the 10 GeV/c angle scan.

Figure 5.14 shows the normalized distribution of recorded hits on one of the PHOTONIS MCP-PMT columns for different angles of incidence of the hadronic beam. Projections of the distribution at an angle of incidence of 7° show a good agreement with Monte Carlo predictions (see Figures 5.15 and 5.16, the fits inside the plots are applied to the data sample). The signal coming from Cherenkov photons is clearly visible and shows a linear dependence between the track angle of incidence and the position on the photon sensor. The mean channel number with respect to the angle is plotted in Figure 5.17. The width of the photon distribution is the single photon resolution, which turns out to be in good agreement with Monte Carlo predictions. In case of the PHOTONIS MCP-PMT it is as low as 5.68 mrad. The geometric acceptance for the Hamamatsu MCP-PMT in the chosen setup was worse which results in a minimal value of single photon resolution of 10.28 mrad.

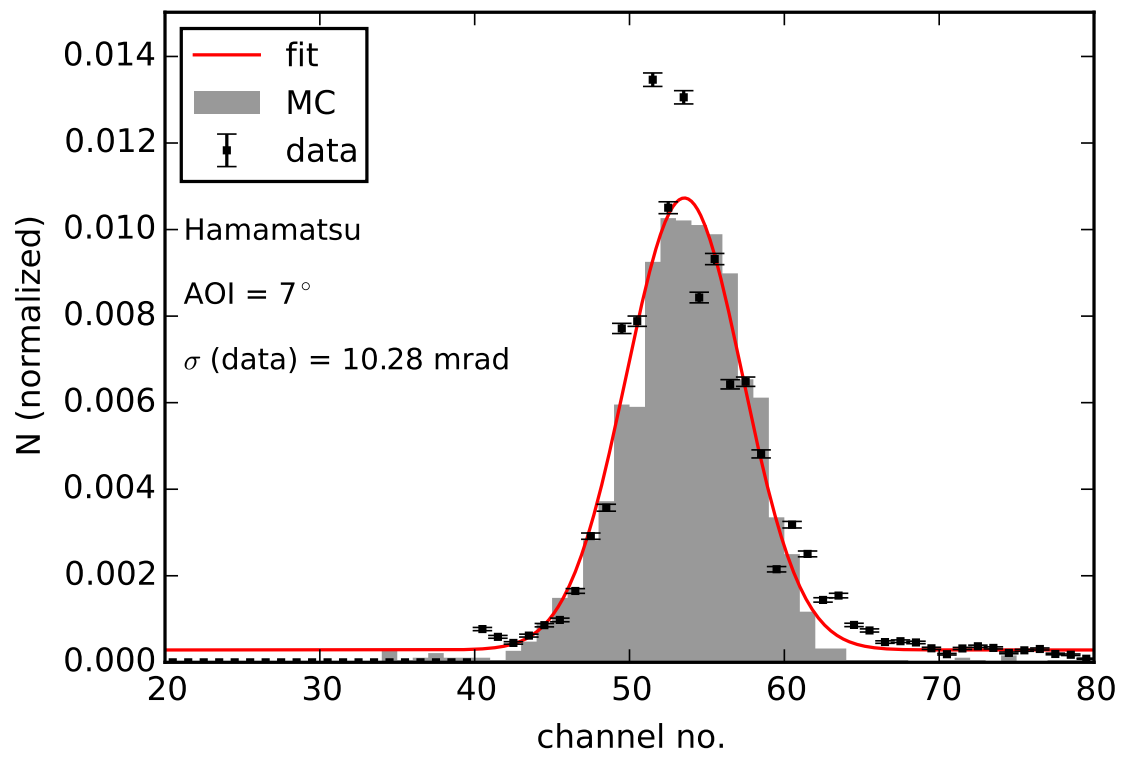

**Figure 5.16:** Single photon resolution on the Hamamatsu MCP-PMT at an angle of incidence of 7° during the 10 GeV/c angle scan.

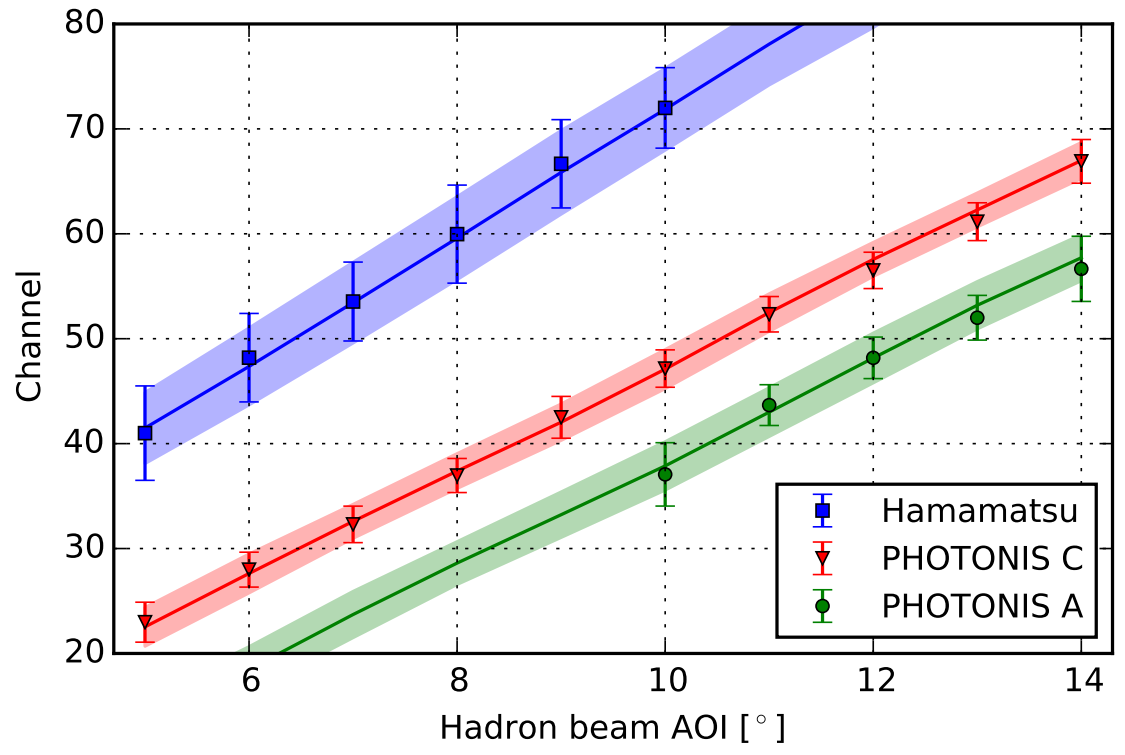

**Figure 5.17:** Linear dependence of the reconstructed Cherenkov light position on the MCP as function of the incident angle of the hadron beam. The error bars denote the width of the measured distributions. The colored bands denote the width and position according to Monte Carlo simulations.

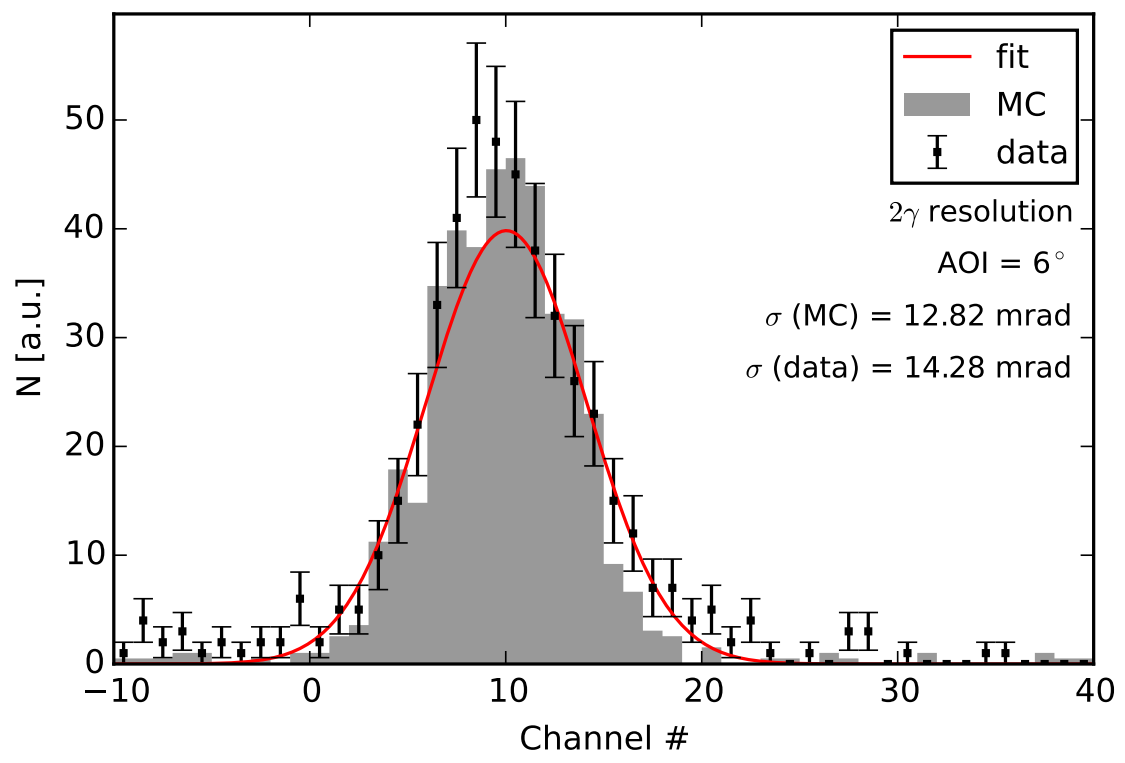

**Figure 5.18:** Two photon resolution between the PHOTONIS and the Hamamatsu MCP-PMT.

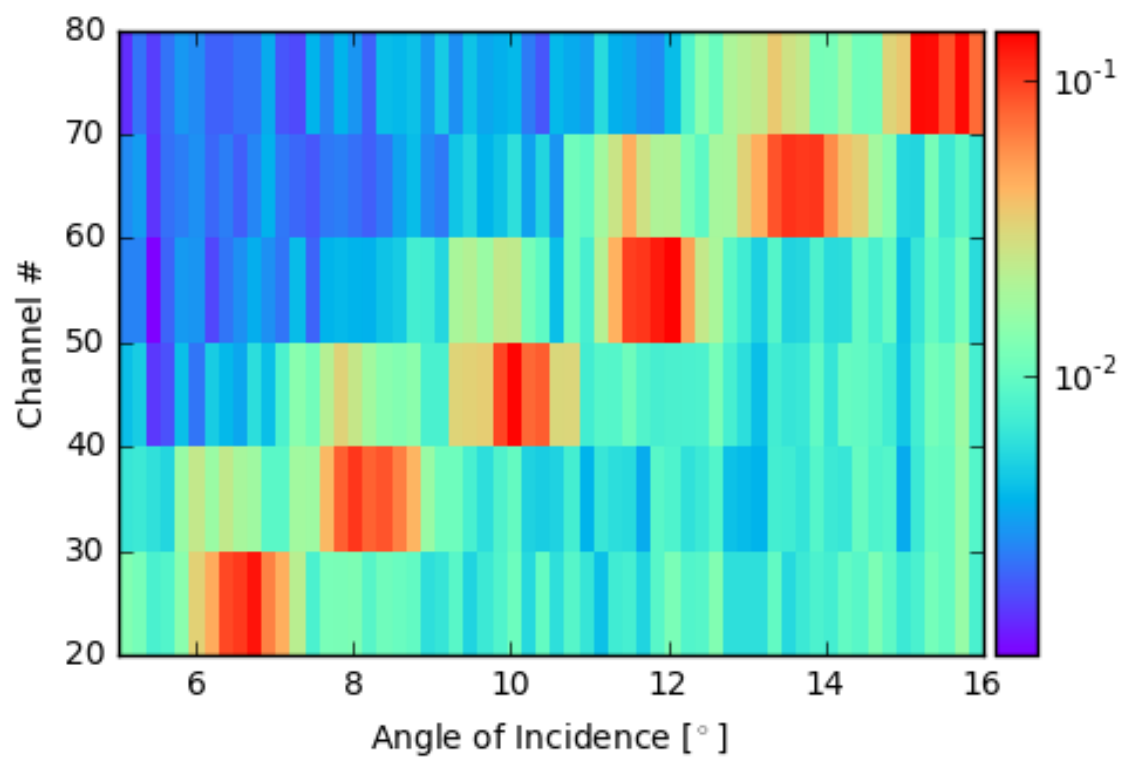

**Figure 5.19:** Hitmap for a 3 GeV/c angle scan. The prominent clusters are caused by Cherenkov photons coming from pions and other light hadrons and leptons. The proton peaks are suppressed and located towards smaller channel numbers.

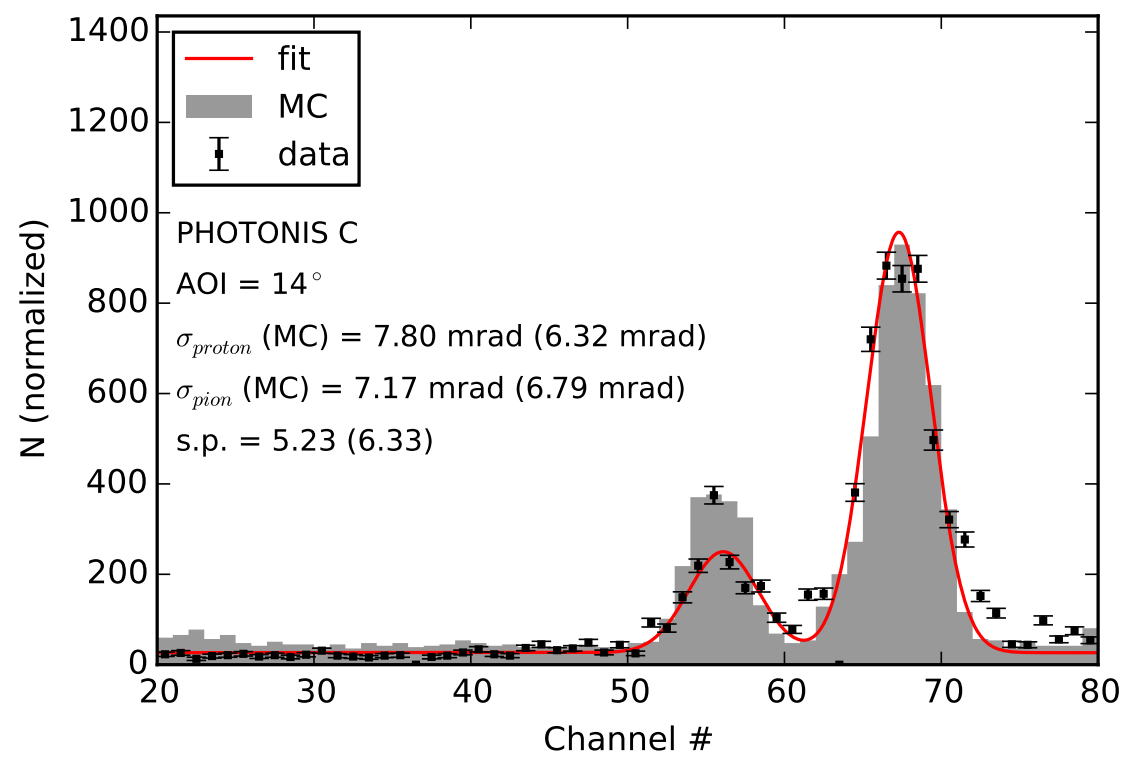

**Figure 5.20:** Cherenkov photon distribution for a PHOTONIS column at a beam angle of incidence of 14° during the angle scan at 3 GeV/c. In the presented case a separation power (s.p.) of better than 5 $\sigma$ was reached. The values in parenthesis are obtained by Monte Carlo simulations.

In addition to the single photon resolution the so-called two photon resolution was determined (Figure 5.18). It is defined by the difference between the hit positions of two different sensor columns during a single event. In comparison to the single photon resolution certain errors cancel due to correlations.

Again the measured distribution is in good agreement with Monte Carlo predictions.

#### Particle Identification at 3 GeV/c

An angle scan at 3 GeV/c was done, as at this energy the Cherenkov distributions for pion and protons are clearly separated. Figure 5.19 shows the hit pattern for different angles of incidence for the PHOTONIS MCP-PMT. Because pions are more than twice as frequent as protons at the CERN beam at this energy, the proton peak is suppressed. Nevertheless as shown in Figure 5.20 the resolution is good enough to identify pions and protons at 3 GeV/c with a separation power of more than $5\,\sigma$ on the single photon level although external preconditions in terms of beam size and position as well as the accepted wavelength interval are not optimal. In case of the Hamamatsu MCP-PMT a separation power of $3.5\,\sigma$ or better was reached. It was limited due to a less favoured geometrical position and acceptance in our prototype setup.

### 5.1.5 Prototype Test with TOFPET Readout at DESY 2016

The 2016 prototype incorporated some major enhancements in comparison to the previous year. The most prominent change was that the now available TOFPET readout was used as DAQ system. In addition the mechanics were rebuild.

The radiator was held by a cross-like structure similar to the final detector design and the FELs together with a PHOTONIS MCP-PMT were positioned inside a 3D-printed case. This setup allowed to fully equip a ROM for the first time. Due to a delayed delivery only the three previously available FELs could be used. The best FEL with optical contact bonding (in terms of this testbeam campaign called FEL2) was positioned at the bottom the ROM. The remaining two (FEL0 which had some inclusions at the glue joints and FEL1 which was bonded with a two component epoxy) were mounted in the top and central position. Three sliding doors and a top cover allowed easy access to the inside of the light-tight prototype box. The analog signals from the MCP-PMTs were forwarded with commercially available shielded coax-cables to PCBs that were integrated into the prototype box cover. Here they were rerouted to match the required channel numbers for the TOFPET readout. Figure 5.22 shows the setup.

The beam test was done in the T24 beam area at DESY which provides a 3 GeV/c electron beam with a beam size of $\approx 5$ mm. The setup is schematically shown in Figure 5.21. The trigger logic included two scintillating counters and a calorimeter behind a shielding wall to avoid the selection of events where the beam electron was scattered and/or lost a significant amount of energy. The data was taken continuously with the TOFPET readout. Different to the laser measurements in the lab, the signal from the MCP-PMTs arrived at the testbeam before the trigger signal from the scintillators, which was taken in to account in the readout mode of the TOFPET that we used.

The goals for this testbeam campaign were to validate the TOFPET readout, get information on the photon yield, and do additional scans to systematically check the behavior of the detector. For this reason a Geant4 stand-alone Monte Carlo simulation was carried out which takes into account all relevant detector and setup parameters such as

- the surface roughness of the prototype radiator,
- transmission values for different optical components like quartz, glue or optical grease,
- the size and position of the collimator inside the T24 beam area including a beam smearing of 1 mrad,
- additional passive material like the prototype box cover,
- rejection of electrons which are scattered out of the calorimeter acceptance,
- the measured quantum efficiency of the MCP-PMT and a collection efficiency of 65 %,
- charge sharing effects.

The amount of charge sharing during the testbeam campaign was determined by looking at clusters of multiple hits on neighboring anode pixels and comparing them to Monte Carlo simulations. In comparison to the lab tests shown in section 4.2 the discriminator thresholds for the individual channels were set to higher values which leads to a relatively small amount of charge sharing.

The measurement program of our test beam campaign consisted of three different types of scans: an scan of the radiator angle with respect to the test beam direction (see Figure 5.23), a horizontal position scan to reduce the distance between the beam and the FELs and different vertical position scans parallel to the row of FELs (see Figure 5.24). In addition to that, high-voltage scans and threshold scans have been carried out to find the best parameters for all runs. During the entire testbeam campaign, the used Photonis MCP-PMT was run with 2550 V that equals a gain of approx. $5 \cdot 10^5$ according to the measurements from the Erlangen group. Further systematic studies regarding the matching of thresholds and gains have to be done at a later stage with a new ASIC version of the TOFPET readout.

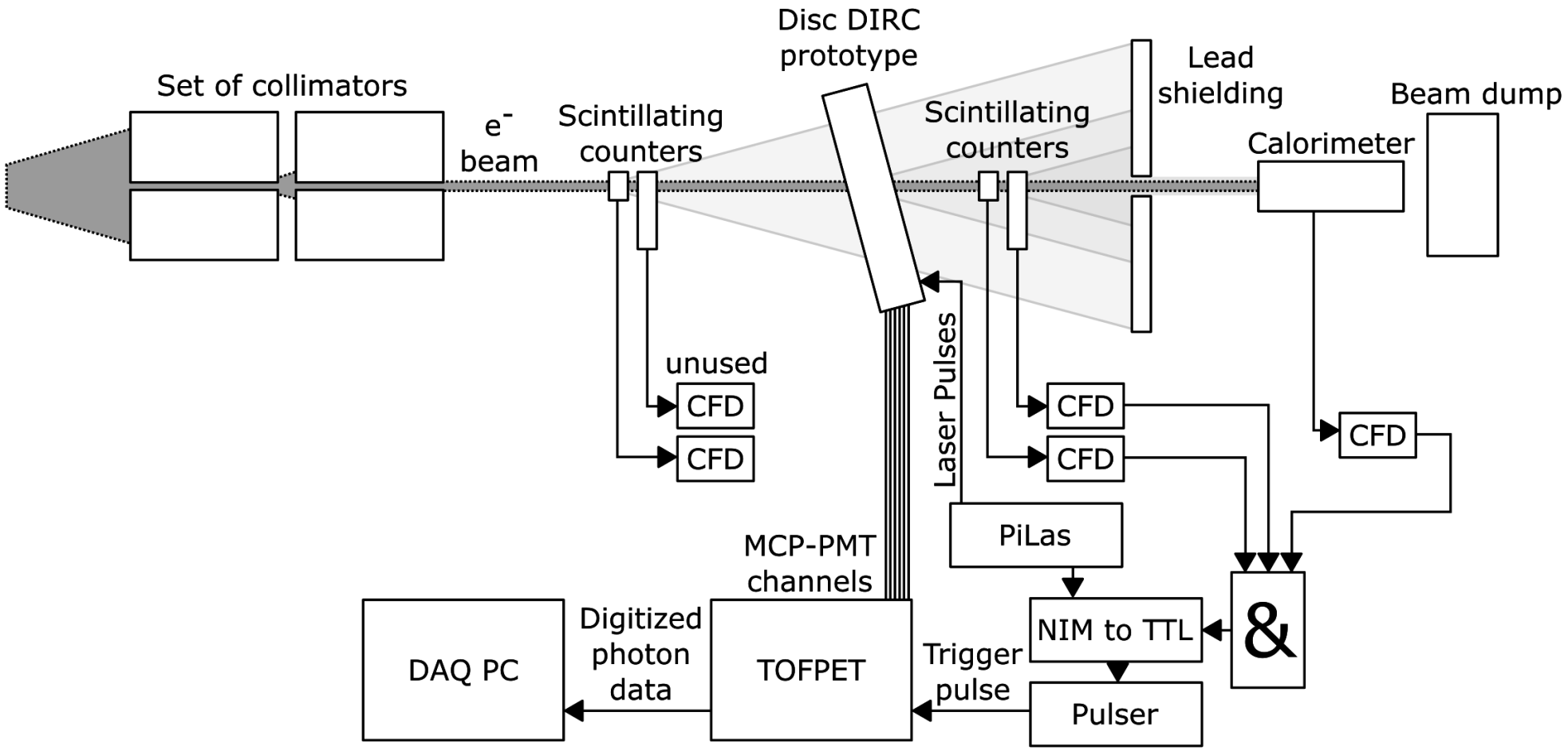

**Figure 5.21:** Setup and trigger/DAQ logic inside the T24 beam area at DESY during the 2016 testbeam [17].

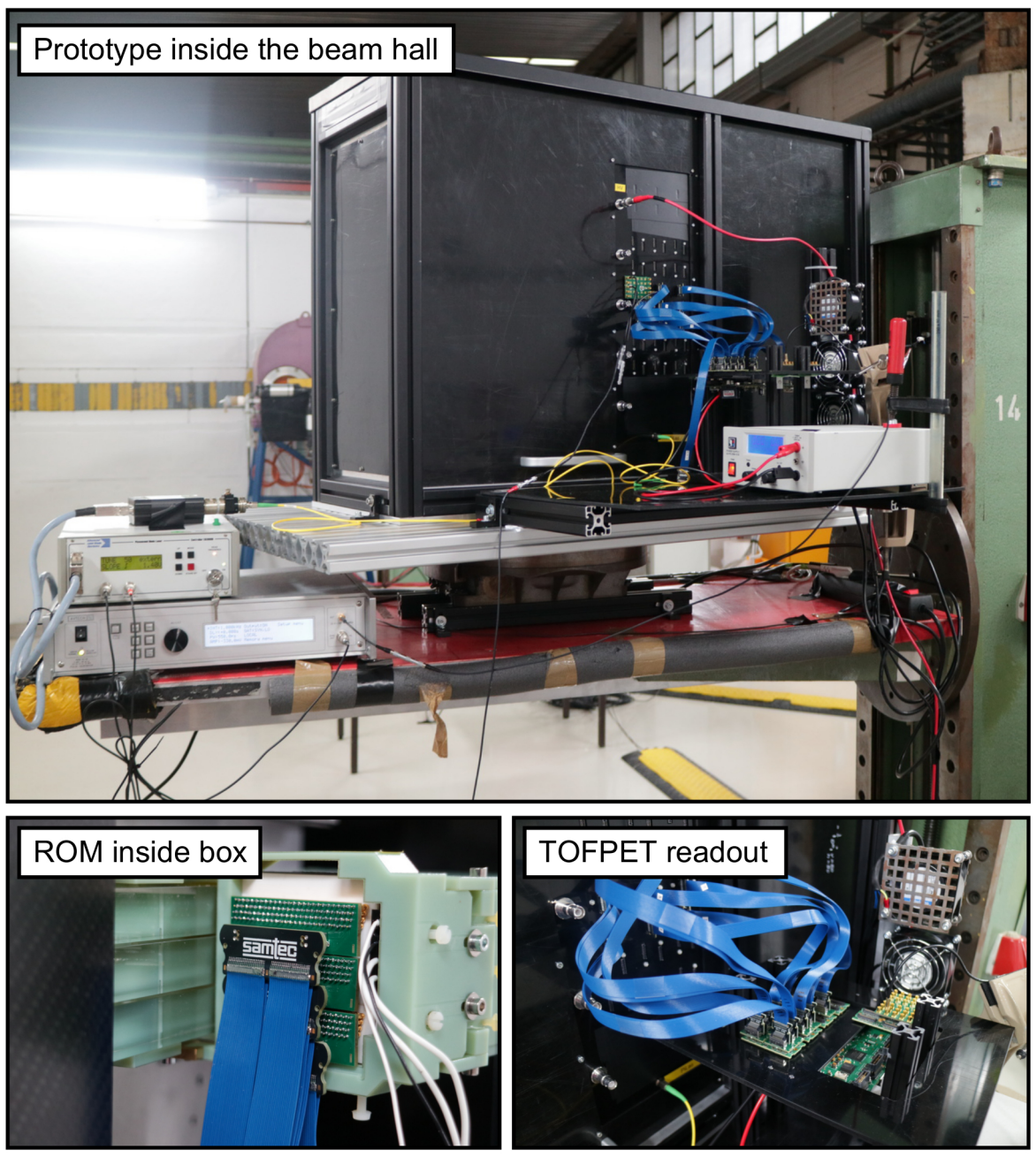

**Figure 5.22:** The 2016 EDD prototype at DESY with TOFPET readout.

Unlike in the 2015 testbeam at CERN, this prototype could be moved such that the whole radiator area was accessible by the electron beam. For the angle scan the beam passed the radiator at the vertical position of FEL1 at a horizontal distance of 450 mm. In addition to the results obtained in the 2015 testbeam campaign an indirect reflection off a side surface can be obtained as seen in Figure 5.24. Due to a different angle between the projection of the Cherenkov photon path and the orientation of the FEL-bar pair ($\alpha_{\mathrm{FEL}}$) these Cherenkov photons hit a higher pixel number and the linear progression with respect to the hadron beam angle is slightly steeper. Therefore also the distance between both lines increases between FEL2 and FEL0.

For each scan consistency checks between the recorded data and MC simulations were performed. For this purpose FEL1 was chosen as a reference as it gives the best results. FEL0 had a problem with air bubbles inside the glue joint which bias the photon yield and also its resolution. FEL2 had a problem due to a small misalignment of 1.5 mrad with respect to the radiator disc, which leads to a slightly broader double image. In addition, as no glue was used to bond the FEL-bar pair of FEL2, the dispersion error is much larger due to photons in the UV range.

### Time Resolution

One of the first measurements was to evaluate the timing performance of the system. Figure 5.25 shows the time of arrival of Cherenkov photons relative to the trigger for all channels for a single run at a defined angle of incidence and position of the electron beam. The relative time shift between the channels has not been calibrated. For each sensor column 96 channels were read out. Besides a prominent cluster each sensor column also contains a second, attenuated cluster which comes from indirect reflections off a side surface of the radiator. The delay of $\approx 2$ ns corresponds to the time of propagation for the additional path length the indirect Cherenkov photons have to travel.

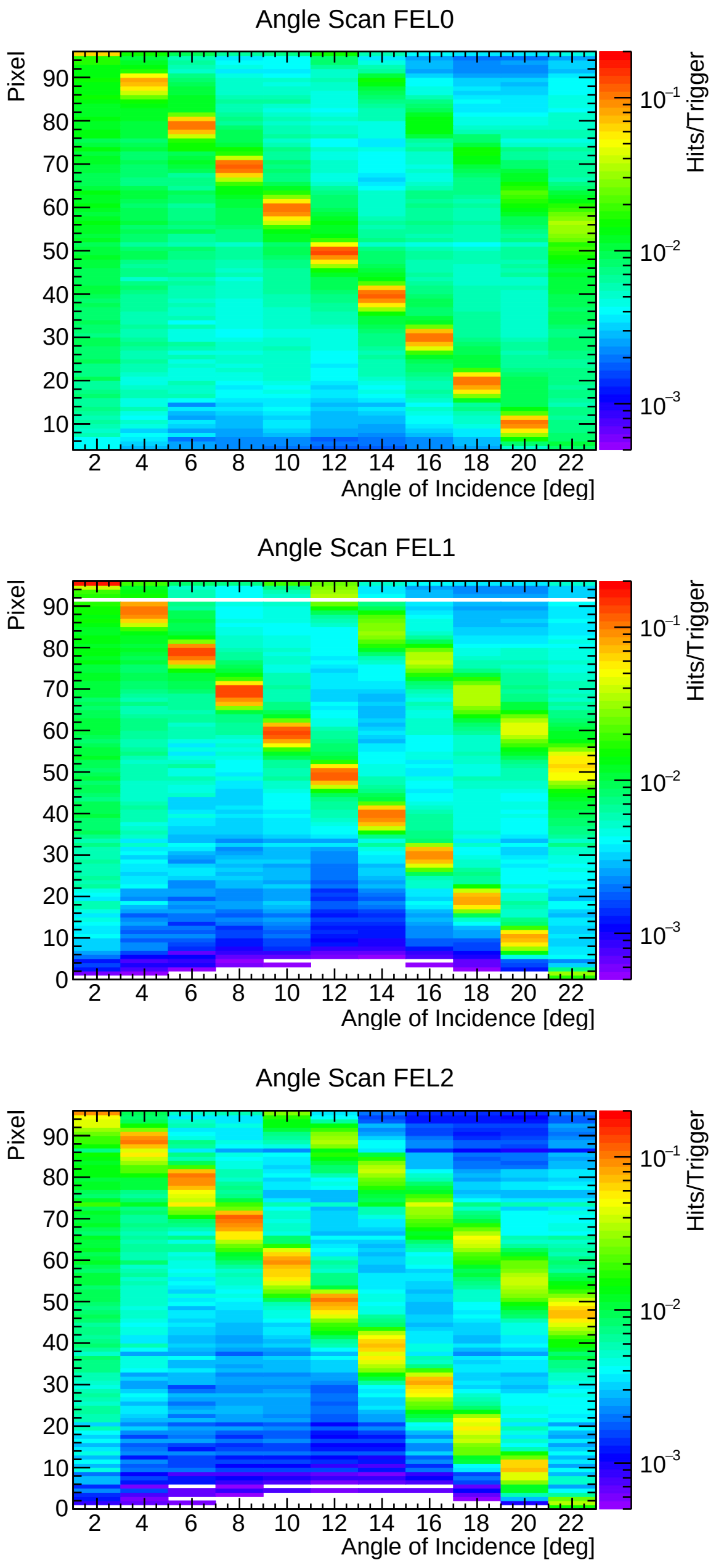

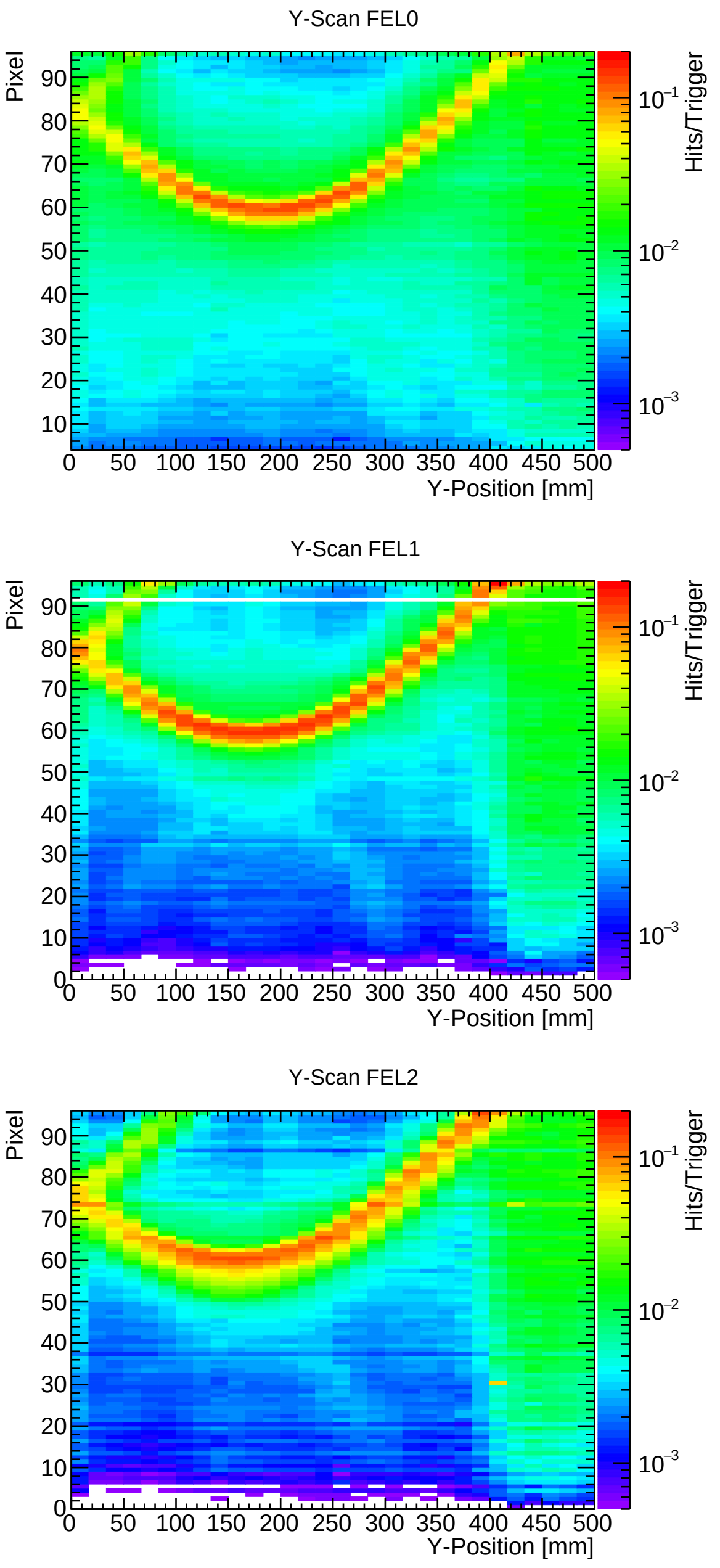

**Figure 5.23:** Angle scan perpendicular to the ROM at a distance of 450 mm. Shown is the distribution of detected photons on the MCP-PMT for each angle setting between 0° and 22°.

**Figure 5.24:** Vertical scan at an angle of incidence of 10°. The center of the beam spot was positioned perpendicular to the ROM at a distance of 347 mm. Shown is the distribution of detected photons on the MCP-PMT for each beam position setting. One vertical bin corresponds to a step size of 17 mm.

The time resolution of the system was obtained by taking the difference of two Cherenkov photons for given pixels in FEL0 and FEL2 (see Figure 5.26) such that charge sharing cannot cause both pixels to fire simultaneously. The distribution was fitted with the sum of two Gaussian functions to include charge sharing from neighboring anodes. By divid-ing the value for the time resolution by $\sqrt{2}$ to account for the independence of both pixels an individual uncertainty of 320 ps is reached. This value will improve when the time-over-threshold information is included and time walk is corrected. Another contribution comes from the chromatic dispersion which will be reduced by an optical filter in front of the MCP-PMT. In the final EDD the resolution is expected to be improved due to the magnetic field in $\overline{\text{P}}$ANDA, that reduces the spread of the electron cloud in the MCPs and its charge charing on neighboring anodes.

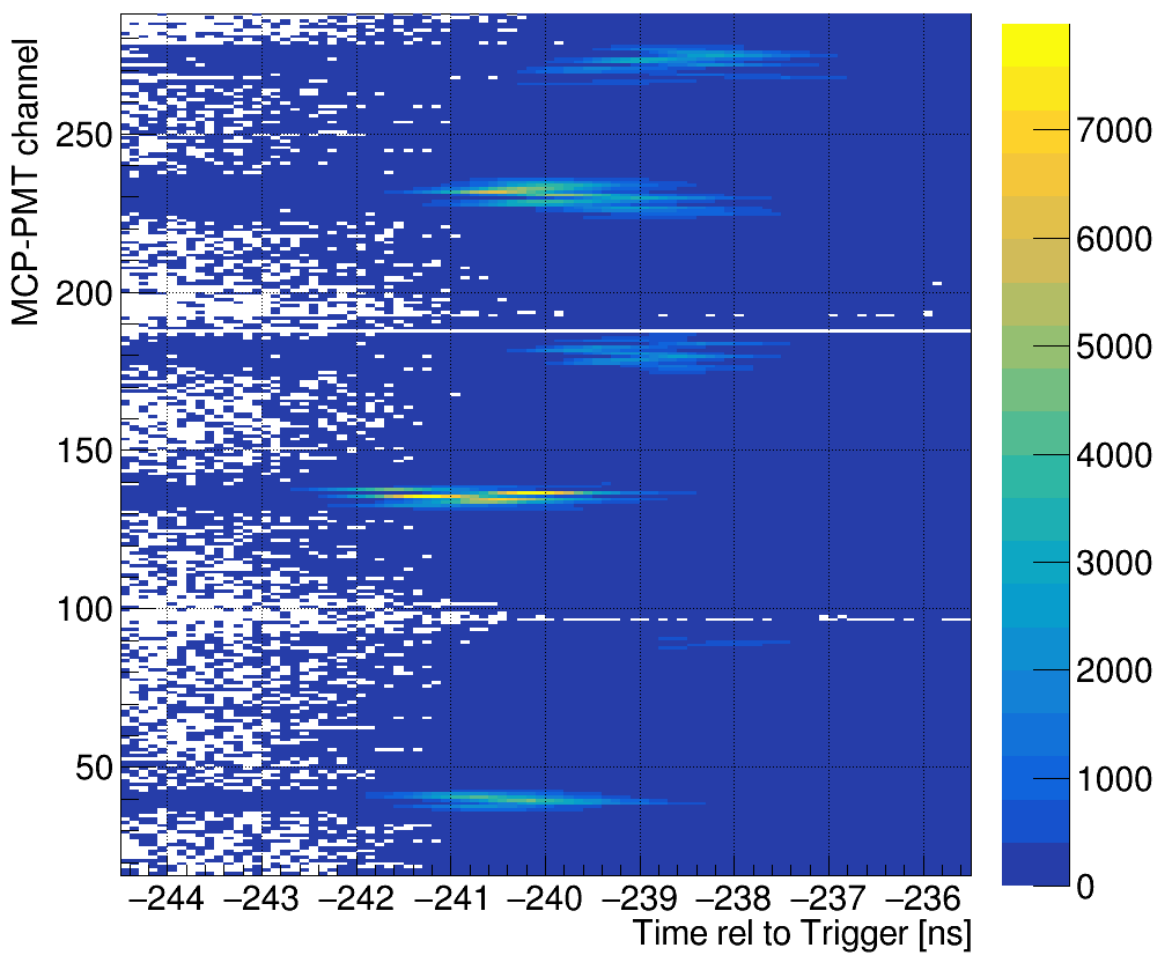

**Figure 5.25:** Position (channel) dependent time of arrival of Cherenkov photons relative to the trigger scintillator signal [17].

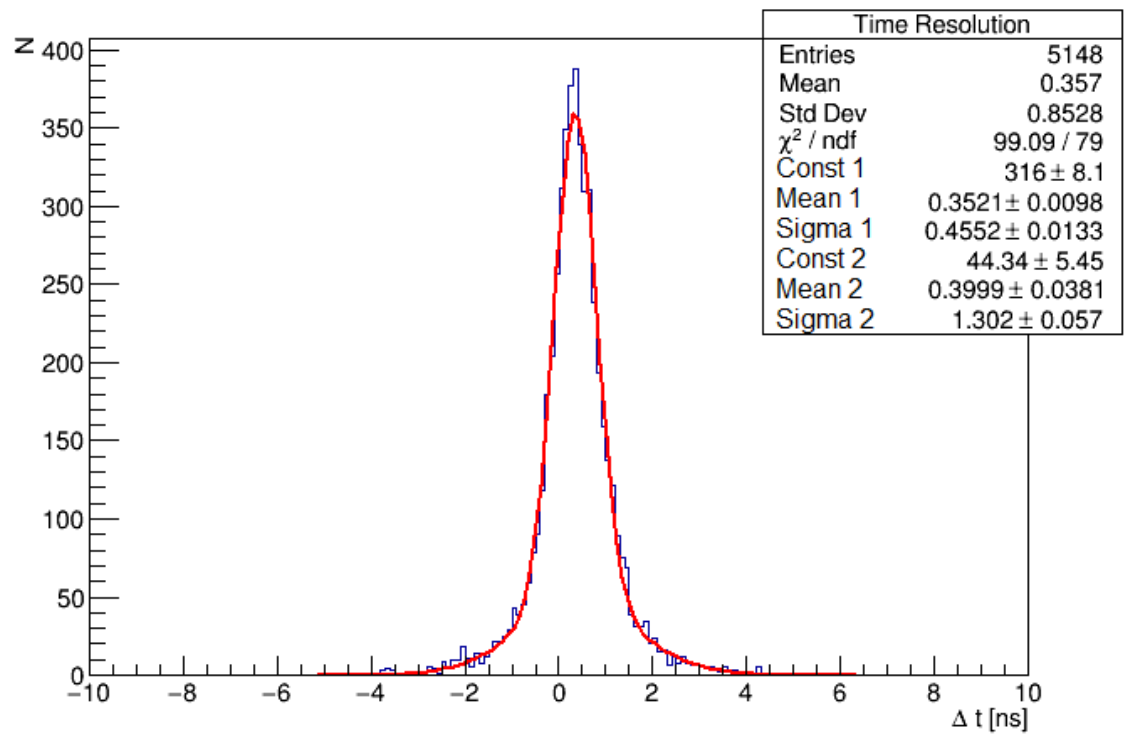

**Figure 5.26:** Time resolution obtained from time differences between the two directly illuminated pixels [17].

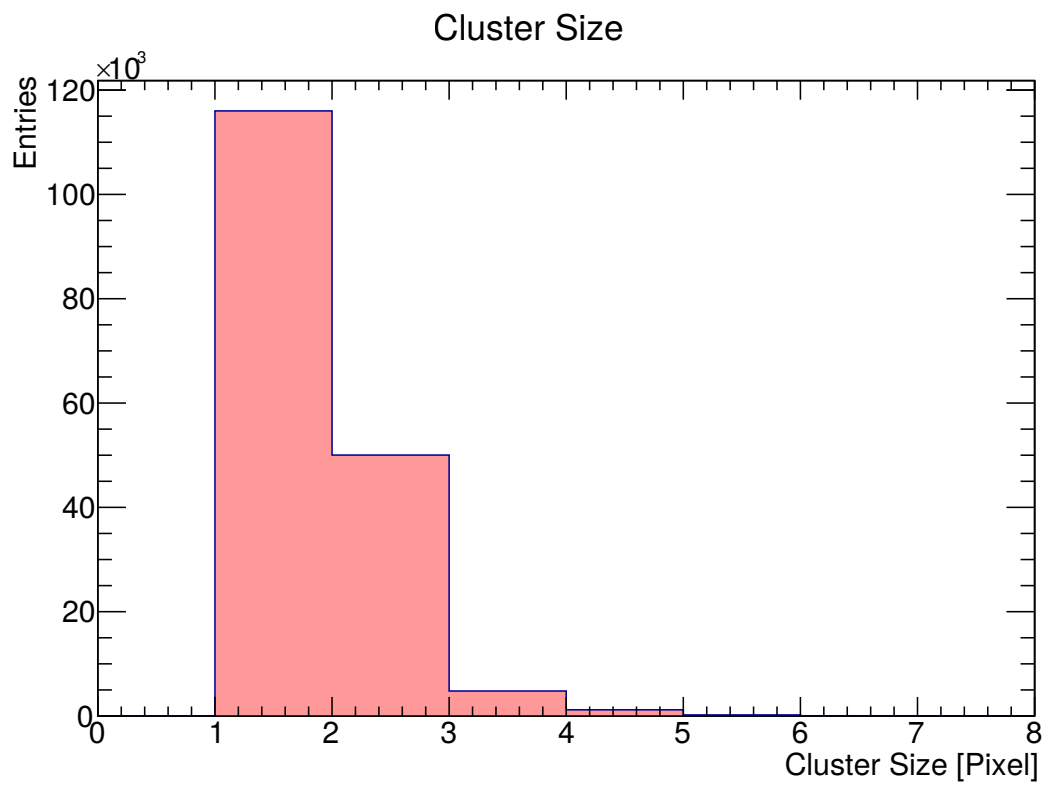

**Figure 5.27:** Cluster sizes for single events. About 30% of all events record hits on two neighboring pixels which is a rough estimate for the the amount of charge-sharing.

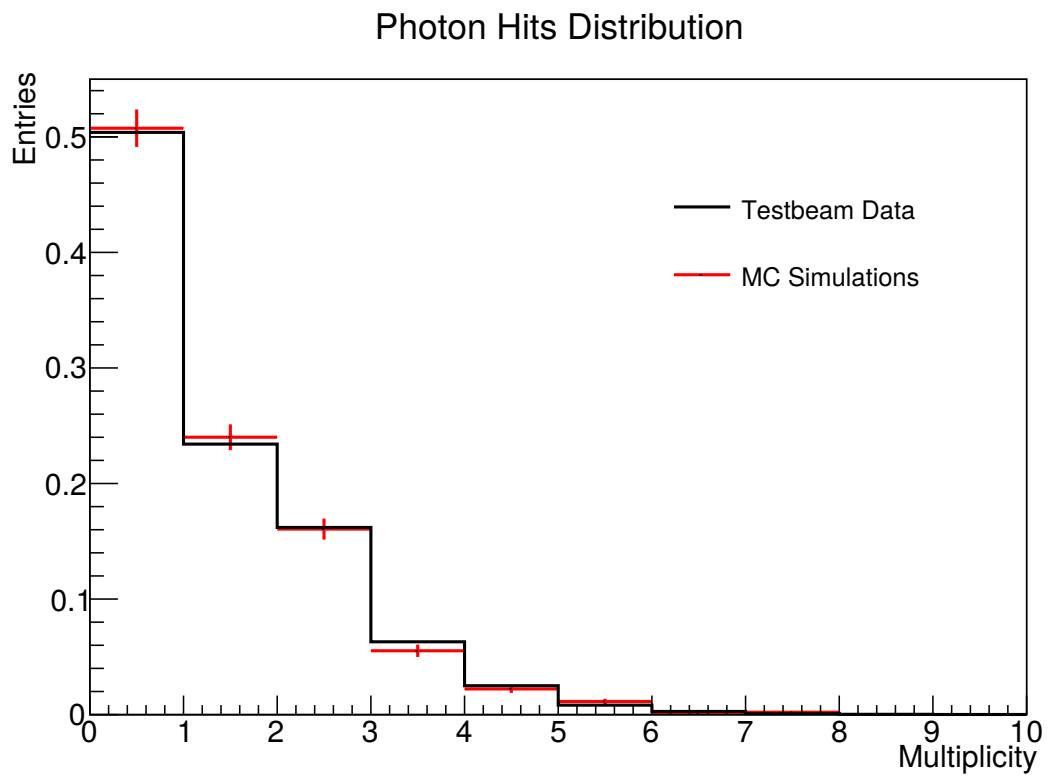

**Figure 5.28:** Cherenkov photon multiplicities at 6°. The Monte Carlo charge-sharing probability is set to 30%.

### Yield and Multiplicity Distributions of Cherenkov Photons

An important parameter for the verification of the detector performance is the photon yield, in other words the number of photons which are detected per charged hadron track. This number depends on various processes and parameters such as the photon loss due to bulk absorption inside the optical material or surface losses. In addition only a fraction of about 30% of the photons is converted into an electron inside the MCP-PMT photocathode and only another 65% hit a microchannel and produce secondary electrons to produce a significant signal for the detector readout. On the other hand hits are recorded without being caused by a Cherenkov photon. Besides dark counts, which have a relatively small effect because they can be reduced by proper time cuts, charge-sharing and electronic cross-talk can produce additional hits.

As described in section 4.2.3.6 charge-sharing is only relevant if no external magnetic field is present. In order to estimate the contribution of charge-sharing in the test experiment, the amount of clus-ters was evaluated. A cluster is defined as neighboring pixels which simultaneously record a hit during a single event. As shown in Figure 5.27 in about 30% of the events such a cluster occurs. Although the probability of two hits during a single event in one MCP-PMT column is about 10%, they do not have to necessarily hit two neighboring anodes, why a charge-sharing effect of 30% is a realistic estimate.

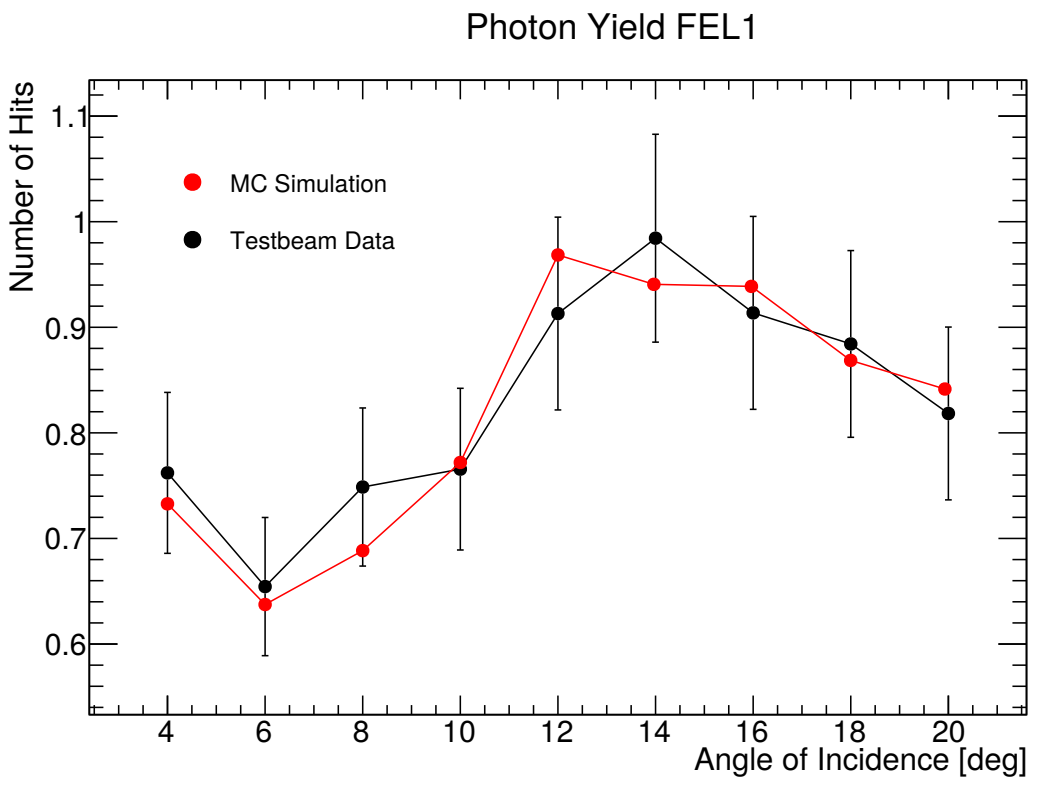

**Figure 5.29:** Photon yield at the central FEL for the angle scan. The measured and simulated data are in good agreement. The FEL and bar were joined by an epoxy resin which cuts off photons below 300 nm.

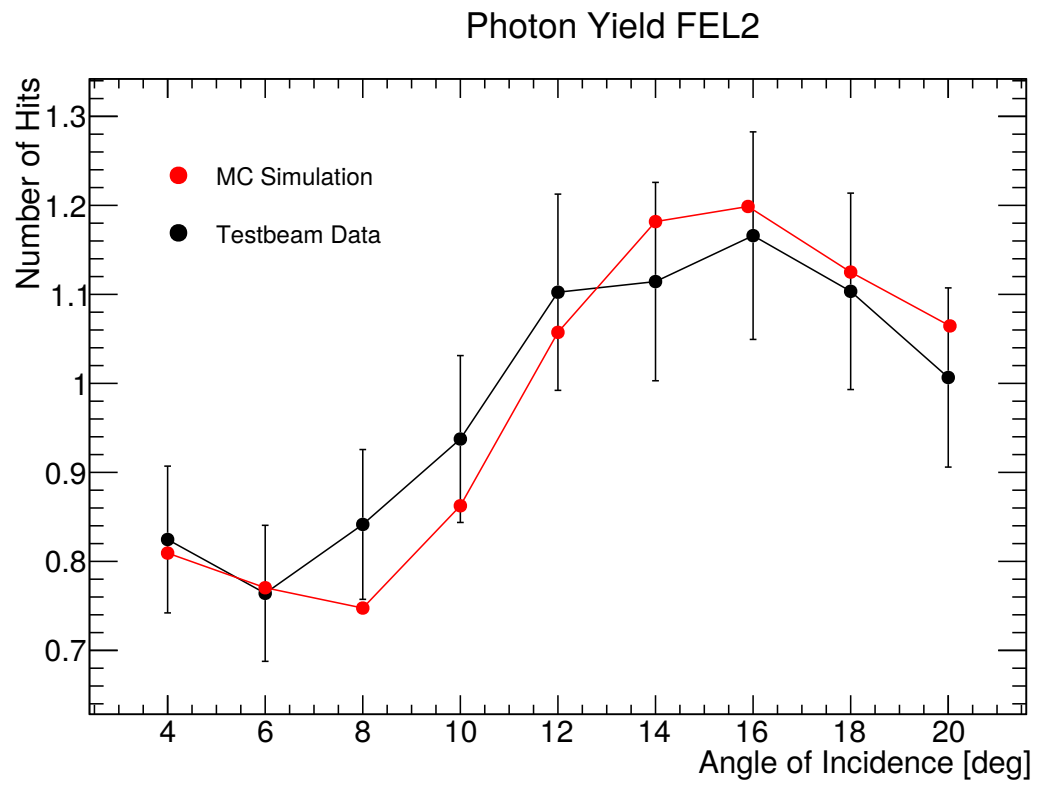

**Figure 5.30:** Photon yield at the lower FEL for the angle scan. The measured and simulated data are in good agreement. The wavelength range is limited by a photon cutoff due to the optical grease at the MCP-PMT surface. FEL and bar are joined via optical contact bonding.

As shown in Figure 5.28 this charge sharing estimate leads to a good agreement between Monte Carlo and the recorded data for the photon multiplicities. This plot shows data from the angle scan at 6° where only direct photons but no reflected photons from the side surfaces are expected.

In Figures 5.29 and 5.30 the average number of photons per track is plotted for the angle scan. Both plots show an increase of the yield at 10°. This effect is due to additional photons which are reflected off the side surfaces of the radiator. A systematic error of 10% was estimated due to fluctuations of the MCP-PMT's QE and the estimate of the charge-sharing effect. The average photon yield in Figure 5.30 is slightly higher as this FEL was contacted via optical contact bonding and hence has no cut-off at 300 nm due to an epoxy bonding. For this reason additional photons in the UV region hit the MCP-PMT. Both plots show a good agreement with Monte Carlo.

### Resolution of the Cherenkov Angle Measurement

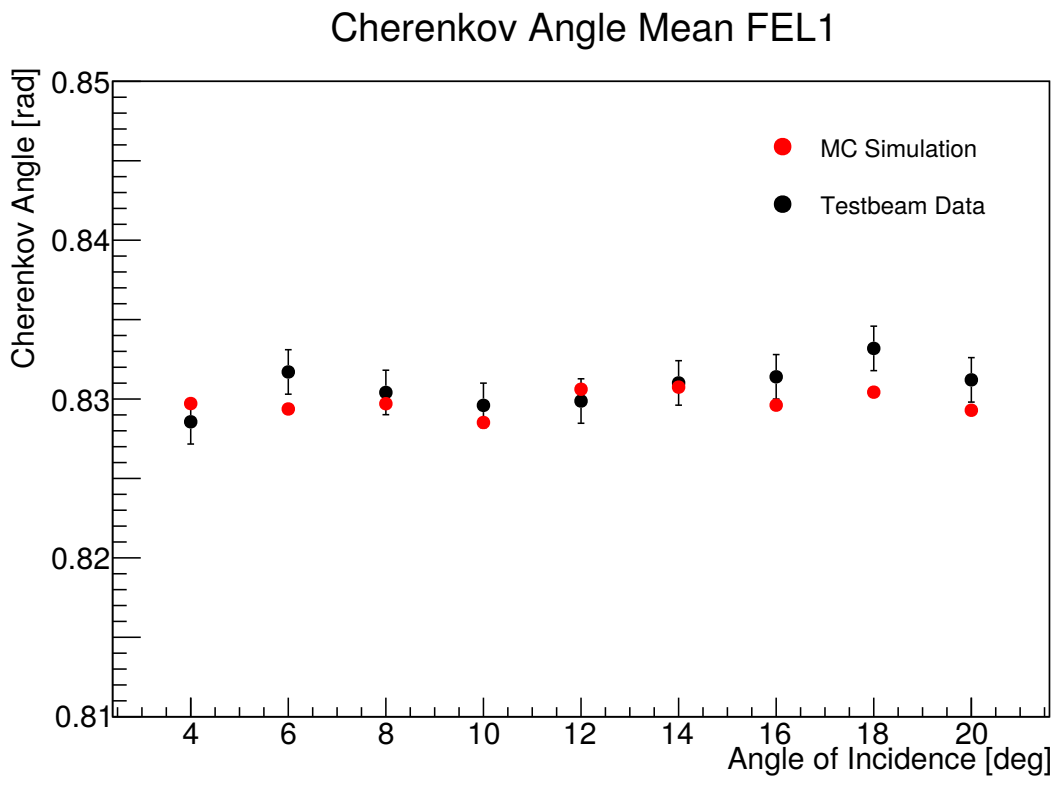

**Figure 5.31:** Measured Cherenkov angle at the central FEL as a function of the incident angle of the beam on the radiator plate. The extracted angle is independent of the beam direction and agrees with MC simulations.

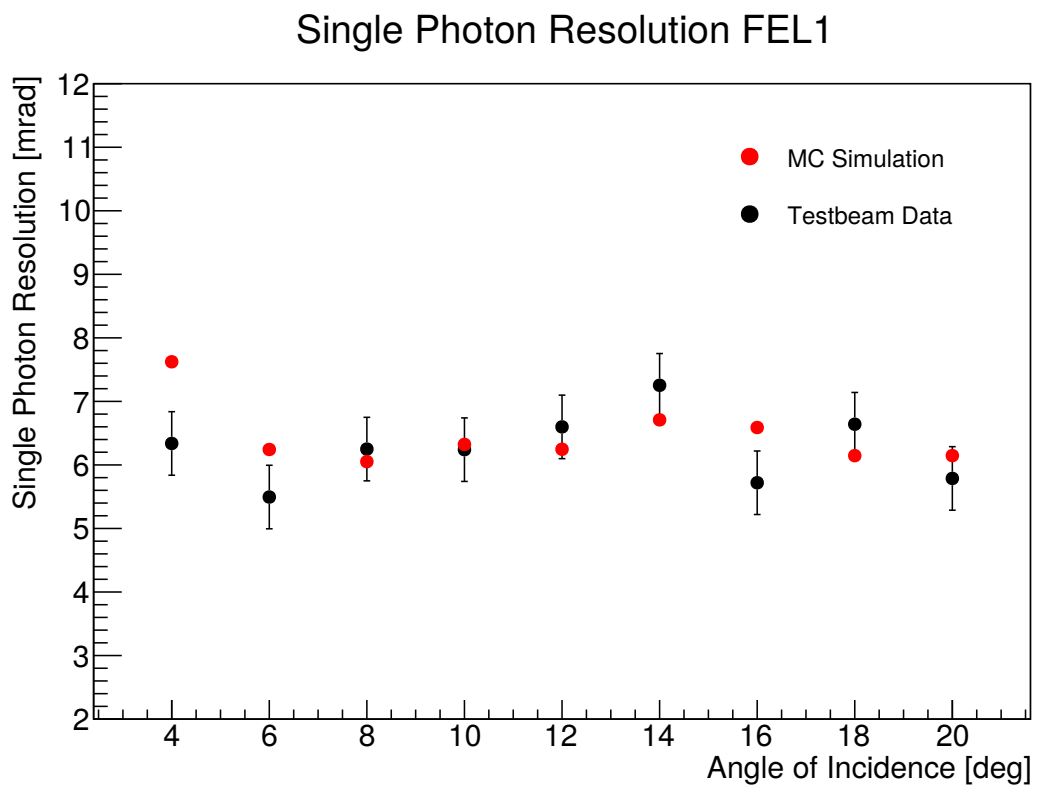

**Figure 5.32:** Measured single photon resolution at the central FEL for the angle scan. It agrees with MC simulations.

Besides the yield of photons, the angle resolution of single Cherenkov photons is a key parameter for the detector performance. The test beam at DESY in 2016 was used to determine this single photon resolution as a function of several experimental parameters, using a well defined beam. Figure 5.23 shows the two dimensional correspondence of the input angle and the measured photon position on the photon sensor. From that position, the Cherenkov angle can be extracted. Figure 5.31 shows that the extracted Cherenkov angle is basically independent of the angle between the radiator plate and the beam direction. The single photon resolution was determined by a Gaussian fit to the corresponding Cherenkov angle distributions.

Figure 5.32 shows the resolution (i.e. the width of the bands in 5.23) as a function of the beam angle with respect to the normal of the radiator plate. The systematic dependence of the single photon resolution arises mainly because at different angles different paths inside the FEL-bar pair are taken. They were determined by a separate MC simulation. The systematic error for the mean Cherenkov angle is mainly driven by the beam angle uncertainty which is taken into account via error propagation. The measurement of the mean of the Cherenkov angle distributions as well as the single photon resolution show a good agreement with the MC predictions.

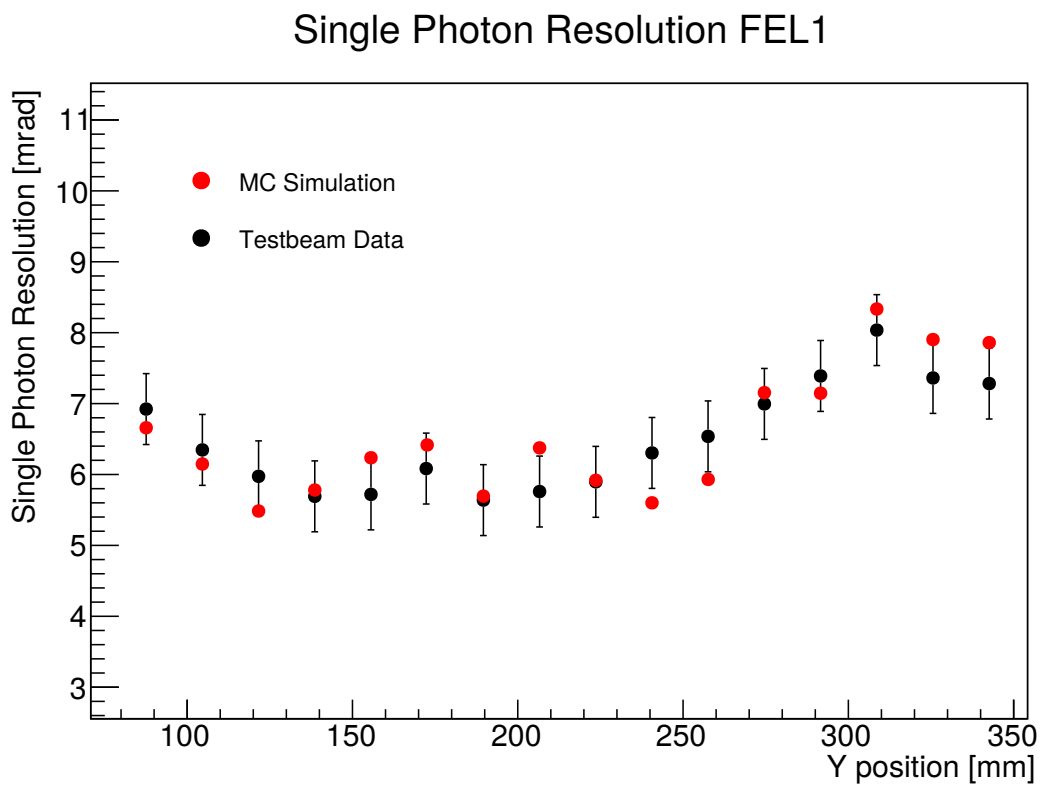

**Figure 5.33:** Measured single photon resolution at the central FEL for the angle scan vs. MC simulations.

Figure 5.24 shows the two dimensional correspondence of the vertical radiator position and the measured photon position on the sensor. This scan with only one ROM simulates the pattern of a fully equipped prototype. The 30 positions of the scan of one sensor translate into an equivalent prototype where one side is equipped with 30 equidistant sensors. The vertical scan changes the geometrical acceptance of the Cherenkov photons and hence the angle $\alpha_{FEL}$. The hit map nicely shows a (distorted) ring pattern with an indirect reflection for low vertical positions coming from the edge of the radiator plate. In Figure 5.33 the single photon resolution of the Cherenkov angle is plotted for the position scan. As expected the single photon resolution becomes slightly larger as the relative vertical distance between FEL1 (placed at the minimum at 170 mm) and the beam increases. The Monte Carlo simulation agrees with the measured results within their uncertainties.

### Combined Event Challenge

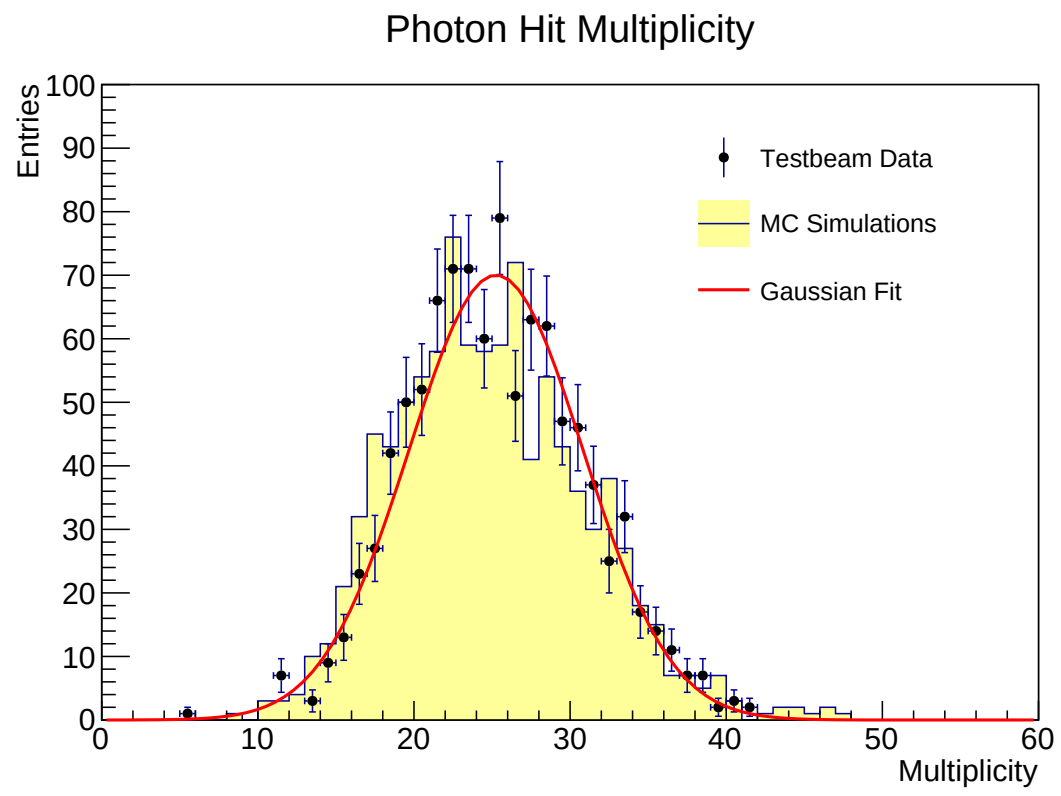

**Figure 5.34:** Number of photons per combined event.

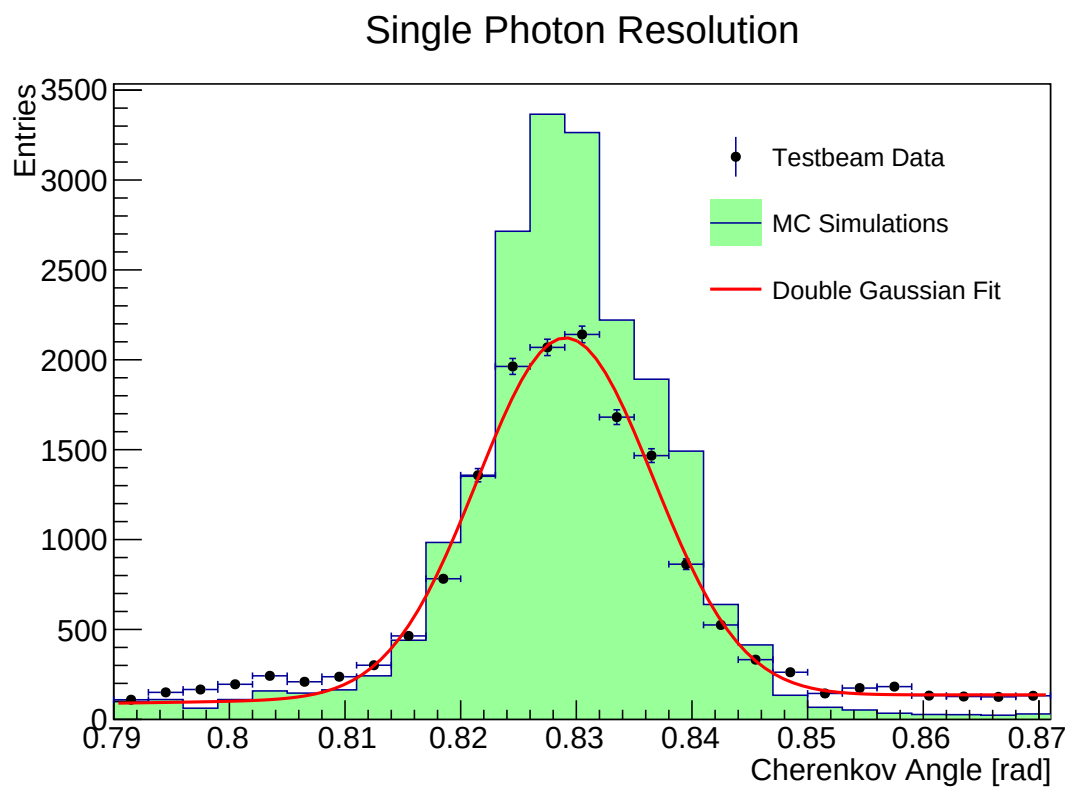

**Figure 5.35:** Single photon resolution of all recorded hits. The data sample has been fitted with two Gaussian functions.

New, so-called "combined events" can be constructed from adding 30 different events from 30 different positions of the vertical scan. This has been done for the whole data set, so that each combined event has a hit multiplicity, that simulates events of a detector with 30 FEL columns, i.e. six fully equipped ROMs. Figure 5.34 shows the distribution of the combined event multiplicity, and a good agreement between Monte Carlo and data can be observed. On average 26 hits (24 hits) per combined event were registered for the test beam (Monte Carlo) data.

The single photon resolution (Figure 5.35) of the full data set is a bit larger than the single photon resolution of the individual FEL columns (compare Figure 5.33) as systematic errors of the mean position add up, but the mean value of 7.57 mrad for the test beam data compare well with the 7.66 mrad of the Monte Carlo sample.

The challenge of the combined event analysis is to show that the combined events can be used to reconstruct the Cherenkov angle with a precision that is much better than the single photon resolution. Figure 5.35 shows significant tails beyond the Gaussian distribution that come from the fact that the combined events contain the sum of background hits from 30 data events. In the ideal case, a calibrated reconstruction program based on likelihood methods would be needed to separate signal from background. For this study, a simplified truncated mean method was applied to reject background events before the individual Cherenkov angles were averaged to calculate the mean Cherenkov angle for each combined event: For each virtual FEL column a mean was calculated and the hit with the largest variance was excluded. This procedure was repeated until all remaining hits were inside an interval around the mean value which roughly corresponds to three times the single photon resolution. The size of this interval was determined by optimizing the mean Cherenkov angle resolution of the test beam data sample and was subsequently also used for the Monte Carlo data sample. In case of the test beam data on average 13 hits passed the selection routine and a mean Cherenkov angle resolution of 2.52 mrad has been obtained. In case of Monte Carlo data 18 hits were selected and the mean Cherenkov angle resolution was measured to be 2.16 mrad. Figure 5.36 shows the result and that a good agreement is reached. The combined event analysis is expected to give results that are worse that the results of a fully equipped detector, as the combined events combine Cherenkov light from 30 different beam particles with different positions and divergence.

To conclude, the combined event challenge showed successfully that an improvement by a factor 3 in resolution can be achieved, using a simplified reconstruction algorithm to combine the photon hits of one event.

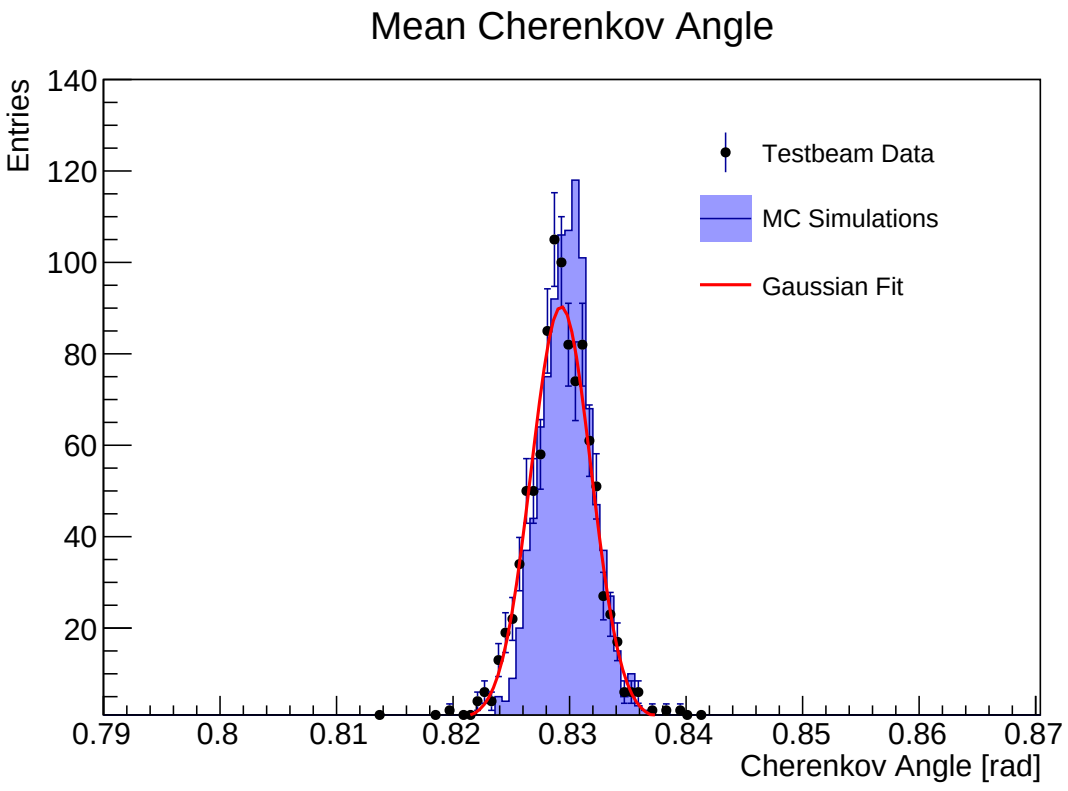

**Figure 5.36:** Mean Cherenkov angle for each combined event. The data sample has been fitted with a single Gaussian to obtain the mean Cherenkov angle resolution.

## 5.2 Final Endcap Disc DIRC

### 5.2.1 Expected Detector Performance

The previous chapter showed that the small prototype detectors do not have the performance which the proposed EDD will have. This is mainly due to four reasons: The available photon detectors do not have the final wavelength sensitivity curves, the MCP-PMTs are not in the final, strong fringe field of the $\overline{\text{P}}$ANDA solenoid, the background conditions in a test beam are much reduced compared to a real high luminosity hadron experiment, and most importantly, our prototype detectors were stripped down versions with only one or a few sensors on a smaller sized radiator plate. Instead, the final EDD will have 24 MCP-PMTs per quadrant. It is planned to build a full size quadrant and test it in a real environment to verify its performance. Until such a full-size quadrant is funded, we rely on Monte Carlo simulations to estimate the performance of the final EDD. However, the previous chapter has shown that our Monte Carlo codes are reliable and reproduce well the properties of the tested Disc DIRC prototypes.

Most important for the EDD is to reach the anticipated resolution in the measurement of the Cherenkov angle. Figure 5.37 summarizes the steps that lead from the limited performance of the tested prototypes to the resolution of the final EDD. Table 5.1 lists the corresponding numbers.

Panel a) shows the simulated distribution of the Cherenkov angle of the 2016 DESY test beam. The beam was placed 450 mm away from the FEL with an angle of incidence of 16°. Results of the actual measurement and of the simulation consistently

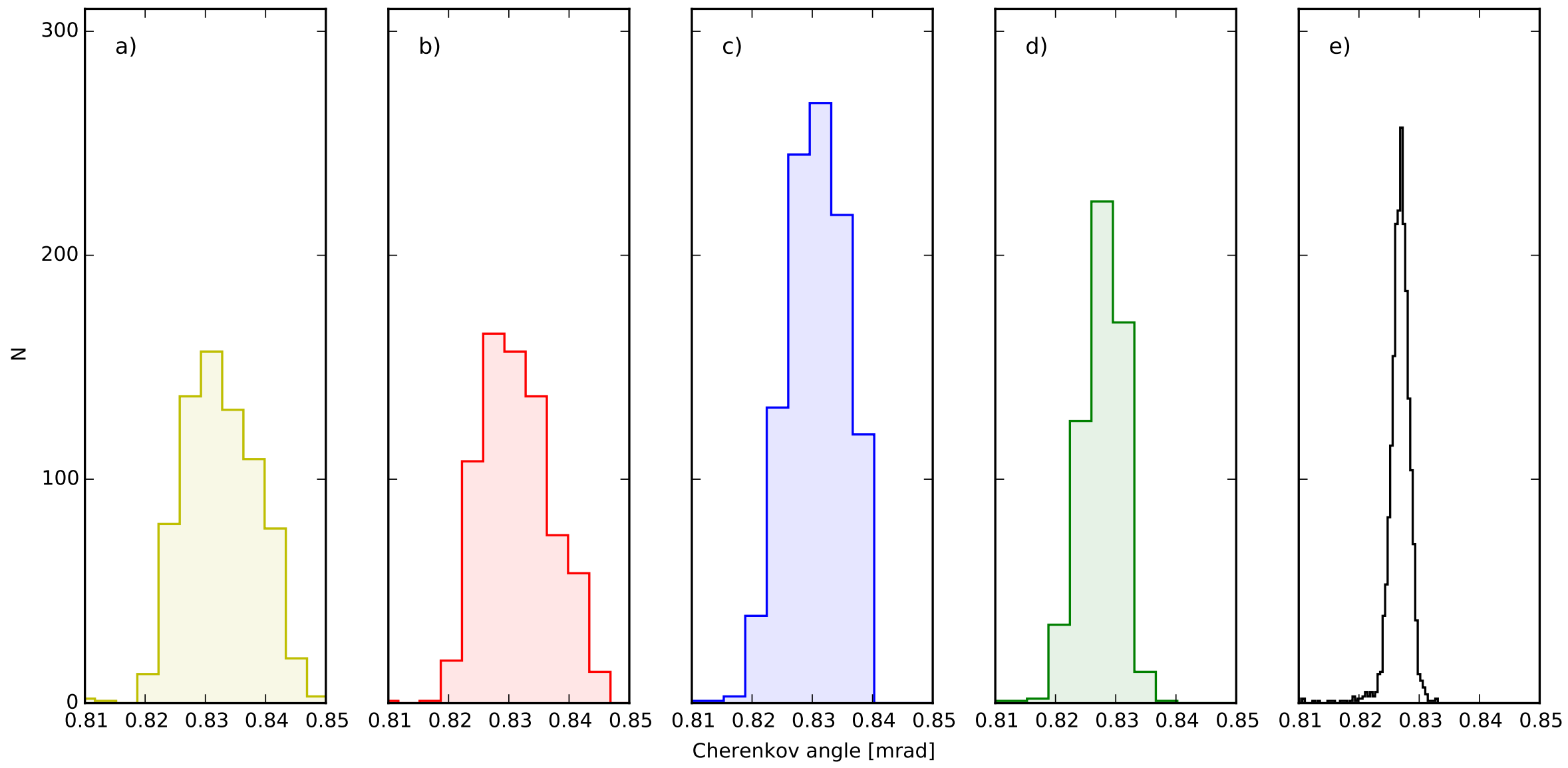

**Figure 5.37:** Improvement of the single photon resolution by a stepwise transfer from testbeam conditions with electrons to a fully equipped EDD inside $\overline{\text{P}}$ANDA: a) electrons for the small prototype with one FEL at the DESY testbeam; b) pions for the small prototype; c) pions inside the $\overline{\text{P}}$ANDA experiment d) adding an edge filter which blocks photons below 355 nm; e) averaging the information of all 72 FELs in a full quadrant. Table 5.1 gives a quantitative overview.

yield a single photon resolution of about 6 mrad.

Panel b) shows the resolution for the case that hadrons (pions) instead of electrons cross the detector. The resolution is slightly reduced as deflections of the hadronic particle by multi Coulomb scattering are reduced compared to electrons.

Panel c) shows an improved situation, which corresponds to the transition from the small testbeam radiator to a full size quadrant which is hit by a pion that originates from the $\overline{\text{P}}$ANDA target vertex at the same angle of incidence. The photon yield is increased due to geometrical acceptance effects and the single photon resolution is improved (4.8 mrad) as $\overline{\text{P}}$ANDA will provide a more precise information on the position and angle of the incoming hadron track compared to the information that was available at the DESY test beam.

Panel d) includes the anticipated wavelength filter which does not transmit photons below 355 nm. This way, the yield is reduced but at the same time the dispersion effect on the single photon Cherenkov angle resolution is reduced and a resolution of 3.5 mrad is obtained. This value also corresponds to the maximum achievable single photon resolution with the chosen detector design.

Panel e) shows the final performance of the EDD. Here, the information of all 72 FELs is combined while before only one FEL was used. The binning in a)-d) corresponds to the segmentation of the photon sensor. Due to the averaging of all measured photons, panel e) allows a finer binning of the measured angle and a resolution as good as 1.8 mrad is obtained. In the presented case of a 3 GeV/c particle this full detector resolution translates into a $\pi/K$ separation power of more than 6 $\sigma$.

**Table 5.1:** Overview of Cherenkov angle resolutions shown in Figure 5.37.

| radiator | particle | filter | #FEL | resolution |
|---|---|---|---|---|
| prototype | e | - | 1 | 6.06 mrad |
| prototype | $\pi$ | - | 1 | 5.95 mrad |
| $\overline{\text{P}}$ANDA | $\pi$ | - | 1 | 4.80 mrad |
| $\overline{\text{P}}$ANDA | $\pi$ | LP | 1 | 3.45 mrad |
| $\overline{\text{P}}$ANDA | $\pi$ | LP | 72 | 1.80 mrad |

### 5.2.2 PID Performance using Two Different Analysis Methods

Detector simulations based on the Geant4 framework [18] have been extensively used during the design of the detector [19]. As previously described these models are able to account for surface losses at the optical interfaces, bulk attenuation, detection efficiencies and the dead time of the front-end electronics. Particle and optical photon transport is completely handled by Geant4 up to the sensi-tive surface of the photodetector (see Figure 5.38). The photon detection process is then handled in a parametrized way by applying efficiencies, dead time filtering and smearing of measured observables. An accumulated hit-pattern for $10^4$ simulated pions at a momentum of 4 GeV/c, polar and azimuth angles $\theta = 16°$, $\phi = 45°$ for the final EDD design using MCP-PMTs is shown in Figure 5.39.

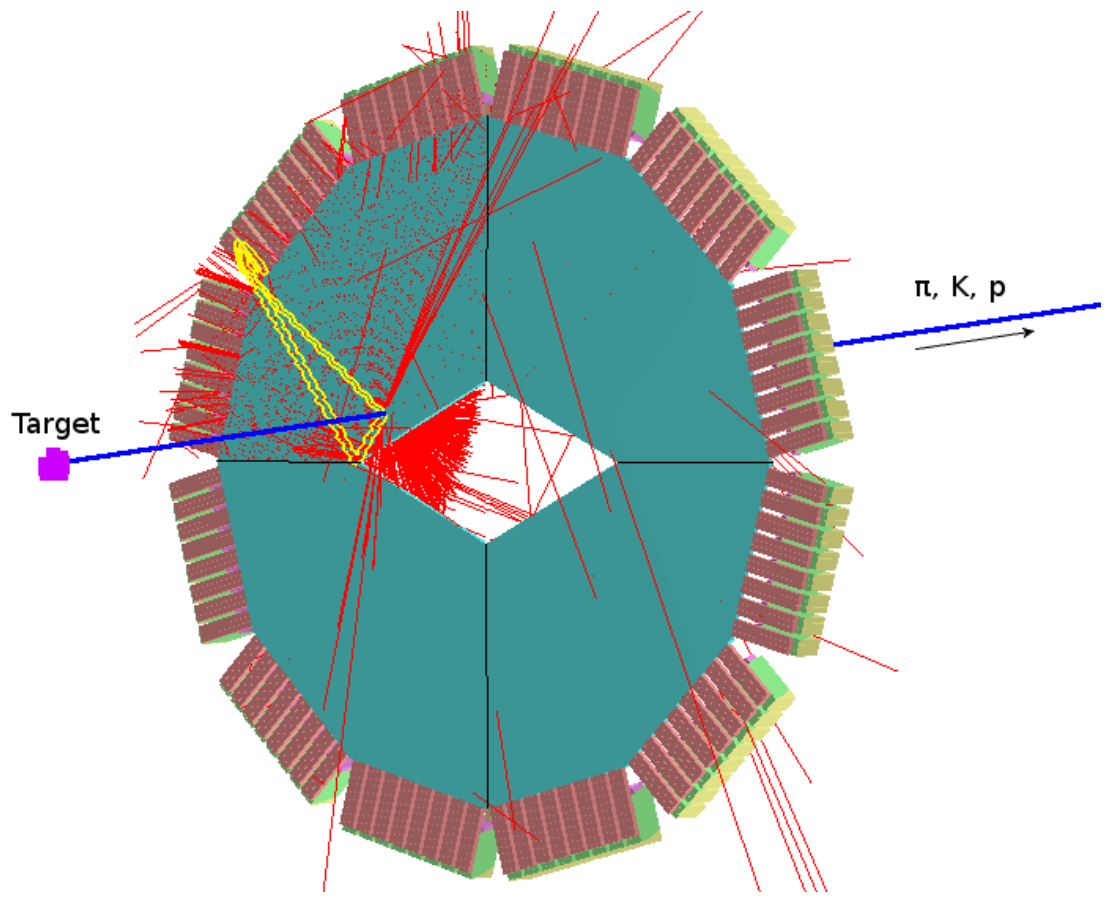

**Figure 5.38:** PandaROOT event display showing photons originating from a charged particle track hitting the Endcap DIRC. Only a few photon tracks are highlighted for clarity.

The reconstructed single photon Cherenkov angle $\theta_c$ and the corresponding reconstructed mean Cherenkov angles per track with an RMS of 1.55 mrad are plotted in Figure 5.40. The peak structure is a result of the discrete pixel structure of the MCP-PMT anode folded with geometric effects and chromatic dispersion. The reconstruction algorithm is purely analytical and does not rely on simulation specific input for calibration. This is an advantage over approaches based on look-up tables where the reconstruction uses Monte Carlo data to predict Monte Carlo results (see also section 3.3).

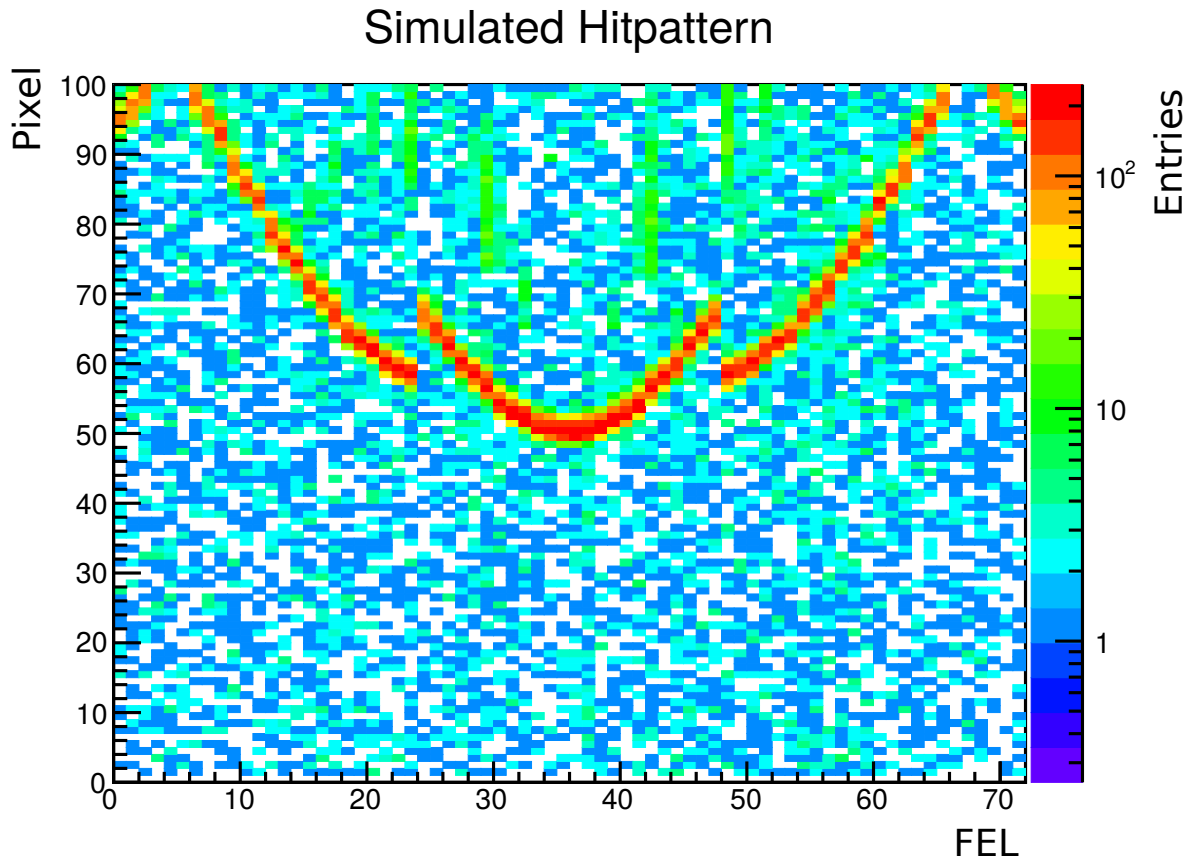

**Figure 5.39:** Accumulated hit patterns (log-scale) without experimental background from the continuous $\bar{p}p$ interaction. The hit pattern was generated with 2000 pions at 3 GeV/c at $\phi = 45°$ and $\theta = 16°$.

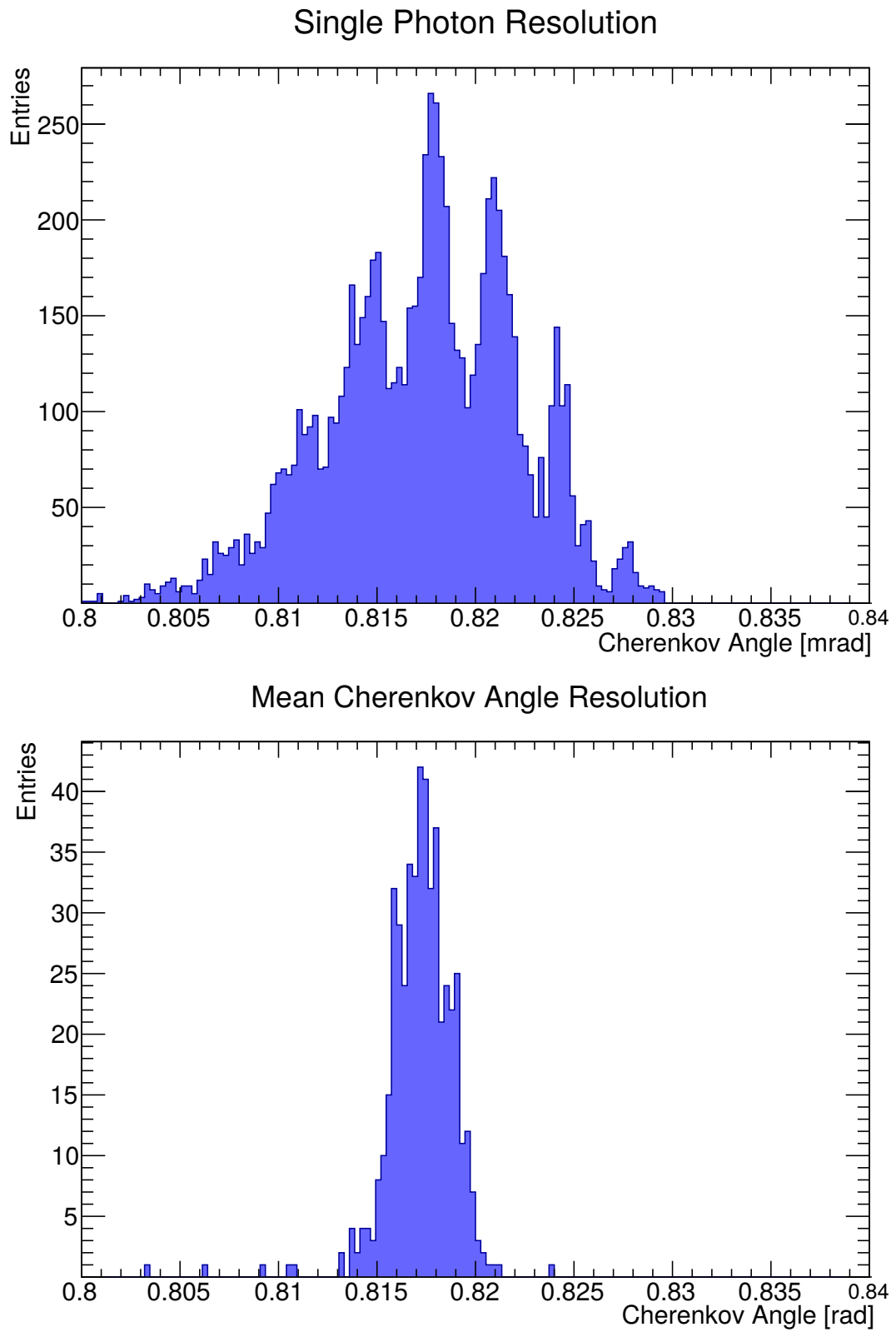

**Figure 5.40:** Distribution of reconstructed single photon Cherenkov angles (top) and the corresponding Cherenkov angles of each track (bottom) reconstructed from $10^4$ simulated patterns at 4 GeV/c.

The experimental background including its time structure as well as the influence of the magnetic field have not been modeled in the original simulations which were used for the development of the detector design. After the design was finished, the Monte Carlo model has been ported to the PandaROOT framework and complete EDD simulations in the $\overline{\text{P}}$ANDA environment have been done, including so-called "time based" simulation which include all subdetectors and hence account for additional showers created in their material [19]. The results demonstrated a stable reconstruction performance and resolutions which are compatible with the event based, single pattern results.

In this time based simulation, the reconstruction of single probe tracks has been performed in presence of the full experimental background at the nominal interaction rate of 20 MHz of a 15 GeV/c beam. This background model includes also multi-track events where more than one track is present in a single EDD quadrant. The background events have been obtained using the DPM event generator included in the PandaROOT package. Hits from background and probe track simulations were then mixed and ordered by time-stamp while taking a dead time of 40 ns per pixel into account. To test the reconstruction, hits were extracted from the resulting time based data stream using a simple time-window approach. All hits in this window were then fed into the reconstruction software which eliminates photon background by means of cuts in time and position. The remaining procedure is analog to the event-based reconstruction.

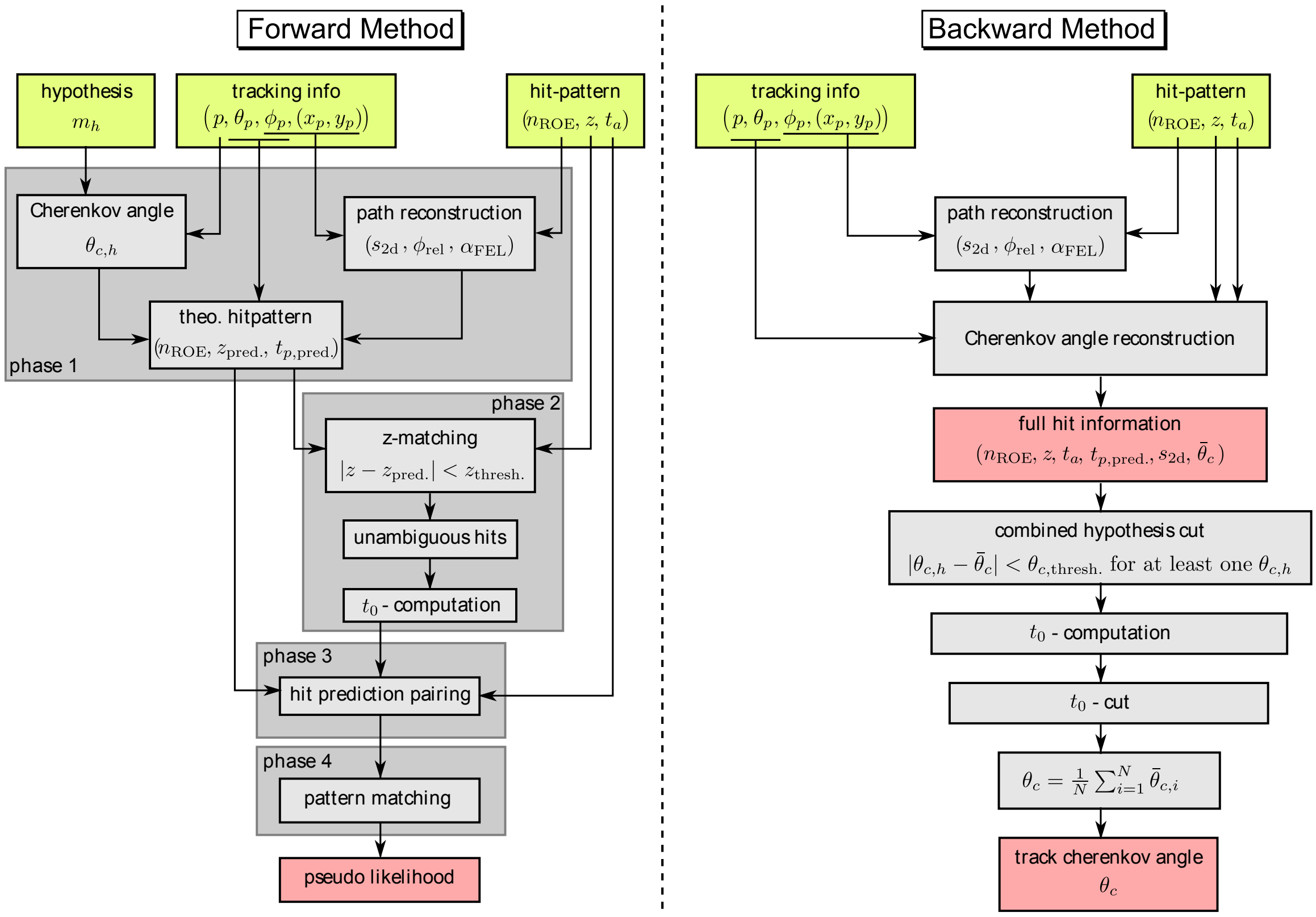

**Figure 5.41:** Flow charts for the foward and backward reconstruction method.

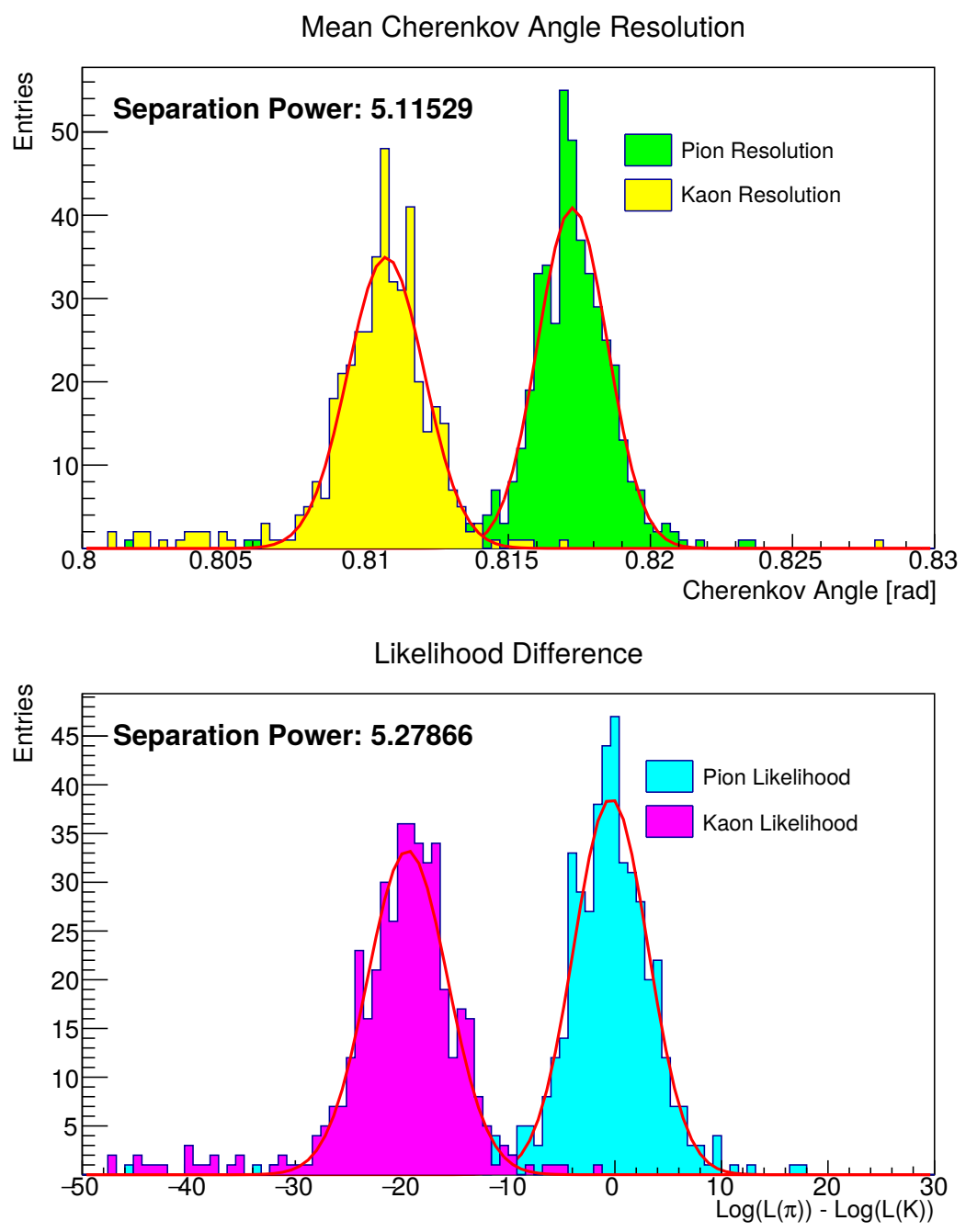

**Figure 5.42:** Separation powers for the backward method (top) and forward method (bottom).

This event-based PID algorithm is similar to the one developed for the SiPM based design [20, 19]. It is a generic algorithm for plate based DIRC counters and works for any radiator shape with the outline of a convex polygon. This algorithm has been further improved by switching from the likelihood based forward method to the direct reconstruction of the Cherenkov angle (backward method).

As input parameters, the tracking system has to provide the momentum vector and position at the track-radiator intersection together with a coarse timestamp. The coarse time is used to define an initial time window for hit selection. The track position is needed to predict all possible projected photon paths (like the purple lines in Figure 3.7) including reflections. These predictions can be used to either compute a photon hit position $z_{\text{theo}}$ on the MCP-PMT from a Cherenkov angle hypothesis (forward method) or a Cherenkov angle candidate $\theta_{c,\text{theo}}$ from the measured hit position $z$ (backward method). Due to multiple possible paths, some of the predictions are ambiguous.

The forward method involves the matching to different hypotheses, one at a time, while the backward method results directly in the Cherenkov angle (see also the flow charts in Figure 5.41). As shown in Figure 5.42 both methods perform equivalently. The latter has been chosen as it reduces the computational load, what is especially important regarding the future online reconstruction. For every measured $z$ a set of triples $(n_{\text{FEL}}, t_a, \theta_{c,\text{theo}})$ is computed, with FEL number $n_{\text{FEL}}$ and time of photon arrival $t_a$. Most of the path ambiguities can be directly resolved by disregarding nonphysical $\theta_{c,\text{theo}}$. After this step, the common emission time $T_0$ is computed from all unambiguous hits, and a time cut can be used to resolve remaining ambiguities. The mean $\theta_{c,\text{theo}}$ of the so cleaned sample is then used for calculating a likelihood value for particle identification.

### 5.2.3 Overall Performance using two different Photocathode Options

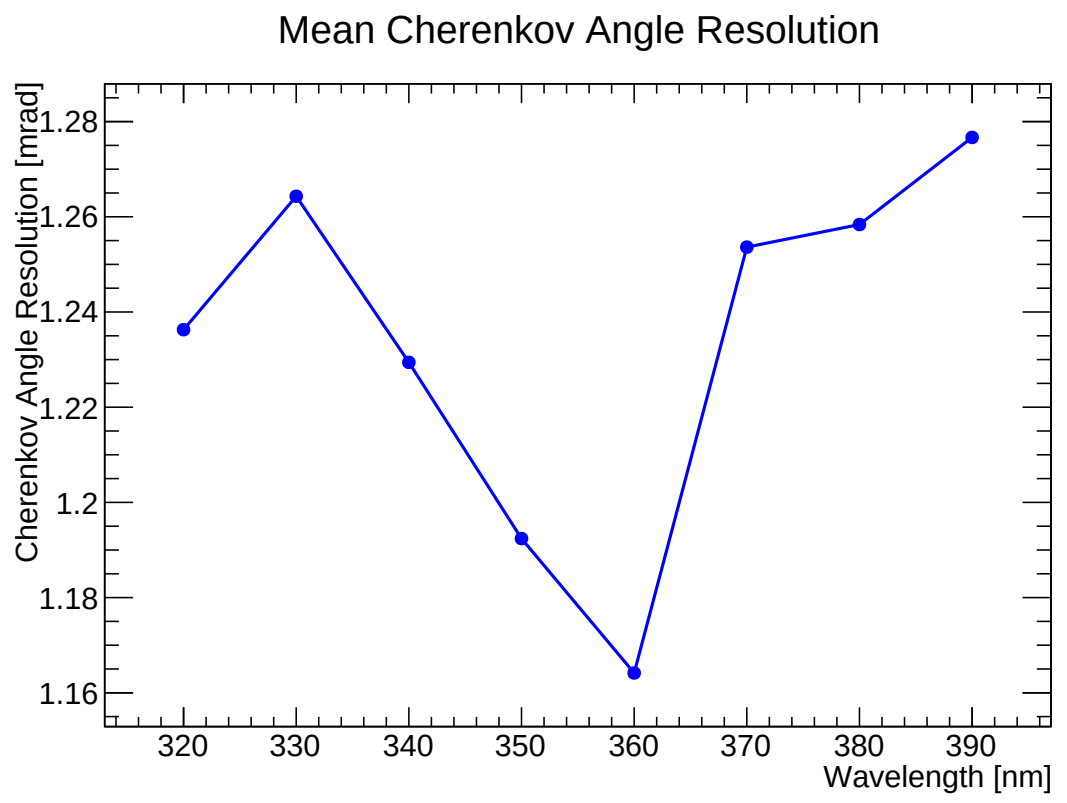

**Figure 5.43:** A long-pass filter with a starting edge at about 355 nm gives the best single photon resolution for a "blue" photocathode.

For the final design of the Endcap Disc DIRC two options are available which differ by the choice of the MCP-PMT photocathode. The first option contains a photocathode with a high quantum efficiency in the blue spectrum. In order to limit the number of photons regarding the restricted lifetime and to limit the chromatic error, an edge filter will be added. This filter can either be placed between the MCP-PMT and the FEL or be included as coating on the MCP-PMT entry window. As shown in Figure 5.43 a long-pass filter with a start value of about 355 nm provides the best performance for this particular photocathode.

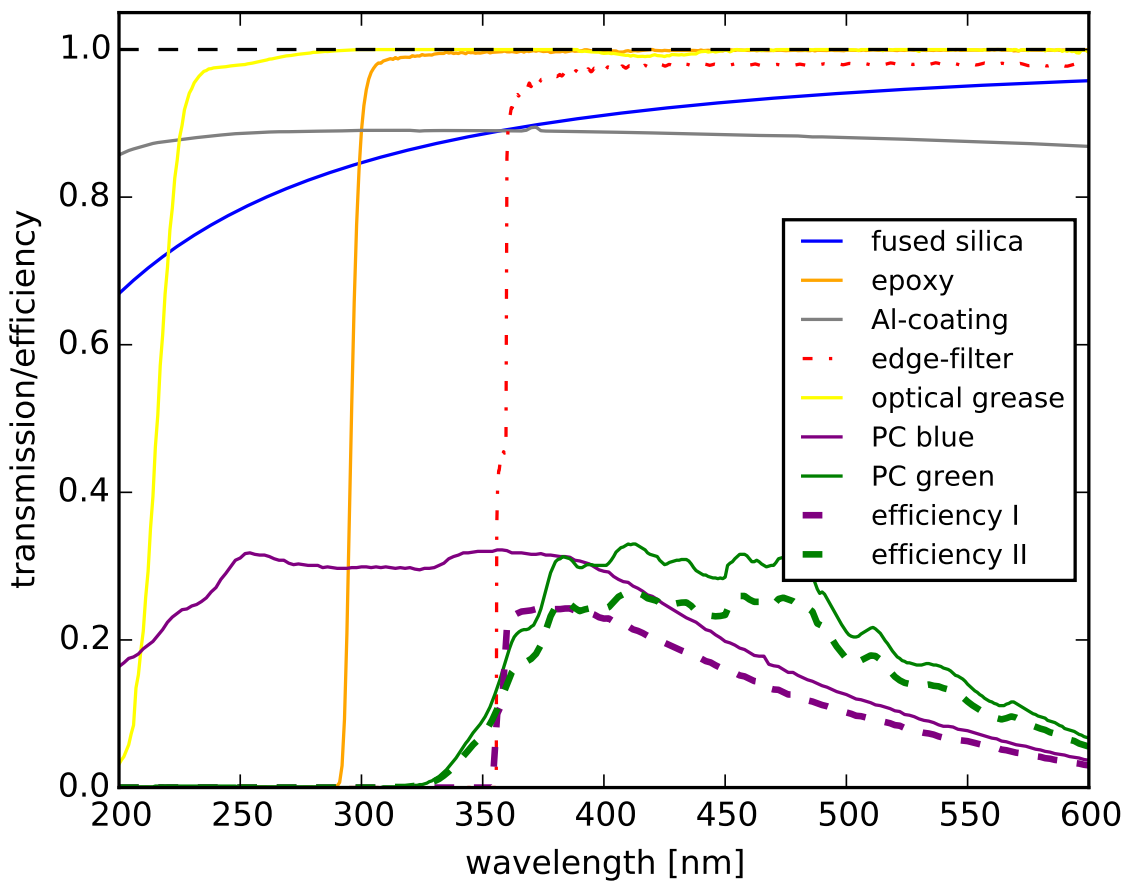

**Figure 5.44:** Transmission values and efficiencies for different optical components that have been used in the subsequent simulations. The transmission for quartz has been calculated for 400 nm photons with at $\phi = 41°$ which travel a projected distance of 1 m. The QE of the blue photocathode has been folded with an edge filter whereas the green photocathode goes without an additional filter. The shown efficiencies do not contain a collection efficiency. The plotted values were provided by [14, 21, 15, 11, 22].

The second option is based on a photocathode where the efficiency peaks in the green visible spectrum. Both, Hamamatsu and PHOTONIS, have such a photocathode on offer but yet need to apply it to a 2-inch tube. The advantage of such a green photocathode is that the photon yield at higher wavelengths, where the dispersion is smaller, is improved and at the same time no additional filter has to be used.

Figure 5.44 shows both options along with transmission values for the individual components. In comparison to the EDD design proposed in [19] the new design reduces the number of ROMs per side from nine to eight to gain space to ease mechanical design and assembly (see chapter 6).

The performance for both MCP-PMT options was mapped out by a Geant4 simulation to retrieve $\pi/K$ separation power, photon yield, mean Cherenkov angle resolution and the reconstruction efficiency at 4 GeV/c. The whole radiator (polar vs. azimuthal angle) has been scanned and for each bin 500 events were generated. The results are summarized on the left side of Figure 5.45 for the "blue" photocathode option which uses a long-pass filter and on the right side of Figure 5.45 for the "green" photocathode option which goes without an additional filter.

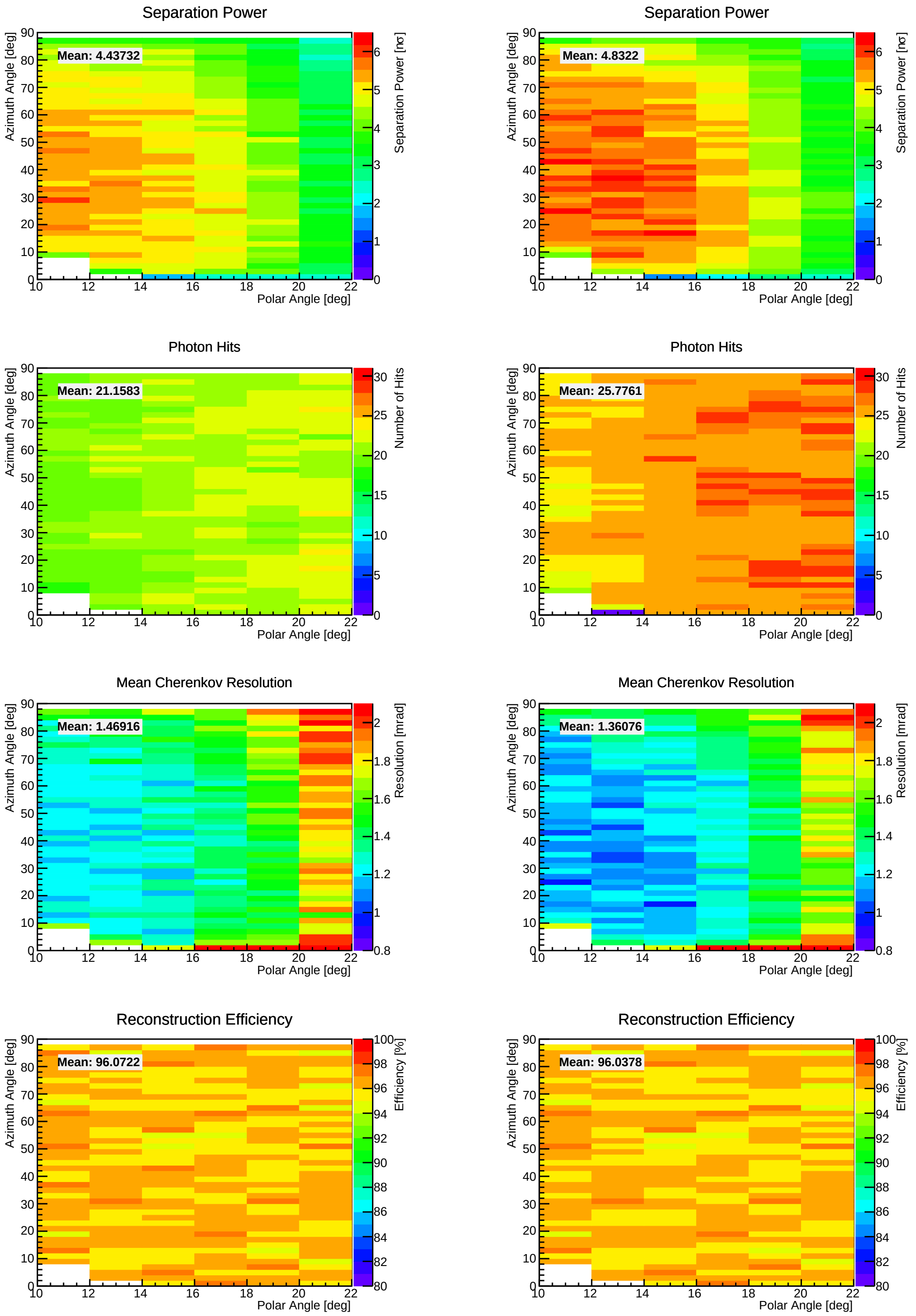

**Figure 5.45:** EDD performance overview plots: blue photocathode plus filter (left) and green photocathode (right).

Both options perform similar in terms of the efficiency which is 97% in both cases. This efficiency value contains also events where the probe track is lost due to inelastic collisions. By suppressing this process the efficiency reaches values above 99%.

The performance regarding the separation power and the mean Cherenkov angle resolution is slightly asymmetric and drops towards very small and large azimuthal angles and large polar angles. The difference in azimuthal direction is caused by overlapping photon hits due to reflections off the small side surfaces. Towards larger polar angles the geometric acceptance due to the width of the FELs changes which leads to a reduced resolution of $\alpha_{FEL}$. However, the mean Cherenkov angle resolution is good enough to provide a separation power of more than $3\,\sigma$ for most of the occupied area. By comparing both options one can obtain an improved performance for the "green" photocathode option which amounts to $0.4\,\sigma$ in terms of separation power.

The downside of the "green" photocathode option is the larger number of photons. On average about 20% more photons are registered for each charged particle track which leads to higher lifetime requirements of the MCP-PMT photocathode and the MCP-PMT's rate capability as described in chapter 4.2.

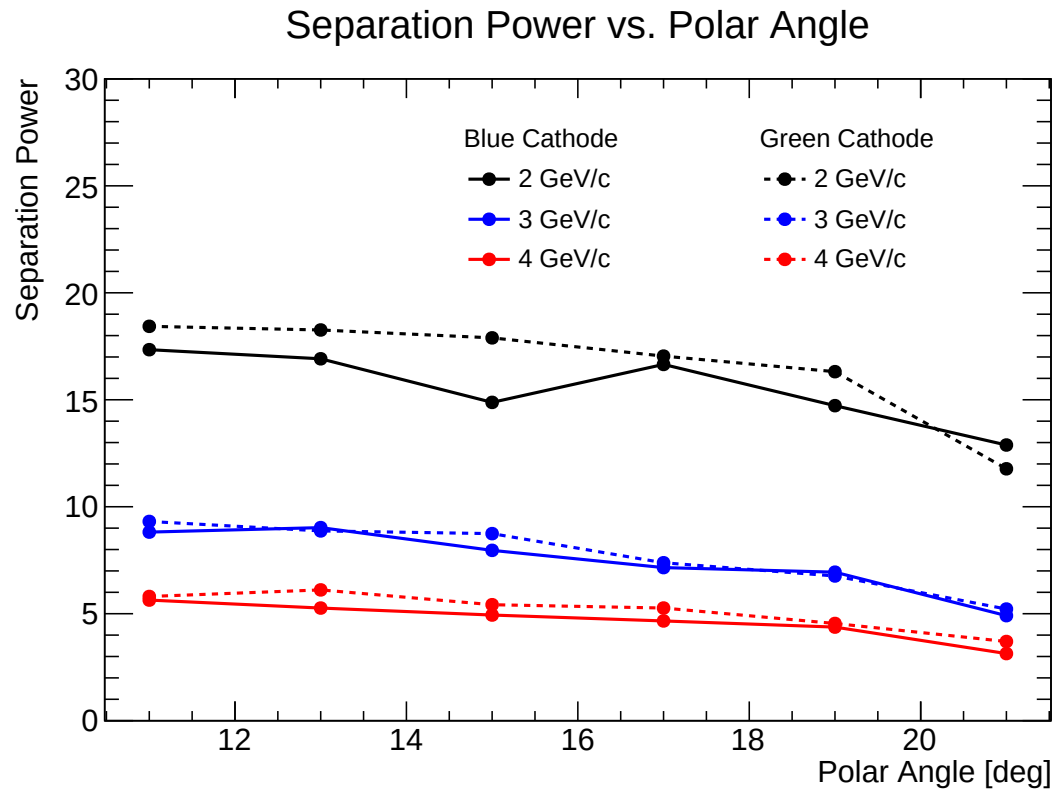

**Figure 5.46:** Separation power for different particle momenta and polar angles at $\phi = 45°$ for both photocathode options.

Both options are once more compared in Figure 5.46 which plots the separation power for both options versus the polar angle for an azimuthal position of 45°. The studies reveal that a photocathode which is sensitive in the green visible spectrum is an interesting choice which does not only make an additional component in terms of a long- or band-pass filter obsolete but also yields and improved performance. However, 2-inch tubes with this photocathode option are not available at present. For this reason the option with an available "blue" photocathode and a long-pass filter with a cut-off wavelength of 355 nm, which is also commercially available, is the baseline option at present. The filter can be applied to the entry window of the MCP-PMT which makes a transition to the "green" photocathode option possible at a later time of the experiment, without changing the FELs.

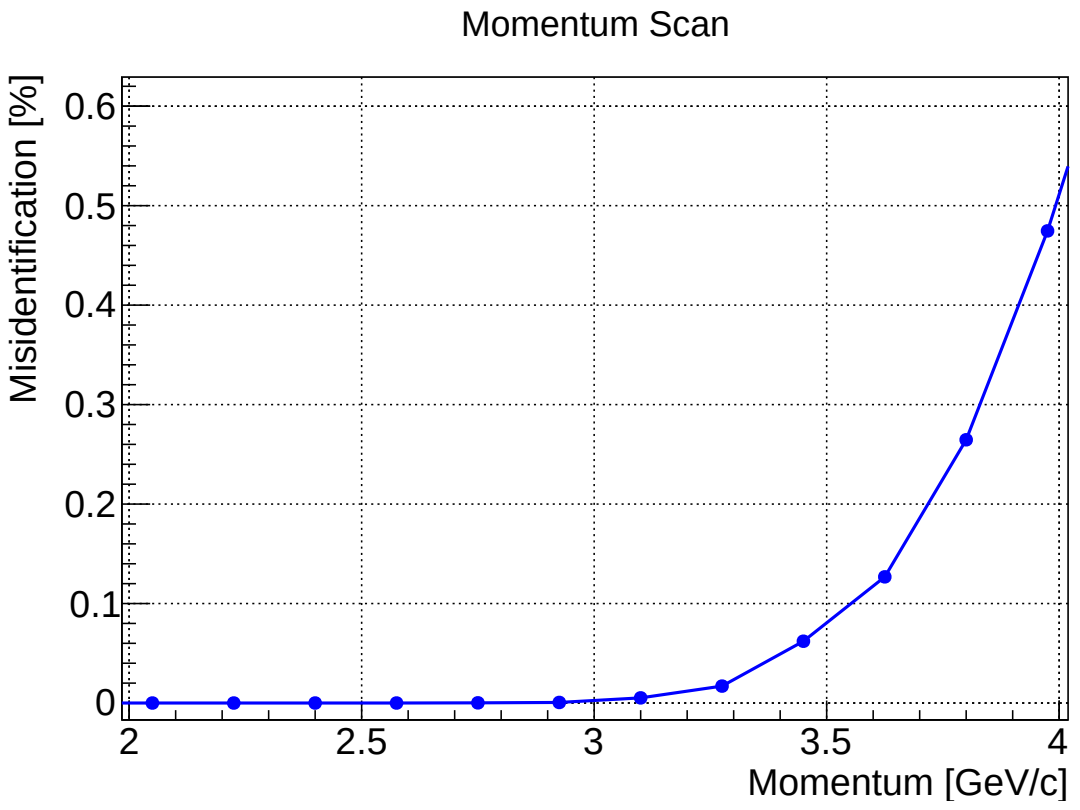

**Figure 5.47:** The simulated misidentification for $\pi/K$ separation as a function of the particle momentum assuming equal particle fluxes.

For comparison, Figure 5.47 illustrates the misidentification probability for $\pi/K$ separation as a function of the particle momentum. For this study, it has been assumed to have equal fluxes of both particle species. At the average separation power for 4 GeV/c momentum it will be possible obtain a misidentification below 1%.

The simulated separation power over the full kaon phase space, covered by the Endcap Disc DIRC and the Barrel DIRC, is shown in Figure 5.48. For both detectors, the performance drops at larger polar angles but stays above the minimum requirement. The vertical line represents the overlap region between them.

## 5.3 Conclusions

Extended testing of various DIRC prototypes and detailed Monte Carlo simulations give consistent results for the photon yield as well as the single photon resolution. The capability of a Disc DIRC to provide identification of charged particles was demonstrated with two different setups. By performing a combined event analysis, where independent events during a vertical scan were combined to simulate a fully equipped prototype, a mean Cherenkov angle resolution of 2.52 mrad was obtained which is already close to the anticipated design goal. Motivated by these measurements the obtained single photon resolution was used to extrapolate the performance of the detector from a single FEL inside the prototype to a fully equipped Endcap Disc DIRC inside P̄ANDA. By adding precise tracking, appropriate wavelength filters and averaging over all available FELs a mean Cherenkov angle resolution of 1.8 mrad was obtained which is sufficient to provide the anticipated particle identification on a 3 $\sigma$ level. Even though the presented predictions are based on solid ground, it is advisable to build and test a full quadrant prototype before the final complete EDD is produced.

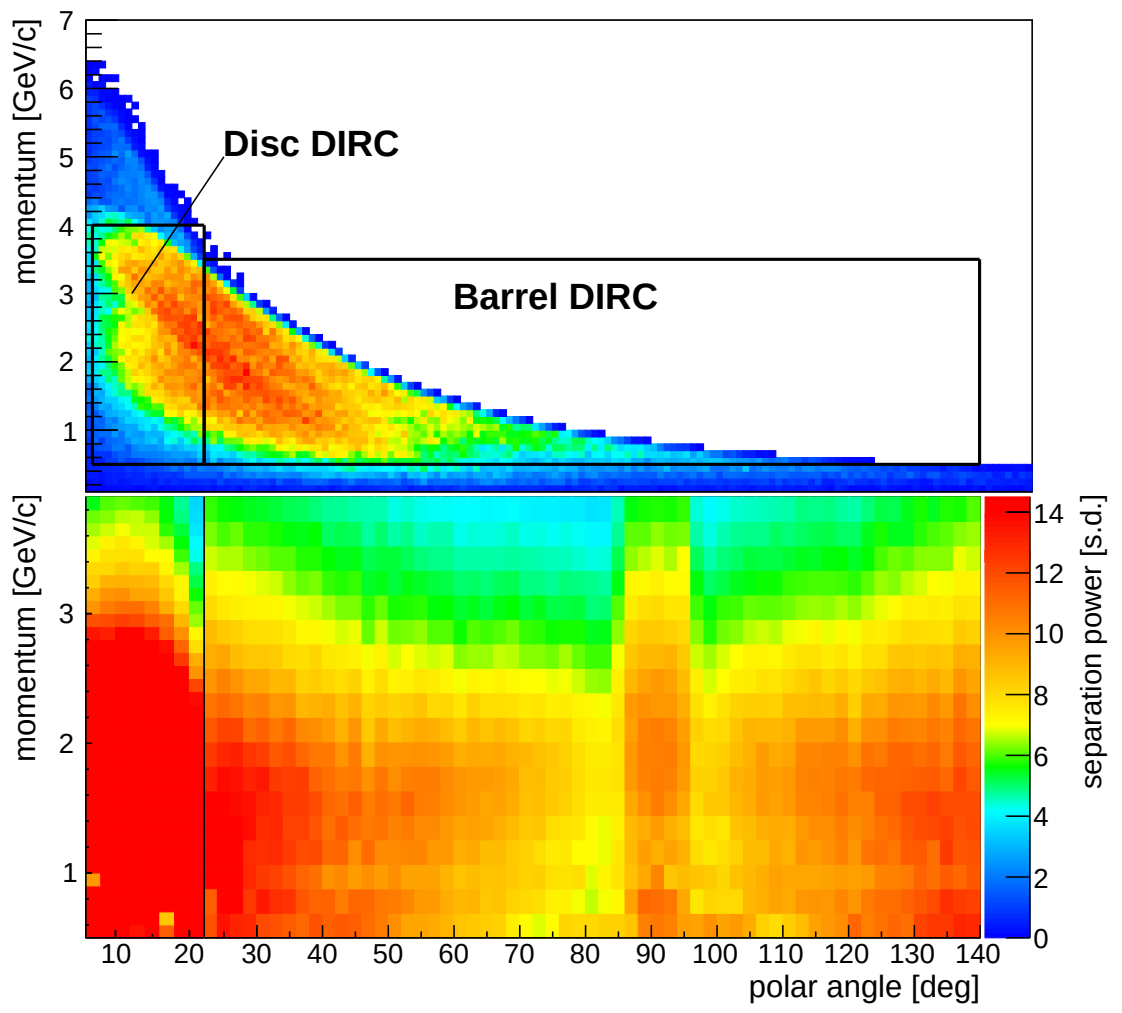

**Figure 5.48:** Top: The kaon phase with the acceptance area of the Disc DIRC and Barrel DIRC. Bottom: The simulated separation power for the full kaon phase space in PANDA for both detectors.

## References

[1] F. Uhlig et al. Breakthrough in the lifetime of microchannel plate photomultipliers. *Nucl. Instr. Meth. A*, 787:105–109, 2015.

[2] M. Düren et al. The PANDA time-of-propagation disc DIRC. *J. Instrum.*, 4:C12013, 2009.

[3] K. Föhl et al. The focussing light guide disc DIRC design. *J. Instrum.*, 4:C11026, 2009.

[4] O. Merle. Development of Reconstruction Methods and Algorithms for the PANDA Disc DIRC. Diploma thesis, Justus-Liebig-Universität Gießen, 2009.

[5] M. Düren et al. The PANDA 3D Disc DIRC. *J. Instrum.*, 7:C01059, 2012.

[6] K. Föhl et al. First particle identification with a Disc-DIRC detector. *Nucl. Instr. and Meth. A*, 732:346–351, 2013.

[7] I. Frohlich et al. A General Purpose Trigger and Readout Board for HADES and FAIR-Experiments. *IEEE Transactions on Nuclear Science*, 55(1):59–66, 2008.

[8] F. Anghinolfi et al. NINO: an ultra-fast and low-power front-end amplifier/discriminator ASIC designed for the multigap resistive plate chamber. *Nucl. Instr. and Meth. A*, 533:183–187, 2004.

[9] M. Mota and J. Christiansen. A high-resolution time interpolator based on a delay locked loop and an RC delay line. *IEEE journal of solid-state circuits*, 34:1360–1366, 1999.

[10] J. Rieke. Analysis of testbeam data taken with a Disc DIRC prototype for PANDA. Master Thesis, Justus-Liebig-Universität Gießen, 2013.

[11] E. Etzelmüller. Developments towards the Technical Design and Prototype Evaluation of the PANDA Endcap Disc DIRC. PhD thesis, Justus-Liebig-Universität Gießen, 2017.

[12] C. Schwarz et al. The PANDA Barrel DIRC. *J. Instrum.*, 11:C05013, 2016.

[13] F. Uhlig. Charakterisierung und Anwendung von schnellen Photosensoren im Hinblick auf ihren Einsatz im PANDA-Experiment. PhD thesis, Friedrich-Alexander-Universität Erlangen-Nürnberg, 2015.

[14] Nikon Corporation Glass Business Unit. 10-1, Asamizodai 1-chome, Minami-ku, Sagamihara, Kanagawa 252-0328, Japan.

[15] Berliner Glas KGaA Herbert Kubatz GmbH & Co. Waldkraiburger Straße 5, 12347 Berlin, Germany.

[16] Cargille Laboratories. 55 Commerce Rd., Cedar Grove, NJ 07009, USA.

[17] J. Rieke. Design of a compact photon detection system for the PANDA Disc DIRC prototype. PhD thesis, Justus-Liebig-Universität Gießen, in preparation.

[18] S. Agostinelli. Geant4—a simulation toolkit. *Nucl. Inst. and Meth. A*, 506:250–303, 2003.

[19] O. Merle. Development, design and optimization of a novel Endcap DIRC for PANDA. PhD thesis, Justus-Liebig-Universität Gießen, 2015.

[20] O. Merle et al. Simulation and reconstruction of photon patterns in the PANDA 3D Disc DIRC. *J. Instrum.*, 7:C01068, 2012.

[21] S. Duarte Pinto. private communications, PHOTONIS Netherlands B.V. 2016.

[22] M. Hoek. private communications, Johannes-Gutenberg-Universität Mainz. 2017.

# 6 Mechanics and Integration

The mechanical design of the $\overline{\text{P}}$ANDA Disc DIRC has to fulfill the following requirements:

- usage of non-magnetic and radiation-hard materials,
- removability of certain components for maintenance (sensors and electronics),
- secure and precise assembly and alignment,
- prevention of instabilities or damage due to thermal expansion,
- placement of sensitive fused silica optics inside hermetically sealed and light-tight compartments,
- minimizing the material budget and radiation length in the acceptance region of subsequent detectors,
- economically priced construction.

These conditions have to be fulfilled in a tight spatial environment as the DIRC is mounted on the support of the Endcap EMC in front of its thermal insulation [1]. The EMC environment will be protected with two layers of superinsulation with a total thickness of 2 cm.

Figure 6.1 gives an overview of the space which is available inside the endcap holding structure (EMC support). Because of the geometry of the Endcap EMC the available space in $z$ for the readout region varies between 148 and 242 mm. In addition the insulation does not have the dodecagonal symmetry of the EDD which further restricts the available space and forces to build ROMs with different bar lengths.

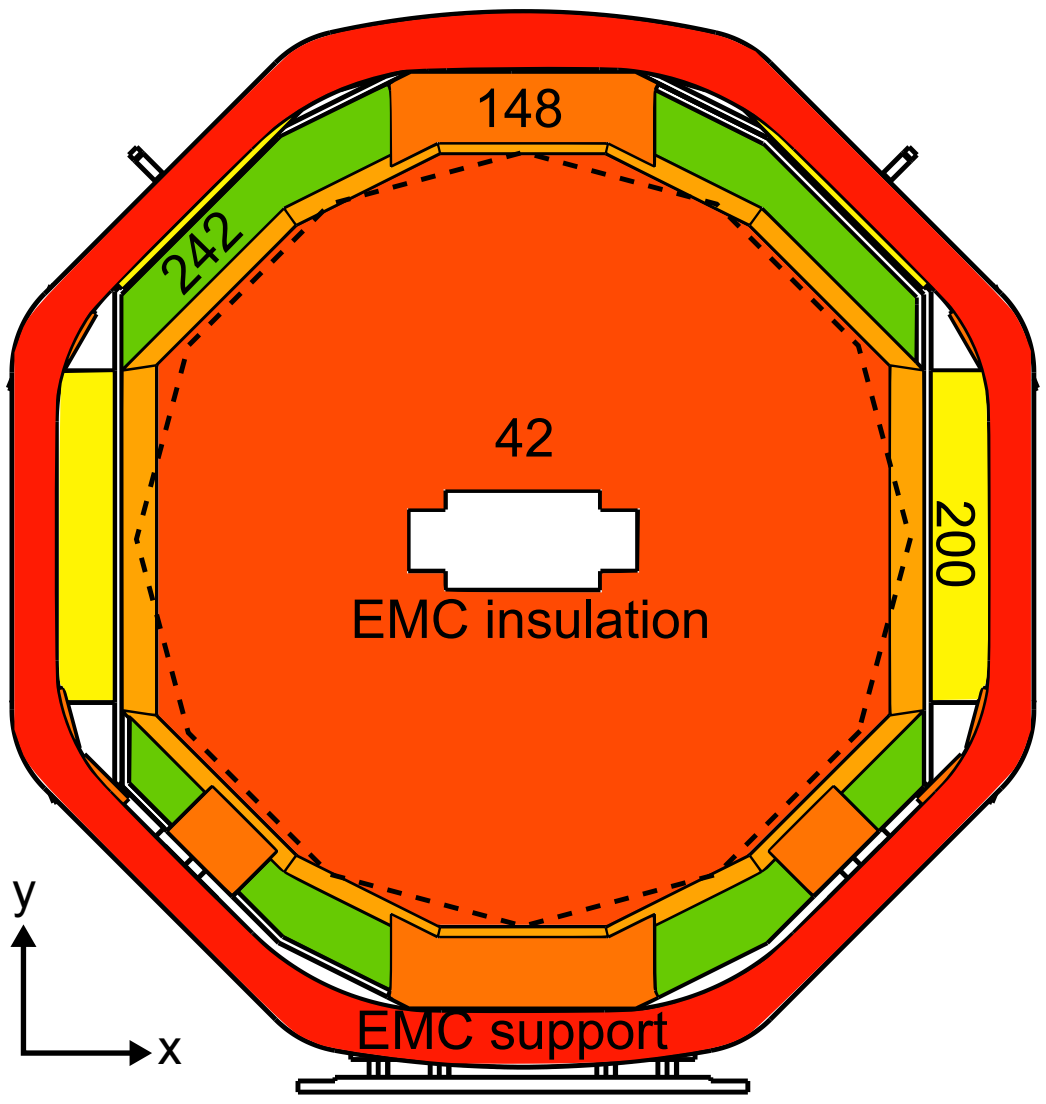

**Figure 6.1:** Downstream view of the EMC endcap with the available space in mm along the z-axis. In the radiator area only 42 mm are available for the radiator and the corresponding mechanics.

## 6.1 Design Approach

For demonstrable reasons not all requirements from the above list can be satisfied at once. The MCP-PMTs, for example, contain Kovar, which is ferromagnetic. The proposed design therefore aims at a reasonable compromise of all constraints which in the end allows a safe operation of the Endcap Disc DIRC. Another challenge comes from the strict requirements of a precision aligned optical system.

The available space is insufficient to use spring-loaded precision tracks which were used for existing prototypes (see section 5.1). Therefore it was decided to permanently glue the bar of the ROMs to the radiator to form a rigid unit which means that the alignment has to be done only once [2]. This approach was studied by FEM simulations to estimate the deformation due to gravity and due to the magnetic field. Figure 6.2 shows the result for a quadrant in a top and bottom position. The stability of the optical system is improved by connecting the eight ROMs on each side of a quadrant to minimize shear strains and torsion.

**Table 6.1:** Dimensions for the $\overline{\text{P}}$ANDA Endcap Disc DIRC mechanical structures.

| Part | Property | Value |
|---|---|---|
| Mounting Plate | Int. Radius | 800 mm |
| | Ext. Radius | <1460 mm |
| | Tot. weight | 66 kg |
| | $\Delta$z | 15 mm |
| Radiator | Int.Radius | >240 mm |
| | Ext. Radius | <1056 mm |
| | Tot. weight | $4 \times 34$ kg |
| | $\Delta$z | 20 mm |
| ROMs | Int. Radius | >1025 mm |
| | Ext. Radius | <1300 mm |
| | Tot. weight | $96 \times 1.9$ kg |
| | $\Delta$z | 137 mm |
| One Quadrant | Tot. weight | $\approx 85$ kg |
| Full Detector | Tot. weight | $\approx 450$ kg |

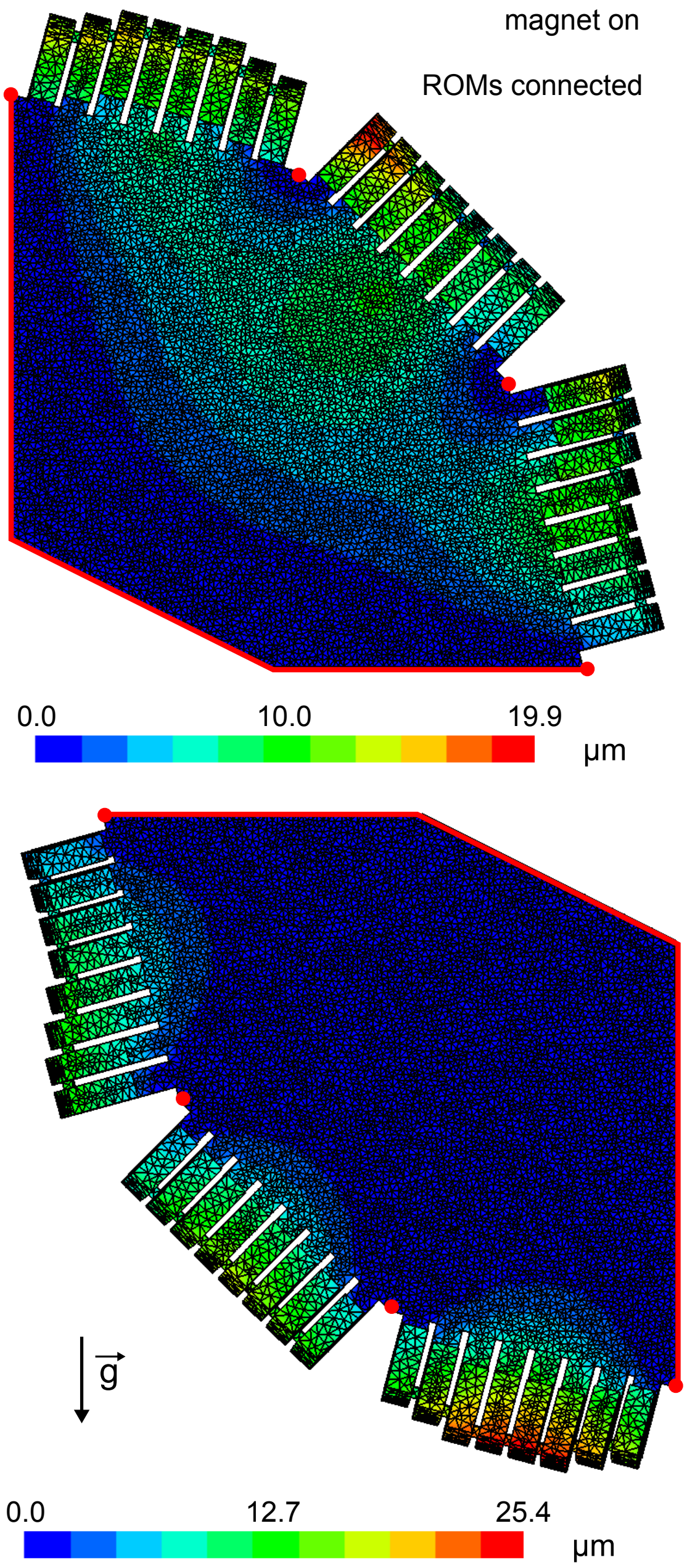

**Figure 6.2:** FEM simulations regarding the displacement of the optical system in the upper and lower positions inside the stabilizing cross. The red lines and dots indicate positions where the optical system is fixed. The simulation includes an estimation of the impact of the inhomogeneous magnetic field on the MCP-PMTs.

The mechanical design roughly differentiates between large-, medium- and small-scale components.

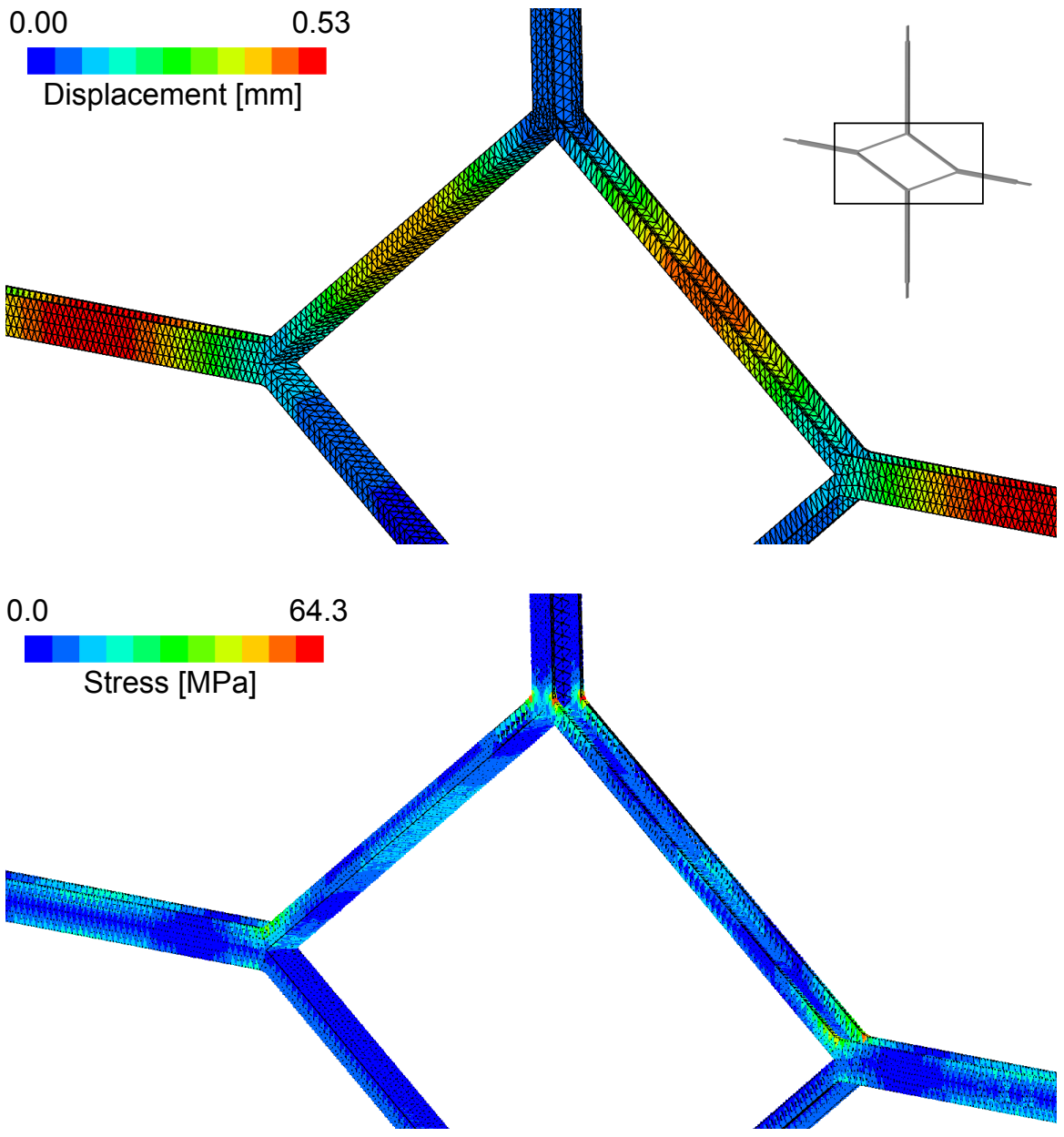

**Figure 6.3:** FEM simulations for the displacement (top) and stresses (bottom) if the two upper positions are loaded by one quadrant each. In the final detector all four positions will be equipped with one quadrant.

Large-scale components resemble the overall mechanical structure above the individual quarters. The largest one is the mounting plate (MP) which is connected with the upstream side of the Endcap EMC support. It has a circular hole with a radius of 800 mm in its center and will carry the entire weight of the Disc DIRC which is about 450 kg (a summary of the most important dimensions and weight loads can be found in table 6.1). Like the endcap holding structure it is made of aluminum. The MP is subdivided into four regions by a stabilizing cross (SC) with a rhombic cutout in its center. Each region will later host one quarter of the EDD. In order to minimize the material budget, the SC will be made out of four 2 mm thin aluminum U-profiles which follow the shape of the inner DIRC radiators and will be connected on their long sides forming an H-profile. This concept of a SC has already been applied to the latest prototype as can bee seen in Figure 6.4. The SC will be clamped tautly to the mounting plate to avoid any shifts in z-direction. Small buttons or nylon stripes will keep the fused silica radiator at distance to the SC. Figure 6.3 shows a FEM calculation for the case of just two quadrants which occupy the top positions. The design goal is to limit the total displacement to less than 0.5 mm which is already the case for this setup. The stress levels are moderate with uncritical stress peaks, far below the permissible elastic limit of a typical aluminum alloy in the corners of the stabilizing cross. In the final detector all four spots inside the SC will be filled with a quadrant which adds additional stability to the mechanical system.

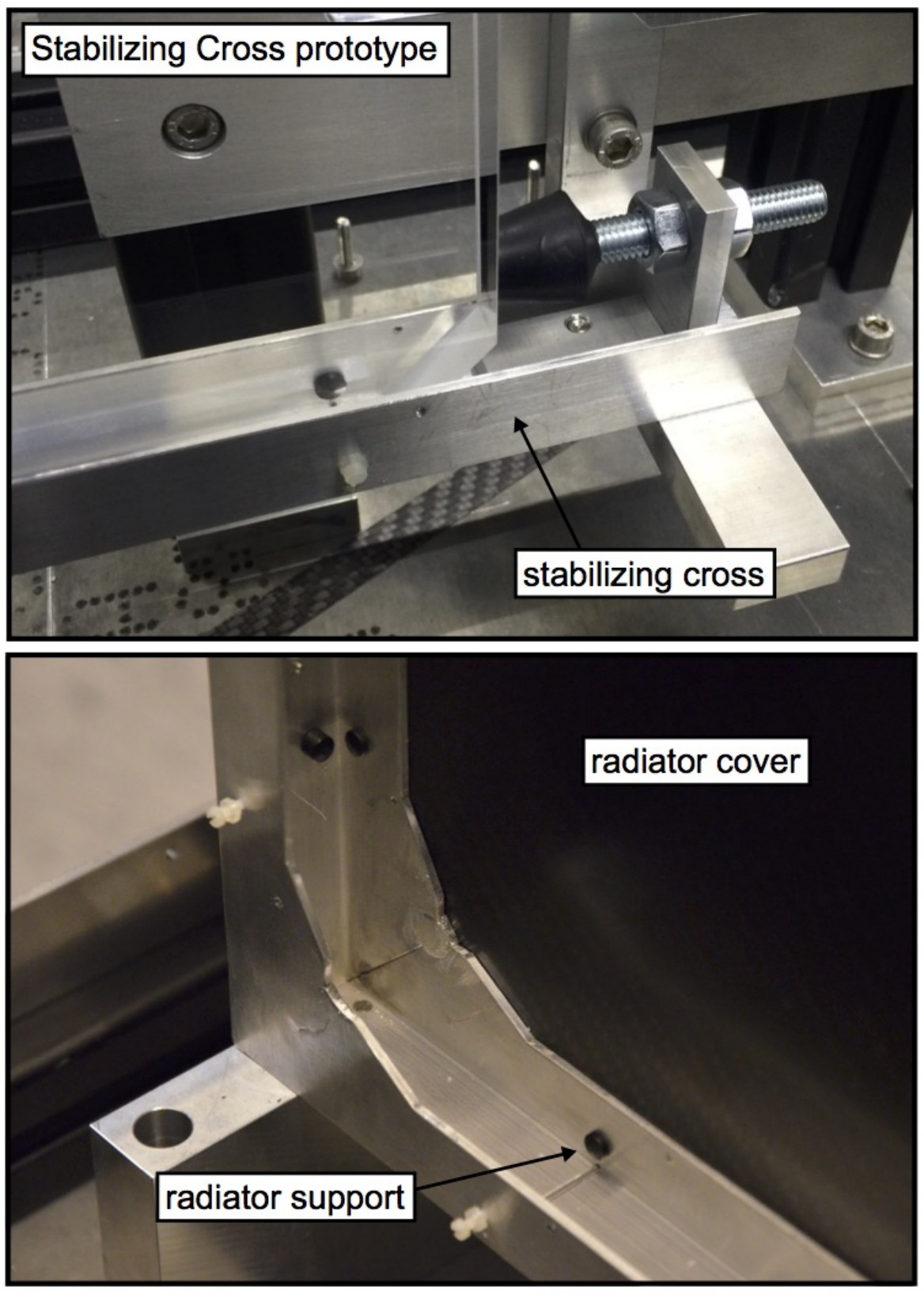

**Figure 6.4:** Prototype of a welded stabilizing cross which includes small areas of support for the fused silica radiator.

Medium-scale components are the housing of the fused silica radiator discs and ROMs. Both have to be operated in a gas- and light-tight environment. The fused silica radiators will be sealed by radiator covers (RC). As the RCs will be in the acceptance region of the Endcap EMC they are required to have a small material budget. Each RC has to cover an area of roughly $0.7\,\mathrm{m}^2$. For this reason a stiff and robust solution is needed. Different options ranging from lightweight honeycomb panels made of aluminum to a plate made of carbon fiber are available. The RCs will be fixed along the SC and connection blocks which will be introduced below. The outer parts of the radiator covers will be overlapped by the ROM housings (RH). These compartments have to be gas- and light-tight as well and will require an individual shape for each side because of the different geometry of the EDD and the insulation of the Endcap EMC. Figure 6.5 shows an example of one side where all ROMs can be placed at the same distance to the radiator edge.

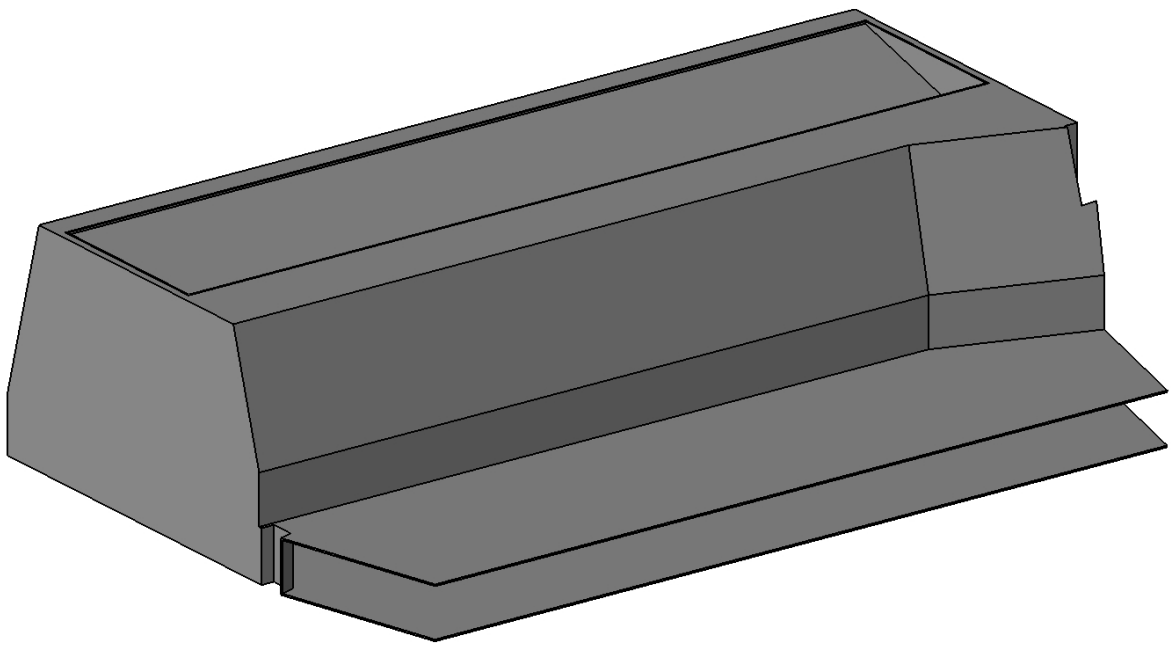

**Figure 6.5:** Isotropic CAD view on the ROM housing at the north-west position. The shape has to adapt to the available space which is restricted by the insulation of the Endcap EMC without interfering with ROM components.

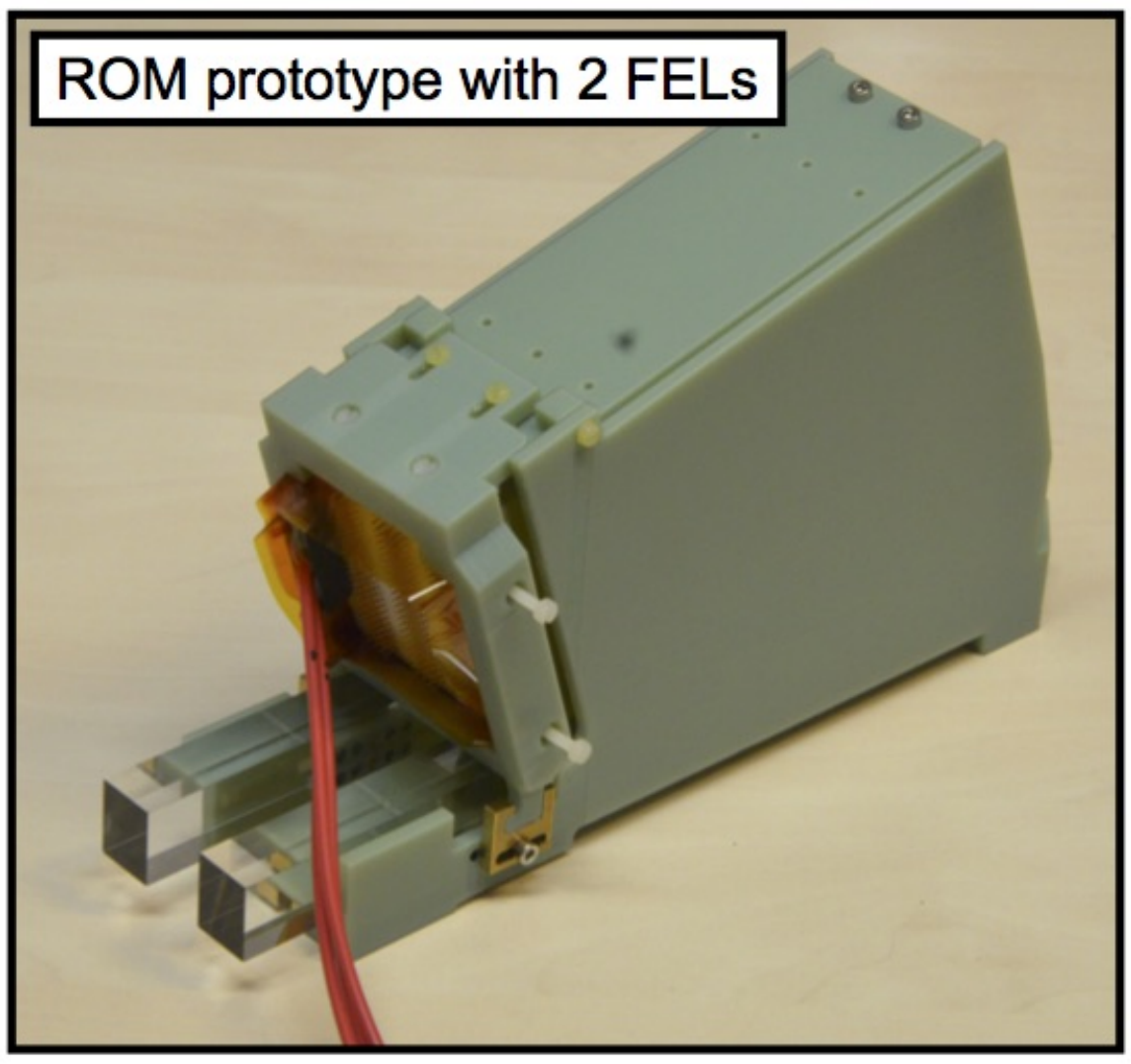

**Figure 6.6:** Equipped ROM prototype for the 2016 testbeam setup at DESY.

The readout modules will be supported by a compact cover (see also figures 6.6 and 6.8. To minimize the contact between optics and mechanics, only the optically irrelevant parts of the light guides and bars will be glued to the ROM case.

Between the twelve RH blocks on each side in total twelve spring-loaded connection blocks (CB) will keep the fused silica radiators in their position inside the SC. The connection blocks will touch the fused silica radiators at their outer corners which are non-critical regarding the path of the Cherenkov photons and press the quadrants towards the center of the SC. At 3, 6, 9 and 12 o'clock the connection blocks will slide over the SC and will touch two radiators at once. This technique will prevent the Disc DIRC from being damaged in the case of unwanted thermal expansion of the material which could lead to a displacement of several hundred $\mu$m.

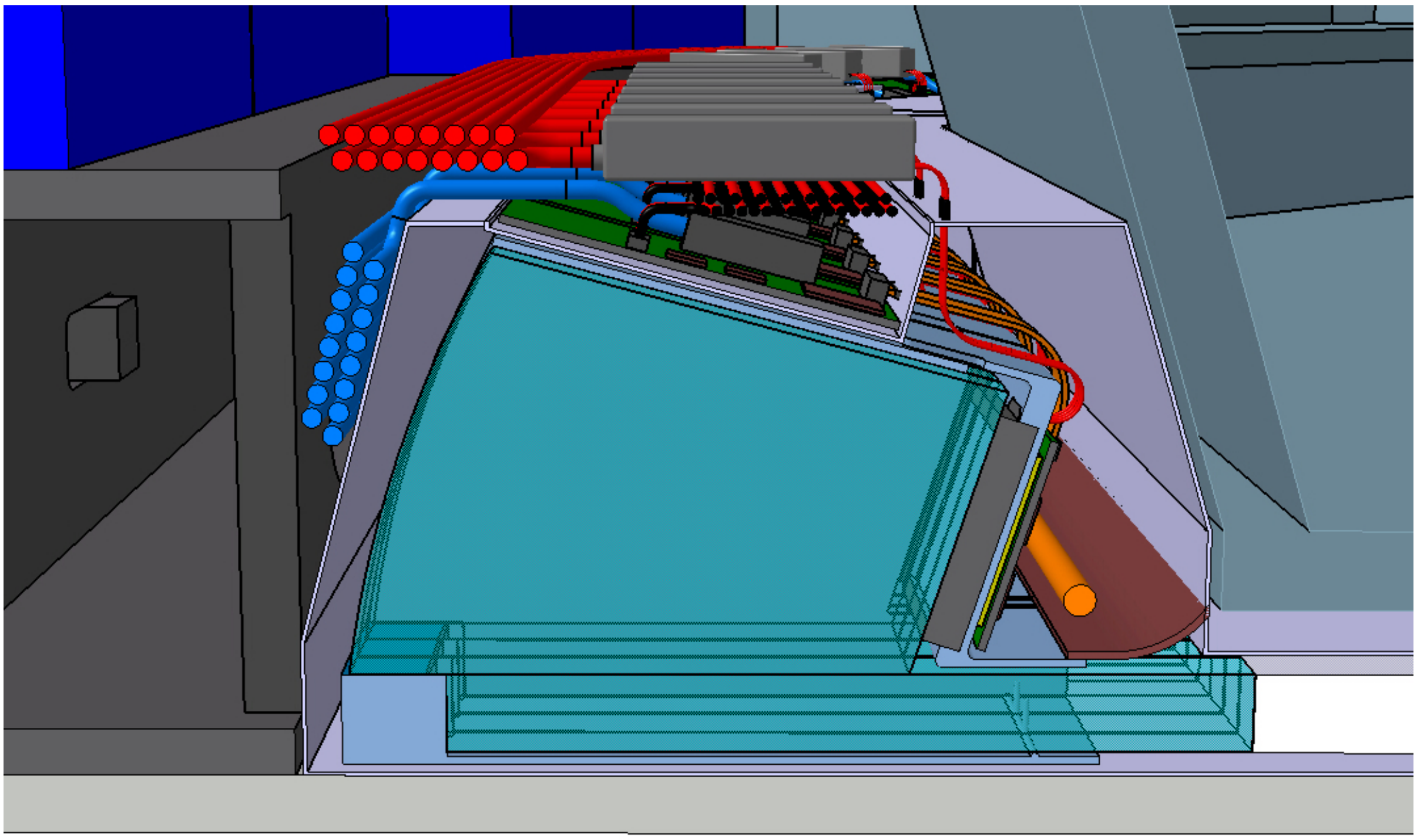

**Figure 6.7:** CAD section view of the readout part at the north-west position where the space in z amounts to 242 mm. The image shows the arrangement of the cables as well as the inside of a ROM case and ROM housing.

## 6.2 Integration into $\overline{\text{P}}$ANDA

### 6.2.1 Neighboring Subdetectors

The EDD will be enclosed by two other subdetectors. Upstream the GEM detector [3] will provide a largely flat boundary with an elevation in the upper region for its cable routing. In the downstream direction the boundary is given by the insulation of the end cap calorimeter [4]. This insulation has the shape of an irregular dodecagon. Therefore the available space for the focusing optics and readout varies for individual ROMs. This requires the introduction of additional light guides (called bars) with different lengths for individual ROMs (see chapter 3). These bars allow to position the FELs at different distances from the radiator. The calorimeter insulation has sloped surfaces on the sides that fall in the opposite direction of the FELs' focal plane which further reduces the available space (see Figure 6.7). At 6 and 12 o'clock, the available height is limited to only 148 mm, whereas other areas provide more space which can be exploited for cable routing and the distribution of the dry nitrogen gas.

The Disc DIRC, including the ROM-Housing, has to fit into the opening of the endcap holding structure which has a diameter of up to 2546 mm. Lateral hollows inside the holding structure provide additional space which is, however, difficult to access.

### 6.2.2 Assembly Procedure

The assembly procedure will consist of four stages:

1. ROM assembly, which prepares the readout modules for the final assembly.
2. Quadrant assembly, which consists of a complete setup of one quadrant.
3. Horizontal assembly, where the complete Disc DIRC is assembled on a horizontal table.
4. Vertical assembly, which includes the joining of all cables and establishing of the connection between the MP and the holding structure of the endcap calorimeter.

#### ROM Assembly

At the beginning of all assembly stages all components have to be cleaned and dust free before they are used. Therefore the assembly will be done in a dedicated clean room. As depicted in Figure 6.8, the ROM assembly contains several steps. In a first step three focusing optics consisting of a FEL and a bar each, will be glued to the ROM case. This will be the only glue joints between optical and mechanical components to minimize any risks due to the different thermal expansion coefficients. The FELs are connected to the bars via optical bonding. We have already good experience with that being done by the company Berliner Glas.

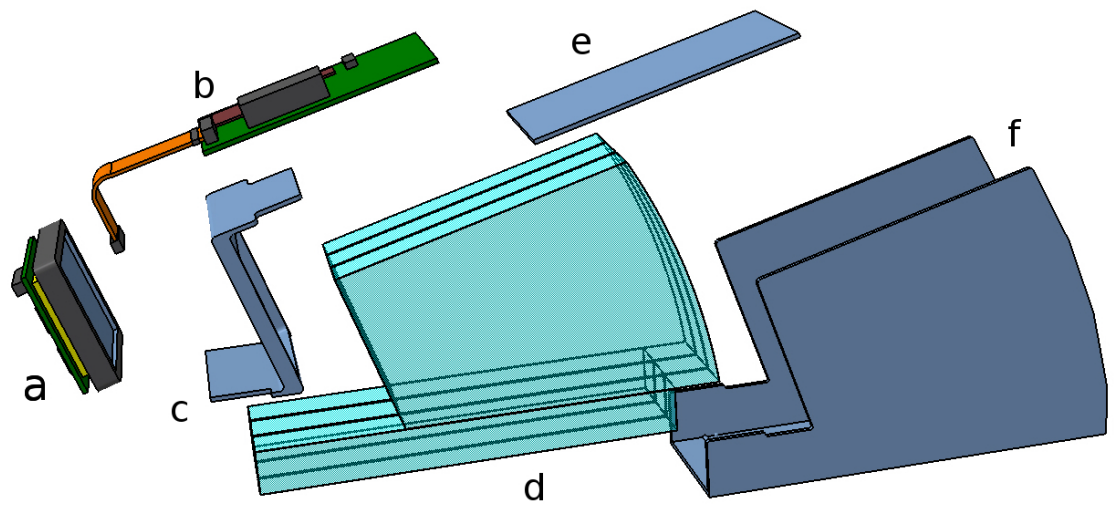

**Figure 6.8:** Exploded view of a ROM: (a) MCP-PMT with PCB, (b) FEE board, (c) MCP-PMT holder, (d) FEL-bar pair, (e) ROM top cover and (f) ROM case.

The distance between the pairs of FELs and bars within a ROM is 1 mm and the bars have to finish parallel towards the radiator within few $\mu$m. The MCP-PMT along with the FEE board and the corresponding mount will be added during a later stage.

The ROM case is going to be designed light- and gas-sealed and includes the optical elements and the sensitive side of the photosensor. It will be connected to the radiator box in a light-tight way. The connectors of the MCP-PMTs as well as the FEE-PCBs are outside of that closed volume, in order to make sure that outgasing and electronics cooling do not affect the fused silica components.

## Quadrant Assembly

A crucial part of the quadrant assembly will be the connection between the ROMs and the radiator. The joint between radiator and bars must be precisely aligned and of high optical quality. In this connection the following requirements have to be fulfilled:

- high transmission for the used photon wavelengths,
- index of refraction similar to fused silica,
- radiation hardness,
- strong bonding.

The most convenient choice would be Epotek 301-2 [5] which has already been used for the BaBar DIRC and the Belle II TOP counter [6]. Due to the size of the quadrant an accelerated curing inside a high temperature environment is not possible which is why several weeks have to be reserved for the complete gluing procedure.

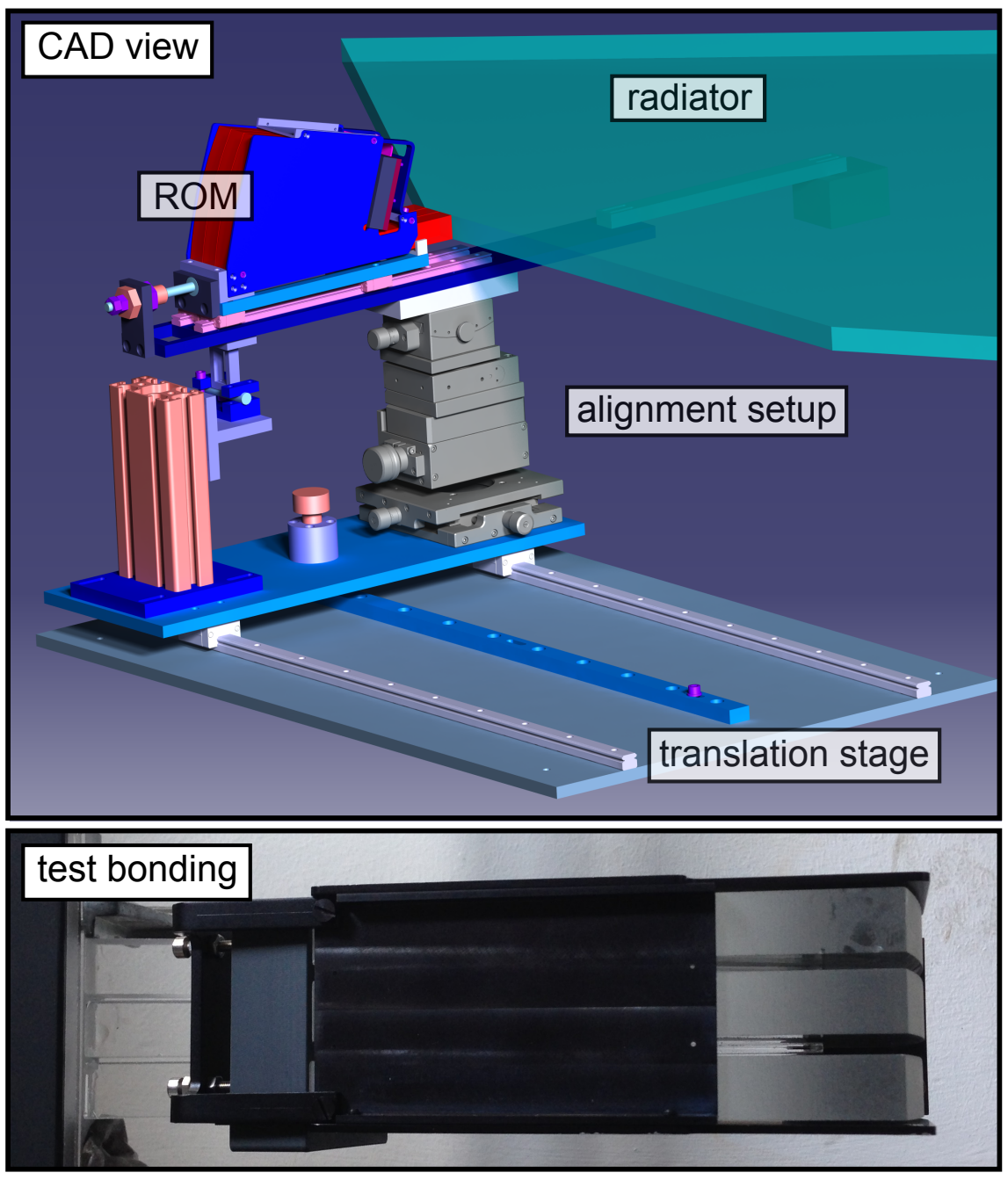

**Figure 6.9:** Gluing station for coupling the ROMs to the fused silica radiator (top) and a gluing test for a ROM prototype at a lateral position (bottom) [2].

The whole procedure will be done in the horizontal position of the radiator plane. Starting at an outward position the ROMs will be bonded to the radiator piece by piece using a dedicated gluing station (see Figure 6.9). This setup will allow a precise positioning of the ROM and a movement perpendicular to the lateral surface of the radiator, i.e. we are able to adjust the ROMs in all 6 degrees of freedom while gluing the bar to the plate. To avoid bending of the fused silica radiator caused by gravity the plate will be placed on a flat surface equipped with small plastic balls to minimize the area of contact. Before the glue is applied the parallel transition between the radiator and bar surface is checked with a calibration laser. After the glue has pre-cured the support will be removed and prepared with the subsequent ROM. This optical system is foreseen to be a rigid unit which is not disassembled any more within the lifetime of the detector. Yellowing of glue is not expected and would have only a minor impact on the transmission as the thickness of the glue joints is only few tens of micrometers. The strengths of the glue joints have already been studied by attaching weights to glued bars with different torques. Nevertheless, further gluing tests with the recently arrived radiator plate from NIKON will be performed.

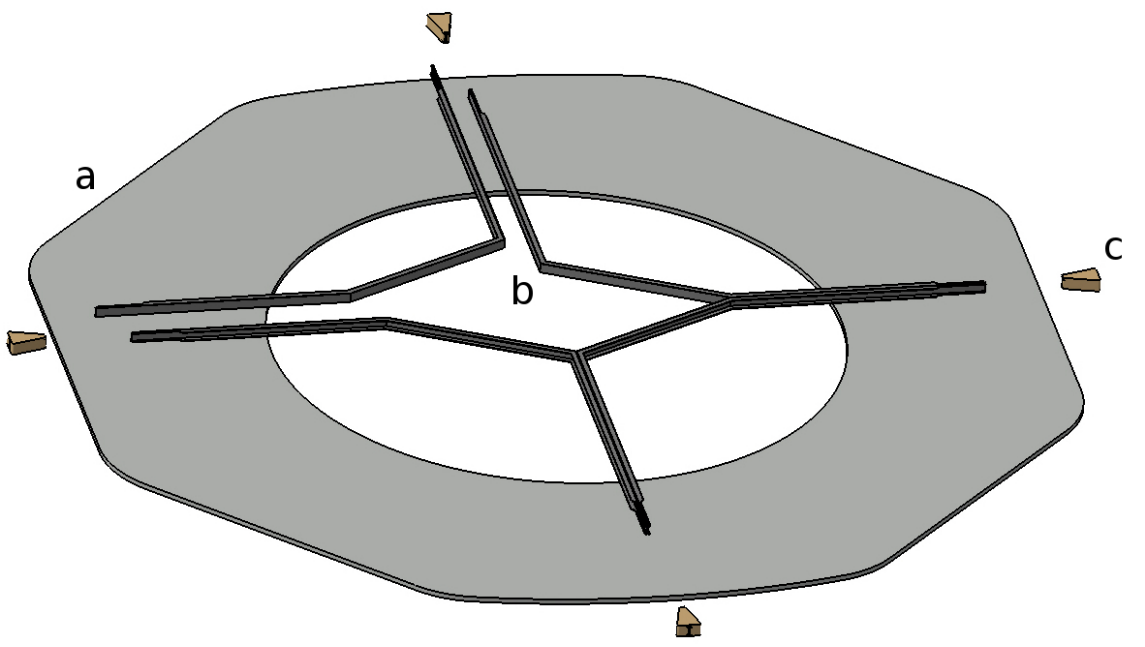

**Figure 6.10:** Assembly of the stabilizing cross (SC, b) which consists of four parts on the mounting plate (MP, a). The stabilizing cross will be clamped tautly by four connection blocks at 0, 3, 6 and 12 o'clock.

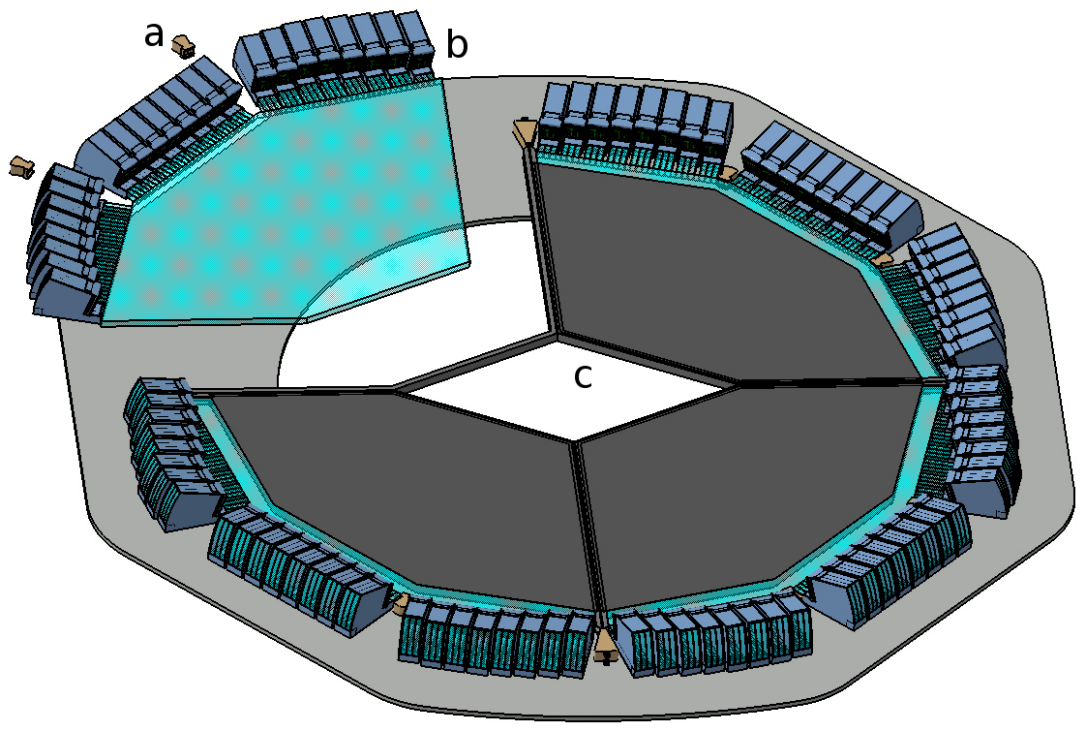

**Figure 6.11:** A fully assembled quadrant (b) will be inserted into the stabilizing cross (c) and mounted via two additional connection blocks (a).

After the quadrant assembly each boxed quadrant can be individually tested and represents a fully functional, independent detector unit. This can be pre-calibrated either by a laser system in a muon test stand or test beam prior to being installed in the $\overline{\text{P}}$ANDA detector.

## Horizontal Assembly

The first step of the horizontal assembly is to mount the stabilizing cross (SC) on to the mounting plate (MP, see Figure 6.10). The four individual SC profiles will be made of aluminum and will be riveted to form one large cross and then be tautly clamped to the MP by four connection blocks.

Afterwards the optical system consisting of one radiator and 24 ROMs will be placed inside the SC (see Figure 6.11). As soon as the quadrant is introduced, spring-loaded connection blocks will be pressed against the outer corners of the radiator to fix them in their position. The radiator covers are than mounted onto the SC and the cables for the laser calibration system and the tubes for the dry nitrogen supply can be connected.

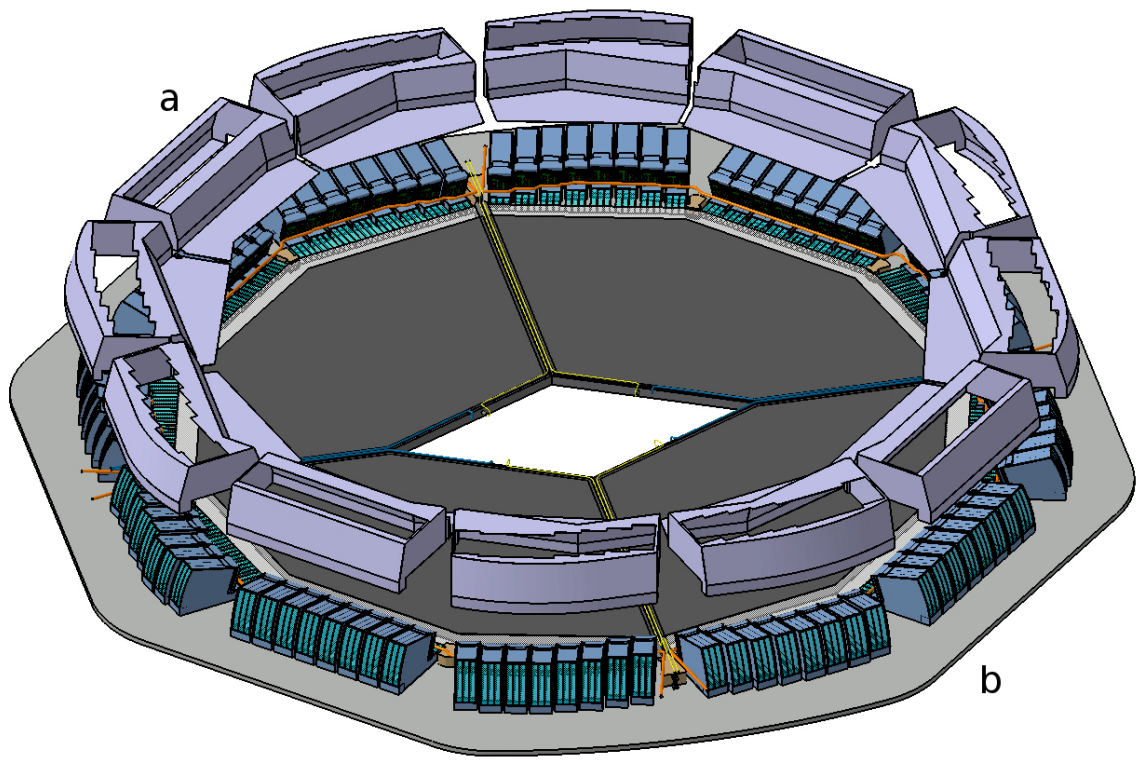

**Figure 6.12:** After the cooling pipes have been put into place the readout part will be sealed by twelve ROM housings (a) which are mounted to the MP (b).

For the horizontal assembly the MP has to be put on a temporary external support structure, as the SC is not designed to stabilize the quadrants in this position.

Before the ROM housings (RH) can be attached, the MCP-PMTs have to be attached to each ROM. In connection with this step also the cabling and cooling inside the RH has to be established and linked to connectors which lead to the outside of the RH. The RH can then be fixed on the MP and to enclose the ROMs. After the RHs are in place their top is closed and the FEE boards can be added and plugged in via a feedthroughs.

## Vertical Assembly

After all components are mounted on the MP the EDD has to be tilted into its vertical position for operation. This task could be done with a dedicated mounting table. The table has to provide shock absorbers and needs to be movable on concrete floor and on the tracks which will later move the endcap frame onto the $\overline{\text{P}}$ANDA target spectrometer. As soon as the Disc DIRC ist put into its vertical position facing the endcap holding structure, it can slowly approach the opening until only few cm are left (see also Figure 6.15). Before the closing, all cables have to be plugged into the dedicated connectors. Then the endcap can be closed up with the Disc DIRC and the MP can be screwed onto the outer frame of the endcap holding structure.

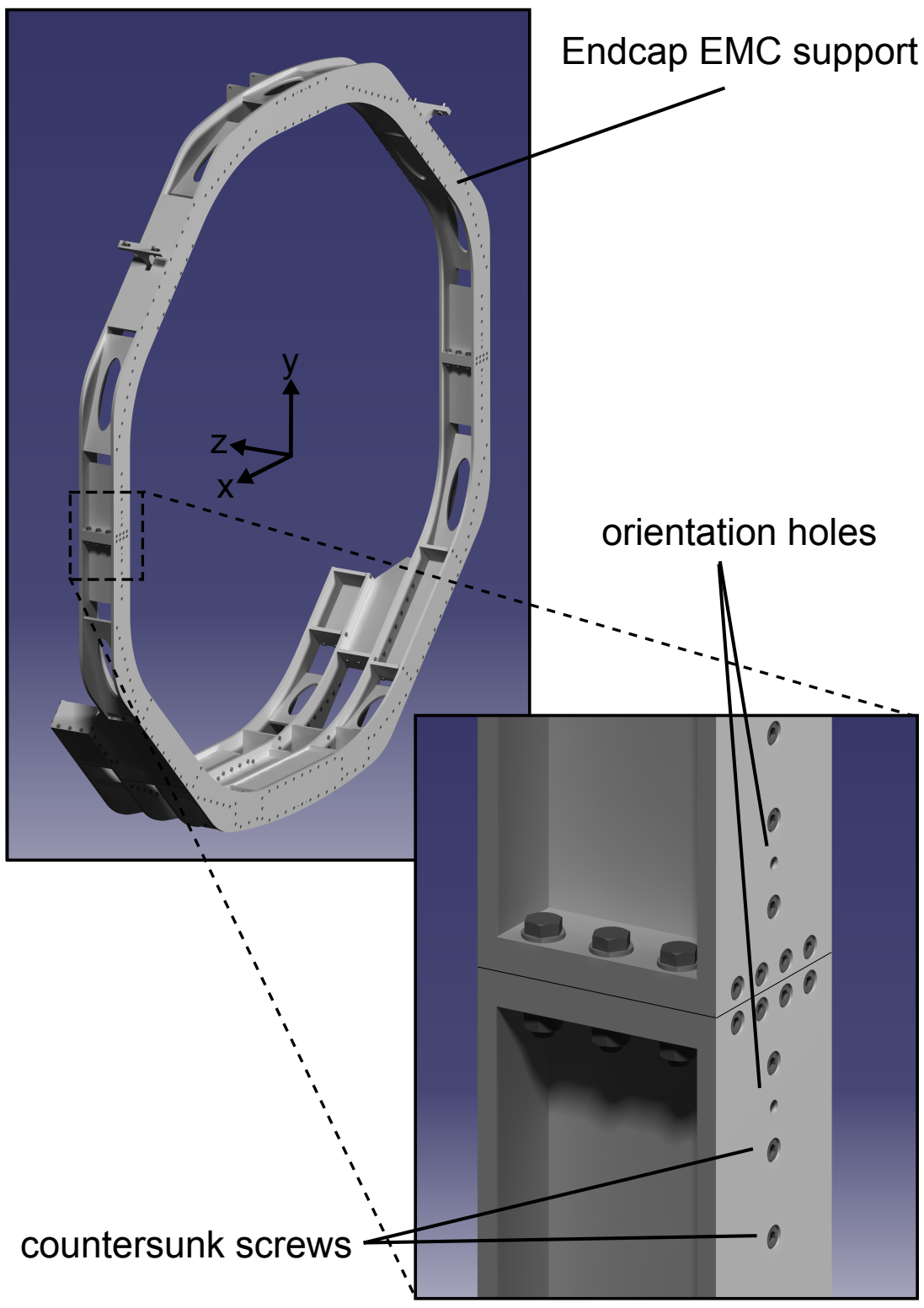

**Figure 6.13:** Endcap EMC holding structure with changes that have been made to mount the EDD.

For this reason all screws on the outer frame have been countersunk. As the production of the endcap holding structure is already in progress at the time this report is written it was decided to add two orientation holes per quadrant (see Figure 6.13). Due to the weight of the endcap calorimeter, the top-to-bottom distance of the opening is expected to be reduced by several mm [4]. The finalization of technical drawings for the MP is planned to be postponed until this is verified to avoid any risks. Additional holes will then be added using these orientation holes and an adjusted drilling stencil. After the MP is mounted on the endcap holding structure, the external support will be removed together with the mounting table. In Figure 6.14 all steps are summarized.

## 6.3 Supply Lines and Cables

All electrical cables of the EDD will be selected in compliance with the FAIR cable rules (fire safety, radiation resistance, bending radius, etc.). The EDD requires cables for HV supply of the MCP-PMTs, LV supply of the front-end electronics, optical data cables and cooling for the DAQ and front-end electronics, gas lines for the gas system and optical fibers for the laser calibration system. Table 6.2 gives an overview of the present status. The listed cables will be divided into two separate cable trays on the east and west-side of the detector. A cross section of about 9,000 $mm^2$ per half will be required. A cross section view of such a cable tray is shown in Figure 6.16. Because the available cable ducts in the top positions (north-north-west NNW and NNE) are partly occupied by the GEM tracker and the central cable ducts (NWN, NEN, SWS and SES) of the endcap are fully reserved for the Endcap EMC the EDD will use the two unoccupied ducts at the positions at SSW and SSE as shown in Figure 6.17. Nevertheless it might be convenient to use the top ducts at SSW and SSE for the cooling lines.

The routing of the cables will be done along the ROM housings and on the backside of the mounting plate. Layers of cables from preceding sides will be stacked (see also Figure 6.18). The HV cables will be connected to voltage dividers which sit on top of the ROMs if the space permits. However at positions with less than 200 mm space in z-direction the voltage dividers have to be redistributed and stacked as well. Figure 6.19 shows the situation without the ROM housing for a side where all ROMs have the same distance.

For a proper functionality of the Disc DIRC it is necessary to maintain the radiator discs, focusing optics and the electronic components in a low moisture environment. This will be ensured by a gas system that flows dry and clean nitrogen over the radiator surfaces. It will be based on the one used at the BaBar DIRC [7]. Thin, about 3 mm thick gas tubes will be laid along the arms of the stabilizing cross and will end inside its hollows towards the center. From there the gas can flow through between radiator and its cover into the ROM Housings and back to the outside of the $\overline{\text{P}}$ANDA detector. The rate will be several liters per hour per quadrant, which results in a light overload pressure. To minimize the impact of outgassing by electronic components which share a volume with the fused silica optics several preventive measures are foreseen. As shown in Figure 6.7 a slice between the ROM cover and the ROM housing guides the gas flow through ROM which it leaves at the back side to enter the ROM housing where the PCB is located. In addition the FEE board is placed outside of the ROM housing.

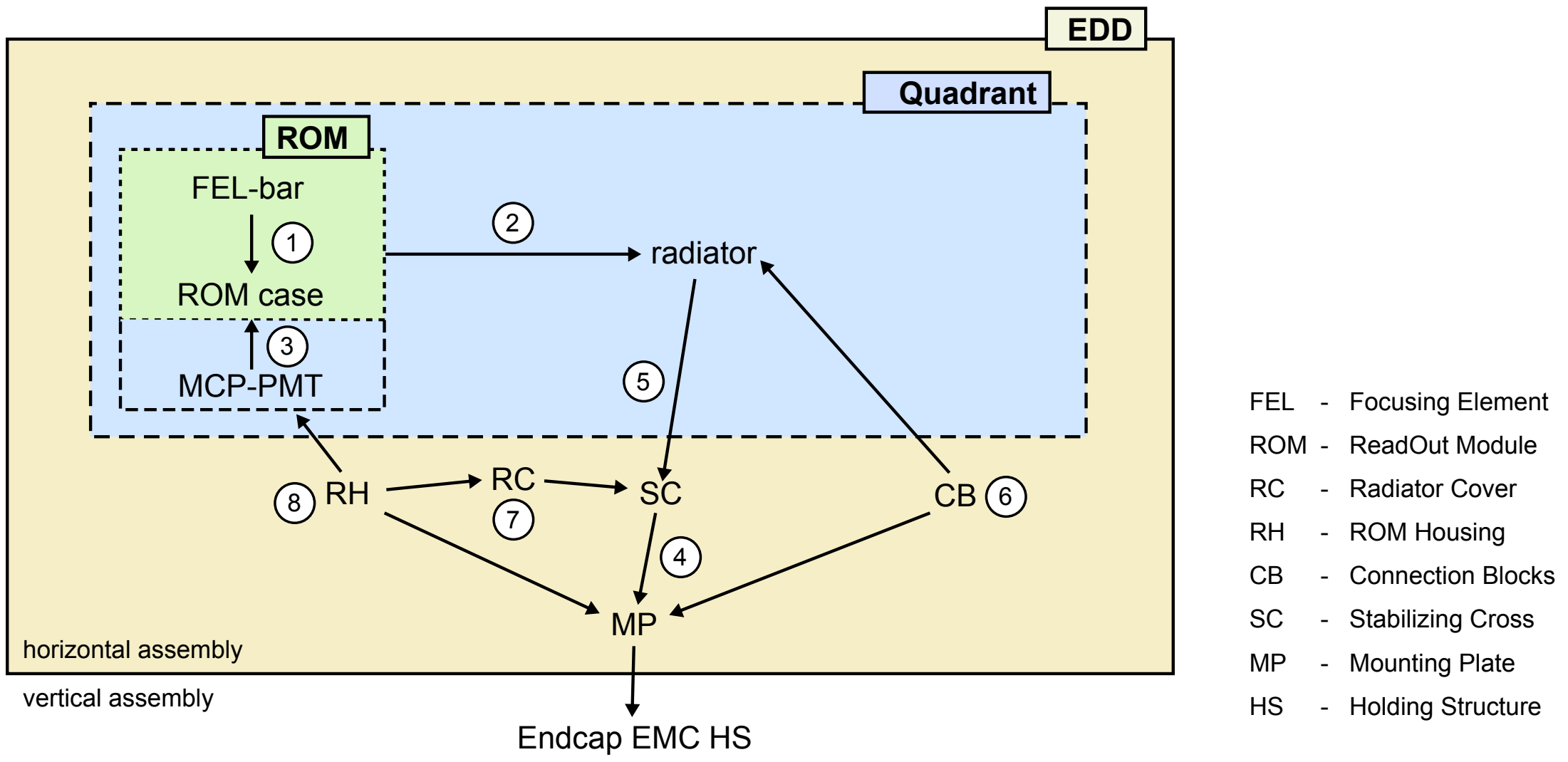

**Figure 6.14:** Schematic of the assembly procedure [2]. Different assembly stages are summarized inside the boxes. The numbers in circles indicate the order of the assembly.

**Table 6.2:** Overview of the total numbers of cables and supply lines. The total cross section amounts to $10,100\,\mathrm{mm}^2$.

| type | connection | number | diameter [mm] | cross section [mm$^2$] |
|---|---|---|---|---|
| HV | 96 MCP-PMTs | 96 coaxial cables | 5 mm | 2,400 |
| LV | 96 ROMs | 192 cables | 2.5 mm | 1,200 |
| data | 96 ROMs | 96 optical fibers | 5 mm | 2,400 |
| gas | 4 quadrants | 8 pipes | 10 mm | 800 |
| cooling | 4 quadrants | 8 pipes | 60 mm | 3,200 |
| laser | 4 quadrants | 4 optical fibers | 5 mm | 100 |

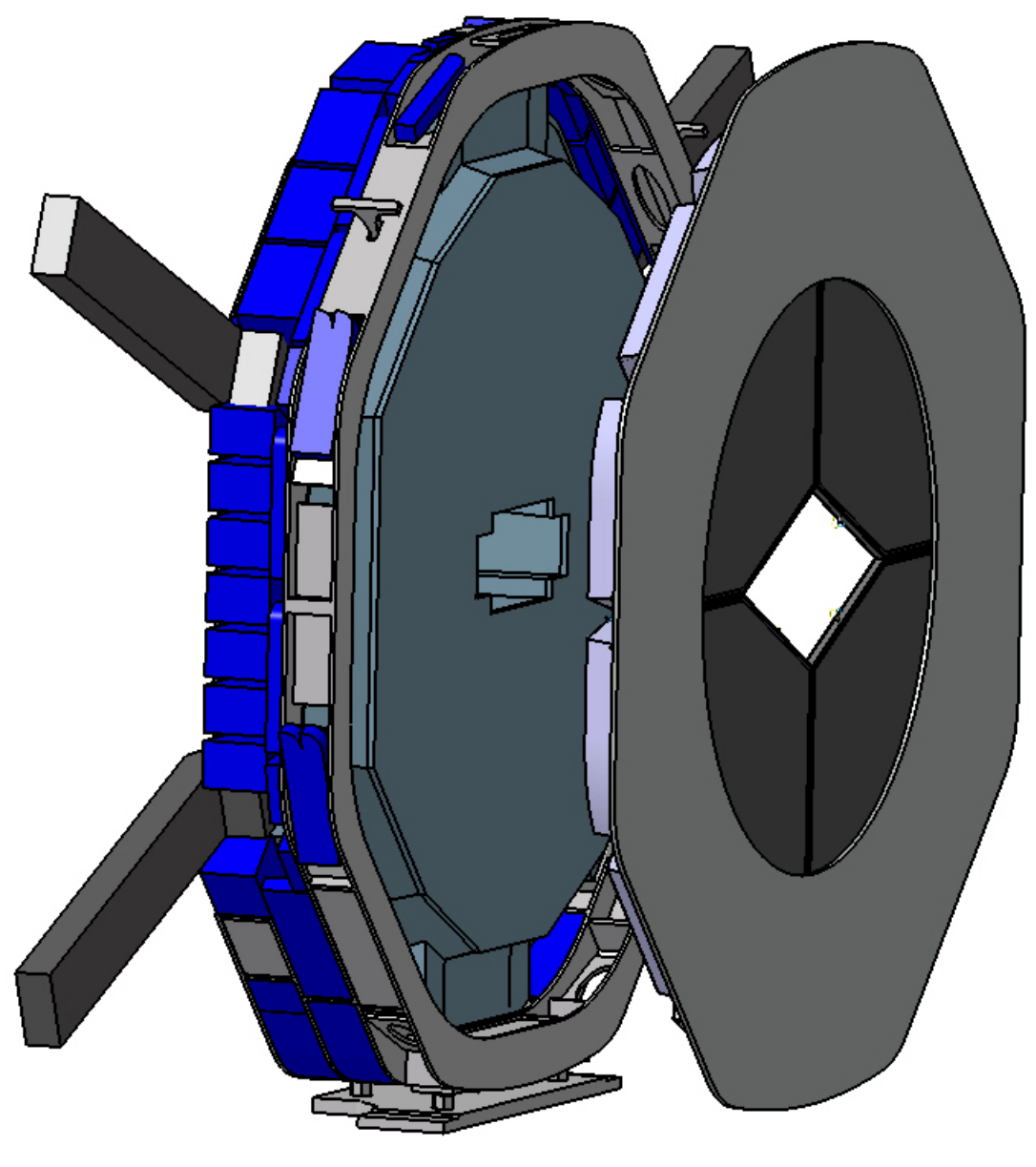

**Figure 6.15:** The mounting plate which carries the EDD will be connected to the Endcap EMC.

## 6.4 Maintenance

The mechanical concept aims at a maintenance-free operation and materials and components are chosen according to this requirement. The optical system of each quadrant will be a single rigid unit which does not require subsequent alignment. If any of the MCP-PMTs or the front-end electronics require intervention these components can be replaced. In order to do this the full EDD has to be removed from the endcap holding structure and the corresponding ROM housing has to be lifted to gain access to the bare ROMs.

A replacement of the MCP-PMTs can only be done in a long shutdown of PANDA, when the magnet doors are opened, and the endcap EMC is moved out and the EDD is fully accessible. The MCP-PMTs can be removed after disconnecting the readout PCBs. Nevertheless, one single faulty MCP-PMT does not worsen the $\pi/K$ separation significantly.

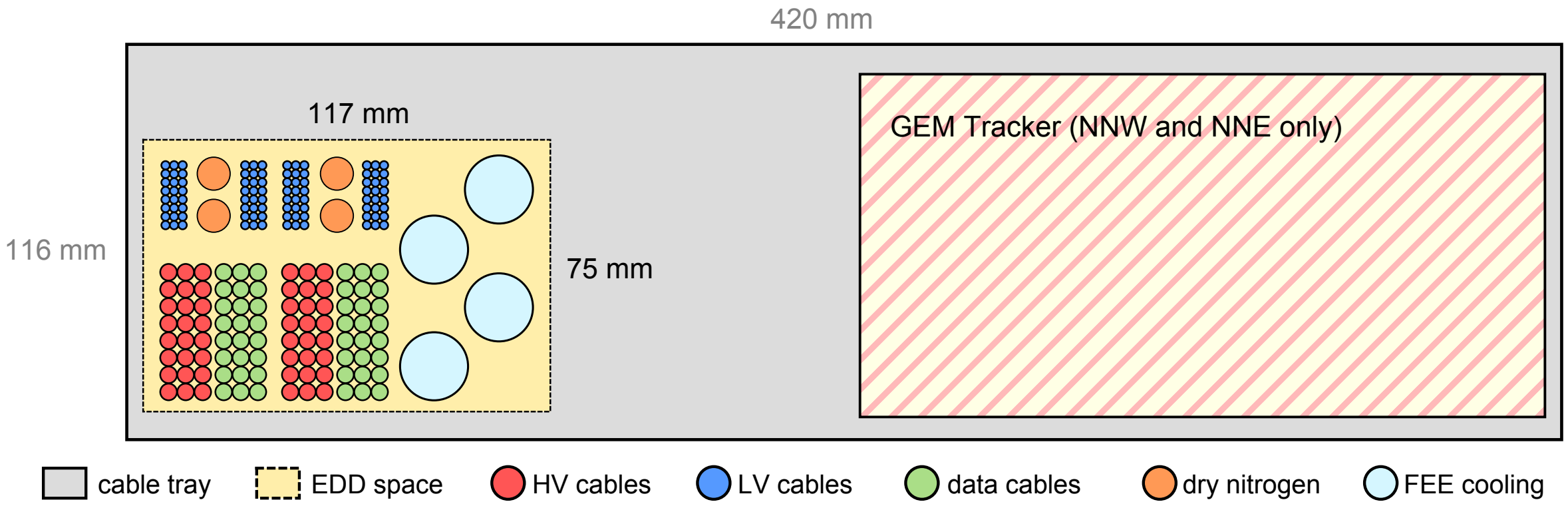

**Figure 6.16:** Cross section of a cable tray which also indicated the required space for two quadrants. The space reserved for GEM trackers only applies to the two upper trays whereas the lower trays are not occupied.

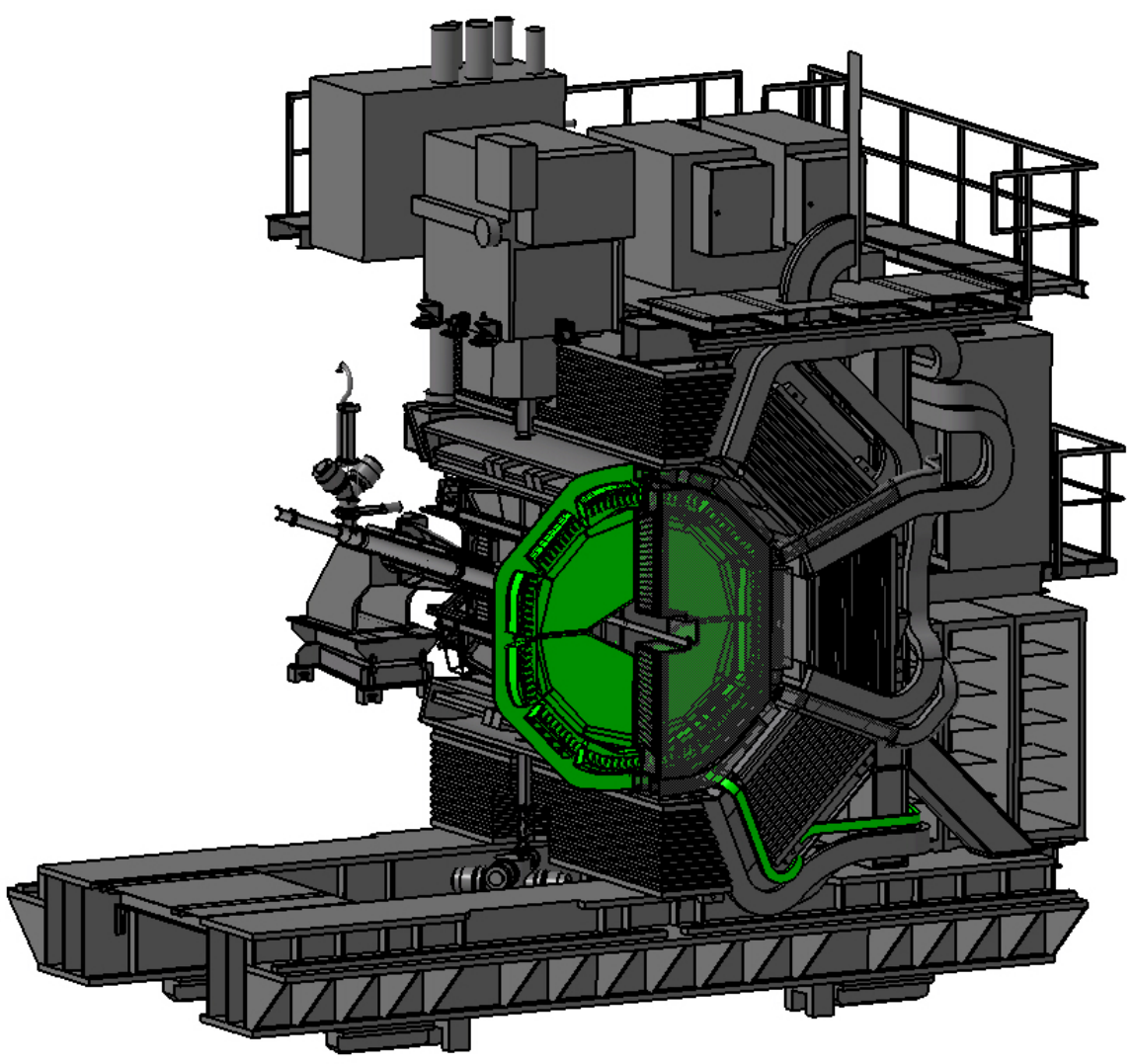

**Figure 6.17:** The EDD inside the $\overline{\text{P}}$ANDA target spectrometer with the anticipated cable tray for one half of the detector (highlighted in green).

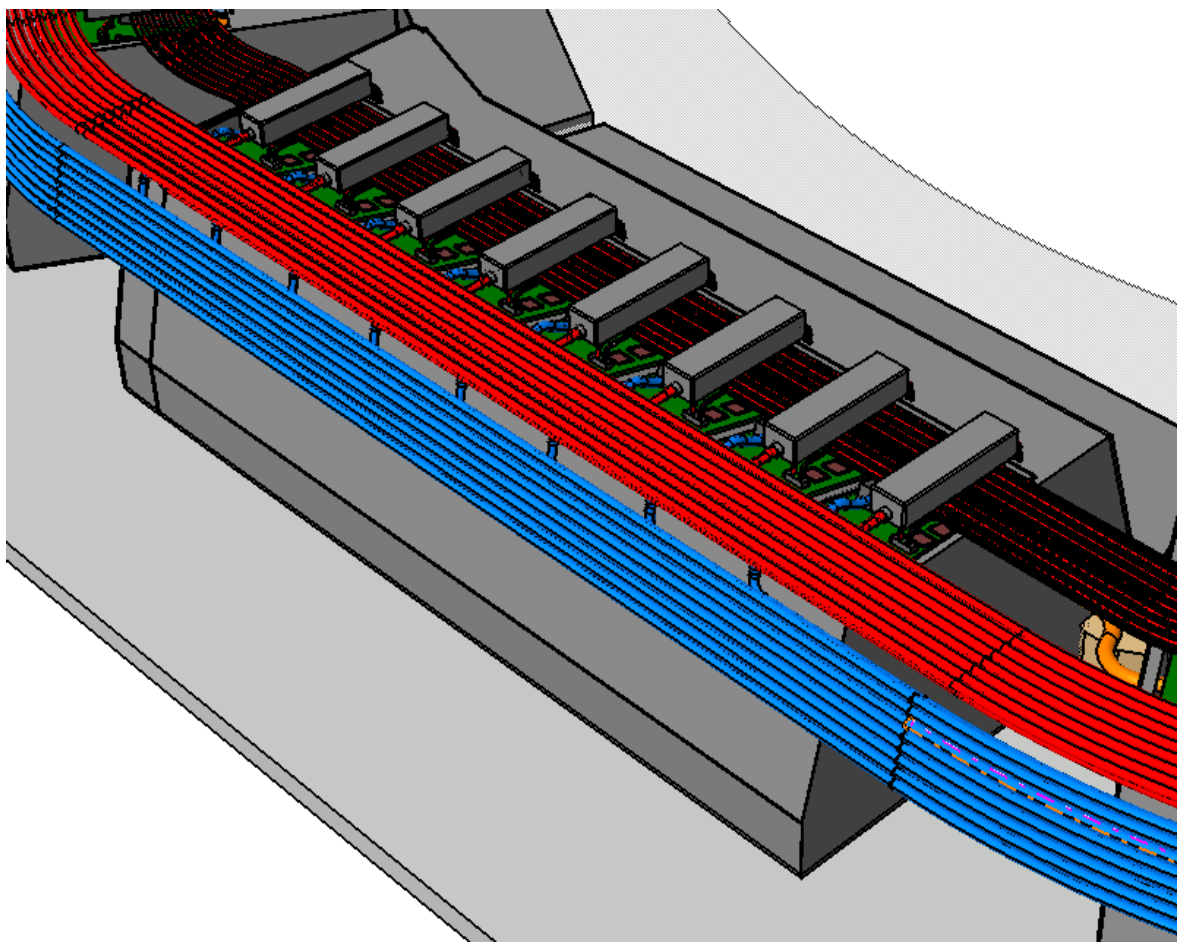

**Figure 6.18:** Backside of ROMs with cabling coming from the preceding side of the quadrant.

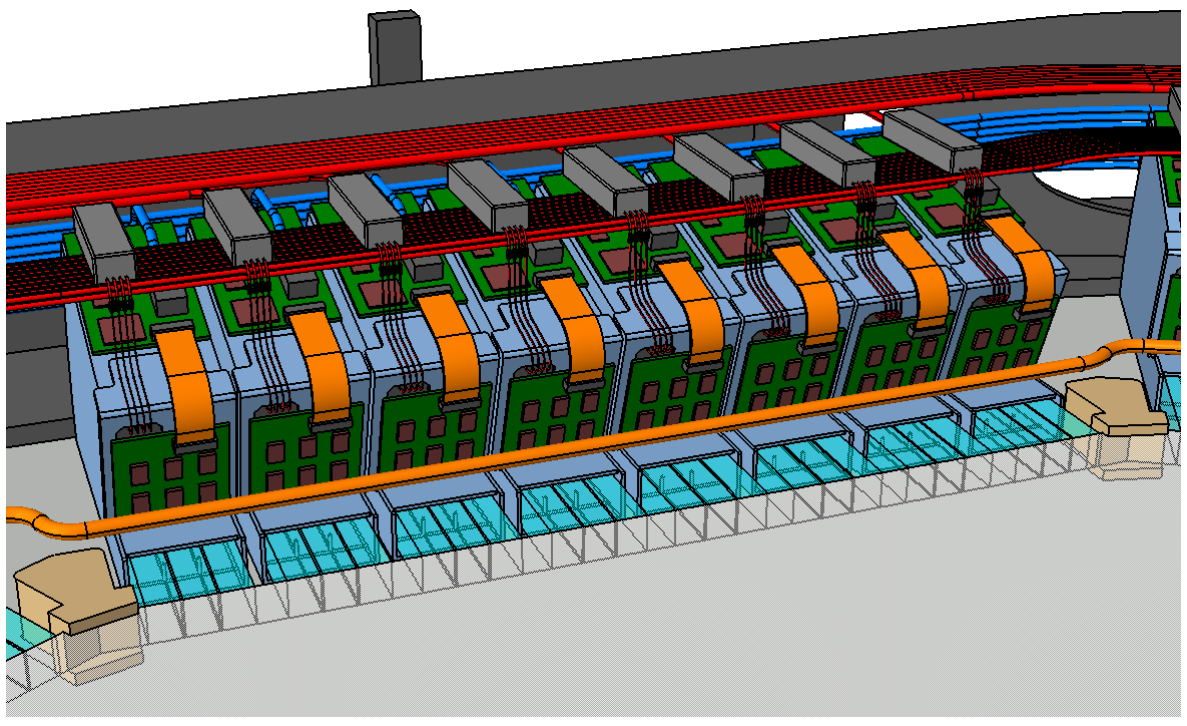

**Figure 6.19:** Cabling without ROM housing for one side of a quadrant.

## References

[1] PANDA Collaboration. Technical Progress Report. 2005.

[2] E. Etzelmüller. Developments towards the Technical Design and Prototype Evaluation of the PANDA Endcap Disc DIRC. PhD thesis, Justus-Liebig-Universität Gießen, 2017.

[3] R. Arora et al. A Large GEM-TPC Prototype Detector for PANDA. *Phys. Procedia*, 37:491–498, 2012.

[4] PANDA Collaboration. Technical Design Report for: PANDA Electromagnetic Calorimeter (EMC). arXiv:0810.1216, 2008.

[5] Epoxy Technology, Inc. 14 Fortune Drive, Billerica, MA 01821, USA.

[6] K. Inami et al. TOP counter for particle identification at the Belle II experiment. *Nucl. Instr. Meth. A*, 766:5–8, 2014.

[7] I. Adam et al. The DIRC particle identification system for the BaBar experiment. *Nucl. Instr. Meth. A*, 538:281–357, 2005.

# 7 Project Management

The participating institutions of the $\overline{\mathsf{P}}$ANDA Cherenkov group are located at the Universities of Erlangen-Nürnberg, Giessen, Glasgow and Mainz as well as BINP Novosibirsk, GSI Darmstadt, JINR Dubna, and SMI Vienna. These institutions share the responsibilities of the Barrel DIRC, the Endcap Disc DIRC and the forward RICH.

## 7.1 Work packages

The working group at the University of Giessen will be responsible for the project management of the first-of-series EDD quadrant and most of its work packages. In case of the optical components (radiator, FELs and filter) the production will be done by external companies whereas the quality assurance will be carried out using existing setups at the University of Giessen. MCP-PMTs will be produced by external companies and qualified by the working group at the University of Erlangen-Nürnberg. The readout is based on existing hardware and the required front-end board design as well as the production will be outsourced to an external company. This process will be monitored by the University of Giessen where the final parts will be tested subsequently. Development of the software, the mechanical design and manufacturing (except large parts) as well as the assembly of the first-of-series EDD quadrant will be done at the University of Giessen.

The modular structure of the detector and the outsourcing of the production of the critical elements allows to cope with the precision requirements of the detector with limited manpower.

In addition to these work packages the expertise on the optical elements, photon sensors, readout devices, and the tests in beam lines are shared within the whole Cherenkov group. An important role has the close cooperation with the Russian groups in PANDA.

It is planned to distribute the work packages for the remaining quadrants 2, 3, and 4 in a similar manner. PANDA is planning to involve additional groups to guarantee the long-term operation and maintenance of the EDD.

## 7.2 Schedule

The project schedule of the Endcap Disc DIRC is shown in Figure 7.1. Assuming that funding for the production will be available by the beginning of 2019 calling for tenders can start. Signed contracts are expected by Q2/2019. The process of producing and assembling the EDD will be done in two steps. A first full quadrant will be completed in Q3/2022. It will be extensively tested to study the final detector performance. In the meantime the development of the front-end electronics will be finalized to meet the spatial requirements for $\overline{\mathsf{P}}$ANDA. According to this time line a fully equipped first-of-series EDD quadrant can be installed inside the $\overline{\mathsf{P}}$ANDA Start Setup to participate in Phase 1. According to the current $\overline{\mathsf{P}}$ANDA schedule, that can be seen in Figure 7.2, the installation will take place between Q3/2021 and Q3/2023 together with the Endcap EMC. Commissioning with protons is planned in Q3/2023.

In a second step the components for the remaining quadrants 2, 3 and 4 will be produced. To profit by future developments regarding lifetime and spectral acceptance the production of MCP-PMTs for the remaining quadrants will start in 2020 and will be completed by Q4/2022. The assembly and testing of the last three quadrants will take place between Q3/2022 and Q3/2025 following an installation into $\overline{\mathsf{P}}$ANDA in Q1/2026.
In the event of a delayed funding the schedule is subject to change. Phase 3 including the Recycling Energy Storage Ring (RESR) is beyond the timescale. The commissioning of the filter for phase 2 has been dropped because it might not be required if MCP-PMT with the desired quantum efficiency spectrum are available at that time. However, we keep it as a backup solution.

## 7.3 Manpower

The manpower required and available is a mixture of staff, postdocs and PhD students as well as master and bachelor students that do R&D, design, assembly and testing. All major items for production are outsourced. The optical tests as well as testing of MCP-PMTs are assumed to be done by experienced people with assistance from students. The workshop of the University of Giessen will do the required production and fitting of small mechanical work. Productions of PCBs are done externally as well as parts that are 3D-printed. All units have limited numbers, so that half-automated testing is feasible for a small team. For the current prototype production, the manpower is sufficient, and the outsourcing of the FEE is fully financed. If the funding of the full EDD is foreseen, the manpower has to be increased.

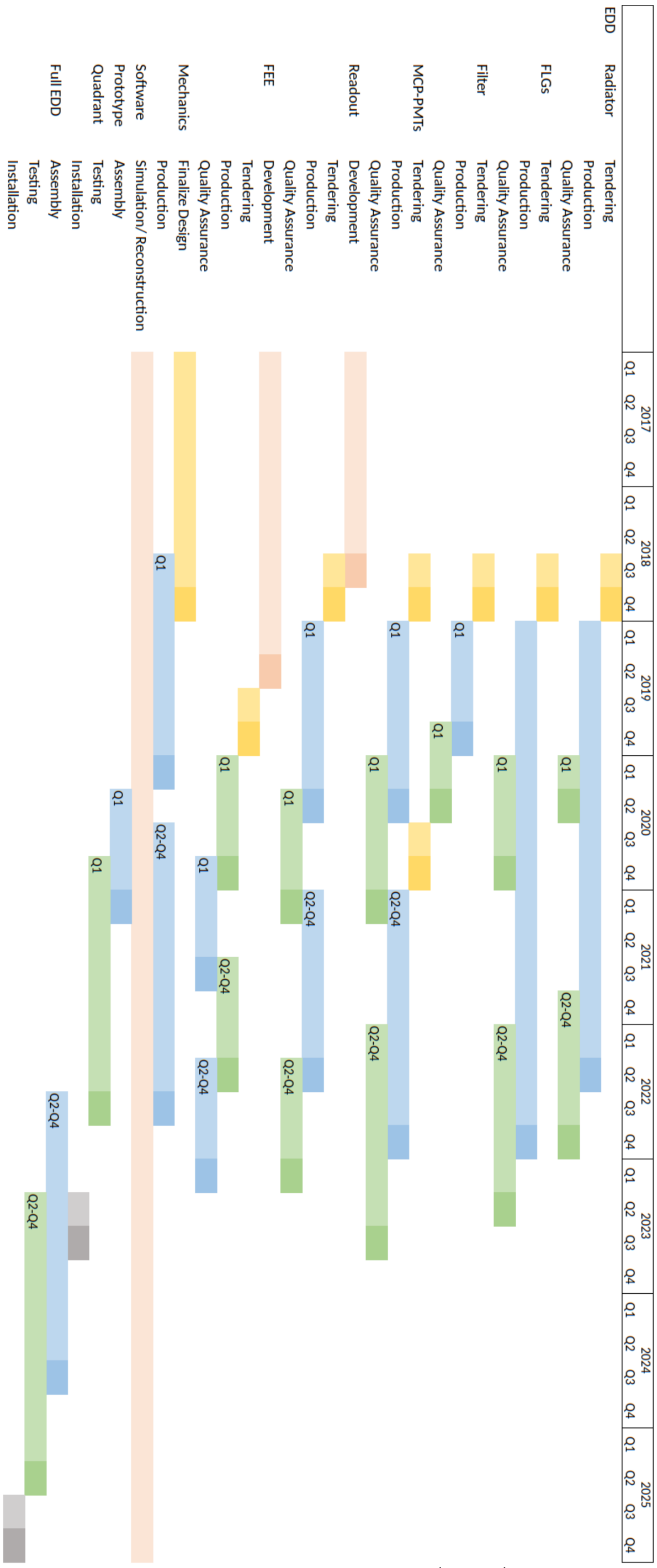

**Figure 7.1:** The projected time line for the Endcap Disc DIRC (EDD) under the assumption that the funding for a first-of-series quadrant is available starting from the beginning of 2019 and the $\overline{\mathsf{P}}$ANDA commissioning starts according to the time line shown in the top part of the overview. The production and assembly will be done in two main steps starting with a first-of-series quadrant. The remaining quadrants will be build subsequently and may incorporate improvements as a result of the experience from the previous quadrant.

## 7.4 Quality Control

### 7.4.1 Optical Elements

The optical elements will be fabricated in external companies according to well defined specifications. The specifications will be guaranteed by the producing companies and will be cross checked in our laser lab and, in case of doubts, by an external company.

The quality control of the radiator plates can be performed in Giessen by several methods. After optical inspection by eye, a 2-dimensional laser scan will measure the two bulk surfaces of each radiator plate for thickness variations of the plate. As a second, overall test of the optical properties of the plate, laser beams of several wavelengths that are fed into the lateral faces of the plate at various angles and positions will be transported to the opposite side face by total reflection. A control of intensity, shape and direction of the transmitted and reflected beams, using a calibrated CCD chip, will allow to verify the main specifications regarding surface roughness and transmission properties.

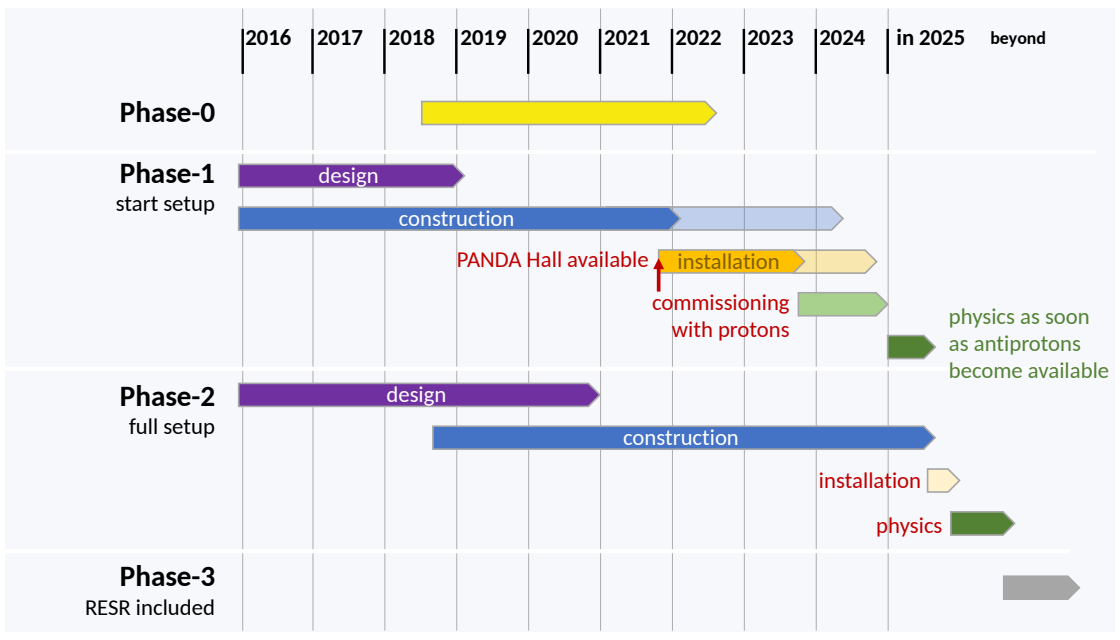

**Figure 7.2:** The PANDA Schedule as published in April 2018.

The assembly and gluing of the optical elements can be done in the Giessen clean room on the laser table using dedicated mechanical supports or, preferably, in a dedicated clean room close to the $\overline{\text{P}}$ANDA detector. In the process of gluing, the alignment of the elements will be controlled by a reflecting laser beam and a CCD chip.

The optical system of each EDD quadrant will be a single, rigid piece, consisting of one radiator and 24 ROMs with 3 FEL each. Its overall performance can be mapped out by coupling a laser into the radiator at a lateral face or via a coupling prism at the plate surface with a calibrated CCD chip that is mounted in the focus area where finally the MCP-PMTs will be mounted.

### 7.4.2 Sensors and Readout

The EDD will consist of 96 identical readout chains, 24 for each quadrant. Each chain consists of one position sensitive MCP-PMT and the corresponding digitizing ASICs on a PCB together with the control electronics, the voltage supply and the outgoing optical fibers and cables.

The ASICs and the PCBs will be produced externally and will be electronically tested on arrival.

The MCP-PMTs will be produced by external companies and will have individual, measured specification sheets. In one of the labs of the Cherenkov group (presumably at Erlangen or Giessen) the amplification, cross talk, time resolution and rate stability will be measured using one of the tested readout chains, and a focused, attenuated, pulsed laser beam in single photon mode.

### 7.4.3 The Quadrant

Once the optical system and the readout chains are tested separately, the complete quadrant will be assembled mechanically in horizontal orientation. The 24 readout chains can be connected to the DAQ system and tested in the dark environment with HV off and on. The light tightness of the stabilizing cross, radiator covers and ROM housings can be tested as well. Using the LED calibration system, light pulses will be used to test all MCP-PMT readout channels for operation.

The performance of a quadrant can be mapped out either using a cosmic trigger system at Giessen which is currently begin upgraded or at a dedicated test beam facility. It is planned to perform extensive studies with the first-of-series quadrant.

### 7.4.4 System Integration

The four quadrants will be shipped to the $\overline{\text{P}}$ANDA hall. The rigid optical systems and the photo sensors and their readout chains will be transported separately. The limited size ($\approx 1.3 \times 1.3$ m$^2$) and weight ($\sim 85$ kg) of the optical system of each quadrant can be handled, however with great care and in special transport boxes.

Optionally, it will also be possible to mount only one, two or three quadrants if the production is staged, e.g. for funding reasons. In this case the missing quadrants are replaced by dummy plates for mechanical stability of the mounting cross.

After connecting all readout systems, cables and cooling pipes, all readout channels can be tested using the LED calibration system and cosmics.

## 7.5 Safety

Both, design and construction of the Endcap Disc DIRC including the infrastructure for its operation will be done according to the safety requirements of FAIR and the European and German safety regulations. Detailed procedures for the assembly, installation, and operation of the EDD will be provided to ensure personal safety as well as the integrity of the EDD components including interference with other parts of the $\overline{\textsf{P}}$ANDA experiment.

Accessing the EDD will only be possible after a cool-down period of the front endcap EMC. Nevertheless, the irradiation of the EDD and its electronics will be less than for several other detectors in the target spectrometer. For accessing the MCP-PMTs and the EDD readout electronics, the magnet has to be opened and the endcap EMC has to be removed.

### 7.5.1 Mechanics

The strength of the EDD support structures have been computed with physical models in the course of the design process. Additional forces during a quench of the superconducting magnet have been taken into account.

### 7.5.2 Electrical Equipment and Cooling

All electrical equipment in $\overline{\textsf{P}}$ANDA will comply to the legally required safety code and concur to standards for large scientific installations following guidelines worked out at CERN to ensure the protection of all personnel working at or close to the components of the $\overline{\textsf{P}}$ANDA system. Power supplies will have safe mountings independent of large mechanical loads. Hazardous voltage supplies and lines will be marked visibly and protected from damage by near-by forces. All supplies will be protected against over-current and over-voltage and have appropriate safety circuits and fuses against shorts. DC-DC converters have to be cooled to avoid overheating and the power supply cables will be dimensioned correctly to prevent overheating. All cabling and optical fiber connections will be executed with non-flammable halogen-free materials according to up-to-date standards. A safe ground scheme will be employed throughout all electrical installations of the experiment. Smoke detectors will be mounted in all appropriate locations.

Lasers or high output LEDs will be employed in the calibration and readout systems and their light is distributed throughout the EDD optical systems. For these devices all necessary precautions like safe housings, color coded protection pipes, interlocks, proper warnings and instructions as well as training of the personnel with access to these components will be taken.

The operation of the quartz plates and the readout electronics require an air conditioning with dry nitrogen and a water cooling system. A control system will be employed to detect any malfunctioning of the gas and cooling system and includes interlocks for electronics and high-voltage supplies. In the case of water loss or in case of abnormal temperature gradients between coolant input and output the control system enacts the appropriate safety procedures. A redundant system of temperature sensors allows the continuous monitoring of the effectiveness of the system.

### 7.5.3 Radiation Aspects

Shielding, operation and maintenance of all $\overline{\textsf{P}}$ANDA components will be planned according to European and German safety regulations to ensure the proper protection of all personnel. The access to the experimental equipment during beam operation will be prohibited and the access during maintenance periods will be cleared after radiation levels are below the allowed thresholds.

The EDD equipment can be activated by radiation leading to low-energy radioactivity of the activated nuclei. Therefore all equipment has to be monitored for radiation before it is taken out of the controlled area.

# 8 Acknowledgements

We acknowledge the support of the Bundesministerium für Bildung und Forschung(BMBF), HGS-HIRe, and HIC for FAIR. We also thank the CERN and DESY staff for the opportunity to use the beam facilities and for their on-site support.

# List of Acronyms

**3D** 3-Dimensional

**ADC** Analog to Digital Converter

**ALD** Atomic Layer Deposition

**APD** Avalanche Phtoto Diode

**ASIC** Application Specific Integrated Circuit

**BINP** Budker Institut of Nuclear Physics

**CCD** Charge Coupled Device

**CERN** Conseil European pour la Recherche Nucleaire

**CIS** Charge Injection System

**CSS** Control System Studio

**CTE** Coefficient of Thermal Expansion

**CTS** Central Trigger System

**DAQ** Data Acquisition

**DC** Direct Current

**DCR** Dark Count Rate

**DCS** Detector Control System

**DESY** Deutsches Elektronensynchrotron

**DIRC** Detector for Internally Reflected Cherenkov Light

**dSiPM** digital Silicon Photomultiplier

**EDD** Endcap Disc DIRC

**EMC** Electromagnetic Calorimeter

**EPICS** Experimental Physics and Industrial Control System

**FAIR** Facility for Antiproton and Ion Research

**FDD** Focusing Disc DIRC

**FEE** Front-End Electronics

**FEL** Focusing Element

**FEM** Finite Element Method

**FPGA** Field Programmable Gate Array

**FS** Forward Spectrometer

**GBT** GigaBit Transceiver

**GBTX** GigaBit Transceiver

**GBT-SCA** GigaBit Slow Control ASIC

**GEM** Gas Electron Multiplier

**GSI** Gesellschaft für Schwerionenforschnung

**GUI** Graphical User Interface

**HADES** High-Acceptance Dielection Spectrometer

**HESR** High Energy Storage Ring

**HPTDC** High Performance Time-to-Digital Converter

**HV** High Voltage

**KVI** Kernfysisch Versneller Instituut

**LED** Light-Emitting Diode

**LHC** Large Hadron Collider

**LMS** Laser Monitoring System

**LSB** Least Significant Bit

**LV** Low Voltage

**LVDS** Low-Voltage Differential Signaling

**MAPMT** Multi-Anode PMT

**MC** Monte Carlo

**MCP PMT** Multi-Channel Plate PMT

**MP** Mounting Plate

**MVD** Micro Vertex Detector

**PANDA** antiProton ANnihilation at DArmstadt

**PC** Photocathode

**PCB** Printed Circuit Board

**PDE** Photon Detection Efficiency

**PET** Positron Emission Tomography

**PHL** Pellet High Luminosity Mode

**PID** Particle Identification

**PMT** Photomultiplier

**PTR** Pellet Tracking Mode

**QE** Quantum Efficiency

**RH** ROM Housing

**RICH** Ring Imaging Cherenkov Counter

**RMS** Root Mean Square

**ROM** Readout Module

**RTV** Room Temperature Vulcanization

**SiPM** Silicon Photomultiplier

**SC** Stabilizing Cross

**SciTil** Scintillating Tile Hodoscope

**SEU** Single Event Upset

**SFT** Silicon Fibre Tracker

**SLHC** Super Large Hadron Collider

**SODA** Synchronisation Of Data Acquisition

**SR** Shift Register

**STIC** Silicon Photomultiplier Timing Chip

**STT** Straw Tube Tracker

**TDC** Time to Digital Converter

**TDR** Technical Design Report

**TOF** Time-of-Flight

**TRB** Time-to-Digital Converter Readout Board

**UV** Ultraviolet